\documentclass[12pt]{report}
\usepackage{array}
\usepackage{citesort}
\usepackage{fancyhdr}
\usepackage{graphicx,psfrag,amsmath}
\DeclareGraphicsExtensions{.pdf}
\usepackage{nccmath}
\usepackage{chapterbib}
\usepackage[usenames,dvipsnames]{color} \usepackage{setspace}
\usepackage{multirow}
\usepackage{tikz}

 \fancyhfoffset[LE]{.5 cm}
\clearpage{\pagestyle{empty}\cleardoublepage} 
\renewcommand{\chaptermark}[1]{\markboth{\chaptername\
\thechapter}{}}

\newcommand{\ta}{\tilde\alpha}
\newcommand{\tb}{\tilde\beta}

\newcommand{\bm}{\bar\mu}

\newcommand{\da}{\dot\alpha}
\newcommand{\db}{\dot\beta}
\newcommand{\la}{\lambda}
\newcommand{\ga}{\gamma}
\newcommand{\al}{\alpha}

\newcommand{\e}{\eta}

\newcommand{\mt}{\tilde m}

\newcommand{\at}{\tilde a}

\newcommand{\om}{\omega}

\newcommand{\Sig}{\bf\Sigma}
\newcommand{\sss}{\sigma}
\newcommand{\ssb}{{\overline{ \sigma}}}
\newcommand{\sq}{\sqrt{2}}

\newcommand{\sqt}{\sqrt{3}}

\newcommand{\os}{\overline\Sigma}
\newcommand{\s}{\Sigma}
\newcommand{\Sigb}{{\overline\Sigma}}
\newcommand{\oot}{\overline {126}}
\newcommand{\boot}{\bf\oot}

\newcommand{\ovl}{\overline}
\newcommand{\nnu}{\nonumber\\}
\def\blfootnote{\xdef\@thefnmark{}\@footnotetext}

\newcommand{\bee}{\begin{eqnarray*}}
\newcommand{\eee}{\end{eqnarray*}}

\newcommand{\ol}[1]{\overline{#1}}
\newcommand{\wt}[1]{\tilde{#1}}

\newcommand{\hc}{{\rm h.\,c.}\,}
\newcommand{\PL}{{ P_L}}
\newcommand{\PR}{{ P_R}}
\newcommand{\Dt}{(4\pi)^2\frac{d}{dt}}

\def\mt{{\ifmmode\td M_t\else $\td M_t$\fi}}
\def\as{{\ifmmode\alpha_s\else$\alpha_s$\fi}}

\let\td=\tilde

\def\co#1{{\ifmmode{\cal O}_{#1}\else${\cal O}_{#1}$\fi}}
\def\cs#1{{\ifmmode{\cal S}_{#1}\else${\cal S}_{#1}$\fi}}
\def\at{{\ifmmode{\tilde A}\else$\tilde A$\fi}}

\def\fr#1.#2.{{#1\over #2}}

\def\mg{{\ifmmode M_{GUT}\else $M_{GUT}$\fi}}

\tikzstyle{startstop} = [rectangle, rounded corners, minimum
width=3.2cm, minimum height=1cm,text centered, draw=black,
fill=red!30] \tikzstyle{io} = [trapezium, trapezium left angle=70,
trapezium right angle=110, minimum width=3.2cm, minimum
height=1cm, text centered, draw=black, fill=blue!30]
\tikzstyle{process} = [rectangle, minimum width=3.5cm, minimum
height=1cm, text centered, draw=black, fill=orange!30]
\tikzstyle{decision} = [diamond, minimum width=2cm, minimum
height=1cm, text centered, draw=black, fill=green!30]
\tikzstyle{arrow} = [thick,->,>=stealth]

\tikzstyle{decision} = [diamond, draw, 
    text width=3.5em, text  centered, node distance=3cm, inner sep=0pt]
\tikzstyle{block} = [rectangle, draw,
    text width=6.5em, text centered, rounded corners, minimum height=1.9em]
\tikzstyle{block5} = [rectangle, draw,
    text width=2.3em, text centered, rounded corners, minimum height=2em]
\tikzstyle{line} = [draw, -latex'] \tikzstyle{cloud} = [draw,
ellipse,fill=red!20, node distance=3cm,
    minimum height=2em]
\tikzstyle{block1} = [rectangle, draw, text width=9em, text
centered, rounded corners, minimum height=2em] \tikzstyle{block4}
= [rectangle, draw, text width=11em, text centered, rounded
corners, minimum height=2em] \tikzstyle{block2} = [rectangle,
draw, text width=9em, text centered, rounded corners, minimum
height=1em] \tikzstyle{block3} = [rectangle, draw, text width=9em,
text centered, rounded corners, minimum height=1em]
\tikzstyle{decision2} = [diamond,minimum width=.5cm, minimum
height=.5cm, draw, text width=1.1cm, text badly centered, node
distance=1.8cm, inner sep=0pt]
\tikzstyle{block11} = [rectangle, draw,
    text width=5.8em, text centered, rounded corners, minimum height=2em]

\tikzstyle{block22} = [rectangle, draw,
    text width=4.0em, text centered, rounded corners, minimum height=2em]
\tikzstyle{block23} = [rectangle, draw,
    text width=3.3em, text centered, rounded corners, minimum height=1.5em]

 \def\fr{\frac} 
 \def\be{\begin{equation}}
\def\ee{\end{equation}} \def\bea{\begin{eqnarray}}
\def\eea{\end{eqnarray}}

\usepackage{amsmath,amssymb,amsfonts}
\usepackage{color}
\usepackage{amsmath}
\usepackage{amsfonts}
\usepackage{amssymb}
\usepackage{graphicx}
\usepackage{slashed}            
\usepackage{bbm}                
\usepackage{xspace}             

\usepackage{tikz}
\usetikzlibrary{arrows,shapes}
\usetikzlibrary{trees}
\usetikzlibrary{matrix,arrows}              
\usetikzlibrary{positioning}                
\usetikzlibrary{calc,through}               
\usetikzlibrary{decorations.pathreplacing}  
\usepackage{pgffor}                         
\usetikzlibrary{decorations.pathmorphing}   
\usetikzlibrary{decorations.markings}
\tikzset{
    vector/.style={decorate, decoration={snake}, draw},
    provector/.style={decorate, decoration={snake,amplitude=2.5pt}, draw},
    antivector/.style={decorate, decoration={snake,amplitude=-2.5pt}, draw},
    fermion/.style={draw=black, postaction={decorate},
        decoration={markings,mark=at position .55 with {\arrow[draw=black]{>}}}},
        fermion1/.style={draw=black, postaction={decorate}},
    fermionbar/.style={draw=black, postaction={decorate},
        decoration={markings,mark=at position .55 with {\arrow[draw=black]{<}}}},
    fermionnoarrow/.style={draw=black},
    gluon/.style={decorate, draw=black,
        decoration={coil,amplitude=4pt, segment length=5pt}},
    scalar/.style={dashed,draw=black, postaction={decorate},
        decoration={markings,mark=at position .55 with {\arrow[draw=black]{>}}}},
    scalar1/.style={dashed,draw=black, postaction={decorate}},
    scalarbar/.style={dashed,draw=black, postaction={decorate},
        decoration={markings,mark=at position .55 with {\arrow[draw=black]{<}}}},
    scalarnoarrow/.style={dashed,draw=black},
    electron/.style={draw=black, postaction={decorate},
        decoration={markings,mark=at position .55 with {\arrow[draw=black]{>}}}},
    bigvector/.style={decorate, decoration={snake,amplitude=4pt}, draw},
}

\begin{document}

\newpage\pagenumbering{roman}\thispagestyle{empty} 
{\Large\baselineskip 27pt
\begin{titlepage}
\begin{center}
 {\bf{\textsf{STUDY OF BARYON NUMBER AND LEPTON FLAVOUR VIOLATION IN
 THE NEW MINIMAL SUPERSYMMETRIC SO(10)GUT}}} \par
\vspace{0.7in} {\bf{\textsf{A THESIS}}}\\ \textsf{ {Submitted to the\\
FACULTY OF SCIENCE\\ PANJAB UNIVERSITY, CHANDIGARH\\ for the
degree of} }\\\par \vspace{0.1in}\textbf{\textsf{DOCTOR OF
PHILOSOPHY}}\\\par \vspace{0.8in}\textsf{\textbf{2014}} \par
\vspace{0.3in} \par \vspace{0.8in} \textbf{\textsf{CHARANJIT
KAUR}}
\par \vspace{0.4in}
 \par
\vspace{0.38in}
{\textsf{DEPARTMENT OF PHYSICS\\CENTRE OF ADVANCED STUDY IN
PHYSICS\\PANJAB UNIVERSITY\\CHANDIGARH, INDIA}}
\end{center}
\end{titlepage}
} \thispagestyle{empty}
 \mbox{}\newpage \thispagestyle{empty} \vspace*{4.in}
\begin{center}  \emph{{\huge\textbf{Dedicated To My Parents}}}
 \end{center}
\newpage\thispagestyle{empty}
\mbox{}\newpage\thispagestyle{empty}
\thispagestyle{empty} \baselineskip 20 pt \begin{center}{\Large
\bf \emph{Acknowledgements}}\end{center} \medskip First and
foremost I would like to express my sincere gratitude to my
advisor Prof. C. S. Aulakh for his guidance, consistent support,
patience and encouragement during the period of my research work.
His tireless efforts to do research acted as an inspiration for
me.

I am indebted to all my teachers who always persuaded me to
acquire higher education. I owe my thanks to my family for their
unconditional support and encouragement throughout my Ph.D.  I
express my special thanks to my brother Gurpal Singh Khosa for
bearing the brunt of my frustration and rages at times, for always
understanding my work priorities, for maintaining my link with
basic physics and for providing me homely environment.

I am  thankful to my fellow labmate Ila Garg for various
stimulating discussions, constructive criticism, companionship  at
the oddest of times.

I am grateful to the chairperson of Department of Physics, P.U.
Chandigarh for providing me required facilities to work. I would
like to acknowledge University Grants Commission, India for
financial support during my Ph.D.

I am thankful to my senior Dr. Rama Gupta for guiding me time to
time. There are countless contributors to whom I am indebted but
few whom I could not skip to mention who anticipated my work and
were always there for me whenever I needed them are Siman Dhillon,
Amanpreet Chahal, Khushwant Chahal, Manbir Kaur, Jagdish Kaur,
Harleen Asees Gill, Rajni Bansal, Arshdeep  and Jashanpreet Kaur
Bath.

Above all I am grateful to Mighty God for blessing me with great
opportunities and keeping me motivated to have the enthusiastic
attitude towards research.\\
\newpage\thispagestyle{empty} \mbox{}\thispagestyle{empty}\newpage\thispagestyle{empty}
\thispagestyle{empty}  \begin{center}{\Large \bf
Abstract}\end{center}We study baryon number (B) and lepton flavour
violation (LFV) in a supersymmetric model based on SO(10) gauge
group called New Minimal Supersymmetric SO(10) Grand Unified
Theory (NMSGUT).

We calculated one loop GUT scale threshold corrections to the
relation between the NMSGUT and effective minimal supersymmetric
standard model (MSSM) Yukawa couplings. Strong renormalization of
the Higgs line entering these vertices, due to the large number of
GUT fields coupled to them allows lowering of the SO(10) couplings
required for fitting MSSM couplings. Since the same SO(10) Yukawas
are responsible for B-violation, proton decay lifetimes compatible
with experimental limits are generically achievable. We
successfully searched the NMSGUT parameter space for values that
allowed accurate fits of known MSSM couplings and have acceptable
dimension five operator mediated B violation rates.

The spectra of sparticles used in the B violation calculations
were improved by including one loop corrections. Searches
including these corrections require careful control to avoid
instability.  We found fits compatible with the MSSM data and B
violation limits even after inclusion of loop corrections.

The effective theory of the NMSGUT includes
 lepton number violating couplings and thus our fits also imply predictions for LFV rates.
 We computed NMSGUT estimates- based on successful fits- for important observables in the lepton sector such as lepton flavour
violating processes (e.g. $l_i \rightarrow l_j \gamma, l_i
\rightarrow 3 l_j$), the muon g-2 anomaly ($a_{\mu}$) and the CP
violation parameter ($\epsilon_{{\small{CP}}}$ ) relevant for high
scale leptogenesis scenarios. For LFV estimation we have included
heavy right handed neutrino thresholds.

We computed the two loop renormalization group evolution equations
of the NMSGUT hard and soft supersymmetry breaking parameters.
These equations are useful for running the parameters from Planck
scale ($M_P$) to the unification scale ($M_X^0$). With a randomly
chosen set of couplings at $M_P$, we run down them to $M_X^0$ and
find a significant variation in the soft parameters. These changes
could explain crucial features such as negative non universal
Higgs mass squared values needed by the NMSGUT for successful
fitting of fermion Yukawas.

We propose generation of the Standard Model fermion hierarchy by
extension of the NMSGUT with $O(N_g)$ family gauge symmetry. In
this scenario Higgs representations of SO(10) also carry family
indices and are called Yukawons. VEVs of these Yukawon fields
break GUT and family symmetry and generate MSSM Yukawa couplings
dynamically. As in the NMSGUT, the effective MSSM matter fermion
couplings to the light Higgs pair are determined by the null
eigenvectors of the MSSM type Higgs doublet superfield mass matrix
$\cal H$. A consistency condition on the doublet $([1,2,\pm 1])$
mass matrix ($Det(\cal{H})=$0) is required to keep one pair of
Higgs doublets light in the effective MSSM. We show that the
Yukawa structure generated by null eigenvectors of $\cal H $ are
of generic kind required by the MSSM. We studied a toy model with
two generations as well as the realistic three generation
($N_g=3$) case. We considered a number of generic possibilities,
with random GUT scale parameters, which produce acceptable Yukawa
eigenvalues and lepton and quark mixing angles, but small neutrino
masses. This justifies  searches for realistic dynamical fermion
fits in the future by generalizing the programs and techniques
used for the NMSGUT. \thispagestyle{empty}
\newpage
\thispagestyle{empty}
\thispagestyle{empty}
\begin{center}{\Large \bf Publications}\end{center} \vspace{.5cm}
\begin{itemize}
\item C. S. Aulakh, I. Garg and \textbf{C. K.
Khosa\footnote{khosacharanjit@gmail.com}},
  \emph{``Baryon stability on the Higgs dissolution edge: threshold corrections and suppression of baryon violation in the NMSGUT"},
  \textbf{Nucl.\ Phys.\ B} {\bf 882}, 397 (2014)
  [arXiv:1311.6100 [hep-ph]].

\vspace{.5cm} \item
  C. S. Aulakh and \textbf{C. K. Khosa},
  \emph{``SO(10) grand unified theories with dynamical Yukawa couplings"},
 \textbf{PRD} {\bf 90}, 045008 (2014)
  [arXiv:1308.5665 [hep-ph]].

 \end{itemize}
\thispagestyle{empty}
\newpage\thispagestyle{empty} \mbox{}\newpage
\clearpage\pagenumbering{roman} \tableofcontents
\mbox{}\newpage
\thispagestyle{empty}\addcontentsline{toc}{chapter}{List of
Figures} \listoffigures \newpage\thispagestyle{empty}
\mbox{}\newpage
  \addcontentsline{toc}{chapter}{List of Tables}
  \listoftables
\newpage\thispagestyle{empty} \mbox{}\newpage
\clearpage\pagenumbering{arabic}
\chapter{Introduction}\label{chapter1}
\section{Standard Model and Beyond }
The Standard Model (SM) supplemented by effective operators
describing small neutrino masses  is the established model of
particle physics whose predictions have been tested experimentally
upto high accuracy. It is a renormalizable, spontaneously broken
chiral Yang Mills quantum field theory describing strong and
electroweak interactions, based upon the principle of local gauge
invariance with the gauge group $ G_{SM} \equiv SU(3)_C \times
SU(2)_L \times U(1)_Y$ and 3 families of 15 chiral fermion fields
describing the known matter particles and antiparticles.
Corresponding to this gauge group there are twelve gauge bosons
$(G^a,W^i,B)$ out of which three become massive after spontaneous
symmetry breaking of the gauge group \[ SU(3)_C\times
SU(2)_L\times U(1)_Y \longrightarrow SU(3)_C\times U(1)_{em} \]
due to the vacuum expectation value (VEV) of Higgs field
$\Phi$(1,2,1) \be \Phi =
\begin{pmatrix} \Phi^{+}\cr \Phi^0 \cr \end{pmatrix}
\quad ;  \quad \langle \Phi\rangle =
\begin{pmatrix} 0 \cr \upsilon/\sqrt{2} \cr \end{pmatrix} \ee
where $\upsilon$=246 GeV.  Three linear combinations of the four
$SU(2)_L \times U(1)_Y$ gauge bosons $(W^i,B)$, $W^{\mp}$
$(\frac{W_1 \pm i W_2}{\sqrt{2}})$ and Z ($c_W W^3-s_WB$)  acquire
mass while the other orthogonal combination of $W_3$ and B- i.e.
$s_WW^3+c_WB$ remains massless and is identified as the photon
(the gauge boson of the electromagnetic interactions). Here $W^i
(i=1,2,3)$ and B denote the gauge boson of $SU(2)_L$ and $U(1)_Y$
respectively, $c_W (s_W)=\cos {\theta_W}(\sin {\theta_W})$,
$\theta_W$ is Weinberg angle, defined as $\tan \theta_W=
\frac{g'}{g}$ ($g'$ and $g$ are gauge couplings of $U(1)_Y$ and
$SU(2)_L$ respectively). 8 gluons ($G^a,a=1...8$) corresponding to
unbroken $SU(3)_{C}$ remain massless. Fermionic matter consists of
three generations, embedded in the SM gauge group with following
quantum numbers :
\[ Q_L=\begin{pmatrix}
u_L  \cr d_L
\end{pmatrix}\quad ; \quad L_L=\begin{pmatrix} \nu_L \cr e_L
\end{pmatrix}\]\be
Q_{LA}(3,2,\frac{1}{3}), u_{RA}(3 ,1 ,\frac{4}{3}) ,
d_{RA}(3,1,-\frac{2}{3}), L_{LA}(1,2,-1), e_{RA}(1,1,-2) \ee where
A= 1, 2, 3 is the generation index. Clearly, all the fermion
fields are present in chiral pairs (i.e. both $ \Psi_L $ and
$\Psi_R $) except the neutral fermion $\nu_{L}$. Dirac mass terms
like : $m_{\psi}\bar\psi{\psi}$ are not allowed by gauge
invariance. However, charged fermion masses are generated via the
Yukawa couplings of the fermions with the Higgs doublet since the
Higgs acquire a VEV. Visible, stable, matter content of the
universe is thought to be made of first generation fermions
(except for $\nu_\mu$, $\nu_\tau$ which persist) since the heavier
generations decay to them with lifetimes shorter than $10^{-10}$s.

Discovery of Higgs boson at Large Hadron Collider (LHC)
\cite{Aad:2012tfa,Chatrchyan:2012ufa} confirmed the existence of
all the SM particles. Although the SM seems to be a very
successful theory but there seems to be no justification for its
basic assumptions like the existence of arbitrary Yukawa
couplings. There is no profound explanation for the origin of
families and their observed mixing structure. Furthermore, charge
quantization remains unexplained. The fourth fundamental
interaction- gravity- is not included. Besides these structural
defects other flaws of the SM are the following unexplained
experimental observations:
 \begin{enumerate}\item Neutrino oscillations indicate non zero
neutrino masses \cite{osi1,osi2,Hirata,Ambrosio,SKC,Ahmad}.
Although neutrino masses can be included via (non-renormalizable)
dimension $\geq$ 5 operators \cite{weinbergd5} in the effective
Lagrangian, there is no way to generate neutrino masses in the
renormalizable SM without any extension. This indicates SM is
incomplete. \item SM does not explain observed baryon asymmetry of
the universe: $n_B/n_\gamma \sim 10^{-10}$. \item Only 4$\%$ of
the universe is visible matter, while the remaining 96$\%$ is dark
matter (DM) and dark energy. SM has no DM candidate. Dark energy
is an even greater mystery.
\end{enumerate}
Answers to these questions require physics beyond SM. Therefore in
spite of its experimental successes, the SM suffers from a number
of limitations, and can't be an ultimate theory of nature.  It is
only an effective theory at low energies ($\leq 200$ GeV) of some
more profound and complete theory. For instance one can generate
tiny neutrino masses through seesaw mechanism by extending the
particle content of SM \cite{MohapatraSenjanovicSeesaw}. Depending
upon the nature of additional particle one can have Type I
\cite{MohapatraSenjanovicSeesaw}, II \cite{TypeII} or III
\cite{TypeIII} seesaw contribution. Type I seesaw requires
addition of a gauge singlet right handed (``sterile") neutrinos.
Type II and III need Higgs and fermion which are triplet
irreducible representations (irreps) (of $SU(2)_L$) respectively.
Several beyond SM approaches like Supersymmetry (Susy), Grand
Unified Theories (GUT), Extra Dimension, String Theory,
Technicolor Model and 4th generation model etc. have been
considered. String theory is a theoretical tool which replaces
point particle by one dimensional string whose excitations are the
usual point particle fields. Extra dimensions models are based on
the assumption that real world is higher dimensional and its extra
spatial dimensions are compact. Susy and GUT together provide an
attractive framework which has explanation of most of the above
mentioned open questions. From now onwards we will focus on this
approach.
\section{Supersymmetric Grand Unification}
\subsection{Supersymmetry}Supersymmetry is the most appealing
extension of the SM. It relates the existence of the fermions with
bosons and vice versa. Each SM particle has its partner called
superpartner which has the same quantum numbers as the particle
except the spin which differs by half. Certain relations among the
allowed coupling constants ensure the invariance of the actions
under transformations with anti-commuting Lorentz spinor parameter
of Susy \cite{Wess}. This leads to conserved supercurrents and
Noether supercurrents that are constants of motion. Higgs mass is
sensitive to radiative corrections from new physics which vary
quadratically with the scale of heavy particles associated with
the new physics. Stabilization of Higgs mass was the main
motivation to introduce Susy. It stabilizes Higgs mass against
radiative corrections by cancelling the loop contribution of a
particle with the contribution from its superpartner
\cite{gaugehierarchy}. Supersymmetric version of the SM is called
Minimal Supersymmetric Standard Model (MSSM). Superpartners of the
SM fermions are scalars, called s-particles while that of gauge
and Higgs bosons are fermions called gaugino and higgsino
respectively. These gauginos and higgsinos mix to form neutralino
and chargino states. The field content of the MSSM is given in
Table \ref{MSSMp}. Renormalization Group (RG) evolved gauge
couplings of the MSSM accurately unify at GUT scale of order of
$10^{16}$GeV which has been a strong motivational hint to study
Susy.

Both fermions and bosons are placed into an irreducible
representation of Susy algebra called supermultiplet.
Supersymmetric gauge theories typically use chiral and vector
supermultiplets. Interaction and mass terms for the various
superfields are described by analytic function (of chiral
superfields ($\Phi$)), called superpotential : \be
W=\frac{M^{ij}}{2} \Phi_i \Phi_j +\frac{y^{ijk}}{6} \Phi_i \Phi_j
\Phi_k \ee where $M^{ij}$ is the mass matrix and $y^{ijk}$ is the
Yukawa coupling. Scalar potential is computed from the
superpotential : \be V=F^{*i}F_i+\frac{1}{2}\sum_a D^a D_a \quad ;
\quad F^i=\frac{\partial W}{\partial \phi_i} \quad ; \quad D^a=g^a
\phi^{\dag} T^a \phi \ee Index $a$ run over the adjoint
representation of the group so the corresponding term exhibit
gauge interactions.

Clearly, Susy is broken in nature because none of the Susy
particles has been observed till date. Even without being exact
symmetry, Susy can solve the hierarchy problem in a elegant way
provided Susy is softly broken. `Soft breaking' means that the
symmetry breaking terms are super-renormalizable (i.e. mass or
scalar trilinear terms). Soft mass terms are introduced by hand in
the Lagrangian ($\mathcal L_{{\scriptsize{\mbox{soft}}}}$) to
distinguish the mass of Susy particles from their SM partners.
Supersymmetric Lagrangian is given by  \be \mathcal{L}=\mathcal
L_{{\tiny{\tiny{\mbox{Susy}}}}}+{\cal
L}_{{\scriptsize{\mbox{soft}}}}\ee Here $\mathcal
L_{{\tiny{\mbox{Susy}}}}(\phi,\psi,A_\mu,\lambda,F,D)$ is globally
supersymmetric action coupling matter and gauge fields to each
other and their superpartners. The generic form is: \bea \mathcal
L_{{\tiny{\mbox{Susy}}}}&=&D^{\mu}\phi^* D_{\mu}\phi+\bar\psi i
\gamma^\mu D_\mu \psi+i g_a \sqrt{2} \phi^* T^a \lambda^a
\psi\nonumber\\&&+\frac{g_a^2}{2}D_a^2 -\biggr|\frac{\partial
W}{\partial \phi}\biggr|^2 -\frac{1}{2}\frac{\partial^2
W}{\partial \phi^2}\psi\psi  +h.c. \eea We follow notation of
\cite{martin,godbole} (see these Refs. for reviews). $\mathcal
L_{{\scriptsize{\mbox{soft}}}}(\phi,\lambda)$ is the Susy
violating part having the generic form of gaugino mass, scalar
bilinear and trilinear terms: \be
\mathcal{L}_{{\scriptsize{\mbox{soft}}}}=M \lambda \lambda
+{m}_\phi^2 |\phi|^2 +(A_0 W(\phi)+h.c.) \ee Note that all the
parameters ($M$, ${m}_\phi$, $A_0$) in $\mathcal
L_{{\scriptsize{\mbox{soft}}}}$ have mass dimensions. If one
consider the different soft masses and trilinear couplings for the
MSSM scalars and gaugino masses then one would end up with 105
free parameters \cite{Dimopoulos105} and this scenario is called
unconstrained MSSM. However it is known that flavour violation in
the soft breaking terms must be nearly absent to avoid disastrous
levels of flavour changing interactions not observed at low energy
\cite{Gabbiani:1996hi}. There are various theoretical mechanisms
that may explain Susy breaking like gravity
\cite{gravity-mediation1,gravity-mediation2}, gauge
\cite{gauge-mediation} and anomaly \cite{anomaly1,anomaly2}
mediation scenario which provide some kind of flavour blind
boundary conditions at GUT scale. As MSSM is an effective theory
of the GUT, soft parameters are evolved from the GUT scale down to
the electroweak scale. Model with universal boundary conditions
for soft sector parameters is called constrained MSSM. Besides
constrained and unconstrained model, phenomenologically more
predictive model based upon the assumptions that the CKM matrix is
the only source of CP violation and flavour mixing, diagonal soft
masses and trilinear couplings with degenerate first two
generation, has been studied extensively. These assumptions reduce
the parameters to 19 and the model is called phenomenological MSSM
\cite{pmssm}.
\begin{table}
\begin{center}$$
 \begin{array}{ |c|c|c|c|c|c| }
\hline \multicolumn{3}{ |c| }{} &\multicolumn{3}{ |c| } {}\vspace{-.3 cm}\\
\multicolumn{3}{ |c| }{\rm {Particle}} &\multicolumn{3}{ |c| } {\rm {Superpartner}}\\\multicolumn{3}{ |c| }{} &\multicolumn{3}{ |c| } {}\vspace{-.3 cm}\\
\hline
\multirow{5}{*}{Fermions} & \multirow{3}{*}{Quarks} & Q=\begin{pmatrix}u_L \cr d_L\end{pmatrix} & \multirow{5}{*}{Scalars} & \multirow{3}{*}{Squarks} & \tilde{Q}=\begin{pmatrix}\tilde{u}_L \cr \tilde{d}_L\end{pmatrix} \\
&  & u_R & &  & \tilde{u}_R \\
 &  & d_R & &  & \tilde{d}_R \\
 & \multirow{2}{*}{Leptons} & L=\begin{pmatrix}\nu_L \cr e_L\end{pmatrix} & & \multirow{2}{*}{Sleptons} & \tilde{L}=\begin{pmatrix}\tilde{\nu}_L \cr \tilde{e}_L\end{pmatrix}\\
 &  & e_R & &  & \tilde{e}_R\\ \hline

\multirow{3}{*}{Gauge Boson} & \rm {Gluon} & g& \multirow{3}{*}{Gaugino} &\rm { Gluino} & \tilde{g}\\
 & \mbox {W Boson }& W^{\pm}, W_3 & & \rm {Wino} & \tilde{W}^{\pm}, \tilde{W_3}\\
 &\rm { Photon} & $ B$ & &\rm { Bino} & \tilde{B}\\ \hline
\multirow{2}{*}{Higgs Boson} &  & H_u& \multirow{2}{*}{Higgsino} &  & \tilde{H}_u\\
 &  & H_d & &  & \tilde{H}_d\\
\hline
\end{array}
$$
\end{center}
\caption{\small{MSSM fields\label{MSSMp}}}
\end{table}
The superpotential of the MSSM is : \vspace{-.1cm}  \be
W_{MSSM}=\bar u Y_u Q H_u+\bar d Y_d Q H_d +\bar e Y_e L H_d +\mu
H_u H_d  \label{super}\ee Here $Q,L, \bar u, \bar d, \bar e$ are
the chiral superfields introduced in the Table \ref{MSSMp}. The
parameters $Y_u$, $Y_d$ and $Y_e$ are Yukawa couplings of up type
quark, down type quark and charge leptons respectively. $H_u$ and
$H_d$ are the Higgs doublets whose
 VEVs ($v_u, v_d, \tan \beta=\frac{v_u}{v_d}$) generate mass for the up and down type fermions. The first three terms are the familiar SM Yukawa interactions while the last term is the Higgs mixing or $\mu$ term. Besides these
terms in the superpotential the following baryon number (B) and
lepton number (L) violating terms are also allowed by gauge
invariance (but are excluded from the MSSM on phenomenological
grounds)\be W_{\Delta B / \Delta L =1}= \frac{\lambda_{ijk}}{2}
L_i L_j \bar e_k +{\lambda'}_{ijk} L_i Q_j \bar d_k+\mu'_i L_i
H_u+\frac{{\lambda''}_{ijk}}{2} \bar u_i \bar d_j \bar d_k  \ee
These operators imply fast proton decay rate unless the couplings
are suppressed (e.g. $|\lambda'\lambda''| \leq
10^{-24}\bigr(\frac{M_{{\tiny{\mbox{Susy}}}}}{100
~{\small{\mbox{GeV}}}}\bigr)^2$). To forbid these $d=4$ operators,
 a discrete symmetry known as R- parity is introduced :
\be R=(-1)^{3(B-L)+2s} \ee Here $s$ is the spin of the particle.
SM particles have $R$=1 and Susy particles have -1 value. Exact
conservation of R-parity require each vertex must have even number
of $R=-1$ particle, so $W_{\Delta B / \Delta L =1}$ is not
allowed. Further, the ``lightest supersymmetric particle (LSP)" is
stable by R-parity conservation  and proves a suitable ``weakly
interacting massive particle (WIMP)" dark matter candidate. This
is a crucial phenomenological virtue of R-parity conserving MSSM.
\subsection{Grand Unified Theories}\vspace{.2cm}
The quest for unification started back in the $19^{th}$ century
with the unification of electric and magnetic forces as
electromagnetic force by Clerk Maxwell. Unification of weak and
electromagnetic interaction is successfully achieved in the SM but
leaves many unanswered questions. Unification of all the known
interactions except gravity is known as Grand Unified Theory.
 It offers a framework to solve many of the shortcomings of the SM.
The first evidence for physics beyond SM came with discovery of
neutrino masses implied by neutrino oscillation data
\cite{osi1,osi2,Hirata,Ambrosio,SKC,Ahmad} which require
theoretical explanation. The seesaw mechanism
\cite{MohapatraSenjanovicSeesaw} is the most appealing mechanism
for explaining smallness of neutrino masses although simply tuning
of Yukawa couplings is also still viable. It relates smallness of
neutrino masses with the existence of heavy particles. It provides
some hint of high scale physics. Essentially the d=5 neutrino mass
or Weinberg operator \cite{weinbergd5} and its higher dimension
generalization are generated when heavy right handed neutrino
($\bar\nu(1,1,0)$) are integrated out (Type I) or a heavy triplet
scalar mass suppress a neutrino mass inducing VEV (Type II) and so
on. Nearly exact unification of the RG  evolved three MSSM gauge
couplings at high scale $\sim$ $10^{16}$ GeV
\cite{dimopoulos,amaldi} is strongest motivation to study GUTs.
All these clues reinforced the proposal \cite{PS,GG} to search for
the larger gauge symmetry of the nature represented by some higher
group whose effective theory is the SM. The basic idea is to unify
gauge as well as matter content to reduce arbitrariness of SM.
$G_{SM}$ $\subset$ $G_{GUT}$, so GUT
group should be of rank $\geq$ 4.\\

In 1974, Pati-Salam proposed the first-ever GUT based upon
$G_{PS}$ $\equiv$ $SU(4) \times SU(2)_L \times SU(2)_R $ gauge
group \cite{PS}. $G_{PS}$ contains the left-right symmetric gauge
group $G_{LR}$ $\equiv$ $SU(3) \times U(1)_{B-L} \times SU(2)_L
\times SU(2)_R$ which breaks to SM. Left-right (LR) symmetric
models offer an appealing  understanding of the origin of parity
violation in the SM. LR symmetry requires existence of the right
handed neutrino. If the Majorana mass of $\bar\nu$ is large a
small neutrino mass is generated through seesaw mechanism
\cite{MohapatraSenjanovicSeesaw}. SM fermion and $\bar\nu$ are
embedded in the LR models as \medskip
\bea &&\psi_{\mu \alpha}(4,2,1)=Q(3,2,1/3)+L(1,2,-1)  \nonumber \\
     \hat\psi^{\mu}_{\dot{\alpha}}(\bar{4},1,\bar{2})&=&\bar d(\bar 3,1,2/3)+\bar u(\bar 3,1,-4/3)+\bar e(1,1,2) +\bar \nu (1,1,0) \label{ps162}\eea
    where L, Q, $\bar e$, $\bar d$, $\bar u$  are written with their SM quantum numbers and the indices $\mu$, $\alpha$, $\dot{\alpha}$ refer to $SU(4)$, $SU(2)_L$ and $SU(2)_R$ respectively. 4-plet of SU(4) treats lepton as a fourth color. Higgs triplets ($\Delta_L(1,3,1)$ and $\Delta_R(1,1,3)$) or doublets ($\chi_L(1,2,1)\oplus \chi_R(1,1,2)$)  and  bidoublet ($\phi(1,2,2)$) can implement symmetry breaking
     \[  G_{LR} \xrightarrow{\text{ $\langle \Delta_R \rangle$ or $\langle \chi_R \rangle$} } G_{SM} \xrightarrow{\text{ $\langle \phi \rangle$ $\langle \chi_L \rangle$} } SU(3)_c \times U(1)_{em} \]$W_L$ and $W_R$ are gauge bosons corresponding to $SU(2)_L$ and $SU(2)_R$ respectively with $m_{W_R}>>m_{W_L}$ as these are missing experimental signature. Electric charge is given by
     \be\vspace{.25cm}  Q_{em}=T_{3L}+T_{3R}+\frac{B-L}{2}\vspace{.25cm} \ee Although $G_{PS}$ unified matter content but still we have 3 gauge couplings, therefore no reduction in number of gauge parameters.\\

Soon after Pati-Salam, Georgi and Glashow proposed single gauge
group GUT SU(5) \cite{GG} which can embed SM gauge group.
      This is the smallest gauge group (rank 4) which provides unification of gauge interactions.  SM fermions are embedded in the 5 dim and
       10- plet (2 index antisymmetric) of SU(5) as
\be \bar{5}=\{L,\bar{d}\} \quad ; \quad 10=\{Q,\bar e,\bar{u}\}\ee
SU(5) group is simple, so it explains charge quantization. It has
24 gauge bosons which transform under the maximal subgroup $SU(3)
\times SU(2) \times U(1)$ as :
 \vspace{.15cm}   \be  24=(8,1,0)+(1,3,0)+(1,1,0)+(3,2,-\frac{5}{3})+(3^*,2,\frac{5}{3}) \vspace{.15cm} \ee  Here (8,1) are the SU(3) gluons, (1,3) are SU(2) gauge bosons,
     (1,1) is gauge boson of U(1) group. Remaining 12 ((3,2)+$(3^*,2)$) are new heavy $X$ and $Y$ gauge bosons (leptoquark) which are responsible
      for proton decay. The experimental lower limit on the life time of proton is more than $10^{33}$ yrs.
       This implies the mass of carriers $X$, $Y$  should be greater than $10^{15}$ GeV. Exchange of $X$, $Y$ or other Higgs leptoquark leads to violation of baryon and lepton number but B-L is conserved.  SU(5) symmetry spontaneously breaks to SM by 24-plet Higgs field ($\Sigma$), while the SM doublets lie in a 5-plet ($H$) of SU(5) :
    \[\vspace{.25cm} SU(5) \xrightarrow{\text{ $\langle \Sigma \rangle $} } SU(3)_c \times SU(2)_L\times U(1)_{Y} \xrightarrow{\text{ $\langle H \rangle$} } SU(3)_c \times U(1)_{em}\vspace{.2cm} \]  SU(5) accommodates SM fermions of the same family in different representations $(\{L,\bar d\}\in 5;(\{Q,\bar e,\bar u\}\in 10 )$. It predicts equal Yukawa couplings for down quark and charged leptons which is not true for first two generations. The original model of Georgi-Glashow  fails to produce neutrino masses.
     Realistic SU(5) models can be build by addition of $24_F$ \cite{Bajc:2006ia,Bajc:2007zf} or $15_H$ \cite{Dorsner:2005ii,Dorsner:2006dj} multiplets as well as a (singlet) right handed neutrino. Since it is a SU(5) singlet the gauge symmetry does not offer any relation between neutrino and charged lepton masses. Then one can produce neutrino masses through seesaw mechanism (Type I and III in case of $24_F$, Type II in $15_H$ and Type I with $\bar\nu$).\\

 SO(10) is a rank 5 GUT \cite{FM} candidate which offers unification of matter as well as of gauge interactions. The embedding chain $SU(5) \times U(1) \subset SO(10)$
  $\rightarrow$ $SO(10) \times U(1) \subset E_6$ shows that $E_6$ is a ``maximal'' GUT gauge group \cite{E6}. 
 This thesis is based upon a successful supersymmetric
   SO(10) model so we will elaborate SO(10) properties in detail in the following section.
\section{SO(10) GUT}
\subsection{Group Theory Essentials}
SO(10) is a special orthogonal group of rank 5 with 45 parameters.
Group elements are generated from generators (J) \be
O=\exp\big(\frac{i}{2}\theta^{ij} J_{ij}\big) \quad ; \quad i,
j=1...10 \ee Antisymmetric generators in the fundamental 10-plet
representation are: \be ( J_{ij})_{kl}=-i \delta_{i[k}
\delta_{l]j} \ee here square bracket represents antisymmetrization
and these generators obey algebra \be [J_{ij},J_{kl}]=i
\delta_{k[i} J_{j]l}-i \delta_{l[i} J_{j]k} \label{algebra} \ee
The fundamental (10-plet) representation $H_i$ transforms as : \be
H'_i=O_{ij}H_j   \ee Taking tensor product of the fundamental
representation one can form higher dimensional symmetric or
anti-symmetric representation. For example 45, 120, 210, 54  are
(2, 3, 4) index antisymmetric and two index symmetric traceless
vector representations respectively. 126 $(\Sigma)$ is self-dual 5
index anti-symmetric representation which requires special
attention because it plays a crucial role in SO(10) model building
specially for neutrino masses. \be \tilde{\Sigma}_{i_1...i_{5}}=
-\frac{i}{5!}\epsilon_{i_1...i_{10}}\Sigma_{i_6...i_{10}} \quad ;
\quad \tilde{\Sigma}= {\Sigma}\ee Similarly one can project out
the anti self-dual $\oot$ :\be \tilde{\bar\Sigma}= -\bar\Sigma \ee
 Apart from tensor representations, orthogonal groups have
spinor representations generated using Clifford algebra of the
$2^N$ (N=5 for SO(10)) dimensional $\Gamma_i$ matrices: \be
\{\Gamma_i,\Gamma_j\}=2 \delta_{ij} \ee From these generators one
constructs : \be \Sigma_{ij}= \frac{[ \Gamma_i,\Gamma_j]}{4i} \ee
The generators $\Sigma_{ij}$ obey the SO(10) commutation algebra
(see Eq. \eqref{algebra}). The explicit form of $\Gamma$ and hence
$\Sigma$ matrices can be found in \cite{wilczekzee,ag1}. Spinor
representation of SO(2N) is $2^N$ dimensional (32 dim for SO(10))
which transform as \be  \Psi'=\exp (-i \theta_{ij} \Sigma_{ij})
\Psi   \ee Irreducible 16($\psi$) dimensional spinor
representation is constructed using projectors: \be
\psi=\frac{1+\Gamma_{{\tiny{\mbox{FIVE}}}}}{2} \Psi \quad ; \quad
\overline {\psi}=\frac{1-\Gamma_{{\tiny{\mbox{FIVE}}}}}{2} \Psi
\quad ; \quad \Gamma_{{\tiny{\mbox{FIVE}}}}=i \Gamma_2 \Gamma_4
....\Gamma_{10} \ee By taking direct products with tensors, spinor
representation lead to an additional class of (``double valued")
representations.

We strictly follow the notations of Ref. \cite{ag1} throughout the
thesis. For completeness we mention the most frequently used ones:
a,b,c(1..6) and $\tilde{\alpha}$, $\tilde{\beta}$,
$\tilde{\gamma}$(1..4) are SO(6) and SO(4) indices respectively,
$\mu, \nu, \lambda $... represent SU(4) indices and run from 1 to
4; $\bar \mu, \bar \nu, \bar\lambda$ run over the color subgroup
(1 to 3) of SU(4); $\alpha$, $\beta$...($\dot{\alpha}$,
$\dot{\beta}$...) denote $SU(2)_L(SU(2)_R)$ doublet indices and
vary from 1 to 2; A, B... and i, j, k.. are SO(10) spinor and
vector indices and run from 1 to 16 and 1 to 10 respectively; A,
B, C=1...3 are also used for family indices.
\subsection{Virtues of Supersymmetric SO(10) GUT}
\begin{itemize}
\item RG evolved gauge couplings of MSSM accurately unify at
$M_{GUT}\sim 10^{16.25}$ GeV. \item It is a natural home for Type
I and Type II seesaw mechanism which generate neutrino masses in
the milli-eV range, via high scale B-L breaking, without any
tuning of Yukawa couplings as is required when the B-L breaking
scale is small. This follows since minimal Susy SO(10) embeds the
minimal Susy LR models \cite{LR1,LR2,rparso10.1,rparso10.2,LR3}
which have high scale breaking of B-L symmetry. \item Large
$\tan\beta$ SO(10) models provide third generation Yukawa
unification \cite{tbtyuk}. Type II seesaw dominated models relate
atmospheric neutrino mixing angle with $b-\tau$ unification
\cite{typeIIdom}.  \item Another important aspect of SO(10) gauge
group is that M-parity ($(-1)^{3(B-L)}$) is effectively part of
SO(10) gauge symmetry since $U(1)_{B-L}$ $\subset$ SO(10). It can
be preserved \cite{rparso10.1,rparso10.2,rparso10} till low energy
with suitable choice of VEV of Higgs field. Using only B-L even
VEVs, R/M- parity preservation ensures stable LSP which can act as
a cold dark matter candidate. \item Observed baryon asymmetry of
the universe can be understood via leptogenesis which explains
baryogenesis through sphaleron processing of a lepton asymmetry
created in L and CP violating decays of heavy neutrino.
\end{itemize}
\subsection{Model Building}
As mentioned spinor representation of an orthogonal group provides
special motivation to study SO(10) GUT.  {\bf{16}}-plet of SO(10)
can accommodate exactly 15 fermions of one SM generation along
with right handed neutrino. {\bf{16}}-plet decompose under two
maximal subgroups of SO(10) : $G_{PS}$ and  $SU(5)\times U(1)$ as
:
     \be 16=\psi_{\mu \alpha}(4,2,1)+\psi^{\mu}_{\dot{\alpha}}(\bar{4},1,2)=10+\bar{5}+1\ee  From Eq. \ref{ps162}, we
      know how SM fermion and right handed neutrino are embedded in $\psi_{\mu \alpha}$ and $\psi^{\mu}_{\dot{\alpha}}$.  As all the matter fields
       are present  in a single irreducible representation thus gauge interactions in SO(10) conserve parity. SO(10) has 45 gauge bosons which decompose under the Pati-Salam group as
     \be 45=(15,1,1)+(1,3,1)+(1,1,3)+(6,2,2)\ee
     An additional 33 gauge bosons (besides the 12 present in the SM) called leptoquarks, mediate B and L violating interactions. One
can have different symmetry breaking chains via the two maximal
subgroups : \be  SO(10) \rightarrow \left\{
    \begin{array}{l l}
    SU(5) \times U(1) \rightarrow SU(5)\rightarrow G_{SM} \\
    G_{PS} \rightarrow G_{LR} \rightarrow G_{SM}
  \end{array} \right.\vspace{.3cm}\ee
These breaking chains proceed via different Higgs sectors. As
clear from the $G_{PS}$ and SU(5) GUT, spontaneous symmetry
breaking  of the larger
      unified group to SM gauge group  requires different higher Higgs representation
       depending upon the group under consideration. Choice of different combination of Higgs irreps give
        different SO(10) models. Possible choices are ${\bf{45}}$, ${\bf{54}}$, ${\bf{126}}$,
${\bf{210}}$ Higgs irreps which contain MSSM singlet.
  Further, Higgs content of the model is chosen not just to break the GUT
symmetry but it should also able to produce realistic fermion mass
mixing data. The tensor product of two {\bf{16}}-plets is
      \be 16 \otimes 16 =10 \oplus
126 \oplus 120  \ee
  Since ${\bf{10}}$, ${\bf{120}}$ irreps are real and ${\bf{126}}$ is complex, the above tensor
   decomposition suggests that only {\bf{10}}, ${\boot}$ and {\bf{120}} Higgs irreps
    can couple to matter bilinear at SO(10) Yukawa vertex.

There are two main classes of SO(10) GUTs distinguished by whether
they use doublets or triplets to break the right handed gauge
group $SU(2)_R$ and whether the seesaw is renormalizable or not.
Model builders considered small representations like
${\bf{10}},{\bf{16}},{\bf{\overline{16}}},{\bf{45}}$
\cite{smallrep}, out of these ${\bf{45}}$, ${\bf{16}}$ and
${\bf{\overline{16}}}$ break SO(10) symmetry to MSSM and
${\bf{10}}$-plet is required for electroweak symmetry breaking.
This model can not produce realistic fermion masses without using
non-renormalizable
          operators.  In the renormalizable regime one needs to use large representations like $\boot$. Further other higher Higgs irreps
          like {\bf{210}} or {\bf{54}} are required for gauge symmetry
          breaking. In this scenario two possible Higgs sets
          sufficient for spontaneous symmetry breaking to MSSM are
          ${\bf{210}}\oplus {\bf{126}} \oplus \boot$ and ${\bf{54}}\oplus {\bf{45}} \oplus {\bf{126}} \oplus
          \boot$. In particular, the model based upon ${\bf{10}}$, ${\bf{126}}$,
$\boot$, ${\bf{210}}$ Higgs representations has minimum number of
parameters and is under development since 1982 \cite{aulmoh,ckn},
was named as $\textbf{Minimal Supersymmetric Grand Unified Theory
(MSGUT)}$ \cite{abmsv}.  ${\bf{126}}$, $\boot$, ${\bf{210}}$ Higgs
irreps break symmetry to the MSSM and  ${\bf{10}}$, $\boot$
generate fermion masses.

Since 2000 \cite{ag1,nathsyed} a
         conversion of SO(10) tensor, spinor representation and their invariants in terms of unitary subgroups
         $G_{PS}$ and $SU(5) \times U(1)$ has facilitated model building in SO(10).
Recently a SO(10) model using only single pair of irreducible
Higgs representation ${\bf{(144+\overline{144})}}$ is proposed
\cite{144-1,144-2}. ${\bf{144}}$ irrep has
      both vector and spinor index. Adjoint and 5-plet of SU(5) contained in ${\bf{144}}$ break the gauge symmetry to $SU(3)_c \times U(1)_{em}$. Quartic
       coupling of Higgs and matter ${\bf{(16.16.144.144)}}$ generates fermion masses.

Babu-Mohapatra proposal that ${\bf{10}},\boot$ Higgs can
completely determine the fermion Yukawa couplings
\cite{Babumohapatra}, triggered
 intense interest in fermion fitting in SO(10) models \cite{typeIIdom,allferm}. MSGUT, a fully specified theory with only 26 real (hard) parameter failed \cite{gmblm,blmdm,bert} to fit
realistic fermion mass mixing data because Type I seesaw
contribution  which dominates over Type II seesaw yields too small
neutrino masses. Faced with this impasse, Aulakh and Garg
\cite{blmdm,nmsgut} investigated the role of ${\bf{120}}$ plet
(which can couple to matter bilinear) in the context of its direct
contribution to fermion masses. Earlier the ${\bf{120}}$-plet was
 considered \cite{bert,moh120,Oshimo-120} mostly as a perturbation to ${\bf{10}}$, ${\bf{\oot}}$, to suppress proton decay or to explain different quark and lepton mixing with arbitrary assumptions.
 The observation that MSGUT accompanied by the $\mathbf{120}$-plet
Higgs (which is the next to minimal candidate) where the
$\mathbf{120}$ and $\mathbf{10}$-plet fit the charged fermion
masses and the $\mathbf{{\overline{126}}} $ is freed to fit
neutrino masses succeeded in achieving a realistic fit. Since the
Type I seesaw neutrino masses are inversely proportional to the
$\mathbf{{\overline{126}}}  $ Yukawa coupling, the freed (to be
small) $\mathbf{126}$-plet coupling enhances the Type I seesaw
masses to viable values (Type II contribution gets further
suppressed) allowing enough freedom to fit all the fermion mass
and mixing data (the d, s quark Yukawa couplings require special
treatment and this yields important information on sparticle
spectra). The small $\mathbf{{\overline{126}}} $ coupling provide
right handed neutrino masses in a leptogenesis \cite{lepto1}
compatible  range ($ 10^8-10^{12} $ GeV). In this way, the GUT
based upon the $ \mathbf{210}\oplus \mathbf{10}\oplus
\mathbf{120}\oplus \mathbf{126}\oplus  \mathbf{{\overline{126}}} $
 Higgs irreps, known as a $\textbf{New Minimal Supersymmetric SO(10) GUT (NMSGUT)}$, emerged as a realistic GUT.
\section{Thesis Outline}
This thesis is a report on development of a realistic Susy SO(10)
model called NMSGUT. The principal new contributions have made are
the following:
\begin{itemize}
\item The NMSGUT \cite{nmsgut} while able to fit the MSSM fermion
hierarchy and predict specific testable super spectra, yields a
proton decay life time $\sim 10^{27}$ yrs if threshold corrections
due to $\sim 700$ superheavy fields  are ignored. We calculated
one loop corrections to the effective MSSM Yukawa vertices. We
found that the tree level relation between MSSM and Yukawa
coupling is strongly renormalized due to the large number of
fields and couplings renormalizing the MSSM Higgs field. This
allows natural suppression to $\tau_p > 10^{34}$ yrs on the
``Higgs dissolution edge" $(Z_{H,\bar H} \approx 0)$ in GUT
parameter space. \item In Ref. \cite{nmsgut} tree level sparticle
spectrum were used. One loop corrections to sparticle masses can
be large specially for the minisplit Susy spectra with large
$A_0$, $\mu$ and $M_A$ parameters found in \cite{nmsgut}. We
incorporate these corrections in the search program.
    \item Since the NMSGUT generates neutrino masses from B-L violating VEVs and the GUT scale slepton soft masses and trilinear couplings are renormalized by loop corrections and right handed neutrino thresholds one expects \cite{Barbieri:1,Barbieri:2,hisano} significant lepton flavour violation (LFV)  in the benchmark observable
        like $l_i \rightarrow l_j \gamma$ rates. We calculate these and other lepton sector predictions of the NMSGUT.
        \item One expects the soft Susy parameters to obey GUT relations at some high scale (e.g. Planck scale or string scale) which need not coincide with the MSSM coupling unification scale $M_X^0=10^{16.33}$ GeV. Thus the RG equations predicting flow of NMSGUT couplings between these scales should be calculated and used to improve the estimate of plausible soft parameter values at $M_X^0$. We calculated the complete two loop NMSGUT RG equations for soft and hard parameters and quote the hard parameter equations in the thesis. The soft parameter RG equations are available in \cite{csaigckkRG,ilathesis}.
            \item The successful fitting of the fermion hierarchy in the NMSGUT naturally motivates attempts at flavour unification which generate the successful NMSGUT and ultimately MSSM fermion couplings dynamically. We have proposed a novel dynamical scenario
                implementing this idea using the experience gained from unifying the MSSM fermion hierarchy in the NMSGUT.
\end{itemize}
Our aim is to check the model compatibility with experimental
data. First of all successful GUT should be able to fit the SM
fermion mass mixing data ($m_{q,l}$,
$\theta_{12,23,13}^{\mbox{{\tiny{CKM}}}}$,
$\delta^{\mbox{{\tiny{CKM}}}}$,
$\theta_{12,23,13}^{\mbox{{\tiny{PMNS}}}}$ and
$\delta^{\mbox{{\tiny{PMNS}}}}$, $\Delta m^2_{\nu}$). Proton decay
is a peculiarity of GUT so it becomes a fundamental test.  In the
lepton sector the experimental upper bound
\cite{MEG,BaBar,SINDRUM} for the branching ratio of lepton flavour
violating decays $\mu \rightarrow e \gamma $, $\tau \rightarrow
\mu \gamma $,  etc. and  the muon anomaly \cite{Stockinger}
require thorough investigation of how these limits constrain the
parameter space of the model. We will discuss all these issues
chapterwise.

In Chapter 2 we review the structure of the model, in particular,
fermion mass generation, effect of superheavy thresholds on gauge
couplings and FORTRAN fitting program algorithm. Tree level Susy
spectrum is presented in the Appendix. In Chapter 3  a generic
mechanism is introduced to suppress fast d=5, baryon decay in
SO(10) GUT with an example solution. Appendix contains detailed
formulae for Higgs renormalization factors.  Loop corrected Susy
spectrum is presented in the Chapter 4 with approximate formulae
which clarify the dominant contributions and fits incorporating
these loop corrections. Chapter 5 is devoted to the lepton sector
phenomenological implications of the model which includes $\Delta
F$=1 LFV processes, $\Delta F$=0 $(a_\mu)$ and calculation of
leptogenesis parameters. In Chapter 6, we present the SO(10)
renormalization group equations for the NMSGUT soft and hard
parameters. In Chapter 7, dynamical Yukawa generation in SO(10)
GUTs extended with the family group $O(N_g)$ is discussed. In
Chapter 8, we summarize our work and conclusions. We also indicate
avenues for further research.
\chapter{New Minimal Supersymmetric SO(10) Grand Unified Theory\label{nmsgutch}}
\section{Introduction}
 The so called ``NMSGUT'' is a renormalizable SO(10) supersymmetric grand
unified theory based upon ${\bf{10(H)}} {\bf\oplus}
{\bf{120(\Theta)}}{\bf\oplus} {\bf{126(\Sigma)}}{\bf\oplus}
{\bf{\oot(\bar\Sigma)}}{\bf\oplus} {\bf{210(\Phi)}}$ Higgs irreps.
All the SM fermions along with the right handed neutrino are
accommodated  in three copies $(\bf{\psi_A})$ of ${\bf{16}}$-plet.
${\bf{126}}$, ${\bf{\oot}}$, ${\bf{210}}$ Higgs irreps participate
in spontaneous symmetry breaking (SSB) from Susy SO(10) to MSSM in
steps or at once and are therefore called adjoint type Higgs
multiplets (AM).  ${\bf{10}}$, ${\bf{\overline{126}}}$,
${\bf{120}}$ Higgs irreps couple to matter bilinear to generate
fermion masses and are hence called \cite{ag2} fermion mass (FM)
type Higgs multiplets. ${\bf{120}}$-plet has no MSSM singlet so it
does not participate in symmetry breaking. Use of ${\bf{\oot}}$
irrep in the GUT scale SSB offers automatic implementation of high
scale Type I and Type II seesaw. ${\bf{126}}$ is introduced to
preserve Susy in the GUT scale SSB which exhibits crucial R-parity
conservation ( only B-L=2 even fields have VEVs). ${\bf{10}}$ and
${\bf{120}}$-plet are mainly responsible for generating charged
fermion masses and small Yukawa coupling of
${\bf{\overline{126}}}$ is crucial for viable neutrino masses. The
heavy
   right handed neutrino in range $10^{8-12}$ GeV (compatible with leptogenesis) and milli-eV neutrino masses
    as required by neutrino oscillations,
   through seesaw mechanism are achievable. Thus this model is capable of
    fitting known SM fermion
mass mixing data.

The structure of the theory includes its mass spectra, RG
evolution, effective MSSM, B violation effective superpotential,
threshold effects, fermion fitting etc. and has already been
elucidated \cite{ag2,abmsv,ag1,gmblm,blmdm,nmsgut}. Extensive
computer codes were developed, incorporating the NMSGUT formulae
to search the GUT parameter space for viable parameter sets
\cite{aulakhgarg}. In
 this chapter we review the structural features and predictions of the
 model. We include the description of how the GUT threshold
 corrections modify MSSM Yukawas that are incorporated in the NMSGUT
 code. The actual threshold corrections are given in the next
 chapter.
\section{Structure}
\subsection{Spontaneous Symmetry Breaking\label{ssb}}
The superpotential of theory which involves Yukawa couplings for
Higgs and matter fermions is given by \bea
W_{{\scriptsize{\mbox{NMSGUT}}}}=&\frac{1}{2}M_H H_i^2+
\frac{m}{4!} \Phi_{ijkl} \Phi_{ijkl} +\frac{\lambda}{4!}
\Phi_{ijkl} \Phi_{klmn}
\Phi_{mnij}+\frac{\eta}{4!}\Phi_{ijkl}\Sigma_{ijmno}
\overline{\Sigma}_{klmno}\nonumber\\& + \frac{M}{5!}\Sigma_{ijklm}
\overline{\Sigma}_{ijklm} +\frac{1}{4!}H_i \Phi_{jklm}(\gamma
\Sigma_{ijklm}+ \overline{\gamma} \overline{\Sigma}_{ijklm}  )
\nonumber\\& +\frac{m_{\Theta}}{2(3!)}\Theta_{ijk}\Theta_{ijk} +
\frac{k}{3!}\Theta_{ijk}H_m \Phi_{mijk}
  +
 \frac{\rho}{4!}\Theta_{ijk}\Theta_{mnk}\Phi_{ijmn} \nonumber\\
  &+  \frac{1}{2(3!)}
 \Theta_{ijk}\Phi_{klmn}(\zeta \Sigma_{lmnij}
  +  \bar\zeta \bar\Sigma_{lmnij})
 +h_{AB} \psi^T_A C^{(5)}_2 \gamma_i \psi_B H_i \nonumber \\& +
\frac{1}{5!}f_{AB} \psi^T_A C^{(5)}_2 \gamma_{i_1}....\gamma_{i_5}
\psi_B\overline{\Sigma}_{i_1...i_5} +  \frac{1}{3!}g_{AB} \psi_A^T
 C_2^{(5)}\gamma_{i}\gamma_{j}\gamma_{k}\psi_B \Theta_{ijk} \eea
Here $h$, $f$ and $g$ are Yukawa couplings of ${\bf{10}}$,
${\bf{\oot}}$, ${\bf{120}}$ Higgs. These are complex symmetric
($h,f$) and anti-symmetric ($g$) matrices in flavour space. We can
diagonalize one out of these by performing U(3) rotations in the
flavour space since the kinetic terms of the three
${\bf{16}}$-plets are invariants under U(3). These Yukawas
contribute 21 real parameters (real diagonal $h$(3)+ complex
symmetric $f$(12) + complex antisymmetric $g$(6)). In addition to
these, superpotential has trilinear couplings ($\lambda, \eta,
\gamma, \overline{\gamma}, k, \rho, \zeta, \bar\zeta $) and masses
($M_H, M,m_{\Theta}, m $) which contribute 24 parameters. In total
model has 45 real parameters out of which 2 can be fixed by using
fine tuning condition for Higgs mass (which we will explain later)
and 5 phases can be removed by redefining Higgs fields (therefore
we choose real ($\gamma, \overline{\gamma}, m_{\Theta}, m, M $)).
M is determined from $\lambda, \eta, m ,\xi$ $(M=\frac{\xi \eta
m}{\lambda})$ and further $\xi$ parameter is determined from $x$
(solution of
Eq. \ref{cubic}).  
We are left with 37 parameters. Although this seems a lot, it is
minimal in comparison to any other SO(10) GUT which provides
realistic fermion mass mixing data and experiment compatible
B-decay rates. The GUT scale (SM neutral) VEVs that break the
gauge symmetry down to the SM symmetry (in the notation of
\cite{ag1}) are \bea
 {\langle(15,1,1)\rangle}_{210} :
\langle{\phi_{abcd}}\rangle&=&{a\over{2}}
\epsilon_{abcdef}\epsilon_{ef}\qquad
\langle(15,1,3)\rangle_{210}~:~\langle\phi_{ab\ta\tb}\rangle=\omega
\epsilon_{ab}\epsilon_{\ta\tb}\nnu \langle(1,1,1)\rangle_{210}~:
~\langle\phi_{ {\tilde \alpha}{\tilde \beta} {\tilde
\gamma}{\tilde \delta}} \rangle &=&p\epsilon_{{\tilde \alpha}
{\tilde \beta} {\tilde \gamma}{\tilde \delta}}\qquad
\langle(10,1,3)\rangle_{\oot} ~:
  \langle{\overline\Sigma}_{\hat{1}\hat{3}\hat{5}
\hat{8}\hat{0}}\rangle= \bar\sigma \nnu &&
\langle({\overline{10}},1,3)\rangle_{126} ~:
\langle{\Sigma}_{\hat{2}\hat{4}\hat{6}\hat{7}\hat{9}}
\rangle=\sigma \eea  As a function of these VEVs the
superpotential becomes \be
W=m(p^2+3a^2+6\omega^2)+2\lambda(a^3+3p\omega^2+6a
\omega^2)+(M+\eta(p+3a-6\omega))\sigma\bar\sigma \ee It is
sufficient to calculate F terms from the above superpotential and
investigate the conditions for them to vanish. The VEVs of
${\bf{210}}$ do not contribute to any D term leaving only $D_{B-L}
\sim (|\sigma|^2-|\bar\sigma|^2)$ contribution to the D-term
potential. Thus vanishing of F and D terms determine MSSM vacuum.
Dimensionless  VEVs (in units of $m/\lambda$) can be ensured by
writing all VEVs in terms of single complex parameter
$x(=-\frac{\lambda \omega}{m}=-\tilde{\omega})$ :\be
\tilde{p}=\frac{x(5 x^2-1)}{(1-x)^2} \quad ; \quad
\tilde{a}=\frac{x^2+2x-1}{1-x} \quad ; \quad
 \tilde{\sigma} \tilde{\bar\sigma} =\frac{2}{\eta}\frac{\lambda x(1+x^2)(1-3x)}{(1-x)^2}\ee
 Note that $\mbox{Arg}(\sigma)+\mbox{Arg}(\bar\sigma)$ is B-L invariants while
$\mbox{Arg}(\sigma)-\mbox{Arg}(\bar\sigma)$ can be set to zero by
a B-L transformation. Thus effectively $\sigma=\bar\sigma$. Then
$F_{p,a,\omega}=0$ and vanishing of $F_{\sigma,\bar\sigma}$
requires \cite{ag2,abmsv,bmsv} : \be
 8 x^3 - 15 x^2 + 14 x -3 +\xi
(1-x)^2 =0\label{cubic} \ee where $\xi ={{ \lambda M}\over {\eta
m}} $. For each value of $\xi$, three solutions of $x$ are
available. The complex parameter $x$ is used for systematic survey
of parameter space of the model \cite{ag2,gmblm,blmdm} since its
variation directly affects the VEVs and thus the masses in the
theory and each value of $x$ fixes a unique $\xi$, whereas solving
(Eq. \ref{cubic}) for $x$ given $\xi$ requires checking three
solutions separately.\vspace{.2cm}
\subsection{Superheavy Spectrum\label{spec}}
SO(10) Higgs representations are decomposed into the SM gauge
group representations by first decomposing into Pati-Salam (PS)
labels. As an example we will discuss splitting of 10-plet. We
first need to decompose 10-plet under $SO(6)\times SO(4)$ as : \be
10(H_i)=6(H_a)+4(H_{\tilde\alpha}) \ee Here $a$, $\tilde\alpha$
are SO(6) and SO(4) indices respectively. $SO(6)\sim SU(4)$ and
$SO(4)\sim \{SU(2),SU(2) \}$ facilitate PS decomposition. Complete
technology of orthogonal to unitary conversion is presented in
\cite{ag1}. PS decomposition of 10-plet is : \be H_i(10)=H_{\mu
\nu}(6,1,1)+H_{\alpha \dot{\alpha}} (1,2,2) \ee 6-plet
($[\mu\nu]$) is two index antisymmetric representation of SU(4).
Second multiplet is SU(4) singlet and doublet of both $SU(2)_L$
and $SU(2)_R$, represented by $\alpha$ and $\dot{\alpha}$
respectively. From the breaking chain
\[ SU(2)_R \times U(1)_{B-L} \rightarrow U(1)_Y  \] Hypercharge is given by \be Y=2 T_{3R}+(B-L) \ee
so that one finds the beautiful LR symmetric electromagnetic
charge formula : \be Q=T_{3L}+T_{3R}+\frac{B-L}{2} \ee PS to the
MSSM decomposition can be easily achieved using SU(4)
decomposition $(\mu=\bar\mu+4,\bar\mu=1,2,3)$ and hypercharge
formulae given above. Under the SM gauge group $(SU(3)_C \times
SU(2)_L \times U(1)_Y)$
 the $\bf{10}$ plet decomposes as : \be
 10=H_\alpha(1,2,1)+\bar{H}_\alpha(1,2,-1)+t_{\bar{\mu}
 }(3,1,-\frac{2}{3})+\bar t^{\bar{\mu}
 }(\bar{3},1,\frac{2}{3}) \label{10decom}\ee
 The $H_\alpha$, $\bar H_{\dot{\alpha}}$ contribute to the MSSM doublets while $t,\bar t$ exchange contribute to proton decay.
Similarly other Higgs irreps are decomposed into the PS labels :
\vspace{.1cm} \be \overline\Sigma(\oot)=
\overline\Sigma^{(s)}_{\mu\nu \dot{\alpha} \dot{\beta}} (10,1,3)+
\overline\Sigma^{\mu\nu}_{ (s)\alpha \beta} (\overline{10},3,1) +
\overline\Sigma^{\nu}_{\mu \alpha \dot{\alpha}}
(15,2,2)+\overline\Sigma^{(a)}_{\mu\nu } (6,1,1)\ee\vspace{.10cm}
\be \Sigma(126)= \Sigma^{(s)}_{\mu\nu \alpha \beta} (10,3,1)+
\Sigma^{\mu\nu}_{ (s)\dot{\alpha} \dot{\beta}} (\overline{10},1,3)
+ \Sigma^{\nu}_{\mu \alpha \dot{\alpha}}
(15,2,2)+\Sigma^{(a)}_{\mu\nu } (6,1,1)\ee \bea
\Phi(210)&=&\Phi_{\mu}^{\nu}(15,1,1)+\Phi(1,1,1)+\Phi_{\mu
\dot{\alpha} \dot{\beta}}^{\nu}(15,1,3)+\Phi_{\mu \alpha
\beta}^{\nu}(15,3,1)\nnu &+&\Phi^{(a)}_{\mu\nu\alpha \dot{\alpha}
}(6,2,2) +\Phi^{(s)}_{\mu\nu\alpha \dot{\alpha}
}(10,2,2)+\Phi^{\mu\nu}_{(s)\alpha \dot{\alpha}}(\bar{10},2,2)
\eea \bea \vspace{1.0cm}\Theta_{ijk}(120)
&=&\Theta^{(s)}_{\mu\nu}({10,1,1})+
{\Theta}^{\mu\nu}_{(s)}(\overline{10},1,1)
+{\Theta_{\nu\alpha\dot\alpha}}^{\mu}(15,2,2)\nonumber\\
&+& \Theta^{(a)}_{\mu\nu\dot\alpha\dot\beta}(6,1,3)+
{\Theta^{(a)}_{\mu\nu}}_{\alpha\beta}(6,3,1)+
\Theta_{\alpha\dot\alpha}(1,2,2) \eea As discussed for
${\bf{10}}$-plet, all Higgs irreps are decomposed into SM labels.
The ${\bf{592}}$ ${\bf{(10+120+126+\oot+210)}}$ fields in the
Higgs sector fall precisely into 26 different types of SM gauge
representations which are labelled by the 26 letters of the
English alphabet \cite{ag2,nmsgut}. The decomposition of SO(10) in
terms of its ``Pati-Salam'' labels (i.e. the maximal subgroup
$SU(4)\times SU(2)_R\times SU(2)_L$) provided a translation manual
\cite{ag1} from SO(10) to unitary group labels. Using this
technology all the invariants of the superpotential are decomposed
into PS labels. Decomposition of invariants corresponding to
${\bf{10+126+\oot+210}}$ are given in \cite{ag2,ag1} and the ones
involving ${\bf{120}}$-plet are given in \cite{nmsgut}. To
illustrate we give decomposition of one invariant $\eta \phi
\Sigma \bar\Sigma$ \vspace{.2cm}\bea {\eta
\over{4!}}\phi\Sigma{\overline{\Sigma}}&=& \eta
[2i\phi_{\mu}^{~\nu}(\s_{\nu}
^{~\lambda\alpha\da}\os_{\lambda\alpha\da}^{~\mu}+\os_{\nu}^
{~\lambda\alpha\da}\s_{\lambda~\alpha\da}^{~\mu}) \\
&+&2i\phi_{\mu}^{~\nu}(\vec{\os}_{\nu\lambda(s)(R)}.\vec{\s}_{(s)(R)}^
{\mu\lambda}+\vec{\s}_{\nu\lambda(s)(L)}.\vec{\os}_{(s)(L)}^{\mu\lambda})
\label{2ndline}\\
&+& \phi(\vec{\os}_{\mu\nu(s)(R)}.\vec{\s}_{(s)(R)}^
{\mu\nu}-\vec{\s}_{\mu\nu(s)(L)}\cdot{\vec{\os}}_{(s)(L)}^{\mu\nu})
\label{3rdline}\\
&+&i\sqrt{2}(-\tilde{\s}^{\mu\nu(a)}\os_{\nu}^{~\lambda~\alpha\da}
\phi_{\mu\lambda(s)\alpha\da}+\s_{\mu\nu(a)}\os_{\lambda}^{~\nu~\alpha\da}
\phi^{\mu\lambda(s)}_{\alpha\da}) \\
&+&i\sqrt{2}(\tilde{\os}^{\mu\nu(a)}\s_{\nu}^{~\lambda~\alpha\da}
\phi_{\mu\lambda(s)\alpha\da} -
\os_{\mu\nu(a)}\s_{\lambda}^{~\nu~\alpha\da}
\phi^{\mu\lambda(s)}_{\alpha\da})\\
&-&2i\s^{~\nu\alpha\da}_{\mu}(\phi_{\nu\lambda(s)\da}^{~\beta}\os^
{\mu\lambda(s)}_{\alpha\beta(L)}+\os_{\nu\lambda\da\db(R)}^{(s)}
\phi^{\mu\lambda(s)\db}
_{\alpha}) \\
&-&2i\os_{\mu}^{~\nu\alpha\da}(\phi_{\nu\lambda(s)\alpha}^{~\db}\s^
{\mu\lambda(s)}_{\da\db(R)}+\s_{\nu\lambda\alpha\beta(L)}^{(s)}\phi^
{\mu\lambda(s)\beta}_{\da})\\
&-&2(\tilde\s^{\mu\nu(a)}\vec{\phi}_{\nu(R)}^{\lambda}.\vec{\os}_
{\mu\lambda(R)}+\s_{\mu\nu(a)}\vec{\phi}_{\lambda(L)}^{\nu}.\vec{\os}^
{\mu\lambda(L)}) \\
&+&2(\tilde{\os}^{\mu\nu(a)}\vec{\phi}_{\nu
(L)}^{\lambda}.\vec{\s}_
{\mu\lambda(L)}+\os_{\mu\nu(a)}\vec{\phi}_{\lambda
(R)}^{\nu}.\vec{\s}^
{\mu\lambda(R)})\\
&-&2\sqrt{2}(
\os_{\mu}^{~\nu\alpha\da}\s_{\nu\alpha}^{~\lambda\db}\phi
_{\lambda(R)\da\db}^{~\mu}+ \s_{\mu}^{~\nu\alpha\da}
\os_{\nu\da}^{~\lambda\beta}
 \phi_{\lambda(L)\alpha\beta}^{\mu} )  \\
&-&\sqrt{2}(\phi_{\nu~(R)}^{~\mu~\da\db}\os_{\mu\lambda\db}
^{(s)(R)\dot\gamma}
\s_{(R)\dot\gamma\da}^{\nu\lambda(s)}+\phi_{\nu~(L)}^{~\mu~\alpha\beta}
\s_{\mu\lambda\beta}^{(s)(L)\gamma}
\os_{(L)\gamma\alpha}^{\nu\lambda(s)})\label{3rdlastline}\\
&+&i\sqrt{2}(-\tilde\phi^{\mu\nu\alpha\da}_{(a)}\os_{\nu~\da}
^{~\lambda~\beta}\s_{\mu\lambda\alpha\beta}^{(s)(L)}+\phi_{\mu\nu(a)}^
{\alpha\da}\os_{\lambda~\alpha}^{~\nu~\db}\s^{\mu\lambda(s)}_{\da\db(R)})
\\
&+&i\sqrt{2}(\tilde\phi^{\mu\nu\alpha\da}_{(a)}\s_{\nu~\alpha}
^{~\lambda~\db}\os_{\mu\lambda\da\db}^{(s)(R)} -\phi_{\mu\nu(a)}^
{~\alpha\da}\s_{\lambda~\da}^{~\nu~\beta}
\os^{\mu\lambda(s)}_{\da\db(L)})]\label{phSSb}\eea The
supermultiplet masses are determined from this decomposition by
using symmetry breaking VEVs. Mass terms are divided into 3 types-
unmixed chiral, mixed pure chiral and mixed chiral-gauge
\cite{ag2}:
\begin{enumerate}
\item Unmixed chiral are those chiral fermions which transform as
SM conjugate pairs and form Dirac fermions. In total there are 13
multiplets of this type- $A[1,1,\pm 4]$, $B[6,2,\pm\frac{5}{3}]$,
$I[3,1,\pm\frac{10}{3}]$, $M[6,1,\pm\frac{8}{3}]$,
$N[6,1,\mp\frac{4}{3}]$, $O[1,3,\mp2]$, $Q[8,3,0]$, $S[1,3,0]$,
$U[3,3,\pm\frac{4}{3}]$, $V[1,2,\mp3]$, $W[6,3,\pm\frac{2}{3}]$,
$Y[6,2,\mp\frac{1}{3}]$, $Z[8,1,\pm2]$. For example $A[1,1,4]$ and
$\bar A [1,1,-4]$ form a Dirac fermion and its mass originate from
$M \Sigma \bar\Sigma$ and using ${\bf{210}}$ VEV in $\eta \phi
\Sigma \bar\Sigma$ superpotential invariants.
    \be \bar A [1,1,-4]=\frac{\bar\Sigma_{44(R-)}}{\sqrt{2}} \quad ; \quad A [1,1,4]=\frac{\Sigma_{(R+)}^{44}}{\sqrt{2}} \ee
    First term of Eqns. \eqref{2ndline}, \eqref{3rdline} and \eqref{3rdlastline} will contribute to $m_A$ :
    \be m_A=2(M+\eta(p+3a+6\omega)) \ee
     \item Mixed pure
chiral scenario correspond to when we have more than one multiplet
from different SO(10) Higgs having same SM quantum numbers. The
model has $C[8,2,\pm 1]$, $D[3,2,\pm\frac{7}{3}]$,
$K[3,1,\mp\frac{8}{3}]$, $L[6,1,\pm\frac{2}{3}]$,
$P[3,3,\mp\frac{2}{3}]$, $R[8,1,0]$, $h[1,2,\pm1]$ and $
t[3,1,\mp\frac{2}{3}]$ multiplets that belong to this category. We
will give explicit form of $h$ mass matrix when we will discuss
the emergence of effective MSSM Higgs. \item Mixed chiral gauge
are the multiplet which mixes among themselves and also with gauge
particles. These multiplets are named as $E[3,2,\pm\frac{1}{3}]$,
J$[3,1,\pm\frac{4}{3}]$, X$[3,2,\mp\frac{5}{3}]$, $F[1,1,\pm2]$,
$G[1,1,0]$. Mass matrix for $E$ is given by :
\end{enumerate}
 \bea 
{\scriptsize \left(
\begin{array}{cccccc}-2(M+\e(a-\om))&0&0&0&0&(i \omega -ip + 2ia )\zeta\\ 0&-2(M+\e(a-3\om))&
-2\sq i\e\sss&2i\e\sss&ig\sq\ssb^*& \hspace{-5mm}  (-3 i\omega +ip + 2ia )\bar\zeta\\
0&2i\sq\e\ssb&\hspace{-5mm}-2(m+\la(a-\om))&-2\sq\la\om&2g(a^*-\om^*)&-\sqrt{2}\bar{\zeta}\bar{\sigma}\\
0&-2i\e\ssb&-2\sq\la\om&-2(m-\la\om)&\sq g(\om^*-p^*)& \bar{\sigma}\bar{\zeta}\\
0&-ig\sq\sss^*&2g(a^*-\om^*)&g\sq(\om^*-p^*)&0&0\\
( ip-i\omega- 2ia )\bar\zeta&(3i\omega -ip - 2ia )\zeta& -\sqrt{2}
\zeta \sigma & \sigma\zeta&0&\hspace{-5mm} -(m_\Theta
+\frac{\rho}{3}a - \frac{2}{3}\rho \omega)\\
\end{array}\right) } \nonumber\eea
\begin{enumerate}\item[]Rows and columns are labelled by
 $(\bar E_1$,$\bar E_2$,$\bar
E_3$,$\bar E_4$,$\bar E_5$,$\bar E_6)$ $[\bar 3,2,-{1\over 3}]$
$\oplus$ $(E_1$,$E_2$,\\$E_3$,$E_4$,$E_5$,$E_6)$$[3,2,{1\over
3}]$$\equiv$$(\Sigma_{4 \dot 1}^{\bar\mu\alpha}$,$\Sigb_{4\dot
1}^{\bm \alpha}$,$\phi^{\bm 4\alpha}_{(s)\dot 2}$,$\phi^{(a) \bm
4\alpha}_{\dot 2}$,$\lambda^{\bm 4\alpha}_{\dot 2}$,$\Theta_{4
\dot{1}}^{\bar\sigma\alpha})$$\oplus$ $(\bar\Sigma_{\bar\mu \alpha
\dot{2}}^{4}\s_{\bm\alpha\dot 2}^4$,$\phi_{\bm 4\alpha\dot 1}^{(s)}$,\\
$\phi_{\bm 4\alpha\dot 1}^{(a)}$,$\lambda_{\bm\alpha\dot
1}$,$\Theta_{\bar\sigma\alpha}^{4 \dot{1}}) $. $5^{th}$ row and
column are gaugino contributions.
\end{enumerate}
As a check, SU(5) irreps mass spectra is generated from above
spectra using special direction \be p=a=\pm \omega  \ee of VEVs
\cite{nmsgut}. The superheavy fields play a crucial role as they
provide threshold corrections to the unification scale, gauge
couplings and Higgs fields etc. that we will explain in the
subsequent sections and chapters. The complete GUT scale spectrum
and couplings of NMSGUT have been given in
\cite{ag2,ag1,nmsgut,bmsv,fuku04}.
\subsection{RG Analysis\label{rga}}
After symmetry breaking large number of fields get mass of order
of GUT scale which give threshold correction to the gauge
couplings \cite{weinberg-RGT,hall}. In \cite{ag1} effect of
superheavy thresholds on $\alpha_G(M_X)$, $\sin^2 \theta_W$ and
$M_X$ is investigated using Weinberg and Hall approach.
Alternatively \cite{nmsgut} one can predict $\alpha_3(M_Z)$
instead of $\sin^2 \theta_W$. Effect of these superheavy
thresholds on $\alpha_3(M_Z)$, $\alpha_G^{-1}(M_X)$ and
$\log_{10}M_{X}$ ($\Delta_G, \Delta_3, \Delta_X$) is calculated in
\cite{csaskga3} using precisely measured value of $\sin^2
\theta_W(M_Z)$. It was shown that although there are large number
of superheavy fields their spectrum spread around $M_X^0$ gives
contribution of both signs so that the sums can be reasonable
modification of the tree level results. This analysis has
disproved the conjecture that large number of fields will give
huge corrections to the untameable observables making Susy SO(10)
unification meaningless \cite{dixitsher}. In
\cite{ag1,nmsgut,csaskga3} mass of the lightest vector particle
mediating proton decay ($X[3,2,\pm \frac{5}{3}]$) is chosen as the
matching scale ($M_X$) between the effective MSSM and GUT scale.
Relation between the gauge couplings of the effective MSSM
($\alpha_i^2=\frac{g_i^2}{4 \pi}$) and GUT ($\alpha_G$) is given
by \cite{weinberg-RGT,hall}\bea {1\over{\alpha_i(M_Z)}}
 ={1\over{\alpha_G(M_X)}} +
 8 \pi b_i \ln{{M_X}\over{M_Z} }
  + 4 \pi \sum_j {{{b_{ij}} \over {b_j}}} \ln X_j -
4\pi\lambda_i(M_X)+.... \label{relation}\eea here second and third
term represent one-loop and two-loop gauge running \be X_j= 1 + 8
\pi b_j \alpha_G(M_X^0)
 ln{{M_X^0}\over{M_Z }}
 \ee
 $b_i$, $b_{ij}$ (i, j=1, 2, 3) are the one-loop and two-loop gauge beta function coefficients :
\be \{b_1,b_2,b_3\}=\frac{1}{16\pi^2} \{{{33}/ 5},1,-3\}\nonumber
\ee \be
 b_{ij}={1\over{(16\pi^2)^2}}
  \left({\begin{array}{ccc}{{199}/{25}}  &
{{27}/{5}}  & {{88}/{5}}\\
 {9/ 5} &25 & 24 \\
 {{ 11}/{5}} & 9& 14
 \end{array}}\right)
\ee Last term of the Eq. \eqref{relation} represent the leading
order effects of superheavy thresholds.
 In the ${\overline{\mbox{MS}}}$ scheme one
has : \be \lambda_i (\mu)=-{2\over {21}} (b_{iV} +
b_{i{\tiny{\mbox{GB}}}})
 + 2(b_{i{\tiny{\mbox{V}}}} + b_{i{\tiny{\mbox{GB}}}})\ln{{M_V}\over{\mu }} +2 b_{i{\tiny{\mbox{S}}}}\ln{{M_V}\over{\mu }}+2
b_{i{\tiny{\mbox{F}}}}\ln{{M_F}\over{\mu }} \ee where
$b_{\tiny{\mbox{V}}}$, $b_{\tiny{\mbox{S}}}$,
$b_{\tiny{\mbox{F}}}$, $b_{\tiny{\mbox{GB}}}$  denote one-loop
beta functions of vectors,
 scalars, fermions
and goldstone bosons respectively with  a sum over heavy  mass
eigenstates. Corrections depend upon the ratio of masses so they
are independent of $m$ (mass of ${\bf{210}}$-plet),
  the overall mass scale parameter. Dots (in Eq. \ref{relation}) represent the two loop
  contribution of matter Yukawa couplings. Three equations (Eq. \ref{relation}) are used to determine
$\alpha_3(M_Z)$, $M_X$ and $\alpha_G(M_X)$. The threshold
correction \cite{ag2,gmblm} formulae are :\bea
\Delta^{(th)}(\ln{M_X}) &=&{{\lambda_1(M_X) - \lambda_2(M_X) }
\over{2(b_1 - b_2)}} \nonumber \\
 \Delta_X &\equiv& \Delta^{(th)}(Log_{10}{\frac{M_X}{1 \mbox{ GeV}}}) + \Delta^{\mbox{(2-loop)}}(Log_{10}{\frac{M_X}{1 \mbox{ GeV}}}) \nonumber\\
   &=&  0.222 + {\frac{5({{\bar b}'}_1 -{{\bar b}'}_2
 )}{56 \pi}}
 Log_{10}{{M'}\over  {M_X}}  \nonumber \\
 \Delta_3 &\equiv &\Delta^{(th)} ({\alpha_3} (M_Z))  \nonumber\\&=&
 {{100 \pi (b_1-b_2)\alpha(M_Z)^2}\over{[(5b_1+3b_2-8b_3)\sin^2\theta_W(M_Z)-3(b_2-b_3)]^2}}
 \sum_{ijk}\epsilon_{ijk}(b_i-b_j)\lambda_k(M_X)\nonumber\\
&=&   .000311667 \sum_{M'} (5 {{\bar b}'}_1
 -12{{\bar b}'}_2 +7{{\bar b}'}_3) \ln{{M'}\over  {M_X
 }} \nonumber \\
 \Delta_G &\equiv &\Delta^{(th)}(\alpha_G^{-1}(M_X)) +\Delta^{\mbox{(2-loop)}}(\alpha_G^{-1}(M_X))  =  \frac{4 \pi(b_1
\lambda_2(M_X)-b_2 \lambda_1(M_X))}{b_1-b_2}\nonumber \\
&=& -1.27 +  {\frac{1}{56 \pi}} \sum_{M'}( 33 {{\bar b}'}_2 - 5
{{\bar b}'}_1) \ln{{M'}\over {M_X }}
  \label{Deltath} \eea
Here ${\bar b'}_i = 16\pi^2 b_i'  $ are  one-loop $\beta$ function
coefficients ($ \beta_i=b_i g_i^3 $)
 for multiplets with  mass $M'$ and $\lambda_i$ are
 the leading contributions of the superheavy thresholds \cite{hall,ag2}.
 Using the experimental values \bea M_Z&=&91.1876\pm .0021~\mbox{GeV} \quad ; \quad
(\alpha(M_Z))^{-1} =127.918\pm .018 \nonumber \\ && \sin^2
\theta_W=0.23122+.00015 \quad ; \quad m^t_{\mbox{pole}}=172.7 \pm
2.9~\mbox{GeV}\eea threshold corrections are estimated. Two
expressions of $M_X= M_X^0 10^{\Delta_X} $ and $M_X=m_{\lambda_X}
= |m/\lambda|{g\sqrt{ 4 |\tilde{a} + \tilde{w}|^2 +
   2 |\tilde{p}+ \tilde{\omega}|^2 }}   $ determine $m$ parameter :\bea \Delta_X &=& \Delta (Log_{10}{{M_X} \over {1 \mbox{GeV}}})\nonumber \\
  | m| &=& M_X^0 10^{  + \Delta_X }
   {{|\lambda|}\over
   {g\sqrt{ 4 |\tilde{a} + \tilde{w}|^2 +
   2 |\tilde{p}+ \tilde{\omega}|^2 }}} \mbox{GeV} \label{mvalue}\eea
   where \vspace{.3cm}
   \be  g=\sqrt{4\pi (25.6+\Delta_G)^{-1}}  \ee is the threshold corrected SO(10) gauge coupling. Theory should remain perturbative after including threshold
effects and mass of the proton decay mediating gauge boson should
not be lowered too much so as not to violate experimental bounds.
These requirements constrain the corrections as :
  \bea
-20.0\leq \Delta_G &\equiv&  \Delta  (\alpha_G^{-1}(M_X))  \leq 25 \nonumber \\
3.0 \geq  \Delta_X &\equiv &\Delta (Log_{10}{M_X}) \geq - 0.3\nonumber \\
-0.017< \Delta_{3} &\equiv & {\hat\alpha_3}(M_Z)  <
-0.004\label{criteria22} \eea
 All the superheavy VEVs and hence masses are determined in
terms of parameter $x$. Solution of this variable depend upon the
superpotential parameters through $\xi=\frac{\lambda  M}{\eta m}$
parameter. Threshold corrections are not very sensitive to
$\lambda,\eta, \gamma, \bar\gamma $ as shown by scanning the
parameter space. Systematic survey of behavior of these
unification stability monitoring parameters versus $x$ and $\xi$
is shown in \cite{ag2,gmblm,blmdm}.
\subsection{MSSM Higgs}
VEVs $p,a,\omega, \sigma, \bar\sigma$ of the multiplets of
${\bf{210,216,\oot}}$ Higgs irreps break the gauge symmetry to the
SM gauge group. ${\bf{10}}$-plet (see Eq. \ref{10decom}) has
$h[1,2,1]$ multiplet having MSSM Higgs quantum numbers. Similarly
other Higgs irreps ${\bf{210,126,\oot,120}}$ also have these
multiplets which participate in electroweak symmetry breaking. Six
such doublets are \vspace{.3cm} \bea h^{(1)}&=&H_{\alpha \dot{1}
}\quad ; \quad h^{(2)}=\bar{\Sigma}_{\alpha \dot{1} }^{(15)} \quad
; \quad h^{(3)}=\Sigma_{\alpha \dot{1} }^{(15)}\nnu h^{(4)}&=&
\frac{\Phi_\alpha^{ 44\dot{1} }}{\sqrt{2}} \quad ; \quad
h_\alpha^{(5)}= \Theta_{\alpha \dot{1}}\quad ; \quad h^{(6)}=
\Theta_{\alpha \dot{1}}^{(15)} \eea where $\Sigma_{\alpha \dot{1}
}^{(15)}$, $ \overline{\Sigma}_{\alpha \dot{1} }^{(15)} $ refer to
singlet inside $(15,2,2)$ submultiplet of the $\bf{126  }$,
$\bf{\oot  }$. $h^{(4)}$ comes from $(10,2,2)$ of the $\bf{210 }$,
$h^{(5)}$ and $h^{(6)}$ refer to the singlet inside the $(1,2,2)$
and $(15,2,2)$ of $\bf{120}$ submultiplet. Similarly the six
doublets which transform as $\bar h[1,2,-1]$ are : \bea
\bar{h}^{(1)}&=&H_{ \dot{2} }^{\alpha}\quad ; \quad
\bar{h}^{(2)}=\bar{\Sigma}_{ \dot{2}}^{(15)\alpha} \quad ; \quad
\bar{h}^{(3)}=\Sigma_{ \dot{2}}^{(15)\alpha} \nnu \bar{h}^{(4)}&=&
\frac{\Phi^{ \dot{2}\alpha}_{ 44 }}{\sqrt{2}} \quad ; \quad
\bar{h}^{(5)}= \Theta_{ \dot{2}}^\alpha \quad ; \quad
\bar{h}^{(6)}= \Theta_{ \dot{2}}^{(15)\alpha} \eea These doublets
mix via a $6\times 6 $ mass matrix $ \cal H $($W=\bar h
\mathcal{H} h +....$) \bea {\cal{H}}= {\scriptsize
  \left( \begin{array}{cccccc}
-M_H & \bar{\gamma}\sqrt{3}(\omega-a) & -\gamma\sqrt{3}(\omega +
a)&
-\bar{\gamma}\bar{\sigma}&kp & -\sqrt{3}ik\omega \\
 -\bar{\gamma}\sqrt{3}(\omega+ a)& 0 & -(2M + 4\eta(a+ \omega))&0 &
 -\sqrt{3}\bar{\zeta}\omega & i(p+2\omega)\bar{\zeta}\\
\gamma\sqrt{3}(\omega-a) & -(2M + 4\eta(a- \omega))&0 & -2\eta
\bar{\sigma}\sqrt{3}& \sqrt{3}\zeta\omega& -i(p-2\omega)\zeta\\
-\sigma\gamma & -2\eta\sigma\sqrt{3}&0 & -2m + 6\lambda(\omega-a)&
\zeta\sigma & \sqrt{3}i\zeta\sigma\\
pk& \sqrt{3}\bar{\zeta}\omega& -\sqrt{3}\omega\zeta&
\bar{\zeta}\bar{\sigma}& -m_{\Theta}&
\frac{\rho}{\sqrt{3}}i\omega\\
\sqrt{3}ik\omega&i(p-2\omega)\bar{\zeta}&
 -i(p+2\omega)\zeta& -\sqrt{3}i\bar{\zeta}\bar{\sigma}&
  -\frac{\rho}{\sqrt{3}}i\omega& -m_\Theta - \frac{2\rho}{3}a\\
\end{array}\right) } \nonumber \eea
Rows and columns of mass matrix are labelled by
 $(\bar{h}_1,\bar{h}_2,$ $\bar{h}_3,\bar{h}_4,\bar{h}_5,\bar{h}_6)[1,2,-1]\oplus$
$(h_1,h_2,h_3,h_4,h_5,h_6)[1,2,1]$. As MSSM is effective theory of
the model so one Higgs should be light. By tuning the parameters
so that $Det {\cal H} =0$ one can keep one pair of Higgs doublets
$H_{(1)}, {\bar H}_{(1)}$ (defined by left and right null
eigenstates of the mass matrix ${\cal H}$) light. We denote the
components of the right (left) null eigenvectors as
$\alpha_i(\bar\alpha_i),i=1...6$, normalized to one and real first
component.  U and $\bar{U}$ are the unitary transformations
 \be h=UH \quad ; \quad \bar h=\bar U\bar H \ee which diagonalize $
{\cal H}^\dag{\cal H}$ and ${\cal H}{\cal H}^\dag $ so that
${\bar{U}}^\dag {\cal H} U $ is diagonal and positive. Since
$U_{i1}=\alpha_i$, $\bar{U}_{i1}=\bar\alpha_i$, in the Dirac mass
matrices of the effective MSSM we can replace $\langle h_i
\rangle$ $\rightarrow \alpha_i v_u, \langle \bar h_i
\rangle\rightarrow \bar\alpha_i v_d$. Thus the ``Higgs fractions$
(\alpha_i,\bar{\alpha}_i )$" (analytical expressions can be found
in \cite{nmsgut}) specify how much the different GUT scale
doublets $h_i,\bar h_i$ contribute to the electroweak (EW)
symmetry breaking by $H=H_{(1)}$, $\bar H=\bar H_{(1)}$.
\subsection{Fermion Masses}
As mentioned the Yukawa couplings of the pairs of  bidoublets
contained in the Higgs set $\bf{10}\oplus \bf{\oot}\oplus \bf{120}
$ give rise to charged fermion masses. Therefore Yukawa coupling
matrices $y_l$, $y_u$, $y_d$  and $y_{\nu}$ at high scale ($M_X$)
are predicted in terms of the $Y_{AB}^{10}=h_{AB}$,
$Y_{AB}^{\overline{126}}=f_{AB}$ and $Y_{AB}^{120}=g_{AB}$ which
specify the Yukawa couplings of the $\bf{10}, \bf{\oot},\bf{120} $
to $\bf{16.16}$. For explicit relations of the fermion Dirac
 masses decomposition of $\bf{16\cdot16\cdot (10 \oplus 120\oplus \oot)}$ is required, which can be found in \cite{ag2,nmsgut}. For illustration we present
 decomposition of $\bf{16\cdot16\cdot 10}$ in terms of the PS and the SM
 labelles:
\bea W_{FM}^H &=& h_{AB} \psi^{T}_{A}{C}_{2}^{(5)}
\gamma_{i}^{(5)}\psi_{B} H_{i} \nonumber\\
&=& \sqrt{2}h_{AB} \big [H_{\mu\nu}\widehat{\psi}^{\mu\da}_{A}
\widehat{\psi}^{\nu}_{B\da} +\widetilde{H}^{\mu\nu}\psi_{\mu
A}^{\alpha} \psi_{\nu\alpha B} - H^{\alpha\dot\alpha}
(\widehat{\psi}^{\mu} _{A\dot\alpha}\psi_{\alpha\mu B}
+\psi_{\alpha\mu A}\widehat{\psi}
_{\dot\alpha B}^{\mu} )\big ]\nonumber\\
&=& 2\sqrt{2} h_{AB}[ \ovl{t}_{1} ({\epsilon} {\bar u}_A {\bar
d}_B+ Q_{A}L_{B})
 + {t}_{1} ( {\epsilon\over 2} Q_{A}Q_{B}+
   \bar u_{A}{\bar e}_{B}-{\bar d}_{A}{\bar {\nu}}_{B})]
\nonumber\\
&-&2\sqrt{2}h_{AB}\bar{h}_{1}[{\bar d}_{A} Q_{B} +\bar e_{A}L_{B}]
+2\sqrt{2}h_{AB}{h}_{1}\big [{\bar u}_{A} Q_{B} +{\bar\nu}_{A}L_B
\big] \eea
 From these invariants one obtains the
Yukawa couplings just by replacing the MSSM Higgs by corresponding
Higgs fractions as  \vspace{.35cm} \bea y^u &=& ( {\hat h} + {\hat
f} + {\hat g} )\quad ;\quad r_1=\frac{\bar\alpha_1}{\alpha_1}
\quad ; \quad r_2=\frac{\bar\alpha_2}{\alpha_2} \nnu
 { y}^d &=& {  ({ {r}}_1} {\hat  h} + { { {r}}_2} {\hat  f}  +
{ {r}}_6 {\hat  g})\quad ; \quad { {r}}_6 =
\frac{{{\bar{\alpha}}_6}+ i \sqrt{3}{{\bar{\alpha}}_5}}{\alpha_6+
i \sqrt{3}\alpha_5} \nnu y^{\nu}&=& ({\hat h} -3 {\hat f}  + (r_5
-3) {\hat{g}})\quad ;\quad {r_5}= \frac{4 i
\sqrt{3}{\alpha_5}}{\alpha_6+ i
   \sqrt{3}\alpha_5}
\nnu { y}^l &=&{ ( { {r}}_1} {\hat  h} - 3 {  { {r}}_2} {\hat  f}
+
   ( { {\bar{r}}_5} -
   3{ {r}}_6){\hat  g})\quad;\quad { {\bar{r}}_5}=
\frac{4 i \sqrt{3}{{\bar{\alpha}}_5}}{\alpha_6+ i
   \sqrt{3}\alpha_5}\nnu
{\hat  g} &=&2i g {\sqrt{\frac{2}{3}}}(\alpha_6 + i\sqt \alpha_5)
\quad;\quad \hat  h = 2 {\sqrt{2}} h \alpha_1 \quad;\quad\hat  f =
-4 {\sqrt{\frac{2}{3}}} i f\alpha_2 \label{yukawa}
\vspace{.25cm}\eea By multiplying these Yukawas with electroweak
VEVs $(v_u,v_d,\tan \beta=\frac{v_u}{v_d})$ one can get fermion
masses. Higgs fractions and SO(10) Yukawas determine MSSM matter
fermion Yukawas which produce the experimental fermion mass mixing
data. To generate Majorana masses $(M \psi \psi )$ for the left
and right handed neutrino we need
  SU(4) 10-plet which contains $L= \pm 2$ components, since the lepton number of the Majorana mass term
 is 2. The multiplets having 10 of SU(4) from $\bf{210}$ and
 $\bf{120}$
 do not have VEVs, thus the only option remaining is $(10,1,3)_{\overline{126}}
 $ and $(\overline{10},3,1)_{\overline{126}}$ that can couple to matter bilinear. Majorana
 mass of the right handed neutrinos is determined by the coupling of the neutrino to the $\bf{\oot } $: \vspace{.3cm} \be M^{\bar\nu}_{AB}=  8 {\sqrt{2}} f_{AB}
 {\bar\sigma} \ee
Majorana mass term and Dirac mass term (which mixes the left and
right neutrinos) give rise to Type I seesaw contribution by
eliminating $\bar\nu_A$
\[ W = {\frac{1}{2}}  M^{\bar\nu}_{AB} \bar\nu_A\bar\nu_B  +
\bar\nu_A m^{\nu}_{AB} \nu_B  + .....\] \[ W_{{\small\mbox{eff}}}=
{\frac{1}{2}} M^{\nu (I)}_{AB} \nu_A\nu_B  + ....\]\be
M^{\nu(I)}_{AB} = -((m^{\nu})^T (M^{\bar\nu})^{-1} m^{\nu})_{AB}
\ee In addition to this there is another contribution to the
neutrino mass known as Type II seesaw. The Type II neutrino mass
is \be M^{\nu}_{AB} =16 i f_{AB} \langle{\bar{O}}_- \rangle \ee
where $\langle{\bar{O}}_- \rangle $ is $SU(2)_L$ triplet VEV ($\in
(\overline{10},3,1)$) of $ \boot$ whose computation
 requires inspection
of relevant terms in the superpotential :
 \be \langle{\bar O}_{-}\rangle = ( i\ga {\sqrt  2} \al_1 +
{ 2 i{\sqrt  6}}\e \al_2- {\sqrt{6}}\zeta \al_6 + i
{\sqrt{2}}\zeta \al_5) \al_4
   {\frac{v_u^2}{M_O}}     \qquad \ee
 and $M_O= 2 (M + \eta (3a-p))$ is mass of superheavy $O$
 multiplet \cite{nmsgut}.
\section{Viable Parameter Space Search}
With appropriate formulae in hand, next task is to check the
compatibility of the model with experimental data. FORTRAN and
Mathematica codes were developed \cite{aulakhgarg} for fermion
fitting along with viable unification, electroweak symmetry
breaking, including Susy threshold corrections and for B-decay
calculations. $\chi^2$ analysis is performed to fit SM mass-mixing
data at two scales- GUT scale ($M_X^0$) and electroweak scale
($M_Z$). GUT scale fitting is based upon the random searches of 37
model parameters (listed in Section \ref{ssb} )  and  $x$
parameter (which is chosen complex and later the phase of
$\lambda$ is fixed to remove this freedom). $M_Z$ scale
calculations involve Susy threshold corrections which require
estimation of Susy spectra. Soft Susy breaking Lagrangian is :
\begin{eqnarray}
  -{\cal L}_{\rm soft}
  &=& m_{\tilde Q}^2 \wt{Q} \wt{Q}^{\dag}
    + m_{\tilde u}^2 \wt{u}^{\dagger} \wt{u}
    + m_{\tilde d}^2 \wt{d}^{\dagger} \wt{d}
 + m_{\tilde L}^2 \wt{L}^{\dagger } \wt{L}
    + m_{\tilde l}^2 \wt{e} \wt{e}^{\dagger}
  \nonumber\\
  &&+ M_{\bar H}^2 H_d^{\dagger} H_{d}
    + M_{ H}^2 H_{u}^\dagger H_u
    - \left( B H_{d} H_u + \hc \right)
  \nonumber\\
  &&+ \left(   A_u
                        \wt{Q} \wt{u} H_u
             + A_d \wt{Q} \wt{d} H_{d}
             + A_l
                        \wt{l} \wt{L}H_{d}
             + \hc \right)
  \nonumber\\
  &&+ \left(   \frac{M_1}{2} \wt{B}\wt{B}
             + \frac{M_2}{2} \wt{W}\wt{W}
             + \frac{M_3}{2} \wt{g}\wt{g} + \hc \right) ~.
  \label{MSSMsoft}
\end{eqnarray}
where $\wt{Q}$, $\wt{d}$, $\wt{u}$, $\wt{e}$ and $\wt{L}$ are
scalar components of $Q$, $d^c$, $u^c$, $e^c$ and $L$
respectively. $\wt{g}$, $\wt{W}$ and $\wt{B}$ are gaugino's of
SU(3), SU(2) and U(1) respectively. SU(2) and generation indices
are suppressed. Tree level Susy spectrum is presented in the
Appendix. The soft Susy breaking parameters at the GUT scale
($M_X$) are described by universal soft Susy breaking parameters-
$m_0$ (universal scalar mass), $m_{\frac{1}{2}}$ (universal
gaugino mass), and $A_0$ (dimensionless universal scalar trilinear
coupling) and Higgs mass parameters $(M^2_H, M^2_{\bar H})$:
\begin{eqnarray}
  m_{\tilde Q}^2 &=& m_{\tilde u}^2 ~=~ m_{\tilde d}^2 ~=~ m_{\tilde L}^2~=~m_{\tilde l}^2 ~=~ m_0^2
\nonumber\\
  M_1 &=& M_2 ~=~ M_3 ~=~ m_{\frac{1}{2}}
\nonumber\\
  A_u &=& A_d ~=~A_l=~A_{\nu} = A_0 ~
\quad ; \quad M^2_H, M^2_{\bar H} \end{eqnarray} This scenario is
called supergravity non universal Higgs mass (SUGRY-NUHM)
parameters at GUT scale. Use of non-universal Higgs masses is
justified as the light Higgs of MSSM is a combination of six
doublets from
$\textbf{10},\textbf{210},\textbf{120},\textbf{126},\boot$ Higgs
irreps. In Chapter 6 we will see how RG flow of the soft
parameters between $M_P$ and $M_X^0=10^{16.33}$ GeV can support
the NUHM assumptions. Low scale fitting is based upon these
parameters. Algorithm of the program is represented by a flowchart
\ref{flowc1}. Task of various functions/subroutine and variables
used in the flowchart is discussed below:
\begin{figure}
 \begin{center}
\begin{tikzpicture}[node
distance=2.6cm, auto]
    \node [block5] (init) {{\small{Start}}};
    \node [block22, below of=init] (identify1) {{\small{Initialize\\ Bestfunk}}};
    \node [block22, below of=identify1] (targ) {{\small{Read Target}}};

    \node [block, below of=targ] (gs1) {{\small{Random Input\\ Generates Simplex}}};
    \node [block, below of=gs1] (amb1) {{\small{AMOEBA1}}};;
    \node [block, below of=amb1] (gf) {{\footnotesize{ GUTTHRESH\\ $\&$\\ FUNKFERM}}};
    \node [block, left of=amb1, node distance=3.5cm] (update1) {{\small{Update Input\\(NMSGUT superpotential parameters)}}};
    \node [block, above of=update1, node distance=3.5cm] (update3) {{\small{Bestfunk=$\chi_{X}$}}};
    \node [decision, below of=gf] (decide) {{\scriptsize{$\chi_{X}< $ Bestfunk?}}};

     \node [block22, right of=decide, node distance=7cm] (init2) {{\small{Initialize\\ Besttune}}};
     \node [block, above of=init2] (gensimp) {{\small{Random Input \\Generates Simplex}}};
      \node [block, above of=gensimp, node distance=3cm] (random2) {{\small{PMX2PMS}}};
      \node [block, above of=random2, node distance=3cm] (amb2) {{\small{AMOEBA2}}};
     \node [block11, above of=amb2] (susythresh) {{\footnotesize{FUNKTUNE}}};
     \node [decision, above of=susythresh] (bestpt) {{\scriptsize{$\chi_{Z}< $ Besttune?}}};
     \node [decision, left=.85cm of bestpt] (mastlp) {{\tiny{miter $\leq$ mitermax}}};
     \node [block5, below of=mastlp, node distance=3cm] (stop2) {{\small{Stop}}};

 \node [block, right of=random2, node distance=3.5cm] (update4) {{\small{Update Input\\(mSUGRY-NUHM parameters)}}};
 \node [block, below of=update4, node distance=3.8cm] (update2) {{\small{Besttune=$\chi_{Z}$}}};
 \node[left of=mastlp, node distance=1.8cm] (dummy) {};

 \node [block23, below of=dummy, node distance=1.0cm] (uptarg) {{\small{Update\\Target}}};
    \path [line] (init) -- (identify1);
    \path [line] (identify1) -- (targ);
    \path [line] (targ) -- node[left] {{\scriptsize{$iter1=iter1+1$}}}(gs1);
    \path [line] (gs1) -- (amb1);
    \path [line] (amb1) -- (gf);
    \path [line] (gf) -- (decide);
    \path [line] (decide) -- node[below] {Yes} node [above]  {{\scriptsize{iter1$\leq$ iter1max}}} (-3.5,-16.0)--(update1);
    \path [line] (-3.5,-8.2) |- (-1.5,-8.2); 
    \path [line] (update1) -- (update3);
   \path [line] (decide.east) -- node[below]{No}(2.0,-16.0)--node[right] {{\scriptsize{iter1$\leq$ iter1max}}}(2.0,-7.8)|- (0.0,-6.1);
     \path [line] (decide.south) -| node[midway,below] {{\scriptsize{Masteriter= Masteriter+1}}}(init2);
    \path [line] (update3) |- (identify1);
\path[line] (gensimp) -- (random2); \path[line] (random2) --
(amb2); \path[line] (amb2) -- (susythresh);
    \path [line] (susythresh) -- (bestpt);
    \path [line] (init2) -- node[right] {{\scriptsize{$iter2=iter2+1$}}}(gensimp);
    \path [line] (bestpt) -- node[above]  {Yes} node[below] {{\scriptsize{iter2$\leq$ iter2max}}}(10.5,-1.8)-- (update4);
    \path [line] (update4) -- (update2);
    \path [line] (update2) |-  (init2);
    \path [line] (mastlp) -- node {No}(stop2);
    \path [line] (bestpt.west) --node[above]  {No}(5.3,-1.8)--(5.3,-2.0) --node[left] {{\scriptsize{iter2$\leq$ iter2max}}} (5.3,-14.5) |- (7.0,-15.1);
     \path [line] (bestpt.north) --   (mastlp.north);
     \path [line] (update4) |-(gensimp);
      \path [line] (mastlp.west) -|node[above]  {Yes} (uptarg);
      \path [line] (uptarg) |- (targ);
\end{tikzpicture}
\caption{Flowchart of FORTRAN search program.}
 \label{flowc1}
\end{center}
\end{figure}
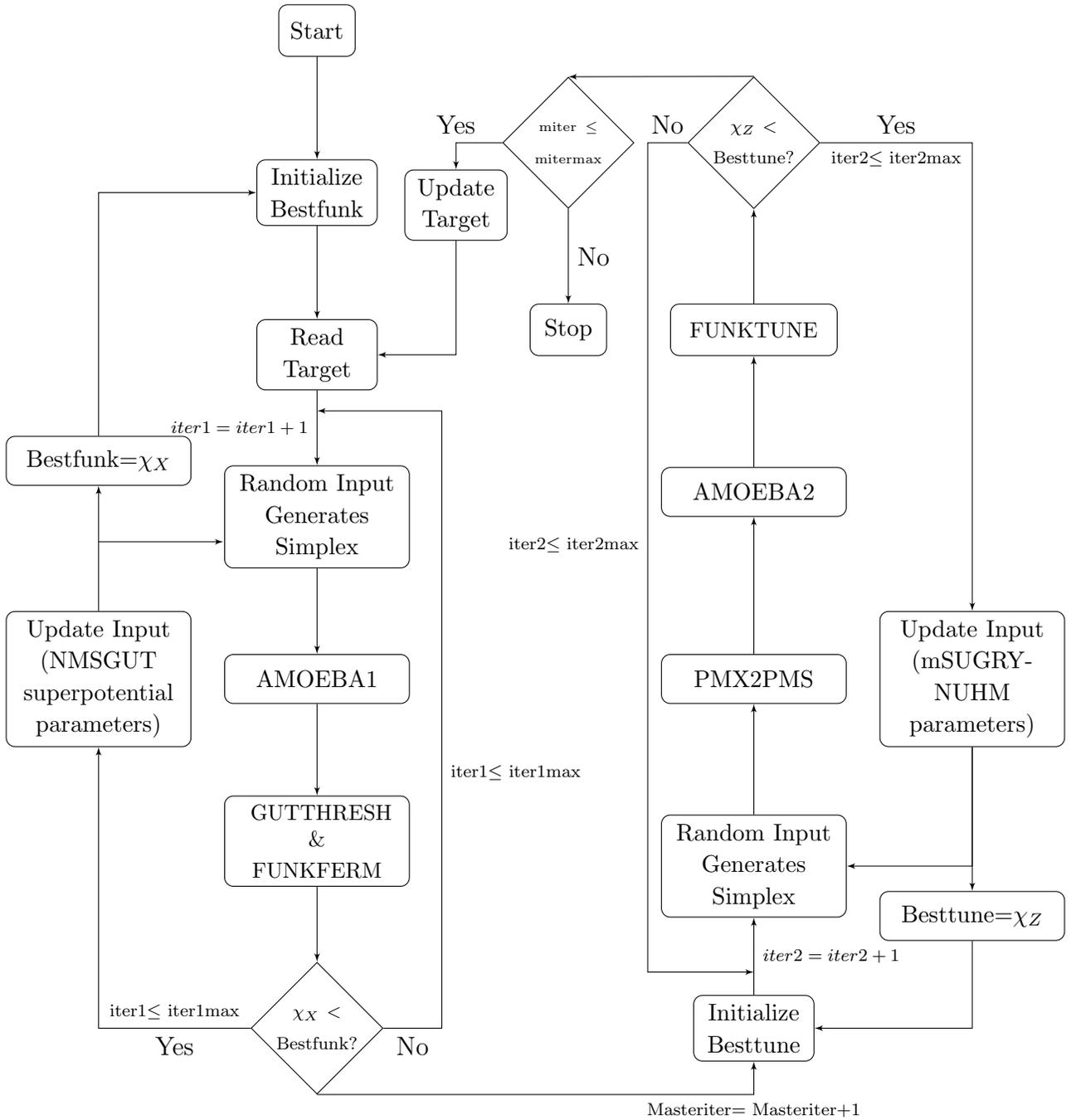
\begin{enumerate}
\item Parameter \emph{Masteriter(miter)} represents the number of
loop iterations from one scale to another. \emph{iter1} and
\emph{iter2} are high scale and low scale iteration parameters of
the search engine. \item Subroutine AMOEBA1/AMOEBA2- There are
various public available packages to find minima of non-linear
functions. Downhill simplex method of Nelder and Mead
\cite{downsim} has been used to find the minima of highly
non-linear function in a multi-dimensional space because it
involves function evaluations only. It is based upon the n-simplex
 having n+1 vertices in n-dimensional space. The AMOEBA subroutine \cite{downsim}
contracts, expands and reflects the simplex so as to converge upon
(local) minima. AMOEBA1 and AMOEBA2 are modified versions of
general subroutine AMOEBA which perform search in different
dimensional space and call separate appropriate functions. AMOEBA1
is used for GUT scale parameter search while AMOEBA2 is used for
low scale searches.

\item Subroutine GUTTHRESH does the calculations discussed in the
Section \ref{spec} and \ref{rga}. First of all it calculates the
superheavy spectrum. Then effect of threshold correction to
unification stability monitoring parameter is checked and overall
mass scale $m$ parameter is fixed. Then coefficients of d=5 proton
decay LLLL and RRRR operator are calculated \cite{ag2,nmsgut}. It
contains penalties to get $\Delta_G$, $\Delta_3$ and $\Delta_X$
within the required range (see Eq. \ref{criteria22}). It provides
$\Delta_G$, $\Delta_3$, $\Delta_X$ and  $m$ parameters as output.

\item Function FUNKFERM- It calculates $\chi^2_{X}$ value by
comparing model predicted values with target (which is run up SM
experimental mass mixing data).  For a given set of values of
superpotential parameters and Yukawa couplings, fermion Yukawas
are calculated using Eq. \eqref{yukawa}. Then it calculates the
eigenvalues and mixing angles in quark and lepton sector. After
that \be \chi^2_{X}=\sum\limits_{i}\bigg(\frac{(O-\bar
O_i)^2}{\delta O_i}\bigg)\ee fitting of SM mass mixing data (18
parameters) is done. $\delta O_i$ is an estimate of the
uncertainty in the GUT scale value based upon extrapolation of
uncertainty at the measured scale (see e.g.
\cite{antuschspinrath}). $\bar O$ and $O$ are experimental and
model predicted values respectively. Here the sum ($i$) run over
the Yukawa couplings (9), quark and lepton mixing angle, CKM phase
and neutrino mass square differences. This function returns
argument \emph{funk} ($\chi^2_{X}$).

   \item Subroutine PMX2PS  uses two-loop MSSM renormalization group equations \cite{Castano,MartinVaughn} to run the hard (Yukawa and gauge coupling) and soft parameters from GUT scale to electroweak scale and vice versa.

    \item Function FUNKTUNE  does $M_Z$ scale calculations. Main task of this function is to perform low scale
     fermion fitting. Fitting of $y_d$ and $y_s$ require inclusion of Susy threshold corrections and which further
      require Susy spectrum  estimation. Yukawa couplings, gauge couplings, scalar masses and gaugino masses at $M_Z$ are input of
       this subroutine. Tree level Susy spectrum is calculated using SPheno
        subroutine (Susy spectrum code) \cite{porod} \emph{TreeMassesMSSM}. Penalties are imposed to
         get positive (and above some lower limit) squark and slepton mass square parameter. Higgs mass is computed using one loop
     effective potential. $\mu$ and $ B$
     parameters are calculated using one loop electroweak symmetry breaking
     conditions and the run down values of the Higgs mass
     parameters at $M_Z$.
\begin{eqnarray}
\mu^2 &=& {1\over2}\biggl[\,
\tan2\beta\biggl(\big(M_{H}^2-\frac{t_2}{v_2}\big)\tan\beta-\big(M_{\bar
H}^2-\frac{t_1}{v_1}\big)\cot\beta\biggr) -M_Z^2\
  \biggr]\  \nonumber\\ B
&=& {1\over 2}\biggl[\tan 2\beta \, \biggl(\big(M_{H}^2 -
\frac{t_2}{v_2}\big)-\big(M_{\bar H}^2 -
\frac{t_1}{v_1}\big)\biggl) -M_Z^2 \sin 2 \beta\biggr]
\label{ourmuB}\end{eqnarray} where $t_{1,2}$ are tadpoles of the
effective potential, calculated using a SPheno subroutine based on
the formulae of \cite{piercebagger}. These can be  extrapolated
back to
     $M_X^0$ to find $\mu$ and $B$ at $M_X^0$ since the RGEs of the other soft masses are independent of these. Both $\mu$ and B are
     assumed real and positive. $M_A$ (pseudo-scalar mass) is calculated  using the
above one-loop corrected value of $B$ parameter \vspace{.3cm} \be
M_A^2= \frac{2B}{\sin2\beta}\vspace{.3cm} \ee Tree level spectrum
is again calculated using updated value of $\mu$ and $B$
parameter. Susy threshold corrections to the Yukawa couplings are
calculated. Then using off-diagonal run down values of Yukawas,
tree
 level spectra is calculated to verify that  ignoring generation mixing does not make much difference. Aim is to find the suitable set of soft parameters which give appropriate corrections to the NMSGUT run down Yukawas to fit them with the SM data.
 Following penalties are imposed upon soft parameters \cite{nmsgut} :
\bea && | {m}_{\tilde{ l},\tilde{L}} /M_1| \geq 0.9
\qquad ;\qquad |{m}_{\tilde{u},\tilde{d},\tilde{Q}} /M_3| \geq 0.75  \quad ;\quad M_{\tilde {u},\tilde {d}} >500 \, \mbox{GeV} \nonumber\\
&&\mu,|A_0| < 150\,\, \mbox{TeV} \quad;\quad m_{\tilde l,H^{\pm}}
> 200\, \mbox{GeV} \nonumber
\\&& \tilde W : 200 \,\, \mbox{GeV}< M_2(M_Z)< 1000.0 \,\,\mbox{GeV} \nonumber
\\ && \tilde g : 500 \,\, \mbox{GeV}< M_3(M_Z)< 1000.0 \,\,\mbox{GeV}\eea
Higgs mass measurements are available since December 2011, in
\cite{nmsgut} SM Higgs was required to be heavier than $114 $ GeV
(LEP limit) and the Bino lighter than the lightest sfermion. Susy
threshold corrected run down Yukawas values are compared with SM
Yukawa via a $\chi^2_{Z}$. \vspace{.3cm} \be \chi^2_{Z}=\sum_i
\bigg(1-\frac{y_i^{\tiny{\mbox{MSSM}}}}{y_i^{\tiny{\mbox{SM}}}}\bigg)^2
\ee Here $y_i^{\tiny{\mbox{MSSM}}}$ are threshold corrected
Yukawas. This function returns $\chi^2_{Z}$.  AMOEBA2 subroutine
  uses this function and calculates its value at each vertex of the simplex.
\end{enumerate}
\vspace{.4cm} Program starts with the GUT scale searches. It
requires target extrapolated SM data. Using two loop RGEs of MSSM
\cite{Castano,MartinVaughn}, central fermion experimental data
(Yukawa + neutrino mass difference and (quark, lepton sector)
mixing data) is extrapolated to one loop unification scale
$M_X^0$=$10^{16.33}$GeV ignoring right handed neutrino thresholds
and assuming normal hierarchy for left handed neutrino masses
(neutrino mass splitting is calculated from extrapolated
coefficient of d=5 operator \cite{weinbergd5,nmass}). At $M_X^0$
canonical parameters are extracted which serves as target for
model calculations. This target file and two other input files
having random set of NMSGUT superpotential couplings and
SUGRY-NUHM parameters are provided. To start with, the value of
variable \emph{bestfunk} (whose role will be explained later) is
fixed along with the initialization of many other parameters. In a
downhill simplex method one needs to provide the initial vertex of
simplex around which it starts searching in the parameter space.
Random input provided is that initial vertex. With slight random
changes from this point all the vertices of the simplex are
generated. At each vertex of simplex GUTTHRESH calculations and
function FUNKFERM is calculated. AMOEBA1 subroutine  also calls
GUTTHRESH and function FUNKFERM for viable unification and
function evaluations and it compares the value of \emph{funk} at
each
 vertex and select the one with minimum \emph{funk}.
 If calculated \emph{funk} at that point is less
than or equal to initial \emph{bestfunk} value then it replaces
that parameter set in the input file. In the next iteration it
start searching around that point. If the lowest funk value is
more than the initial \emph{bestfunk} then in the next iteration
program starts searching around the old point, but as the vertices
of simplex are randomly generated so simplex will be different
from the earlier one. Once all the iterations at high scale are
completed set of hard parameters (Yukawa and gauge couplings) at
GUT scale is obtained.

Then program starts searching for soft parameters. It reads
initially provided random SUGRY-NUHM parameters and generates
simplex as discussed for GUT scale searches. Notice that now
simplex dimensionality is different. Using MSSM RGEs, the diagonal
Yukawa couplings and scalar masses are run down to $M_Z$ scale.
With the fixed value of hard couplings and random soft couplings,
program calls FUNKTUNE at each vertex of simplex and calculates
$\chi^2_{Z}$. Again like $M_X$ scale, AMOEBA2 select the point
with minimum value of $\chi^2_{Z}$ and compare that $\chi^2_{Z}$
with the initially chosen \emph{besttune} value. Soft parameter
input is replaced if the selected point is better than initially
provided. After completing all the low scale iterations, program
provides a set of SUGRY-NUHM parameters. Then new target set of
(Susy threshold) corrected MSSM couplings is provided for next
iteration high scale calculations. Procedure is repeated
\emph{Masteriter} number of times to get reasonable fitting of
MSSM data at high and low scale. At the end the program prints
\emph{bestfunk}  and \emph{besttune} value representing
$\chi_{X},\chi_{Z}$  and stores corresponding parameters in the
input/output files.
\section{Distinct Predictions}
\subsection{Normal s-hierarchy}
NMSGUT fits \cite{nmsgut} prefer large negative values of Higgs
mass squared soft parameters $M_{H,\bar H}^2$ $\simeq$ $(100
~\mbox{TeV})^2$. One loop $\beta$ function of scalar's soft mass
RGEs contains terms proportional to $M_{H,\bar H} Y_f^{\dag} Y_f$.
Due to the large third generation Yukawa couplings, this term
dominates for the third generation evolution and their masses
evolve to large values compared to the first and second
generation. So the model predicts normal s-hierarchy opposite to
the common wisdom. The RG flow of Yukawa coupling and scalar
masses exhibiting this behaviour is given in \cite{gutupend}.
\vspace{.15cm}\subsection{Large $A_0$ and $\mu$
Parameter}\vspace{.35cm}If only ${\bf{10}}$ and ${\bf{120}}$-plet
fit charged fermion masses with $\boot$ Yukawa coupling ($f_{AB}$)
 chosen tiny the NMSGUT can only generate $y_{d,s}(M_X^0)$ values which are smaller by a factor of 3-5 than the
 extrapolated values of SM Yukawas at $M_X^0$. Therefore SM down and strange quark Yukawa require lowering by a factor
of 5 to fit these with run down values of Yukawas. This lowering
is achieved by large $\tan\beta$ (preferred by SO(10) GUTs  for
third generation Yukawa unification )  driven threshold
corrections which require specified soft Susy breaking parameters
\cite{nmsgut}.
  Gluino contribution is a dominant one
loop correction which is proportional to Susy breaking parameter
$\mu$, so large $\mu$ parameter provides significant Susy
threshold corrections to $y_d$ and $y_s$ to match it with GUT
renormalized value down to $M_Z$ scale. For third generation,
bottom quark, slight raising is required, so Susy threshold
corrections should not change it too much. This cancellation can
be achieved by  large $A_0$ (soft trilinear couplings) parameter.
After the Higgs discovery large $A_0$ is favoured in Susy-GUTs for
Higgs mass of 126 GeV \cite{minisplit1}. Heavy third s-generation
and large $A_0$ raise the tree level Higgs mass from 91 GeV to 126
GeV. \vspace{.15cm}\subsection{Bino LSP and Light
Smuon}\vspace{.35cm} Solutions found have  pure Bino LSP and the
light charginos are pure Winos. A striking feature is that there
are solutions with next to LSP (NLSP) as light smuon which can
generate a significant corrections to the muon g-2 and thus remove
the observed anomaly $a_\mu \sim 10^{-9}$. Moreover a light smuon
provides DM co-annihilation channel to get the acceptable relic
density.
\section{Discussion}
NMSGUT superpotential parameter and SUGRY-NUHM type soft
supersymmetry breaking parameters $\{ m_0,$ $m_{1/2},$ $A_0,$ $B$,
$M^2_{H,\bar H} \}$ along with $\mu $
 specified at the MSSM one loop unification
scale $M_X^0=10^{16.33} $ GeV can fit the fermion mass mixing data
as shown in \cite{ag2,nmsgut}. The parameter  $\{ m_0,$ $m_{1/2},$
$A_0,$  $M^2_{H,\bar H} \}$ are randomly chosen by the search
program while $\mu$
 and $B$ are fixed from electroweak symmetry breaking conditions. Moreover  the
constraints from fermion fitting are combined with the
requirements of unification  as well as electroweak symmetry
breaking conditions. Solution sets presented in \cite{nmsgut} have
large proton decay rates of order of $10^{-27}$ $\mbox{yrs}^{-1}$.
However optimized search with respect to baryon decay, including
GUT scale threshold correction will be discussed in the next
chapter.

Among the superpotential parameters $h_{33}$ (largest of all
elements of Yukawas- $h$, $f$, $g$  ) is crucial for fitting
fermion mass-mixing data, it alone can fit third generation within
$5\%$ error. Next relevant parameter is $g_{23}$. $f_{AB}$  is
irrelevant for charge fermion masses and its small value is
crucial for neutrino masses as it enhances Type I and suppress
Type II seesaw contribution. Fits yield right handed neutrino mass
in $10^8$ - $10^{13}$ GeV range which is compatible with
leptogenesis. Heavy right handed neutrino and small $f_{AB}$
generates neutrino masses of order of meV. Fermion Yukawas obey
$b-\tau$ unification ($|\frac{y_b-y_{\tau}}{y_s -y_{\mu}}|\approx
1 $) noted in
\cite{Bajc:SO(10)fitting,core,msgreb,Grimus:1,Grimus:2} based upon
the ${\bf{10-120}}$-plet FM Higgs system. In most of solutions,
superheavy thresholds raise unification scale $M_X$ closer to
$M_P$. Superheavy spectrum varies from $10^{15}$ GeV to $10^{19}$
GeV.

 In addition to the superpotential couplings, $\tan \beta$ is also a crucial
  parameter for realistic fermion mass generation by the GUT Yukawas. For third generation Yukawa unification ($y_t$ $\approx$ $y_b$ $\approx$ $y_{\tau}$), NMSGUT prefer large $\tan \beta$ ($\sim$ 50). As discussed in \cite{nmsgut} large $\tan \beta$ driven Susy
threshold corrections provides a route for successful fitting of
fermion masses at $M_Z$. Fermion fitting and experiment compatible
MSSM Higgs mass require decoupled/mini-split Susy spectrum
\cite{minisplit1,minisplit2} : $|M|_{H,\bar H}$ $\sim$ $100$ TeV,
heavy CP-odd Higgs $M_A$, large $\mu$ and trilinear coupling
$A_0$, light gaugino ($<$ 1 TeV) with pure Bino as LSP, normal
s-hierarchy with $m_{\tilde f_3}$ $\sim$ 50 TeV and degenerate
first two generations.  With this kind of soft spectrum gaugino
mass deviate from the Susy-GUT ratio 1:2:7 operative at one loop
level. Fits prefer large $A_0$ and $\mu$ parameter  $\sim$ 100 TeV
which suppress flavour changing neutral current processes and
avoid problem with charge and color breaking/unbounded
 from below vacua \cite{kuslangseg}.

Besides realistic B-decay rates, other improvements in the fitting
programs which are part of the thesis are loop corrected Susy
spectrum, inclusion of right handed neutrino thresholds for LFV
estimation and to consider the effect of soft parameter RG running
from $M_{P}$ to $M_X^0$. This will be discussed in Chapter 4, 5
and 6 respectively. \eject
\section*{Appendix: MSSM Tree Level Spectrum\label{ap2}}
The MSSM is an effective theory of NMSGUT. Its particle content is
given in the Table \ref{MSSMp} and corresponding
 superpotential and soft Lagrangian ($\mathcal{L}_{{\scriptsize{\mbox{soft}}}}$) are given by Eq. \eqref{super}  and \eqref{MSSMsoft} respectively. Here we
discuss tree level spectrum of the MSSM.
\subsection*{Sparticle Masses}
Squark and slepton mass term in the Lagrangian is given by \be
\mathcal{L}_{m_{\tilde f}}=-\frac{1}{2} \begin{pmatrix} \tilde
{f_L}^\dag & \tilde {f_R}^\dag \end{pmatrix} M_{\tilde
f}^2\begin{pmatrix} \tilde f_L  \cr \tilde f_R \end{pmatrix}\ee
where $\tilde{f}$ represents $\tilde{u}$, $\tilde{d}$, $\tilde{l}$
and $\tilde{\nu}$. Mass matrices are : \be M_{\tilde
e}^2=\begin{pmatrix}
m_{\tilde{L}}^2+m_e^2-(\frac{1}{2}-S_W^2)M_Z^2 C_{2 \beta} &
\frac{1}{\sqrt{2}}(v_1 A_e^*- \mu Y_e v_2) \cr
 \frac{1}{\sqrt{2}}(v_1 A_e- (\mu Y_e)^* v_2) & m_{\tilde{l}}^2+m_e^2-S_W^2 M_Z^2 C_{2 \beta}\end{pmatrix} \ee
\be M_{\tilde \nu}^2=\begin{pmatrix} m_{\tilde{
L}}^2+\frac{1}{2}M_Z^2 C_{2 \beta} & 0\cr
 0 & 0\end{pmatrix} \ee
\be M_{\tilde u}^2=\begin{pmatrix} m_{\tilde{
Q}}^2+m_u^2-(\frac{1}{2}-\frac{2}{3}S_W^2)M_Z^2 C_{2 \beta} &
\frac{1}{\sqrt{2}}(v_2 A_u^*- \mu Y_u v_1) \cr
 \frac{1}{\sqrt{2}}(v_2 A_u- (\mu Y_u)^* v_1) & m_{\tilde{ u}}^2+m_u^2+\frac{2}{3}S_W^2 M_Z^2 C_{2 \beta}\end{pmatrix} \ee

\be M_{\tilde d}^2=\begin{pmatrix} m_{\tilde{
Q}}^2+m_d^2-(\frac{1}{2}-\frac{1}{3}S_W^2)M_Z^2 C_{2 \beta} &
\frac{1}{\sqrt{2}}(v_1 A_d^*- \mu Y_d v_2) \cr
 \frac{1}{\sqrt{2}}(v_1 A_d- (\mu Y_d)^* v_2) & m_{\tilde{ d}}^2+m_d^2-\frac{1}{3}S_W^2 M_Z^2 C_{2 \beta}\end{pmatrix} \ee
 \[ S_W^2=\sin^2\theta_W \qquad ; \qquad C_{2 \beta}=\cos 2\beta \]
Here $m_{\tilde{ L}}^2$, $m_{\tilde{ Q}}^2$, $m_{\tilde u}^2$,
$m_{\tilde d}^2$ and $m_{\tilde l}^2$ are soft mass parameters.
$A_f$, $Y_f$ and $m_f$ are the soft trilinear couplings, fermion
Yukawa couplings and masses. Tree level sparticle masses are
obtained by diagonalizing above (Hermitian) mass matrices via
unitary transformations :- \be  \tilde{U}_u M_{\tilde u}^2
\tilde{U}^\dag_u=\Lambda_{\tilde u}^2 \quad ; \quad  \tilde{U}_d
M_{\tilde d}^2 \tilde{U}^\dag_d=\Lambda_{\tilde d}^2  \ee \be
\tilde{U}_e M_{\tilde e}^2 \tilde{U}^\dag_e=\Lambda_{\tilde e}^2
\quad ; \quad \tilde{U}_\nu M_{\tilde \nu}^2
\tilde{U}^\dag_\nu=\Lambda_{\tilde \nu}^2  \ee where
$\Lambda_{\tilde u}^2$, $\Lambda_{\tilde d}^2$, $\Lambda_{\tilde
e}^2$ and $\Lambda_{\tilde \nu}^2$ are positive definite mass
square parameter.
\subsection*{Higgs Masses}
Physical Higgs particles of the MSSM are : CP-odd neutral $A$,
CP-even $h$ and $H$, and charged Higgs $H^\pm$. Masses of  these
particles can be computed from $M_A$( and $\tan\beta$) which
itself is determined from $B$ parameter : \be
M_A=B(\tan\beta+\cot\beta) \ee $B$ is calculated using EW symmetry
breaking conditions. \be M^2_{H,h}=\frac{1}{2}(M_A^2+M_Z^2\pm
\sqrt{(M_A^2+M_Z^2)^2-4 M_A^2 M_Z^2 C_{2\beta}^2)}  \ee \be
M^2_{H^\pm}=M_A^2+M_W^2 \ee
\subsection*{Gaugino and Higgsino Masses}
Gauginos and Higgsino mix to form chargino and neutralino
eigenstates. \be \mathcal
L_{{\small{\mbox{chargino}}}}=-\tilde\chi^{-T} \mathcal M_{\tilde
\chi^+}\tilde \chi^+ +h.c.    \ee

 \[  \tilde\chi^+=( -i
\tilde W^+ ,\tilde H_u^+)^T     \quad      \tilde\chi^-=( -i
\tilde W^- ,\tilde H_d^-)^T  \]

 \be \mathcal
L_{{\small{\mbox{neutralino}}}}=-\frac{1}{2}\tilde\chi^{0T}
\mathcal M_{\tilde \chi^0}\tilde \chi^0 +h.c.  \ee
\[  \tilde\chi^0=( -i \tilde B ,
-i\tilde W_3, \tilde H_d,\tilde H_u )^T  \]
 Here $\tilde B$, $\tilde W_3$, $\tilde H_d$ and $\tilde H_u$ are Bino, Wino and Higgs components. Neutralino and chargino mass matrices are obtained from $\mathcal{L}_{{\scriptsize{\mbox{soft}}}}$, $\mathcal{L}_{{\scriptsize{\mbox{int}}}}$ (matter-gauge-Higgs) and superpotential : \vspace{.25cm}
 \be \mathcal M_{\tilde
\chi^+}=\begin{pmatrix} M_2 & \sqrt{2} M_W \sin \beta \cr \sqrt{2}
M_W \cos \beta & \mu \end{pmatrix}\label{chargmass}\ee \be
\mathcal M_{\tilde \chi}=\begin{pmatrix} M_1 &0& -M_Z \cos\beta
S_W  &  M_Z \sin \beta S_W  \cr 0 & M_2 &
 M_Z \cos \beta C_W & -M_Z \sin \beta C_W  \cr  -M_Z \cos\beta S_W  &  M_Z \cos \beta C_W  & 0 & \mu \cr
 M_Z \sin \beta S_W & -M_Z \sin \beta C_W  & -\mu & 0 \end{pmatrix} \label{neutmass}\ee
Matrices $\mathcal M_{\tilde \chi^\pm}$ and $\mathcal M_{\tilde
\chi^0}$ are diagonalized to get masses : \be U_-^\dag \mathcal
M_{\tilde \chi^\pm}  U_+=\Lambda_C \quad ; \quad  N^\dag  \mathcal
M_{\tilde \chi^0} N=\Lambda_N \ee Here $U_-$, $U_+$ and $N$ are
unitary matrices and $\Lambda_C$, $\Lambda_N$ are positive
definite masses.
\newpage\thispagestyle{empty} \mbox{}\newpage
\chapter{Baryon Decay and GUT Scale Threshold Corrections} \label{chgut}
\section{Introduction}
GUTs place quarks and leptons in common irreducible
representations. Quarks can transform into leptons by exchanging
gauge and Higgs leptoquarks, thus GUTs predict baryon violating
processes such as proton decay e.g. (p $\rightarrow \pi^0 e^+$).
However non observation of proton decay has put a stringent lower
limit \cite{superk1} on its life time
\[ \tau(p\rightarrow e^+ \pi^0) > 8.2 \times 10^{33} \mbox{yrs} \]
\[  \tau(p\rightarrow K^+ \bar{\nu}) > 2.3 \times 10^{33} \mbox{yrs}  \]
and this contradicts the simplest  models. Hence one must
investigate more refined models, among which the most appealing
are supersymmetric GUTs. In Susy GUTs B and L are violated by the
exchange of superheavy color triplets. In SO(10) GUTs, B-L is
preserved by all the vertices since it is part of gauge symmetry.
Gauge mediated dimension 6 operator proton decay rate is estimated
as \be \Gamma_p \approx \frac{\alpha^2_{\tiny{\mbox{GUT}}}
m_p^5}{M_G^4}  \ee  Here $M_G$ is mass of superheavy gauge boson.
Even with $M_G \sim 10^{16.25} $ GeV this gives $\tau_p \sim
10^{36}$yrs. Threshold corrections can raise the unification scale
near to the Planck scale so this contribution can be even more
strongly suppressed. R-parity forbids fast d=4 baryon number
violating operators. The remaining contribution is d=5 operators
(involving two fermion and two scalars exchanging triplet
Higgsino). Scalars are converted into fermions via gaugino or
higgsino dressing \cite{barondecay1,barondecay2,barondecay3}. In
Susy GUTs dimension 5 operators thus give leading contribution to
proton decay \cite{Sakai,Weinberg-proton} as these are suppressed
only by $\frac{1}{M_H M_{\tiny{\mbox{Susy}}}}$, where $M_H$ is the
mass of triplet Higgsino. Experimental limits put the stringent
constraint on the model parameters \cite{Murayama}.  In this
chapter we will investigate d=5 operator baryon decay and uncover
a generic and natural mechanism to suppress these.
\section{Dimension 5, Baryon Decay Operators}
SO(10) Yukawa interaction include many superheavy Higgs-fermion
interactions. By using the superpotential equations of motion for
the heavy fields
 (just as we eliminate $W^{\pm},Z$ to get the Fermi effective theory of weak interactions from the SM), we obtain two types
  of  d=5 operator which lead
  to proton decay. The effective
    superpotential has generic form : \bea  W_{eff}^{\Delta B\neq  0} = -{ L}_{ABCD}
({1\over 2}\epsilon { Q}_A { Q}_B { Q}_C { L}_D) -{ R}_{ABCD}
(\epsilon {\bar{ e}}_A {\bar{ u}}_B { \bar{ u}}_C {\bar{ d}}_D)
\eea where the first and second term represent contribution of
$SU(2)_L$ doublets and singlets, therefore are called LLLL and RRRR operator respectively. 
In NMSGUT ${\bf{10}}$, ${\bf{\oot}}$ and ${\bf{120}}$ irreps have
$ t[ 3,1,
    {\pm\frac{2}{3}}] \oplus P[3,3,\pm{\frac{2}{3}}] \oplus
     K[3,1,\pm{\frac{8}{3}}]$ multiplets that
couple to fermion and violate B+L. From PS decomposition of the
relevant superpotential invariants $L_{ABCD}$ and $R_{ABCD}$ are
obtained \cite{ag1,ag2,nmsgut}: \bea  L_{ABCD} &=& {\cal S}_1^{~1}
{\tilde h}_{AB} {\tilde h}_{CD} + {\cal S}_1^{~2} {\tilde h}_{AB}
{\tilde f}_{CD} +
 {\cal S}_2^{~1}  {\tilde f}_{AB} {\tilde h}_{CD} + {\cal S}_2^{~2}  {\tilde f}_{AB} {\tilde
 f}_{CD}
-  {\cal S}_1^{~6}  {\tilde h}_{AB} {\tilde g}_{CD}\nnu &-&
 {\cal S}_2^{~6}  {\tilde f}_{AB} {\tilde g}_{CD}
 +  \sqrt{2}({\cal P}^{-1})_2^{~1} {\tilde g}_{AC}{\tilde f}_{BD}-  ({\cal P}^{-1})_2^{~2} {\tilde g}_{AC}{\tilde g}_{BD}
 \eea
and \bea  R_{ABCD} &=&{\cal S}_1^{~1} {\tilde h}_{AB} {\tilde
h}_{CD}
 - {\cal S}_1^{~2}  {\tilde h}_{AB} {\tilde f}_{CD} -
 {\cal S}_2^{~1}  {\tilde f}_{AB} {\tilde h}_{CD} + {\cal S}_2^{~2}  {\tilde f}_{AB} {\tilde f}_{CD}
 - i{\sqrt 2} {\cal S}_4 ^{~1} {\tilde f}_{AB} {\tilde h}_{CD}
\nonumber \\ &+&i {\sqrt 2} {\cal S}_4 ^{~2} {\tilde f}_{AB}
{\tilde f}_{CD}
 + {\cal S}_6 ^{~1} {\tilde g}_{AB} {\tilde h}_{CD} - i
{\cal S}_7 ^{~1} {\tilde g}_{AB} {\tilde h}_{CD} -  {\cal S}_6
^{~2} {\tilde g}_{AB} {  {\tilde f}}_{CD}+ i   {\cal S}_7^{~2}
{\tilde g}_{AB} { {\tilde f}}_{CD}
\nonumber \\
&+&   i{\cal S}_1 ^{~7} {\tilde h}_{AB} {\tilde g}_{CD} -i  {\cal
S}_2 ^{~7} {{\tilde  f}}_{AB} {\tilde g}_{CD}+ \sqrt{2} {\cal S}_4
^{~7} {{\tilde  f}}_{AB} {\tilde g}_{CD}+  i {\cal S}_6 ^{~7}
{\tilde g}_{AB} {\tilde g}_{CD}  +{\cal S}_7 ^{~7} {\tilde g}_{AB}
{\tilde g}_{CD}\nonumber
\\ &-& \sqrt{2} ({\cal
K}^{-1})_1^{~2} {{\tilde  f}}_{AD}{\tilde g}_{BC}-  ({\cal
K}^{-1})_2^{~2} {\tilde g}_{AD}{\tilde g}_{BC}
 \eea
where ${\cal S}= {\cal T}^{-1} $, ${\cal T} $ is $t[3,1,\pm 2/3]$
multiplets mass matrix and
 \bea {\tilde h}_{AB} = 2 {\sqrt 2} h_{AB}  \quad ; \quad {\tilde f}_{AB} = 4
{\sqrt 2} f_{AB} \quad ; \quad {\tilde g_{AB}} = 4 g_{AB} \eea
where $\alpha$, $\beta$ and $\gamma$ are the colour indices and
$SU(2)$ indices are suppressed. Different Susy GUTs will furnish
different coefficient arrays  $L_{ABCD}$, $R_{ABCD}$ and the task
is to convert this information together with the assumptions for
the soft Susy breaking terms (till the superpartners are
discovered) into predictions for the baryon decay rate into
different channels.
\section{Baryon Decay Rate}
We calculate proton decay rates due to d=5 ($\Delta B=\pm 1$)
operators using formulas of \cite{barondecay1}. As a check we
compare the result calculated by using the formalism of
\cite{barondecay2} separately and verify they are the same. The
calculation of baryon decay rates is done in steps as follows.
\begin{itemize}
\item Firstly one has to renormalize the $2\times3^4=162$
component arrays $L_{ABCD}$ and $R_{ABCD}$ from $M_X$ down to
$M_S$ $\sim$ $M_Z$ using the MSSM RG equations supplemented by the
RGEs for the coefficients $L_{ABCD}$,$R_{ABCD}$. The RGE for
$L_{ABCD}$ \cite{barondecay1} is \vspace{-.5cm}
\begin{eqnarray}
\Dt L_{ABCD} &=&
  \left(
    -8g_3^2 -6g_2^2 -\frac{2}{5}g_1^2
  \right)L_{ABCD}
\nonumber\\&&
 + ( Y_u^T Y_u^\ast+ Y_d^T Y_d^\ast )_{~AA'}L_{A'BCD}
( Y_uT Y_u^\ast+ Y_d^T Y_d\ast )_{~BB'}L_{AB'CD}\nonumber\\&&
   +
     ( Y_u^T Y_u^\ast+ Y_d^T Y_d^\ast )_{~CC'}L_{ABC'D}+
     (  Y_l^T Y_l^\ast ) _{~DD'}L_{ABCD'}
     \end{eqnarray}
It is easy to see that last four terms correspond to
     $SU(2)$ invariant corrections to the three $ Q_L$ and one $ L_L$
     external lines of the LLLL operator by a loop involving Higgs
     exchange  while the first term counts the dressing by gauge
     particle exchange on  an external line. A similar equation
     governs the RRRR evolution \cite{barondecay1}. In the combined
     system of the MSSM, soft Susy, LLLL and RRRR- 447 RGEs must be
     integrated down from the scale $M_X$ to
     $M_Z$. Below that scale one has to treat fermions and bosons
     differently and thus one must pass to a component field
     description (instead of superfield).
     \def\tri #1,#2,#3,#4,#5,#6,#7.{
\begin{picture}(150,150)(0,-75)
\thicklines \put(0,50){\line(1,0){50}}\put(25,50){\vector(1,0){0}}
\put(0,-50){\line(1,0){50}} \put(25,-50){\vector(1,0){0}}
\put(150,50){\line(-1,-1){50}} \put(150,-50){\line(-1,1){50}}
\multiput(50,50)(20,-20){3}{\line(1,-1){10}}
\multiput(50,-50)(20,20){3}{\line(1,1){10}}
\put(50,50){\line(0,-1){100}} \put(50,25){\vector(0,1){0}}
\put(50,-25){\vector(0,-1){0}} \put(78,22){\vector(1,-1){0}}
\put(78,-22){\vector(1,1){0}} \put(125,25){\vector(-1,-1){0}}
\put(125,-25){\vector(-1,1){0}} \put(18,60){$#1$}
\put(18,-70){$#2$} \put(82,32){$#5$} \put(82,-40){$#6$}
\put(135,-25){$#4$} \put(135,17){$#3$} \put(32,-6){$#7$}
\end{picture}}
\begin{figure}
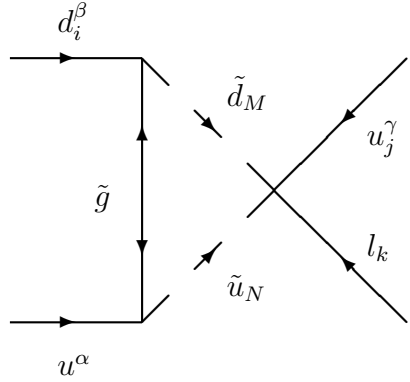

\begin{center}
 \vbox{  \tri
d_i^\beta,u^\alpha,u_j^\gamma,l_k,\tilde d_M, \tilde u_N,\tilde
g.} \caption{Gluino dressing\label{f:1}}
\end{center}
\end{figure}
\item Next the d=4 superpotential is converted into the d=5
effective Lagrangian at $M_Z$ involving 2 scalar and 2 fermion
fields. In order to determine the effective Lagrangian that
governs
     the nucleon decay we must evaluate the dressing diagrams that
     convert the two scalar (A) fields into the corresponding
     fermi fields by exchange of a gluino, chargino or neutralino
     field. For example gluino exchange is governed by the diagrams
     shown in Fig \ref{f:1}. Calculation of a 1-loop
diagram is simplified by assuming that the momenta of the external
fermion lines are negligible compared to superpartner masses. It
is clear that in order to calculate the
     loop we must write the Lagrangian in the mass diagonal basis
     so that the propagators can be easily inserted. Thus we
     not only diagonalize the $3\times3$ Yukawa coupling matrices
     by the usual biunitary transformations.\be
     \hat{f}_{weak}=U^\dagger f_{mass}\quad, \quad \hat{\bar f}_{weak}=V^\dagger
     \bar f_{mass} \ee but also diagonalize the $6\times6$ charged
     sfermion mass squared matrices given in the Appendix of Chapter 2. These are written in the weak
     basis (denoted by hat on $\tilde{F}$)
     \be(\tilde{\hat{f}},{\tilde{\bar{f}}}^\ast)^T=\hat{\tilde{F}}\ee
     or the mass diagonal basis \be
\mathcal {L}=-{\hat{\tilde{F}}}^\dagger
     M^2_{\hat{F}}\hat{\tilde{F}}=-{\tilde{F}}^\dagger{\Lambda}^2_{\tilde{F}}\tilde{F} \quad ; \quad  \tilde{F}=\tilde{U}_{\tilde{F}}\hat{\tilde{F}} \ee
     where ${\Lambda}^2_{\tilde{F}}$ is diagonal positive
     definite. Similarly one must also diagonalize the $4\times4$
     neutralino  and $2\times2$ chargino
     mass matrix by means of
     a symmetric unitary ($N$) and a biunitary ($U^+$, $U^-$) transformations
     respectively. Due to the diagonalization the Yukawa couplings
     in the theory when rewritten in the mass diagonal basis
     become quite complicated.
\item The superpotential
     \be W=\phi_1\phi_2\phi_3\phi_4 \ee will yield 2 fermion, 2 scalar
     terms in the Lagrangian : \be
     \mathcal {L}=-\frac{1}{2}(\psi_1\psi_2)A_3A_4+\mbox{permutations}\ee
     So from $W_{\Delta B=\pm 1}$, one obtain the following d=5
     terms:
     \begin{eqnarray}
  {\cal L}_5 &=& \epsilon_{abc}
 \{
      C(\wt{u}\wt{d}ul_L)^{MNij}
      \wt{u}^{a}_M
      \wt{d}^{b}_N
      ( u_{Li}^{c}l_{Lj} )+  C(\wt{u}\wt{d}ul_R)^{MNij}
      \wt{u}^{a}_M
      \wt{d}^{b}_N
      ( u_{Ri}^{c}l_{Rj} )
\nonumber\\&&
 + C(\wt{u}\wt{l}ud_L)^{MNij}
      \wt{u}^{a}_M
      \wt{l}^{       }_N
      ( u_{Li}^{b}d_{Lj}^{c} )
  +
C(\wt{u}\wt{l}ud_R)^{MNij}
      \wt{u}^{a}_M
      \wt{l}^{       }_N
      ( u_{Ri}^{b}d_{Rj}^{c} )
\nonumber\\&&
 +  C(\wt{d}\wt{\nu}ud_L)^{Mijk}
      \wt{d}^{a}_M
      \wt{\nu}^{     }_i
      ( u_{Lj}^{b}d_{Lk}^{c} )
 \} ~
\label{eq:dim5mass}
\end{eqnarray}
We have mentioned the coefficients corresponding to channel:
proton decays into the charged lepton. Contribution for the other
decay channels can be found in \cite{barondecay1}. Here $a$, $b$
and $c$ are SU(3) fundamental indices. Subscript L and R represent
fermion chirality. The coefficients
$C(\tilde{f}\tilde{f'}f''f''')$ are defined in terms of $L_{ABCD}$
and $R_{ABCD}$ as follows: \vspace{-.5cm}
\begin{eqnarray} &&
C(\wt{u}\wt{d}ul_L)^{MNij}=
  -\sum_{A,B,C,D=1}^3( L^{ABCD} - L^{CBAD} )
  ( \wt{U}_{\tilde{u}}^\dagger )_{A}^{~M}
  ( \wt{U}_{\tilde{d}}^\dagger)_{B}^{~N}
  ( U_u^\dagger )_{C}^{~i}
  ( U_e^\dagger )_{D}^{~j}
~ \nonumber\\&&
 C(\wt{d}\wt{\nu}ud_L)^{Mijk} =
  -\sum_{A,B,C,D=1}^3( L^{ABCD} - L^{CBAD} )
  ( \wt{U}_{\tilde{d}}^\dagger )_{C}^{~M}
  ( \wt{U}_{\tilde{\nu}} ^\dagger )_{D}^{~i}
  ( U_u^\dagger )_{B}^{~j}
  ( U_d^\dagger )_{A}^{~k}
\nonumber\\&& C(\wt{u}\wt{l}ud_L)^{MNij} =
  -\sum_{A,B,C,D=1}^3( L^{ABCD} - L^{CBAD} )
  ( \wt{U}_{\tilde{u}}^\dagger )_{C}^{~M}
  ( \wt{U}_{\tilde{l}}^\dagger)_{D}^{~N}
  ( U_u^\dagger )_{A}^{~i}
  ( U_d^\dagger )_{B}^{~k}
\nonumber\\&& C(\wt{u}\wt{d}ul_R)^{MNij} =
  -2\sum_{A,B,C,D=1}^3 R^{ABCD}
  ( \wt{U}_{\tilde{u}}^\dagger )_{C+3}^{~M}
  ( \wt{U}_{\tilde{d}}^\dagger)_{D+3}^{~N}
  ( V_u^\dagger )_{B}^{~i}
  ( V_e^\dagger )_{A}^{~j}
 \nonumber \\&&
C(\wt{u}\wt{l}ud_R)^{MNij} =
  -2\sum_{A,B,C,D=1}^3 R^{ABCD}
  ( \wt{U}_{\tilde{u}}^\dagger )_{B+3}^{~M}
  ( \wt{U}_{\tilde{l}}^\dagger)_{A+3}^{~N}
  ( V_u^\dagger )_{C}^{~i}
  ( V_d^\dagger )_{D}^{~j}~~~~~~~~
\end{eqnarray}
Here the indices M, N, i, j, k run from 1 to 3 and
$\tilde{U}_{\tilde{f}}$, $U_f$, $V_f$ are the unitary matrices
which diagonalize scalars and fermions respectively.\vspace{.2cm}
\item After redefining the fields to  diagonalize the mass
matrices, we can write the interaction Lagrangian in mass basis.
The quark (lepton)-squark (slepton)- gaugino/higgsino (gluino,
chargino and neutralino) interaction terms are given by:
\vspace{.3cm}\be {\cal L}_{\rm gauge-Yukawa} =
  {\cal L}_{\rm int}(\wt{g})
+ {\cal L}_{\rm int}(\chi^\pm) + {\cal L}_{\rm int}(\chi^0) \ee
\bea
  {\cal L}_{\rm int}(\wt{g}) &=&
  -i\sqrt{2} g_3 \wt{d}^{* I} \ol{\wt{g}} \left[
    \left( \Gamma_{gL}^{(d)} \right)_{I}^{j} \PL
  + \left( \Gamma_{gR}^{(d)} \right)_{I}^{j} \PR
  \right] d_j
\nonumber\\&&
  -i\sqrt{2} g_3 \wt{u}^{* I} \ol{\wt{g}} \left[
    \left( \Gamma_{gL}^{(u)} \right)_{I}^{j} \PL
  + \left( \Gamma_{gR}^{(u)} \right)_{I}^{j} \PR
  \right] u_j  + \hc \eea

\begin{eqnarray}{\cal L}_{\rm int}(\chi^\pm) &=& \phantom{+}
  g_2 \ol{\chi}^-_\alpha \left[
    \left( \Gamma_{CL}^{(d)} \right)_{I}^{\alpha j} \PL
  + \left( \Gamma_{CR}^{(d)} \right)_{I}^{\alpha j} \PR
  \right] d_j \wt{u}^{* I}
 \nonumber\\&&+
  g_2 \ol{\chi}^+_\alpha \left[
    \left( \Gamma_{CL}^{(u)} \right)_{I}^{\alpha j} \PL
  + \left( \Gamma_{CR}^{(u)} \right)_{I}^{\alpha j} \PR
  \right] u_j \wt{d}^{* I}
\nonumber\\&& +
  g_2 \ol{\chi}^-_\alpha \left[
    \left( \Gamma_{CL}^{(l)} \right)_{i}^{\alpha j} \PL
  + \left( \Gamma_{CR}^{(l)} \right)_{i}^{\alpha j} \PR
  \right] l_j \wt{\nu}^{* i}
 \nonumber\\&&+
  g_2 \ol{\chi}^+_\alpha
    \left( \Gamma_{CL}^{(\nu)} \right)_{I}^{\alpha j} \PL
  \nu_j \wt{l}^{* I} + \hc ~
\end{eqnarray}
\begin{eqnarray}
{\cal L}_{\rm int}(\chi^0) &=& \phantom{+}
  g_2 \ol{\chi}^0_{\ol{\alpha}} \left[
    \left( \Gamma_{NL}^{(d)} \right)_{I}^{\ol{\alpha} j} \PL
  + \left( \Gamma_{NR}^{(d)} \right)_{I}^{\ol{\alpha} j} \PR
  \right] d_j \wt{d}^{* I}
 \nonumber\\&&+
  g_2 \ol{\chi}^0_{\ol{\alpha}} \left[
    \left( \Gamma_{NL}^{(u)} \right)_{I}^{\ol{\alpha} j} \PL
  + \left( \Gamma_{NR}^{(u)} \right)_{I}^{\ol{\alpha} j} \PR
  \right] u_j \wt{u}^{* I}
\nonumber\\&& +
  g_2 \ol{\chi}^0_{\ol{\alpha}} \left[
    \left( \Gamma_{NL}^{(l)} \right)_{I}^{\ol{\alpha} j} \PL
  + \left( \Gamma_{NR}^{(l)} \right)_{I}^{\ol{\alpha} j} \PR
  \right] l_j \wt{l}^{* I}
\nonumber\\&& +
  g_2 \ol{\chi}^0_{\ol{\alpha}}
    \left( \Gamma_{NL}^{(\nu)} \right)_{i}^{\ol{\alpha} j} \PL
  \nu_j \wt{\nu}^{* i} + \hc ~
\end{eqnarray}
where $P_{L/R} = \frac{1}{2}(1\mp\gamma_5)$, $g_2$ and $g_3$ are
gauge couplings of SU(2) and SU(3) respectively, $I = 1, 2, \cdots
6$- represents squarks and charged sleptons, $i, j, k = 1, 2, 3$
refer to fermions and sneutrinos, $\alpha$ ($= 1, 2$) and
$\ol{\alpha}$ (= $1, 2, 3, 4$) denote chargino and  neutralino.
Mixing factors ($\Gamma$s) involve unitary matrices
$\wt{U}_{\tilde{f}}$, ${U}_f$, $V_f$, $U_\pm$ and $N$  which
diagonalize scalars, fermion (bi-unitary transformations),
chargino and neutralino mass matrices.
\begin{eqnarray}
    \left( \Gamma_{gL}^{(d)} \right)_{I}^{j} &=&
      \sum_{k=1}^3 \left( \wt{U}_{\tilde{d}}   \right)_I^{~k}
                   \left( U_d^\dagger \right)_k^{~j}
\quad ; \quad
    \left( \Gamma_{gR}^{(d)} \right)_{I}^{j} =\sum_{k=1}^3
      \left( \wt{U}_{\tilde{d}}   \right)_I^{~k+3}\left(V_{\bar d}^T \right)_k^{~j}
\nonumber\\
    \left( \Gamma_{gL}^{(u)} \right)_{I}^{j} &=& \sum_{k=1}^3
      \left( \wt{U}_{\tilde{u}}   \right)_I^{~k}\left (U_u^\dagger \right)_k^{~j}
\quad ; \quad
    \left( \Gamma_{gR}^{(u)} \right)_{I}^{j} = \sum_{k=1}^3
      \left( \wt{U}_{\tilde{u}}   \right)_I^{~k+3}\left(V_{\bar u}^T \right)_k^{~j} ~
\end{eqnarray}
\begin{eqnarray}
    \left( \Gamma_{CL}^{(d)} \right)_{I}^{\alpha j} &=&
      \sum_{k=1}^3 \biggl\{-
        \left( \wt{U}_{\tilde{u}} \right)_I^{~k}
        \left( U_+ \right)_1^{~\alpha}
        \left( U_d^\dagger \right)_k^{~j}
\nonumber\\&&+ \frac{1}{g_{2}} \sum_{l=1}^3\left(
\wt{U}_{\tilde{u}} \right)_I^{~k+3}
 \left( U_+ \right)_2^{~\alpha}
        \left(Y_u \right)_k^{~l}
        \left(U_d^\dagger \right)_l^{~j}
        \biggr\}  ~
\nonumber\\
    \left( \Gamma_{CR}^{(d)} \right)_{I}^{\alpha j} &=&
    \frac{1}{g_{2}} \sum_{k,l=1}^3
        \left( \wt{U}_{\tilde{u}} \right)_I^{~k}
        \left( Y_d^\ast \right)_k^{~l}
        \left( V_d \right)_l^{~j}
\left( U_- \right)_2^{~\alpha}  ~
\nonumber\\
    \left( \Gamma_{CL}^{(u)} \right)_{I}^{\alpha j} &=&
    \sum_{l=1}^3 \biggl\{-
        \left( \wt{U}_{\tilde{d}} \right)_I^{~l}
        \left( U_u^\dagger \right)_l^{~j}
        \left( U_-^\dagger \right)_1^{~\alpha}
\nonumber\\&&+\frac{1}{g_{2}} \sum_{k=1}^3
        \left( \wt{U}_{\tilde{d}} \right)_I^{~k+3}
\left( Y_d \right)_k^{~l} \left(U_u^\dagger \right)_l^{~j}
        \left( U_-^\dagger \right)_2^{~\alpha} \biggr\}  ~
\nonumber\\
    \left( \Gamma_{CR}^{(u)} \right)_{I}^{\alpha j} &=&-\frac{1}{g_{2}} \sum_{k,l=1}^3
        \left( \wt{U}_{\tilde{d}} \right)_I^{~k}
\left( Y_u \right)_k^{~l} \left( V_u \right)_l^{~j}
        \left( U_+^\dagger \right)_2^{~\alpha}    ~
\nonumber\end{eqnarray}
\begin{eqnarray}
    \left( \Gamma_{CL}^{(l)} \right)_{I}^{\alpha j} &=&\sum_{k=1}^3
     \left( \wt{U}_{\tilde{\nu}} \right)_I^{~k}
     \left( U_l \right)_k^{~j}
      \left( U_+^\dagger \right)_1^{~\alpha}
\nonumber\\
    \left( \Gamma_{CR}^{(l)} \right)_{I}^{\alpha j} &=&\frac{1}{g_{2}}\sum_{k,l=1}^3
\left( \wt{U}_{\tilde{\nu}} \right)_I^{~k}
 \left( Y_{e} \right)_k^{~l}
      \left( V_e \right)_l^{~j}
      \left( U_- \right)_2^{~\alpha} ~
\end{eqnarray}
\begin{eqnarray}
\left( \Gamma_{NL}^{(d)} \right)_{M}^{\eta B} &=&
      \sum_{A=1}^3 \biggl\{\frac{1}{\sqrt{2}}
        \left( \wt{U}_{\tilde{d}} \right)_M^{~A}
        \left( U_d^{\dagger} \right)_A^{~B}
       \bigg(\left(N^\dagger
        \right)_2^{~\eta}-\frac{g_1}{3\sqrt{2}g_{2}}\left(N^\dagger
        \right)_1^{~\eta}\bigg)
\nonumber\\&&
      -  \frac{1}{g_{2}} \sum_{C=1}^3\left( \wt{U}_{\tilde{d}} \right)_M^{~A+3}
\left(Y_{d} \right)_A^{~C}
        \left(U_d^\dagger \right)_C^{~B}
         \left(N^\dagger \right)_3^{~\eta}
      \biggr\}  ~
 \nonumber     \\
      \left( \Gamma_{NR}^{(d)} \right)_{M}^{\eta B} &=&
      \sum_{A=1}^3 \biggl\{-\frac{\sqrt{2} g_1}{3 g_{2} }
        \left( \wt{U}_{\tilde{d}} \right)_M^{~A+3}
        \left( V_d \right)_A^{~B}
       \left(N^\dagger \right)_1^{~\eta}
 \nonumber\\&&-  \frac{1}{g_{2}}\sum_{C=1}^3\left( \wt{U}_{\tilde{d}} \right)_M^{~C}
\left(Y_{d} \right)_C^{~A}
        \left(V_d \right)_A^{~B}
        \left(N^\dagger \right)_3^{~\eta}
      \biggr\}  ~
 \nonumber     \\
      \left( \Gamma_{NL}^{(u)} \right)_{M}^{\eta B} &=&
      \sum_{A=1}^3 \biggl\{\frac{1}{\sqrt{2}}
        \left( \wt{U}_{\tilde{u}} \right)_M^{~A}
        \left( U_u^{\dagger} \right)_A^{~B}
       \bigg(-\left(N^\dagger
        \right)_2^{~\eta}-\frac{g_1}{3\sqrt{2} g_{2}}\left(N^\dagger
        \right)_1^{~\eta}\bigg)
\nonumber\\&&
      - \sum_{C=1}^3\left( \wt{U}_{\tilde{u}} \right)_M^{~A+3}
\left(Y_{u} \right)_A^{~C}
        \left(U_u^\dagger \right)_C^{~B}
\left(N^\dagger \right)_4^{~\eta}
      \biggr\}  ~
  \nonumber    \\
      \left( \Gamma_{NR}^{(u)} \right)_{M}^{\eta B} &=&
      \sum_{A=1}^3 \biggl\{\frac{2 \sqrt{2} g_1}{3 g_{2}}
        \left( \wt{U}_{\tilde{u}} \right)_M^{~A+3}
        \left( V_u \right)_A^{~B}
       \left(N^\dagger \right)_1^{~\eta}
\nonumber\\&&- \sum_{C=1}^3\left( \wt{U}_{\tilde{u}}
\right)_M^{~A} \left(Y_{u}^{\dagger} \right)_A^{~C}
        \left(V_u \right)_C^{~B}
 \left(N^\dagger \right)_4^{~\eta}
      \biggr\}
 \nonumber     \\
      \left( \Gamma_{NL}^{(e)} \right)_{M}^{\eta B} &=&
      \sum_{A=1}^3 \biggl\{\frac{1}{\sqrt{2}}
        \left( \wt{U}_{\tilde{l}} \right)_M^{~A}
        \left( U_e^{\dagger} \right)_A^{~B}
       \bigg(\left(N^\dagger
        \right)_2^{~\eta}+\frac{g_1}{\sqrt{2} g_{2}}\left(N^\dagger
        \right)_1^{~\eta}\bigg)
\nonumber\\&&
      -\frac{1}{g_{2}}  \sum_{C=1}^3\left( \wt{U}_{\tilde{l}} \right)_M^{~A+3}
\left(Y_{d} \right)_A^{~C}
        \left(U_e^\dagger \right)_C^{~B}
          \left(N^\dagger \right)_3^{~\eta}
      \biggr\}  ~
   \nonumber   \\
 \left( \Gamma_{NR}^{(e)} \right)_{M}^{\eta B} &=&
      -\sum_{A=1}^3 \biggl\{\frac{ g_1}{\sqrt{2} g_{2}}
        \left( \wt{U}_{\tilde{l}} \right)_M^{~A+3}
        \left( V_e \right)_A^{~B}
       \left(N^\dagger \right)_1^{~\eta}
\nonumber\\&&+\frac{1}{g_{2}} \sum_{C=1}^3\left(
\wt{U}_{\tilde{l}} \right)_M^{~C} \left(Y_{e}^{\dagger}
\right)_C^{~A} \left(V_e \right)_A^{~B}
 \left(N \right)_3^{~\eta}
      \biggr\}  ~~~~~ \qquad
\end{eqnarray}
The
  gluino mass matrix does not need diagonalization so one can
  evaluate the diagram in Fig \ref{f:1} to give a contribution to
  $\mathcal {L}_{eff}$
\bea
\mathcal{L}_{eff}&=&\frac{4}{3i}\frac{g_3^2}{M_{\wt{g}}}\frac{1}{16\pi^2}(
-2 i
L_{mljk})({\wt{U}}_{\tilde{u}}^\dagger)_{mN}({\wt{U}}_{\tilde{d}}^\dagger)_{lM}
\left( \Gamma_{gL}^{(u)} \right)_{N}^{1}
  \left( \Gamma_{gL}^{(d)} \right)_{M}^{i}
  \nonumber \\&& \times H( {\wt{u}}_N, {\wt{d}}_M ){\epsilon_{abc}}(u_L^{a}d_{Li}^{b})(u_j^{c}l_{Lk})\eea
Here the function $ H( {\wt{u}}_N, {\wt{d}}_M )$ arises from the
standard loop integration, having form :
\begin{eqnarray}
  H(x,y) &=&
  \frac{1}{x-y}
  \left(
      \frac{x\log x}{x-1} - \frac{y\log y}{y-1}
  \right) ~
\end{eqnarray}
and the arguments of loop function $H$ are Susy particle mass
squared ratios:
\begin{eqnarray}
  \tilde{d}_{M} &=&
  \frac{ m_{\wt{d}_M}^2 }{ M_{\wt{g}}^2 } ~
~~~
  \tilde{u}_{M} ~=~
  \frac{ m_{\wt{u}_M}^2 }{ M_{\wt{g}}^2 } ~\end{eqnarray}

\item After  calculating one-loop (gluino, neutralino and
chargino) dressing diagrams, effective Lagrangian containing
four-fermi interaction terms relevant to the proton decay into
charged lepton channels is given by:
\begin{eqnarray}
  {\cal L}_{\Delta {B}\neq 0} &=& \frac{1}{(4\pi)^2}
  \epsilon_{abc}
  \{
      \wt{C}_{LL}(udul)^{ik}
      (u_L^{a}d_{Li}^{b})(u_L^{c}l_{Lk})
   +  \wt{C}_{RL}(udul)^{ik}
      (u_R^{a}d_{Ri}^{b})(u_L^{c}l_{Lk})
  \nonumber\\&&
  +  \wt{C}_{LR}(udul)^{ik}
      (u_L^{a}d_{Li}^{b})(u_R^{c}l_{Rk})
   +  \wt{C}_{RR}(udul)^{ik}
      (u_R^{a}d_{Ri}^{b})(u_R^{c}l_{Rk})
  \} ~~~~~
\end{eqnarray}
$\wt{C}$ coefficients can be found in \cite{barondecay1}. \item
Finally matrix elements of the four Fermi operators involving
quark and lepton fields must be evaluated between the baryon and
meson initial and final states to obtain the amplitude for baryon
decay in any channel (e.g. p$\longrightarrow e^+ \pi^0 $). Chiral
Lagrangian technique \cite{Chiraltech} is used to convert the
effective quark Lagrangian to the effective hadronic Lagrangian.
Then partial decay widths of the nucleon are given as
\begin{eqnarray}
  \Gamma( B_i \rightarrow M_j l_k ) &=&
  \frac{m_i}{32\pi}\left( 1 - \frac{m_j^2}{m_i^2} \right)^2
  \frac{1}{f_\pi^2}(A^2_{Long})
  \left(
    \left| A_L^{ijk} \right|^2
  + \left| A_R^{ijk} \right|^2
  \right)
~ \label{eq:pdecayrate}
\end{eqnarray}
where $m_i$ and $m_j$ are the masses of baryon and meson
respectively. ${f_\pi}$ is the pion decay constant having value
139 MeV. A factor of $A_{Long}$ $\approx$ .22 is used to take into
account the renormalization from $M_Z$ to 1 GeV. The explicit
expressions for $A_{L,R}^{ijk}$ (defined in terms of $\tilde{C}$
coefficients) can be found in \cite{barondecay1}
\end{itemize}
As shown in \cite{nmsgut} using the tree level Yukawas successful
fitting of fermion mass mixing data is obtained in the NMSGUT but
proton life time is 6-7 order of magnitude smaller than the
experimental limit. In the literature particular textures of
Yukawa couplings and discrete symmetries are considered to
suppress fast B-decay rates \cite{moh120}. In the next section we
will discuss a generic mechanism to suppress fast B-decay rates in
Susy-GUTs.
\section{GUT Scale Threshold Corrections}
We have computed one-loop GUT scale threshold correction to a
Yukawa coupling of matter field due to heavy fields running in
self energy loops on  lines leading into the Yukawa vertex when
the external light Higgs comes from any of the 6 possible
components \cite{nmsgut} using technique of \cite{wright}. These
threshold corrections are very significant due to the large Higgs
representation used and also play a crucial role in obtaining
parameter sets compatible with constraints on B violation.
\subsection{Formalism}\vspace{.2cm}
In supersymmetric theories, non-renormalization theorem
\cite{nonrenorth} implies that superpotential  couplings are
modified only by wave function renormalization. We have calculated
the large number of the NMSGUT Yukawa vertices that couple light
fermions and Higgs field. To calculate corrections we need to move
into the basis  where mass matrices of heavy fields are diagonal.
We can redefine heavy field $\Phi$ to diagonalize the mass term in
the superpotential \vspace{.3cm}
 \bea{\overline{\Phi }}= U^{\Phi}{\overline{\Phi' }} \quad ;
 \qquad {\Phi } = V^{\Phi}\Phi'\quad \Rightarrow \quad
 {\overline{\Phi}}^T M \Phi ={\overline{\Phi'}}^T M_{{\scriptsize{\mbox{Diag}}}}
 \Phi'\eea
 \begin{figure}[!ht]
\centering
\includegraphics[scale=0.3]{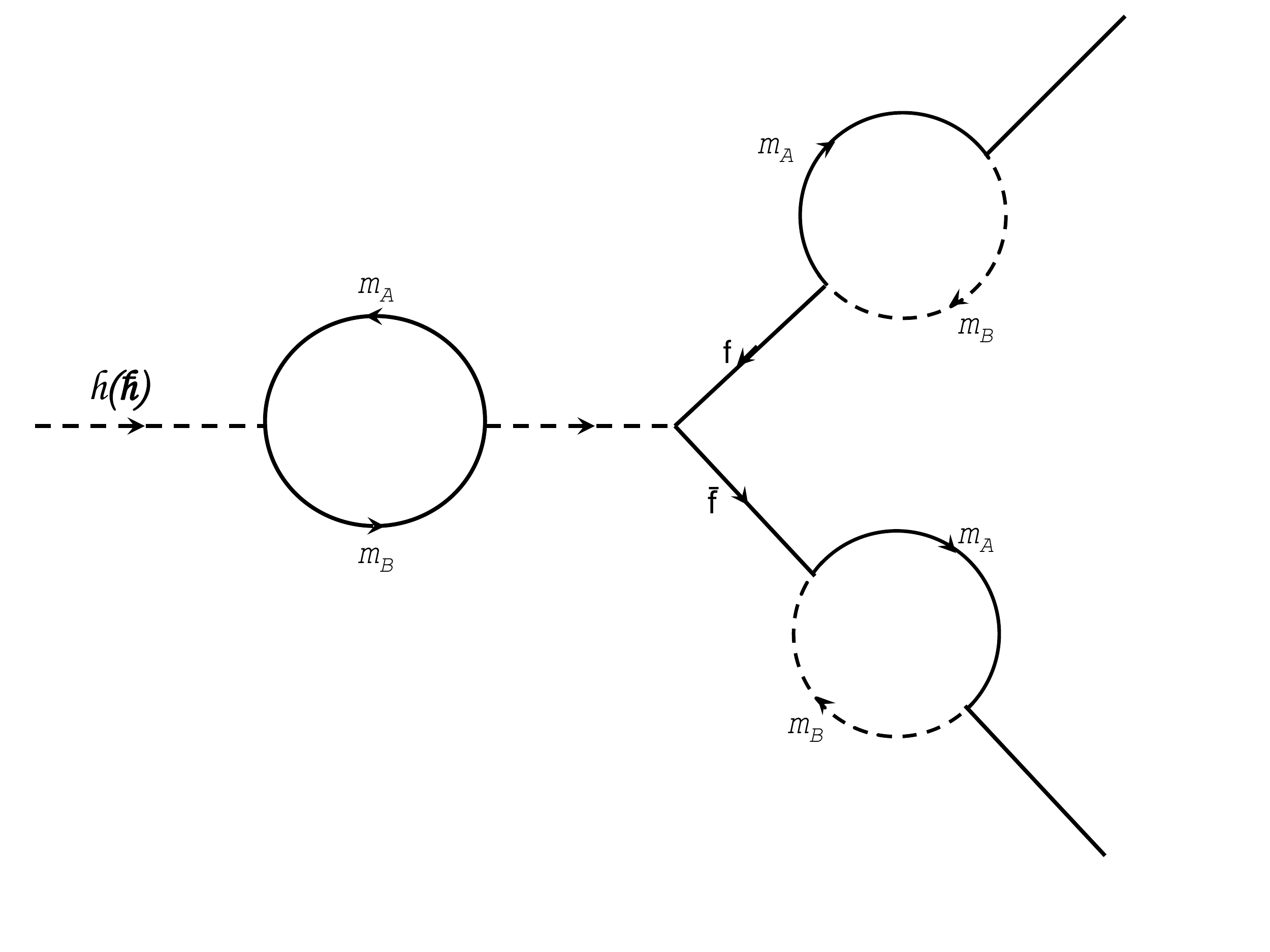}
\caption{Loop corrections to fermion, antifermion and Higgs
line}\label{loops}
\end{figure}
Matter Yukawa vertices : \bea {\cal L} = [f_c^T Y_f f H_f]_F +
h.c. +....\eea As shown in Fig.\ref{loops} heavy superfields can
circulate on any of the three chiral superfields which give  wave
function renormalization in the kinetic terms : \bea {\cal
L}=\bigg[\sum_{A,B}( {\bar f}_A^\dagger (Z_{\bar f})_A^B {\bar
f}_B +{f}_A^\dagger (Z_{f})_A^B {f}_B ) + H^\dagger Z_H H + {\ovl
H}^\dagger Z_{\ovl H} {\ovl H}\bigg]_D +..\eea Here A, B=1, 2, 3
are the generation indices and $H$ and $\bar H $ are the MSSM
light Higgs doublets. Light Higgs fields are the combinations of
all the heavy Higgs $h_i$, 1=1...6 fields of GUT \bea H=\sum_i
\alpha_i^* h_i \qquad ;\qquad
 \ovl H=\sum_i \bar \alpha_i^* \bar h_i \eea
Here $\alpha_i$ and $\bar \alpha_i$ are the Higgs fractions which
describe the contribution of different Higgs fields to light Higgs
and are first columns of the unitary matrices which diagonalize
the Higgs mass matrix. One needs to define a new basis to write
kinetic terms of light matter and Higgs fields in canonical form
as :
\[ f = U_{Z_f} \Lambda^{-\frac{1}{2}}_{Z_f} \tilde f =\tilde
U_{Z_f} \tilde f\qquad ; \qquad \bar f= U_{Z_{\bar f}}
\Lambda^{-\frac{1}{2}}_{Z_{\bar f}} \tilde{\bar f}=\tilde
U_{Z_{\bar {f}}} \tilde {\bar f}\] \be H=\frac{\widetilde
H}{\sqrt{Z_H}}\qquad ; \qquad \ovl H=\frac{\widetilde {\ovl
H}}{\sqrt{Z_{\ovl H}}}\label{f2ftilde}\ee Here $U_{Z_f}$,
$\bar{U}_{Z_{\bar f}}$  are the unitary matrices that diagonalize
$Z_{f, \bar f}$ to positive definite form $\Lambda_{f, \bar f}$.
Then \vspace{.2cm}  \bea {\cal L}&=&\bigg[\sum_A( {\tilde{\bar
f}}_A^\dagger \tilde{\bar f}_A + {\tilde{f}}_A^\dagger \tilde{f}_A
) + \widetilde{H}^\dagger \widetilde{H} + \widetilde{{\ovl
H}}^{\dagger} \widetilde{{\ovl H}}\bigg]_D + \big[\tilde{\bar f}
^T \tilde{Y}_{f} \tilde{f} \widetilde{H}_ {f}\big]_F + h.c. +..
\eea \vspace{.2cm} As a result the MSSM Yukawa couplings in terms
of the tree level (SO(10) determined) Yukawas change
as\vspace{.2cm}  \be\tilde Y_f= \Lambda_{Z_{\bar
f}}^{-\frac{1}{2}} U_{Z_{\bar f}}^T {\frac{Y_f}{\sqrt{Z_{H_f}}}}
U_{Z_f} \Lambda_{Z_f}^{-\frac{1}{2}} = \tilde{U}_{Z_{\bar f}}^T
{\frac{Y_f}{\sqrt{Z_{H_f}}}} \tilde{U}_{Z_f} \label{Ynutilde}\ee
\vspace{.2cm}Generic form of correction factor for any chiral
field $\Phi_i$ is ($Z= 1 -{\cal K}$) :\vspace{.2cm} \bea{\cal
K}_i^j=- {\frac{g_{10}^2}{8 \pi^2}} \sum_\alpha {Q^\alpha_{ik}}^*
{Q^\alpha_{kj}} F(m_\alpha,m_k) +{\frac{1}{32 \pi^2}}\sum_{kl}
Y_{ikl} Y_{jkl}^* F(m_k,m_l) \eea \vspace{.2cm} Here first term is
the contribution of coupling of $\Phi_i$ to gauge field
($A_{\mu}$) in ${\cal L}= g_{10} \, Q^\alpha_{ik} \psi^\dagger_i
{\gamma^\mu A_\mu}^\alpha \psi_k$ ($g_{10}$ is SO(10) gauge
coupling) and second term is Yukawa contribution
($W=\frac{1}{6}Y_{ijk}$$\Phi_i\Phi_j\Phi_k$). $F$ is symmetric
Passarino-Veltman loop function. When both the fields running in
the loop are heavy
 fields then $F(m_1,m_2)$
 has the form
\bea F_{12}(M_A,M_B,Q)={1\over {(M_A^2- M_B^2)}}( M_A^2\ln
{M_A^2\over Q^2} -M_B^2\ln    {M_B^2\over Q^2} )- 1 \eea which
reduces to just \bea F_{11}(M_A,Q)=F_{12}(M_A,0,Q)=  \ln
{M_A^2\over Q^2} - 1 \eea \vspace{.2cm}  when one field is light
($M_B\rightarrow 0)$. One should avoid the sum over light index
when both the fields running in the loop are Higgs fields.
\subsection{Explicit Form of Correction Factors}
In the NMSGUT, right handed neutrino Majorana masses are 3-4 order
of magnitude smaller than the GUT scale. Therefore while
calculating GUT scale threshold correction to the Yukawa coupling
we treat right handed neutrino as light particle like other SM
fermions. The calculation for the corrections to the light Higgs
doublet lines  $H,{\overline H}$ is  much more complicated than
the matter lines  since these are mixtures of pairs of doublets
from the $\mathbf{10,120}$(2 pairs),$\mathbf{ \oot,126,210}$
SO(10) Higgs multiplets: \vspace{.2cm} \be  H= (V^H)^\dagger h
\qquad ;\qquad {\overline H}= (U^H)^\dagger {\overline
h}\vspace{.3cm} \ee Here $V^H$ and $U^H$ are the unitary matrices
which diagonalize Higgs mass matrix ($\mathcal{H}$). The couplings
of the GUT field doublets $h_a,{\bar h}_a, =1,2...6 $  to various
\emph{pairs} of the 26 different MSSM irrep-types (labelled
conveniently by the letters of the  alphabet : see
\cite{ag2,nmsgut}) that occur in this theory is worked out using
the technology \cite{ag1} of SO(10) decomposition via the PS
group. There are again precisely 26 different combinations of GUT
multiplets (labelled by the letter pairs for irreps) which can
combine to give operators that can form singlets with the MSSM $H
[1,2,1]$ (their conjugates gave singlets with ${\overline
H}[1,2,-1]$). Then we get\bea (16 \pi^2){\cal K}_{H} &=&
 3K_{J\bar D}+8K_{R\bar C}+ 9K_{X \bar P}+ K_{VF}+3K_{E\bar J}
 + 9K_{P\bar E} + 6K_{B\bar M}+3K_{X\bar T}
\nonumber\\&& + 3K_{D\bar I}
 +24K_{Q\bar C}  +3K_{T\bar E}
 + 6K_{Y\bar L}+ 18K_{W\bar B}
 + 8K_{C\bar Z}
+ 9K_{E\bar U}\nonumber\\&&
 + 9K_{U\bar D}
  + 3K_{HO}
 + K_{\bar V\bar A}
 + 3K_{K \bar X}
  + K_{H\bar F} + 6K_{N\bar Y}
+ 18K_{Y\bar W}\nonumber\\&&
  + 3K_{V\bar O}
+ 6K_{L\bar B}
 + 3K_{S\bar H}
+ K_{G\bar H}\label{hline}\eea
 To
illustrate the correction factor from the $J\bar{D}$  channel on
Higgs line is given by: \vspace{.25cm} \bea
 K_{J\bar D}
& = & \sum_ {a =1}^{\mbox{d(J)}}\sum_ {a' =1}^{\mbox{d(D)}}
\biggr| \biggr(\gamma V^J_{2a} U^D_{1a'} -\frac {\gamma}{\sqrt
{2}}
  V^J_{3a}U^D_{1a'} +
 \bar\gamma V^J_{2a} U^D_{2a'} +
 \frac { \bar\gamma} {\sqrt {2}} V^J_{3a}U^D_{2a'} - \frac {ik}{\sqrt {2}} V^J_{3a} U^D_{3a'}\biggr) V^H_{11}
 \nonumber \\ && + \biggr( \frac {2\eta } {\sqrt {3}}V^J_{2a} U^D_{1a'} - \sqrt {6}\eta  V^J_{3a} U^D_{1a'} -
  \frac {2 i\bar\zeta } {\sqrt {3}}V^J_{2a} U^D_{3a'} + \sqrt{\frac
{3} {2}}i\bar\zeta V^J_{3a} U^D_{3a'}\biggr) V^H_{21} +
\biggr(\frac{-i}{\sqrt{6}} \zeta V^J_{3a} U^D_{3a'}\nonumber
 \eea \vspace{.4cm}\bea &&-\frac {2i\zeta} {\sqrt{3}}V^J_{2a}U^D_{3a'} +
  \frac{2\eta}{\sqrt{3}} V^J_{2a} U^D_{2a'} - \sqrt{\frac{2}{3}}\eta V^J_{3a} U^D_{2a'}\biggr) V^H_{31} -\biggr(\frac{i \rho}{3} V^J_{5a} U^D_{3a'} +
  4\eta V^J_{1a} U^D_{1a'} \nonumber\\&&+
  2 i\bar\zeta V^J_{1a} U^D_{3a'} +
  2\bar\zeta V^J_{5a} U^D_{2a'}\biggr) V^H_{41} +\biggr(\frac{i \rho}{3\sqrt{2}} V^J_{3a} U^D_{3a'} -
  \frac{\bar\zeta}{ \sqrt{2}} V^J_{3a} U^D_{2a'} -
  \frac {\zeta}{\sqrt {2}} V^J_{3a} U^D_{1a'}\biggr) V^H_{51} \nonumber \\ && +\biggr( \frac{2i\zeta}{\sqrt{3}} V^J_{2a} U^D_{1a'}-\sqrt{\frac{3}{2}}i\zeta V^J_{3a} U^D_{1a'}
    +
    \frac{i\bar\zeta}{\sqrt{6}}V^J_{3a} U^D_{2a'}+
   \frac{2i\bar\zeta}{ \sqrt{3}} V^J_{2a} U^D_{2a'} +
   \frac {\rho}{3\sqrt{3}}V^J_{2a} U^D_{3a'} \nonumber \\ &&- \frac { \sqrt{2}\rho}{3\sqrt{3}} V^J_{3a} U^D_{3a'}\biggr) V^H_{61} \biggr| ^2 F_{12}(m^J_a, m^D_{a'},
   Q) \nonumber \\ &&-2{g_{10}^2} \biggr|\frac{-2i}{\sqrt{3}}\biggr(V^{D*}_{1a'}V^H_{21}+V^{D*}_{2a'}V^H_{31}+V^{D*}_{3a'}V^H_{61}\biggr)\biggr|^2 F_{12}(m_{\lambda_J}, m^D_{a'},
   Q)\eea
It has both Yukawa (first five lines) and gauge (last line)
contribution. V and U (with subscript J, D or H ) are the
 unitary matrices which diagonalize the mass matrix of the respective multiplet. Indices $a$ and $a'$ represent the multiplicity
  of J and D multiplet respectively. In the Appendix we give
explicit formulae for the Higgs correction factor\footnote{These
calculations were done in collaboration with Prof. C. S. Aulakh
and Ila Garg}. The corrections for the fermion lines can be found
in \cite{ilathesis,aulakhgargkhosa}.
\section{Threshold Effects on $\Gamma_{d=5}^{\Delta B\neq 0}$ }
As discussed  SO(10) Yukawa couplings ($h$, $f$, $g$) and heavy
masses $M_i$  determine both fermion masses and
    coefficient of d=5 baryon decay operator
      \[   L_{ABCD}(\{h,g,f\}_{AB}, M_i),R_{ABCD}(\{h,g,f\}_{AB},M_i)
         \]
Canonical kinetic terms after including wavefunction
renormalization factor require transformation of fields to the
tilde basis  (Eq. \ref{f2ftilde}) using bi-unitary transformation.
Then fermion Yukawas ($\tilde Y_f$) are diagonalized to mass basis
(denoted by primes) via the unitary matrices $(U_f^{L,R})$ made up
of the left and right eigenvectors of $\tilde Y_f$.  Phases of
unitary matrices are fixed by the requirement that $(U^L_f)^T
{\tilde Y}_f U^R_f =\Lambda_f$ should yield positive definite
$\Lambda_f$ :
\[ W = (\bar f')^T \Lambda_f f' {\tilde H_f}\]
\[ f= \tilde U_{Z_f} U^R_f f'= {\tilde U_{f}}' f'\qquad ; \qquad
 \bar{f}= \tilde U_{Z_{\bar{f}}} U^L_f \bar{f}'= {\tilde U_{{\bar{f}}}}' \bar{f}'\]
 \be \Lambda_f=(U^L_f)^T
\tilde{U}_{Z_{\bar f}}^T {\frac{Y_f}{\sqrt{Z_{H_f}}}}
\tilde{U}_{Z_f} U^R_f  \label{yukend} \ee As a result the
coefficient $L_{ABCD}$, $R_{ABCD}$ of d=5, $\Delta B =\pm 1$ decay
operator in terms of the Yukawa eigenstate basis transform as \bea
L_{ABCD}' &&=\sum_{a,b,c,d} L_{abcd} (\tilde U_Q')_{aA} (\tilde
U_Q')_{bB}(\tilde U_Q')_{cC} (\tilde U_L')_{dD} \nnu R_{ABCD}'
&&=\sum_{a,b,c,d} R_{abcd} (\tilde U_{\bar e}')_{aA} (\tilde
U_{\bar u}')_{bB}(\tilde U_{\bar u}')_{cC} (\tilde U_{\bar
d}')_{dD} \eea MSSM Higgs ($H,{\bar{H}}$) are mixtures of 6 pairs
of doublets
    from NMSGUT Higgs irreps so Higgs lines have contribution from all the invariants (couplings) of
     superpotential. Although these couplings are small but large number of terms add up to an appreciable correction.
     \emph{Imposing unitarity and perturbativity $Z>0$ one can find the regions of the parameter space where couplings are small
      but $|Z_{H,\bar H}| \approx 0$. Thus the factor $  1/{\sqrt{Z_{H,{\bar H}}}}$ (Eq. \ref{yukend}) will lower the magnitude of the SO(10) Yukawas required to match MSSM
    data by a factor of $10^{-1}$ to $10^{-2}$ and still maintain perturbativity. d=5 operators have no external Higgs line so
    lowered SO(10) couplings will suppress d=5 operators by a factor of $10^{-4} $ to $10^{-8}$.}
\section{Fits Including Threshold Corrections}
Besides d=5, B decay operator coefficients, wavefunction
renormalization also modifies the relation between other
 GUT and MSSM parameters. MSSM $\mu$ and $B$ parameters are larger than the same GUT parameters by the
  factor of $(Z_H Z_{\bar H})^{-1/2}$. Scalar soft masses and soft Higgs masses will be modified by a
   factor of $Z_f^{-1}$ and $Z_{H/\bar{H}}^{-1}$ respectively. Trilinear soft parameters will remain same as wavefunction renormalization factors are absorbed by the Yukawa couplings ($A=A_0 \tilde{Y}_f$). These thresholds redefine $f_{AB}$ as
\be  \tilde{f}_{AB}=(\tilde{U}^T_{\bar\nu} f
\tilde{U}_{\bar\nu})_{AB} \ee This changes the right handed
neutrino Majorana masses ($(M_{\bar\nu})_{AB}\sim f_{AB}
\bar\sigma$). $Y_\nu$ and Higgs field redefinition modify the Type
I seesaw formula.

If we use the NMSGUT parameters for the solutions presented in
\cite{nmsgut} and calculate threshold correction factors for
fermion and Higgs lines $(Z_H, Z_{\bar H}, Z_f, Z_{\bar f})$ we
notice that both the solutions have large negative value of $Z_H,
Z_{\bar H}$ and second one even has negative eigenvalues for
$Z_{f,\bar f}$ as shown in Table \ref{nmsgutsol}. So we performed
a fresh search including the GUT scale threshold corrections to
Yukawas. Our basic search criteria is same as described in the
previous chapter but now include an additional subroutine
implementing threshold corrections to fermion and Higgs fields.
Other improvements we implemented relative to \cite{nmsgut} are:
\begin{table}
 $$
 \begin{array}{cccc}
\hline \hline&&&\vspace{-.3 cm}\\
 \multicolumn{4}{c}{\mbox{SOLUTION 1}}\vspace{0.1 cm} \\
 \hline \hline
 \mbox{Eigenvalues}(Z_{\bar u})& 0.928326& 0.930946& 1.031795\\
 \mbox{Eigenvalues}(Z_{\bar d})& 0.915317& 0.917464& 0.979132\\
 \mbox{Eigenvalues}(Z_{\bar \nu})& 0.870911& 0.873470& 0.975019\\
 \mbox{Eigenvalues}(Z_{\bar e})& 0.904179& 0.908973& 0.971322\\
 \mbox{Eigenvalues}(Z_{Q})& 0.942772& 0.946127& 1.027745\\
 \mbox{Eigenvalues}(Z_{L})& 0.911375& 0.916329& 0.997229\\
 Z_{\bar H},Z_{H}& -109.367 & -193.755 & \\
\hline \hline &&&\vspace{-.3 cm}\\\multicolumn{4}{c}{\mbox{SOLUTION 2}}\vspace{0.1 cm}\\
 \hline \hline
 \mbox{Eigenvalues}(Z_{\bar u})& -7.526729& -7.416343& 1.192789\\
 \mbox{Eigenvalues}(Z_{\bar d})& -7.845885& -7.738424& 1.191023\\
 \mbox{Eigenvalues}(Z_{\bar \nu})& -8.830309& -8.681419& 1.234923\\
 \mbox{Eigenvalues}(Z_{\bar e})& -7.880892& -7.716853& 1.238144\\
 \mbox{Eigenvalues}(Z_{Q})& -9.203739& -9.109832& 1.171956\\
 \mbox{Eigenvalues}(Z_{L})& -9.797736& -9.698265& 1.217620\\
 Z_{\bar H},Z_{H}& -264.776 & -386.534 & \\
 \hline \hline
 \end{array}
 $$
 \caption{\small{Eigenvalues of the wavefunction
 renormalization matrices $Z_f$ for fermion lines and for
MSSM Higgs ($Z_{H,\overline H}$) for solutions presented in
\cite{nmsgut}. \label{nmsgutsol} }}
 \end{table}
 \begin{enumerate}
 \item We imposed the strict unitarity and perturbativity
\[ Z_{f, \bar f, H, \bar H} >0 \]
so searches now prefer smaller values of superpotential parameters
as compared to the case without GUT scale threshold corrections
(see Table 2 \cite{nmsgut} for details).
 \item Including GUT scale threshold corrections we
searched for fits of fermion mass-mixing data in terms  of NMSGUT
superpotential parameters that are compatible with B decay limits.
We have constrained the B decay rates while searching :
\[ \mbox{Max}(L'_{ABCD},R'_{ABCD})< 10^{-22}\,\mbox{GeV}^{-1} \] to get proton life time above $10^{34}$ Yrs. This constraint forces the search towards the regions of parameter space which produce $Z_{H,\bar{H}}\ll 1$
\item Range of soft Susy parameters is almost same as
\cite{nmsgut} except gluino mass$(M_{\tilde g})$ which is kept
greater than $1000$ GeV in accordance with the latest LHC results.
Loop corrected Higgs mass is required to be in the experimentally
indicated range
    \[ 124\, \mbox{GeV} <M_h <126\, \mbox{GeV} \]
\item Another improvement is inclusion of Susy threshold effects
on gauge unification parameters $\alpha_3(M_Z),M_X,\alpha(M_X)$ to
take into account  the spread out  spectrum of supersymmetric
masses. A weighted sum over all the Susy particles
($M_{{\tiny{\mbox{Susy}}}}$) is used in
$\Delta^{{\tiny{\mbox{Susy}}}}_{\alpha_s}$ as given in
\cite{langpol}. \vspace{.2cm} \bea
\Delta^{{\tiny{\mbox{Susy}}}}_{\alpha_s} & \approx &
\frac{-19\alpha_s^2}{28\pi}
\ln\frac{M_{{\tiny{\mbox{Susy}}}}}{M_Z}\nnu
M_{{\tiny{\mbox{Susy}}}} &=&  \prod_i m_i^{-{\frac{ 5}{38}} (4
b_i^1 -9.6 b_i^2 +5.6 b_i^3)} \eea \be
\Delta_X^{{\tiny{\mbox{Susy}}}} = \frac{1}{11.2\pi } \sum_i (b_1 -
b_2)Log_{10}\frac{m_i}{M_Z} \ee \be
\Delta_G^{{\tiny{\mbox{Susy}}}} = \frac{1}{11.2\pi}\sum_i (6.6\,
b_2 - b_1) \ln\frac{m_i}{M_Z}\vspace{.2cm}\ee Here $b_1$,
$b_2,b_3$ are the 1-loop $\beta$ function coefficient of U(1),
SU(2), SU(3) in the MSSM  respectively.
$\Delta^{{\tiny{\mbox{Susy}}}}_{\alpha_s}$ can
 be significant  so  it changes the allowed range at GUT scale.
  We considered the following limits for $\Delta^{{\tiny{\mbox{Susy}}}}_{\alpha_s}$ in the search program.
\be -.0146< \Delta^{{\tiny{\mbox{Susy}}}}_{\alpha_s}  <-0.0102 \ee
Typically we find $M_{{\tiny{\mbox{Susy}}}}$ $\approx$ 10 TeV. Our
constraints on
 the gauge unification parameters are thus : \vspace{-.3cm}\bea
-22.0\leq \Delta_G &\equiv&  \Delta  (\alpha_G^{-1}(M_X))  \leq 25 \nonumber \\
3.0 \geq  \Delta_X &\equiv &\Delta (Log_{10}{M_X}) \geq - 0.03\nonumber \\
-.0126< \Delta_{3} &\equiv & \Delta\alpha_3(M_Z)  <
-.0122\label{criteria} \eea
\end{enumerate}
\subsection{Example Fit}
\begin{enumerate}
\item In Table \ref{table1} we give the values of the NMSGUT
superpotential parameters, changes in gauge unification
parameters- $\Delta_x$, $\Delta_{\alpha_3}$, $\Delta_{\alpha_G}$
(from GUT and Susy), $x$ parameter, the superheavy spectrum and
the mSUGRY-NUHM parameters preferred by the fitting search
program. We also give heavy right handed neutrino masses along
with Type I and II contribution to light neutrino masses. All
parameters are modified (by GUT scale thresholds) parameters. Tree
level relation $\frac{y_b-y_{\tau}}{y_s-y_{\mu}} \sim 1$
\cite{Bajc:SO(10)fitting,core,msgreb,Grimus:1,Grimus:2} is no
longer applicable.
 \item
  Table \ref{table2} shows the successful fitting of extrapolated fermion mass mixing data at GUT scale. Column 2 contains the values
   achieved by the model. Column 3 shows experimental error. In the central block eigenvalues of the fermion correction
    factors ($Z_f,Z_{\bar f}$), $Z_{H, \bar H }$ are given and in the lower block Higgs fractions $\alpha_i, \bar\alpha_i$ are
     presented (which along with the SO(10) Yukawas determine fermion masses). These parameters are determined by the fine tuning
      condition to keep one pair of Higgs doublets light. As discussed $Z_{H, \bar H } \ll 1$  lowers the SO(10) Yukawa couplings therefore $Z_f,Z_{\bar f}$ are close to unity since these are determined by the lowered Yukawas. In Table \ref{table3} (column 2) we show, run down to $M_Z$, values of the fermion masses generated by GUT. Notice that $y_d$ and $y_s$ are smaller by a factor of 3 as compared to their SM values. Fermion masses including large $\tan\beta$ driven Susy threshold corrections are given in $4^{th}$ column.
 \item In Table \ref{table4} we have run down values
  of soft Susy masses (including $M_{H,\bar{H}}^2$) and trilinear couplings. These parameters determine Susy threshold corrections to fermion Yukawas. $\mu$ and $B$ are determined by electroweak symmetry breaking conditions (their values at $M_X$ were obtained by running backup to $M_X$).
 We used the tree level formulae given in the Appendix (of Chapter 2), to calculate Susy spectra. Tables \ref{table5} and \ref{table6} show
  Susy spectra ignoring and including generation mixing. Off-diagonal running changes the spectra marginally.
     \item Here the spectrum has same characteristic as shown in \cite{nmsgut} and mentioned in the previous
     chapter:
      $\mu , A_0, B >$ 100 TeV and normal s-hierarchy. Sometimes we get light smuon having mass 100-200 GeV along with
       Bino (LSP). These special solutions are crucial for model phenomenology as they offer a co-annihilation channel to LSP and can predict appropriate contribution to $(g-2)_{\mu}$. Unoptimized values of important beyond SM (BSM) observables are presented in Table \ref{table7}.
         \item Using the formalism of \cite{barondecay1,barondecay2} we have calculated the decay rate of proton and
          neutron to the different channel as shown in Table \ref{decayrate}. Clearly, B-decay rates are compatible with
           the experiment. In Table \ref{charglu} we consider the chargino and gluino contribution separately and it shows chargino dominance.
           \item Other solution with large value of $M_{\tilde
           \mu}$ are also obtained. They may be found in
           \cite{ilathesis,aulakhgargkhosa}.
 \end{enumerate}
\begin{table}
\vspace{-1.0cm} \addtolength{\tabcolsep}{-5pt}
 $$
 \begin{array}{cccc}
 \hline\hline&&&\vspace{-.4 cm}\\
 {\rm Parameter }&{\rm Value} &{\rm  Field }&{\rm Masses}\\
 &&\hspace{-10mm}{\rm{\small{[SU(3),SU(2),Y]}}} &({\rm Units~of 10^{16} GeV })\vspace{0.1 cm}\\ \hline\hline
       \chi_{X}&  0.3988           &A[1,1,4]&      1.68 \\ \chi_{Z}&
    0.1168
                &B[6,2,{5/3}]&            0.0718\\
           h_{11}/10^{-6}&  3.4611         &C[8,2,1]&{      0.94,      2.41,      5.15 }\\
           h_{22}/10^{-4}&  3.0937    &D[3,2,{7/ 3}]&{      0.08,      3.39,      6.02 }\\
                   h_{33}&  0.0230     &E[3,2,{1/3}]&{      0.09,      0.71,      1.85 }\\
 f_{11}/10^{-6}&
  0.0038+  0.2167i
                      &&{     1.854,      2.65,      5.33 }\\
 f_{12}/10^{-6}&
 -1.0760-  2.0474i
          &F[1,1,2]&      0.29,      0.57
 \\f_{13}/10^{-5}&
  0.0632+  0.1223i
                  &&      0.57,      3.33  \\
 f_{22}/10^{-5}&
  5.0702+  3.6293i
              &G[1,1,0]&{     0.015,      0.14,      0.50 }\\
 f_{23}/10^{-4}&
 -0.3765+  1.7999i
                      &&{     0.498,      0.65,      0.68 }\\
 f_{33}/10^{-3}&
 -0.9059+  0.2815i
              &h[1,2,1]&{     0.291,      2.32,      3.41 }\\
 g_{12}/10^{-4}&
  0.1310+  0.1177i
                 &&{      4.89,     23.26 }\\
 g_{13}/10^{-5}&
 -8.5199+  6.9958i
     &I[3,1,{10/3}]&      0.23\\
 g_{23}/10^{-4}&
 -3.1937-  1.2230i
          &J[3,1,{4/3}]&{     0.201,      0.65,      1.21 }\\
 \lambda/10^{-2}&
 -3.8826+  1.0500i
                 &&{      1.21,      3.83 }\\
 \eta&
 -0.3134+  0.1210i
   &K[3,1, {8/ 3}]&{      1.86,      3.84 }\\
 \rho&
  0.6305-  0.5268i
    &L[6,1,{2/ 3}]&{      1.93,      2.56 }\\
 k&
  0.1926+  0.2311i
     &M[6,1,{8/ 3}]&      2.17\\
 \zeta&
  0.9082+  0.8524i
     &N[6,1,{4/ 3}]&      2.04\\
 \bar\zeta &
  0.2737+  0.6140i
          &O[1,3,2]&      2.77\\
       m/10^{16} \mbox{GeV}&  0.0086    &P[3,3,{2/ 3}]&{      0.64,      3.56 }\\
     m_\Theta/10^{16} \mbox{GeV}&  -2.375e^{-iArg(\lambda)}     &Q[8,3,0]&     0.181\\
             \gamma&  0.3234        &R[8,1, 0]&{      0.08,      0.24 }\\
              \bar\gamma& -3.6166     &S[1,3,0]&    0.2828\\
 x&
  0.78+  0.58i
         &t[3,1,{2/ 3}]&{      0.16,      0.45,      0.9,      2.52  }\\\Delta_X^{\footnotesize{\mbox{Tot}}},\Delta_X^{{\scriptsize{\mbox{GUT}}}}&      0.67,      0.74 &&{      4.08,      4.37,     25.68 }\\
                                 \Delta_{G}^{\footnotesize{\mbox{Tot}}},\Delta_{G}^{{\scriptsize{\mbox{GUT}}}}&-20.46,-23.49           &U[3,3,{4/3}]&     0.238\\
       \{\Delta\alpha_{3}^{\footnotesize{\mbox{Tot}}},\Delta\alpha_{3}^{{\scriptsize{\mbox{GUT}}}}\}{\footnotesize{(M_{Z})}}& -0.0126,  0.0020               &V[1,2,3]&     0.187\\
    M^{\nu^c}/10^{12} \mbox{GeV}&{0.000648,    0.99,   37.28    }&W[6,3,{2/ 3}]&              1.95  \\
 M^{\nu}_{ II}/10^{-9} \mbox{eV}&  .24, 370.1,       13882.34               &X[3,2,{5/ 3}]&     0.063,     2.068,     2.07\\
                  M_\nu(\mbox{meV})&{1.169109,    7.32,   41.46    }&Y[6,2, {1/3}]&              0.08  \\
  \{\rm{Evals[f]}\}/ 10^{-6}&{0.01714,   26.28,  985.21         }&Z[8,1,2]&              0.24  \\
 \hline\hline
 \mbox{Soft parameters}&{\rm m_{\frac{1}{2}}}=
          -152.899
 &{\rm m_{0}}=
         11400.99
 &{\rm A_{0}}=
         -2.00 \times 10^{   5}
 \\
 \mbox{at $M_{X}$}&\hspace{-10mm}\mu=
          1.597 \times 10^{   5}
 &\hspace{-7mm}{\rm B}=
         -1.74 \times 10^{  10}
  &{\rm tan{\beta}}=           51.00\\
 &\hspace{-10mm}{\rm M^2_{\bar H}}=
         -2.07 \times 10^{  10}
 &\hspace{-7mm}{\rm M^2_{  H} }=
         -1.8 \times 10^{  10}
 &
 {\rm R_{\frac{b\tau}{s\mu}}}=
  0.1998
  \\
 \rm{Max(|L_{ABCD}|,|R_{ABCD}|)}&
          8.1104 \times 10^{ -22}
  {\,\rm{GeV^{-1}}}&& \\
 \hline\hline
 \mbox{Susy contribution to}&M_{{\tiny{\mbox{Susy}}}}=12.6~{\small{\mbox{TeV}}}&&
 \\
 {\rm \Delta_{X,G,3}}&{\rm \Delta_X^{{\tiny{\mbox{Susy}}}}}=
            -0.070
 &{\rm \Delta_G^{{\tiny{\mbox{Susy}}}}}=
             3.04
 &{\rm \Delta\alpha_3^{{\tiny{\mbox{Susy}}}}}=
            -0.015
 \\
 \hline\hline\end{array}
 $$
 \caption{\small{NMSGUT superpotential couplings and SUGRY-NUHM soft parameters at $M_X$
  which accurately fit fermion mass-mixing data respecting RG constraints.
 Unification parameters and mass spectrum of superheavy and superlight fields are  also given.
 \label{table1} }}\end{table}
\begin{table}
 $$
 \begin{array}{ccccc}
 \hline\hline
 &&&&\vspace{-.3 cm}\\
 {\rm  Parameter }&{\rm Target} =\bar O_i & {\rm Uncert.= \delta_i  }  & {\rm Achieved= O_i} & {\rm Pull =\frac{(O_i-\bar O_i)}{\delta_i}}\vspace{0.1 cm}\\
 \hline\hline
    y_u/10^{-6}&  2.062837&  0.788004&  2.066323&  0.004424\\
    y_c/10^{-3}&  1.005548&  0.165915&  1.010599&  0.030440\\
            y_t&  0.369885&  0.014795&  0.369792& -0.006256\\
    y_d/10^{-5}& 11.438266&  6.668509& 12.421488&  0.147443\\
    y_s/10^{-3}&  2.169195&  1.023860&  2.189195&  0.019534\\
            y_b&  0.456797&  0.237078&  0.527664&  0.298917\\
    y_e/10^{-4}&  1.240696&  0.186104&  1.224753& -0.085665\\
  y_\mu/10^{-2}&  2.589364&  0.388405&  2.603313&  0.035911\\
         y_\tau&  0.543441&  0.103254&  0.532427& -0.106669\\
             \sin\theta^q_{12}&    0.2210&  0.001600&    0.2210&           -0.0003\\
     \sin\theta^q_{13}/10^{-4}&   29.1907&  5.000000&   29.0755&           -0.0230\\
     \sin\theta^q_{23}/10^{-3}&   34.3461&  1.300000&   34.3574&            0.0087\\
                      \delta^q&   60.0212& 14.000000&   59.7774&           -0.0174\\
    (m^2_{12})/10^{-5}(\mbox{eV})^{2}&    5.2115&  0.552419&    5.2189&            0.0133\\
    (m^2_{23})/10^{-3}(\mbox{eV})^{2}&    1.6647&  0.332930&    1.6650&            0.0011\\
           \sin^2\theta^L_{12}&    0.2935&  0.058706&    0.2926&           -0.0152\\
           \sin^2\theta^L_{23}&    0.4594&  0.137809&    0.4412&           -0.1317\\
           \sin^2\theta^L_{13}&    0.0250&  0.019000&    0.0267&            0.0892\\

  \hline\hline
                    (Z_{\bar u})&   0.957467&   0.957908&   0.957908&\\
                  (Z_{\bar d})&   0.950892&   0.951332&   0.951333&\\
                (Z_{\bar \nu})&   0.925116&   0.925579&   0.925580&\\
                  (Z_{\bar e})&   0.944853&   0.945306&   0.945308&\\
                       (Z_{Q})&   0.968740&   0.969189&   0.969190&\\
                       (Z_{L})&   0.949564&   0.950011&   0.950013&\\
              Z_{\bar H},Z_{H}&        0.000273   &        0.001151    &{}&\\
   \hline\hline
 \alpha_1 &
  0.1609-  0.0000i
 & {\bar \alpha}_1 &
  0.1188-  0.0000i
 &\\
 \alpha_2&
 -0.3140-  0.6026i
 & {\bar \alpha}_2 &
 -0.4802-  0.2961i
 &\\
 \alpha_3 &
 -0.0477-  0.4786i
 & {\bar \alpha}_3 &
 -0.4842-  0.2469i
 &\\
 \alpha_4 &
  0.3903-  0.1942i
 & {\bar \alpha}_4 &
  0.5795+  0.0171i
 &\\
 \alpha_5 &
 -0.0449+  0.0061i
 & {\bar \alpha}_5 &
 -0.0415-  0.1241i
 &\\
 \alpha_6 &
 -0.0071-  0.2982i
 & {\bar \alpha}_6 &
  0.0274-  0.1349i
 &\\
 \hline\hline
 \end{array}
 $$
   \caption{\small{Fit with $\chi_X=\sqrt{ \sum\limits_{i=1}^{17}
 \frac{(O_i-\bar O_i)^2}{\delta_i^2}}=
    0.3988$. Target values,  at $M_X$ of the fermion Yukawa
 couplings and mixing parameters, together with the estimated uncertainties, achieved values and pulls. The eigenvalues of the wavefunction renormalization for fermion and Higgs lines are
 given with Higgs fractions $\alpha_i,{\bar{\alpha_i}}$ which control the MSSM
fermion Yukawa couplings.\label{table2} }}
 \end{table}
 \begin{table}
 $$
 \begin{array}{cccc}
 \hline \hline&&&\vspace{-.3 cm}\\ {\rm  Parameter }&{\rm SM(M_Z) }&{\rm m^{GUT}(M_Z)} & {\rm m^{MSSM}=(m+\Delta m)^{GUT}(M_Z) }\vspace{0.1 cm}\\
 \hline\hline
    m_d/10^{-3}&   2.90000&   1.08183&   3.01515\\
    m_s/10^{-3}&  55.00000&  19.06631&  53.14737\\
            m_b&   2.90000&   3.17508&   3.05602\\
    m_e/10^{-3}&   0.48657&   0.45157&   0.45925\\
         m_\mu &   0.10272&   0.09594&   0.09902\\
         m_\tau&   1.74624&   1.65725&   1.65734\\
    m_u/10^{-3}&   1.27000&   1.10509&   1.27687\\
            m_c&   0.61900&   0.54048&   0.62449\\
            m_t& 172.50000& 145.99987& 170.88573\\
 \hline\hline
 \end{array}
 $$
  \caption{\small{Values of the SM
 fermion masses in GeV at $M_Z$ compared with the masses obtained from
 values of GUT derived  Yukawa couplings  run down from $M_X$ to
 $M_Z$  both before and after threshold corrections.
  Fit with $\chi_Z=\sqrt{ \sum\limits_{i=1}^{9} \frac{(m_i^{{\tiny{\mbox{MSSM}}}}- m_i^{{\tiny{\mbox{SM}}}})^2}{ (m_i^{{\tiny{\mbox{MSSM}}}})^2} }=
0.1153$.\label{table3} }}
 \end{table}
 \begin{table}
 $$
 \begin{array}{cccc}
 \hline\hline&&&\vspace{-.3 cm}\\
 {\rm  Parameter}  &{\rm Value}&  {\rm  Parameter}&{\rm Value}\vspace{0.1 cm} \\
 \hline\hline
                       M_{1}&            210.10&   m_{{\tilde {\bar {u}}_1}}&          14446.81\\
                       M_{2}&            569.81&   m_{{\tilde {\bar {u}}_2}}&          14445.85\\
                       M_{3}&           1000.14&   m_{{\tilde {\bar {u}}_3}}&          24609.79\\
     m_{{\tilde {\bar l}_1}}&           1761.31&               A^{0(l)}_{11}&        -121907.75\\
     m_{{\tilde {\bar l}_2}}&            210.71&               A^{0(l)}_{22}&        -121757.58\\
     m_{{\tilde {\bar l}_3}}&          20777.09&               A^{0(l)}_{33}&         -77289.04\\
        m_{{\tilde {L}_{1}}}&          15308.21&               A^{0(u)}_{11}&        -148456.63\\
        m_{{\tilde {L}_{2}}}&          15258.47&               A^{0(u)}_{22}&        -148455.19\\
        m_{{\tilde {L}_{3}}}&          21320.16&               A^{0(u)}_{33}&         -76985.25\\
     m_{{\tilde {\bar d}_1}}&           8402.95&               A^{0(d)}_{11}&        -122521.00\\
     m_{{\tilde {\bar d}_2}}&           8401.45&               A^{0(d)}_{22}&        -122518.53\\
     m_{{\tilde {\bar d}_3}}&          51842.14&               A^{0(d)}_{33}&         -44046.92\\
          m_{{\tilde {Q}_1}}&          11271.93&                   \tan\beta&             51.00\\
          m_{{\tilde {Q}_2}}&          11270.77&                    \mu(M_Z)&         125591.16\\
          m_{{\tilde {Q}_3}}&          40274.01&                      B(M_Z)&
          2.7861 \times 10^{   9}
 \\
 M_{\bar {H}}^2&
         -1.6336 \times 10^{  10}
 &M_{H}^2&
         -1.7391 \times 10^{  10}
 \\
 \hline\hline
 \end{array}
 $$
  \caption{ \small {Values (in GeV) of the soft Susy parameters  at $M_Z$
 (evolved from the soft SUGRY-NUHM parameters at $M_X$) ($M_{{\tiny{\mbox{Susy}}}}=12.6~{\small{\mbox{TeV}}}$).
 \label{table4}  }}
 \end{table}
 \begin{table}
 $$
 \begin{array}{cc}
 \hline\hline &\vspace{-.3 cm}\\{\mbox {Field } }&{\rm Mass(GeV)}\vspace{0.1 cm}\\
 \hline\hline
                M_{\tilde{g}}&           1000.14\\
               M_{\chi^{\pm}}&            569.81,         125591.22\\
       M_{\chi^{0}}&            210.10,            569.81,         125591.20    ,         125591.20\\
              M_{\tilde{\nu}}&         15308.069,         15258.322,         21320.059\\
                M_{\tilde{e}}&           1761.89,          15308.29,            211.57   ,          15258.60,          20674.72,          21419.56  \\
                M_{\tilde{u}}&          11271.80,          14446.76,          11270.63   ,          14445.80,          24607.51,          40275.87  \\
                M_{\tilde{d}}&           8402.99,          11272.10,           8401.48   ,          11270.95,          40269.19,          51845.93  \\
                        M_{A}&         377025.29\\
                  M_{H^{\pm}}&         377025.30\\
                    M_{H}&         377025.28\\
                    M_{h}&            124.00\\
 \hline\hline
 \end{array}
 $$
  \caption{\small{Susy spectrum calculated ignoring generation mixing effects.
\label{table5} }}\end{table}

\begin{table}
 $$
 \begin{array}{cc}
 \hline \hline&\vspace{-.3 cm}\\{\mbox {Field } }&{\rm Mass(GeV)}\vspace{0.1 cm}\\
 \hline\hline
                M_{\tilde{g}}&           1000.72\\
               M_{\chi^{\pm}}&            570.11,         125537.00\\
       M_{\chi^{0}}&            210.22,            570.11,         125536.98    ,         125536.98\\
              M_{\tilde{\nu}}&          15257.98,          15307.71,         21350.169\\
                M_{\tilde{e}}&            242.61,           1765.59,          15258.25   ,          15307.93,          20733.03,          21453.81  \\
                M_{\tilde{u}}&          11258.18,          11270.54,          14444.57   ,          14445.53,          24609.90,          40301.29  \\
                M_{\tilde{d}}&           8400.19,           8401.71,          11258.52   ,          11270.84,          40294.63,          51879.28  \\
                        M_{A}&         377430.83\\
                  M_{H^{\pm}}&         377430.84\\
                    M_{H}&         377430.82\\
                    M_{h}&            124.13\\
 \hline\hline
 \end{array}
 $$
  \caption{\small{Susy spectrum calculated including  generation mixing effects.\label{table6} }}\end{table}
  \begin{table}
 $$
 \begin{array}{cc}
 \hline\hline &\vspace{-.3 cm}\\ {\rm Parameter }& \mbox{ Value } \vspace{0.1 cm}\\
  \hline\hline
 {\rm{ BR(b\rightarrow s\gamma)} }& 3.294 \times 10^{ -4} \\
\Delta a_{\mu} &
   1.06 \times 10^{ -9}\\
\Delta \rho &
   6.03 \times 10^{ -7}\\
 \epsilon/10^{-7}  &0.12 \\
  \delta_{PMNS} & 6.21^\circ\\
  \hline\hline \end{array}
 $$
 \caption{\small{Unoptimized  values for the solution presented.\label{table7}}}
 \end{table}

\begin{table}
  $$
  \begin{array}{cc}
  \hline\hline&\vspace{-.3 cm}\\ {\rm Parameter }& \mbox{ Value } \vspace{0.1 cm}\\
  \hline\hline
    \tau_p(M^+ \bar\nu)&
                 9.63 \times 10^{  34}

 \\\hline
     \Gamma(p\rightarrow \pi^+ \bar\nu)&
                 4.32 \times 10^{ -37}

 \\
            \mbox{BR}( p\rightarrow\pi^+\bar \nu_{e,\mu,\tau})&
 \{
                 1.3 \times 10^{  -3}
 ,
                 0.34
 ,
                 0.66
 \}
\\
     \Gamma(p\rightarrow K^+ \bar\nu)&
                 9.95 \times 10^{ -36}
 \\
         \mbox{BR}( p\rightarrow K^+ \bar\nu_{e,\mu,\tau}) &
 \{
                 4.6 \times 10^{  -4}
 ,
                 0.15
 ,
                 0.85
 \}\\\hline \hline
    \tau_p(M^0 l^+)&
                 1.053 \times 10^{  36}
 \\\hline
     \Gamma(p\rightarrow \pi^0 l^+ )&
                 4.239 \times 10^{ -37}
 \\
             \mbox{BR}( p\rightarrow\pi^0 \{e^+,\mu^+,\tau^+\})&
 \{
                 2.241 \times 10^{  -4}
 ,
                 0.057
 ,
                 0.943
 \}\\
     \Gamma(p\rightarrow K^0 l^+)&
                 3.532 \times 10^{ -37}
 \\
         \mbox{BR}( p\rightarrow K^0 \{e^+,\mu^+,\tau^+\}) &
 \{
                 5.544 \times 10^{  -4}
 ,
                 0.103
 ,
                 0.897
 \}\\
         \Gamma(p\rightarrow \eta^0 l^+)&
                 1.724 \times 10^{ -37}
  \\
    \mbox{BR}( p\rightarrow \eta^0 \{e^+,\mu^+,\tau^+\})&
 \{
                 2.232 \times 10^{  -4}
 ,
                 0.057
 ,
                 0.943
 \}\\\hline \hline
  \tau_n(M^0 \bar\nu)&
                1.084 \times 10^{  35}
 \\  \hline
     \Gamma(n\rightarrow \pi^0 \bar\nu )&
                 2.170 \times 10^{ -37}
 \\
             \mbox{BR}( n\rightarrow\pi^0 \bar\nu_{e,\mu,\tau})&
 \{
                 1.321 \times 10^{  -3}
 ,
                 0.341
 ,
                 0.658
 \}\\
     \Gamma(n\rightarrow K^0 \bar\nu)&
                 8.79 \times 10^{ -36}
 \\
         \mbox{BR}( n\rightarrow K^0 \bar\nu_{e,\mu,\tau}) &
 \{
                 1.33 \times 10^{  -3}
 ,
                 0.202
 ,
                 0.797
 \}\\
         \Gamma(n\rightarrow \eta^0 \bar\nu)&
                 2.177 \times 10^{ -37}
\\
    \mbox{BR}( n\rightarrow \eta^0 \bar\nu_{e,\mu,\tau})&
 \{
                 5.39 \times 10^{  -4}
 ,
                 0.128
 ,
                 0.871
 \}\\   \hline \hline
 \end{array}
 $$
 \label{table c}
 \caption{\small{d=5 operator mediated nucleon lifetimes $\tau_{p,n}$(yrs), decay rates
  $ \Gamma$(yr$^{-1}$)  and branching ratios in the different channels.\label{decayrate}}}
 \end{table}

 \begin{table}
  $$
  \begin{array}{ccc}
  \hline \hline&&\vspace{-.3 cm}\\{\rm Parameter }  & \rm{\frac{\Gamma^{chargino}}{ \Gamma^{total}}}  &  \rm{ \frac{\Gamma^{gluino}}{ \Gamma^{total}}} \vspace{0.1 cm} \\
  \hline \hline
   p\rightarrow K^+ \bar\nu& 0.795  &  0.029
\\
 p\rightarrow\pi^+ \bar\nu&  1.123&  0.011
\\
p\rightarrow K^0 l^+&  0.794 & 0.029
\\
 p\rightarrow\pi^0 l^+&  1.255 &  0.023
\\
p\rightarrow \eta^0 l^+&  0.61  & 0.015
\\

n\rightarrow \pi^0 \nu & 1.255& 0.023
\\
 n\rightarrow K^0 \nu & 1.111   &  0.015
\\
n\rightarrow \eta^0 \nu &  0.969 & 0.009
\\\hline \hline
\end{array}
 $$
 \label{table c}
 \caption{\small{First and second column contain ratio of nucleon
  decay considering only chargino and gluino contribution to the total
  decay rate.\label{charglu}}}
 \end{table}

\section{Conclusions and Outlook}
We have computed the one-loop GUT scale threshold corrections
\cite{aulakhgargkhosa} to the tree level Yukawas at an SO(10)
Yukawa vertex. Threshold corrections at $M_X$ to the Higgs lines
are very significant due to the large Higgs representation used.
There exist regions in parameter space where the effective MSSM
Higgs renormalization factor can have very
 small value ($Z_{H, \bar{H}}$ $\approx$ 0). These corrections lower the SO(10) Yukawas required to match
MSSM fermion data. The same Yukawas  determine coefficients of d=5
baryon violation operators. The lowered SO(10) Yukawas solved the
problem of fast B decay in NMSGUT \cite{nmsgut}. We have shown
example solutions of NMSGUT parameters and soft Susy breaking
parameters which accurately fit fermion mass mixing data and are
compatible with B decay rates. Solutions found have not only the
Yukawa couplings but also the superpotential parameters
significantly lowered in magnitude as compared to the tree level
solutions. We have not optimized our fits for different
phenomenological constraints from quark and lepton sector. However
we have calculated some important beyond SM (BSM) parameters such
as $a_\mu$, $\rho$, BR($b \rightarrow s \gamma$) which respect the
experiment limits. One of the example fits with a light smuon
($\tilde{\mu}_R$) shows significant Susy contribution to
$a_{\mu}$.

We have also implemented the effects of Susy spectra on gauge
unification parameters $\alpha_3(M_Z)$, $M_X$ and $\alpha(M_X)$.
Susy spectra of the fits including superheavy threshold
corrections have same characteristics as the one without GUT scale
threshold corrections (exhibited in \cite{nmsgut}). Thus they have
large $A_0,\mu \sim 100$ TeV (for realistic fermion mass and
mixing data and $M_h \sim 125$ GeV) parameters and heavy
s-particle spectra which seems to be a likely scenario after Higgs
discovery. Fits obtained deviate significantly from
$\frac{y_b-y_{\tau}}{y_s-y_{\mu}}\approx 1$ which is
characteristic of ${\bf{10}}-{\bf{120}}$-plet generated fermion
fits \cite{Bajc:SO(10)fitting,core,msgreb,Grimus:1,Grimus:2}.

This mechanism of suppressing fast dimension 5 proton decay rates
including GUT scale threshold to light fields, is generic for all
realistic Susy GUTs in which the light MSSM Higgs arise from a
mixture of GUT Higgs doublets coupled to a large number of
superheavy fields. The effect of d=6 B violation operator with one
external Higgs line remains to be checked. \clearpage
\section*{Appendix: Higgs Field Correction Factors}
In this Appendix, we give correction factors from all type of GUT
multiplets to the $H$ and $\bar{H}$ (see Eq. \ref{hline}) : \bea
  K_{R\bar C}
  & = &\sum_ {a = 1}^{\mbox{d(R)}}\sum_ {a' =
  1}^{\mbox{d(C)}} \biggr| \biggr(
  \frac {ik} {\sqrt {2}} V^R_{2a} U^C_{3a'}-\gamma V^R_{1a} U^C_{2a'} + \frac {\gamma } {\sqrt {2}}V^R_{2a} U^C_{2a'} -
  \bar\gamma V^R_{1a}U^C_{1a'} - \frac {\bar\gamma } {\sqrt{2}}V^R_{2a}U^C_{1a'} \biggr) V^H_{11} \nonumber \\ &&
   +\biggr(\frac {2\eta } {\sqrt{3}} V^R_{1a}U^C_{2a'}- \sqrt{\frac{2}{3}}\eta V^R_{2a} U^C_{2a'} +
    \frac {i \bar\zeta} {\sqrt{6}} V^R_{2a} U^C_{3a'}\biggr) V^H_{21} \nonumber \\ &&+\biggr(\frac {2 \eta} {\sqrt{3}} V^R_{1a}U^C_{1a'} +
    \sqrt{\frac{2}{3}}\eta V^R_{2a} U^C_{1a'} + \frac {i\zeta}{\sqrt{6}} V^R_{2a} U^C_{3a'}\biggr) V^H_{31}\nonumber \\ &&
    +\biggr( \frac {\zeta} {\sqrt{2}} V^R_{2a} U^C_{2a'}-\frac {i \rho} {3\sqrt {2}} V^R_{2a} U^C_{3a'} +
   \frac {\bar\zeta}{\sqrt{2}} V^R_{2a} U^C_{1a'}\biggr) V^H_{51}\nonumber \\ && -\biggr(\frac {i\zeta} {\sqrt {6}} V^R_{2a} U^C_{2a'} +
    \frac {i\bar\zeta} {\sqrt{6}} V^R_{2a} U^C_{1a'} -
  \frac {\rho} {3\sqrt{3}}V^R_{1a} U^C_{3a'}\biggr) V^H_{61} \biggr| ^2 F_{12}(m^R_{a}, m^C_{a'}, Q)~~~~~~~~~\,\,\,\,\,\,\eea
\vspace{.2cm}

\bea K_{X\bar P} &= &\sum_ {a = 1}^{\mbox{d(X)}} \sum_ {a' =
  1}^{\mbox{d(P)}} \biggr|\biggr(\bar\gamma V^X_{1a} U^P_{1a'} - \frac{k}{\sqrt{2}} V^X_{2a} U^P_{2a'}\biggr) V^H_{11} -
  \biggr(\frac{2 \bar\zeta}{\sqrt{3}} V^X_{1a} U^P_{2a'} + \frac{\bar\zeta}{\sqrt{6}} V^X_{2a} U^P_{2a'}\biggr) V^H_{21}\nonumber\\&&
  + \biggr(\frac{\zeta}{\sqrt{6}} V^X_{2a} U^P_{2a'} +
\frac{2 \eta}{\sqrt{3}} V^X_{1a} U^P_{1a'} - \frac{2 \sqrt{2}
\eta}{\sqrt{3}} V^X_{2a} U^P_{1a'}\biggr) V^H_{31}\nonumber\\&&+
\biggr(\frac{\rho}{3\sqrt{2}} V^X_{2a} U^P_{2a'} + \bar\zeta
V^X_{1a} U^P_{1a'}\biggr) V^H_{51} +
  \frac{i}{\sqrt{3}} \biggr(\sqrt{2}\bar\zeta V^X_{2a} U^P_{1a'} \nonumber\\&&- \bar\zeta V^X_{1a}
  U^P_{1a'}
  + \frac{\rho}{3} V^X_{1a} U^P_{2a'}- \frac{\rho}{3 \sqrt{2}} V^X_{2a} U^P_{2a'}\biggr) V^H_{61} \biggr| ^2 F_{12} (m^X_a, m^P_{a'},
  Q)\nonumber\\&&
  -2{g_{10}^2} \biggr|i\sqrt{\frac{2}{3}}V^{P*}_{1a'}V^H_{31}-\frac{V^{P*}_{2a'}}{\sqrt{2}}V^H_{51}
  +\frac{i}{\sqrt{6}}V^{P*}_{2a'}V^H_{61}\biggr|^2 F_{12}(m_{\lambda_X}, m^P_{a'}, Q)
  \qquad \qquad~~
\eea

\vspace{.2cm} \bea K_{VF} &= &\sum_{a=1}^{\mbox{d(F)}}
 \biggr|\biggr( k V^F_{4a}-i \gamma V^F_{1a} \biggr)
  V^H_{11}- \biggr(2 \sqrt{3}i \eta V^F_{1a}
  +\sqrt{3}\bar\zeta V^F_{4a}\biggr)
  V^H_{21} -\sqrt{3} \zeta V^F_{4a}
  V^H_{31} \nonumber\\&&- 2\sqrt{3} \lambda V^F_{2a}
  V^H_{41} + i \zeta V^F_{1a}
  V^H_{51} -\biggr(\sqrt{3} \zeta V^F_{1a} -
  \frac{i\rho}{\sqrt{3} } V^F_{4a}\biggr) V^H_{61}
\biggr|^2 F_{12}(m^V, m^F_a, Q)\nonumber\\&&-2{g_{10}^2}
\biggr|iV^H_{41} \biggr|^2F_{12}(m^V, m_{\lambda_F}, Q) \qquad
\qquad ~~~~\eea

\bea K_{E \bar J} & = & \sum_ {a =1}^{\mbox{d(E)}}\sum_ {a'
=1}^{\mbox{d(J)}}\biggr| \biggr(\gamma V^E_{2a} U^J_{2a'} +
\sqrt{2} \gamma i V^E_{3a} U^J_{1a'} - \frac{\gamma}{\sqrt{2}}
V^E_{2a} U^J_{3a'} + \bar\gamma V^E_{1a} U^J_{2a'} +
\frac{\bar\gamma}{\sqrt{2}}V^E_{1a}U^J_{3a'}\nonumber \\&& +
   i k V^E_{4a} U^J_{5a'} - \frac{i k}{\sqrt{2}}V^E_{6a} U^J_{3a'}\biggr) V^H_{11}
    +\biggr( \frac{2 \eta}{\sqrt{3}} V^E_{2a} U^J_{2a'}+
   2 \sqrt{\frac{2}{3}}i\eta V^E_{3a} U^J_{1a'} +
   \sqrt{\frac{2}{3}}\eta V^E_{2a} U^J_{3a'}\nonumber \\ && + \frac{4i \eta }{\sqrt{3}} V^E_{4a} U^J_{1a'}
   -\frac{i\bar\zeta}{\sqrt{6}} V^E_{6a} U^J_{3a'} + \frac {2 i\bar\zeta }{\sqrt{3}} V^E_{6a} U^J_{2a'} -\frac {i\bar\zeta }{\sqrt{3}}V^E_{4a} U^J_{5a'}\biggr) V^H_{21} +\biggr( \frac {2i \zeta}{\sqrt{3} } V^E_{6a} U^J_{2a'}\nonumber \\ && +\sqrt{\frac{3 }{2}}i \zeta V^E_{6a} U^J_{3a'} -
   \frac {2\sqrt{2}i \zeta}{\sqrt{3} } V^E_{3a} U^J_{5a'} + \frac {i \zeta}{\sqrt{3}} V^E_{4a} U^J_{5a'}
   + \sqrt{6}\eta V^E_{1a} U^J_{3a'} + \frac {2\eta}{\sqrt{3}} V^E_{1a} U^J_{2a'}\biggr) V^H_{31}
    \nonumber \\ && + \sqrt{2}i \lambda \biggr(2 V^E_{3a} U^J_{2a'} -
   V^E_{4a} U^J_{3a'} - \sqrt{2} V^E_{3a} U^J_{3a'}\biggr) V^H_{41}
   +\biggr(\frac{i \rho }{3\sqrt{2} }V^E_{6a} U^J_{3a'} +
   \frac {i \rho}{3} V^E_{4a} U^J_{5a'}\nonumber \\ && - \frac{\zeta }{\sqrt{2}}V^E_{2a} U^J_{3a'}
   -\sqrt{2} i\zeta V^E_{3a} U^J_{1a'} - \frac {\bar\zeta}{\sqrt{2} } V^E_{1a} U^J_{3a'}\biggr) V^H_{51} +\biggr( \frac{\sqrt{2} \rho}{3 \sqrt{3}} V^E_{6a} U^J_{3a'} -\frac{\sqrt{2}\rho}{3\sqrt{3}} V^E_{3a} U^J_{5a'}\nonumber \\ && - \frac{\rho }{3\sqrt{3}}V^E_{4a} U^J_{5a'} + \frac{\rho }{3 \sqrt{3}}V^E_{6a} U^J_{2a'} + \frac {2 \zeta}{\sqrt{3} } V^E_{4a} U^J_{1a'} -
   \frac {2i \zeta}{\sqrt{3} } V^E_{2a} U^J_{2a'}+ \frac {\sqrt{2} \zeta}{\sqrt{3} } V^E_{3a} U^J_{1a'}\nonumber \\ && +
    \frac {i\zeta }{\sqrt{6} }V^E_{2a} U^J_{3a'} -
   \sqrt{\frac{3}{2}}i\bar\zeta V^E_{1a} U^J_{3a'} - \frac{2i\bar\zeta}{\sqrt{3}} V^E_{1a} U^J_{2a'}\biggr) V^H_{61} \biggr| ^2 F_{12} (m^E_a, m^J_{a'},
   Q)\nonumber \\ &&-2{g_{10}^2} \biggr|\frac{2i}{\sqrt{3}}U^{E*}_{2a}V^H_{21}+\frac{2i}{\sqrt{3}}
   U^{E*}_{1a}V^H_{31}-\sqrt{2}U^{E*}_{3a}V^H_{41}+\frac{2i}{\sqrt{3}}U^{E*}_{6a}V^H_{61}\biggr|^2
   F_{12}(m^E _a, m_{\lambda_J},Q)\nonumber \\ &&
   -2{g_{10}^2} \biggr|\frac{2}{\sqrt{3}}V^{J*}_{1a'}V^H_{21}-iV^{J*}_{2a'}V^H_{41}
   +\frac{i}{\sqrt{2}}V^{J*}_{3a'}V^H_{41}+iV^{J*}_{5a'}V^H_{51}
   -\frac{1}{\sqrt{3}}V^{J*}_{5a'}V^H_{61}\biggr|^2\nonumber \\ && F_{12} ( m_{\lambda_E},m^J_{a'}, Q)\eea

\vspace{.2cm} \bea K_{P \bar E} &=& \sum_ {a =
1}^{\mbox{d(P)}}\sum_ {a' =
  1}^{\mbox{d(E)}} \biggr| \biggr(\gamma V^P_{1a} U^E_{3a'} - \frac {k}{\sqrt{2}} V^P_{2a} U^E_{4a'}\biggr) V^H_{11}\nonumber\\&& +
  \biggr( \frac{2\eta}{\sqrt{3}}V^P_{1a}( U^E_{3a'} -
\sqrt{2} U^E_{4a'})
  + \frac {\bar\zeta}{\sqrt{6}} V^P_{2a} U^E_{4a'}\biggr) V^H_{21}\nonumber\\&& - \frac{\zeta}{\sqrt{3}}\biggr(2
  V^P_{2a}U^E_{3a'}
  + \frac{V^P_{2a} }{\sqrt{2}} U^E_{4a'} \biggr) V^H_{31} + \biggr(2 \sqrt{2}\eta i V^P_{1a} U^E_{2a'} -
  \frac{\rho V^P_{2a}}{3\sqrt{2}} U^E_{6a'}- \sqrt{2} \zeta V^P_{1a} U^E_{6a'} \nonumber\\&& + \sqrt{2} i \zeta V^P_{2a} U^E_{1a'}\biggr) V^H_{41}
 + \biggr(\frac{\rho}{3\sqrt{2}} V^P_{2a} U^E_{4a'} +
\zeta V^P_{1a} U^E_{3a'}\biggr) V^H_{51} +
 \biggr(\frac {i\zeta }{\sqrt{3}}
  V^P_{1a} U^E_{3a'}\nonumber\\&&- \frac {\sqrt{2}i\zeta }{\sqrt{3}} V^P_{1a} U^E_{4a'}
  - \frac {i\rho}{3\sqrt{3}}V^P_{2a} U^E_{3a'} +
  \frac {i\rho}{3\sqrt{6}}V^P_{2a}U^E_{4a'}\biggr) V^H_{61}\biggr| ^2 F_{12} (m^P_a, m^E_{a'}, Q)
  \nonumber\\&&-2{g_{10}^2} \biggr|-i \sqrt{\frac{2}{3}}U^{P*}_{1a} V^H_{21}-\frac{U^{P*}_{2a}}{\sqrt{2}}V^H_{51}
  -\frac{iU^{P*}_{2a}}{\sqrt{6}}V^H_{61}\biggr|^2 F_{12}(m^P_a, m_{\lambda_{E}}, Q) \qquad \quad~~~\eea

\bea K_{B \bar M} &=& \biggr|\sqrt{2} i \gamma V^H_{11} -
2\sqrt{\frac{2}{3}} i \eta V^H_{21} - \sqrt{2} i \zeta V^H_{51} -
\sqrt{\frac{2}{3}} \zeta V^H_{61}\biggr | ^2 F_{12}(m^B, m^M,Q)
~~~~~~\eea \vspace{-.8cm} \bea K_{X\bar T} &= & \sum_{ a = 1}^{
\mbox{d(X)}} \sum_{ a' = 1}^{
 \mbox{d(T)}} \biggr|\biggr( k V^X_{1a} U^T_{6a'}-\gamma V^X_{2a} U^T_{3a'} -
  i\gamma V^X_{1a} U^T_{4a'} - \bar\gamma V^X_{2a} U^T_{2a'}  -
  \frac{ik}{ \sqrt{2}} V^X_{2a} U^T_{7a'}\biggr) V^H_{11}\nonumber\\&&+ \biggr( \frac {\sqrt{2} i \bar\gamma}{\sqrt{3}} V^X_{1a} U^T_{1a'} -\frac {\bar\gamma}{\sqrt{3}}
  V^X_{2a} U^T_{1a'}- 2\sqrt{\frac{2}{3}}\eta V^X_{1a} U^T_{3a'}- \frac{2i\eta}{\sqrt{3}} V^X_{1a} U^T_{4a'}-
  2\sqrt{\frac{2}{3}}i\eta V^X_{2a} U^T_{4a'}\nonumber\\&& -\frac {\bar\zeta}{\sqrt{3}}V^X_{1a} U^T_{6a'}  - \frac {\sqrt{2}\bar\zeta}{\sqrt{3}} V^X_{2a} U^T_{6a'} +
 \frac{ i\bar\zeta}{\sqrt{6} } V^X_{2a} U^T_{7a'}\biggr) V^H_{21} + \biggr(2 \sqrt{\frac{2}{3}}\eta V^X_{1a} U^T_{2a'} -\frac {\zeta}{\sqrt{3}}V^X_{1a}
 U^T_{6a'} \nonumber\\&&+ \frac {\sqrt{2}\zeta}{\sqrt{3}} V^X_{2a} U^T_{6a'} +\frac {2i\zeta}{\sqrt{3}} V^X_{1a} U^T_{7a'}
- \frac{i\zeta}{\sqrt{6}} V^X_{2a} U^T_{7a'} + \frac
{\sqrt{2}i\gamma }{\sqrt{3}} V^X_{1a} U^T_{1a'} +
  \frac
{\gamma}{\sqrt{3}} V^X_{2a} U^T_{1a'}\biggr) V^H_{31}
\nonumber\\&&-2 i \lambda \biggr(V^X_{2a} U^T_{5a'} + \sqrt{2}
V^X_{1a} U^T_{5a'}\biggr) V^H_{41}  + \biggr( i \zeta V^X_{1a}
U^T_{4a'}-k V^X_{2a} U^T_{1a'} - \frac {i \rho}{3\sqrt{2}}
V^X_{2a} U^T_{7a'} \biggr) V^H_{51}\nonumber\\&& + \biggr( \frac {
\sqrt{2} i \zeta}{\sqrt{3}} V^X_{1a} U^T_{3a'} - \frac
{\sqrt{2}\zeta}{\sqrt{3}} V^X_{2a} U^T_{4a'} + \frac
{i\zeta}{\sqrt{3}} V^X_{2a} U^T_{3a'} -
  \frac
{\zeta}{\sqrt{3}} V^X_{1a} U^T_{4a'}
 -\frac{\sqrt{2} i}{ \sqrt{3}} \bar\zeta V^X_{1a}U^T_{2a'}\nonumber\\&& +\frac{ i \bar\zeta}{ \sqrt{3}}
  V^X_{2a}U^T_{2a'} + \frac{i \rho}{3\sqrt{3}} V^X_{1a} U^T_{6a'} +\frac{ \rho}{3\sqrt{3}} V^X_{1a} U^T_{7a'} +
  \frac { \rho }{3\sqrt{6}}V^X_{2a} U^T_{7a'}-
\sqrt{\frac{2}{3}}ik V^X_{1a} U^T_{1a'} \biggr)\nonumber\\&&
V^H_{61} \biggr| ^2 F_{12} (m^X_{a}, m^T_{a'}, Q)
  -2{g_{10}^2} \biggr|-V^{T*}_{1a'}V^H_{11}-\biggr(\frac{i}{\sqrt{3}}V^{T*}_{2a'}+\sqrt{\frac{2}{3}}V^{T*}_{4a'}\biggr)V^H_{21}
  -\frac{i}{\sqrt{3}}V^{T*}_{3a'}V^H_{31}\nonumber\\&&+i V^{T*}_{5a'}V^H_{41}-\frac{i}{\sqrt{2}}V^{T*}_{7a'}V^H_{51}
  +\biggr(i\sqrt{\frac{2}{3}}V^{T*}_{6a'}+\frac{V^{T*}_{7a'}}{\sqrt{6}} \biggr)V^H_{61}\biggr|^2F_{12}(m_{\lambda_X}, m^T_{a'}, Q) \eea
\vspace{-.8cm}\bea K_{D\bar I} &=& \sum_ {a =
  1}^{\mbox{d(D)}} \biggr| \biggr(\gamma V^D_{2a}- \bar\gamma V^D_{1a} +
  i k V^D_{3a}\biggr) V^H_{11} + \biggr( \frac{i \bar\zeta }{\sqrt{3}} V^D_{3a}-\frac{2 \eta}{\sqrt{3}} V^D_{2a} \biggr)
V^H_{21}\nonumber\\&& + \biggr(-i\zeta \sqrt{3} V^D_{3a} -
2\sqrt{3}\eta V^D_{1a}\biggr) V^H_{31} + \biggr( \zeta
V^D_{2a}-\frac{i\rho }{3} V^D_{3a} + \bar\zeta V^D_{1a}\biggr)
V^H_{51} \nonumber\\&&-\frac{1}{\sqrt{3}} \biggr(i\zeta
 V^D_{2a} -3 i\bar\zeta V^D_{1a}
+\frac {2 \rho }{3} V^D_{3a}\biggr) V^H_{61} \biggr| ^
 2 F_{12}(m^I, m^D_a, Q)
\qquad \quad  \eea  \vspace{-.4cm}
 \bea  K_{Q\bar C} &=& \sum_ {a =
  1}^{\mbox{d(C)}} \biggr| \biggr( \frac{i\bar\gamma }{\sqrt{2}} U^C_{1a}-\frac{i\gamma }{\sqrt{2}} U^C_{2a} -\frac{k}{\sqrt{2}} U^C_{3a}\biggr) V^H_{11} + \biggr(\sqrt{\frac{2}{3}} i \eta U^C_{2a} -
\frac{\bar\zeta}{\sqrt{6}} U^C_{3a}\biggr) V^H_{21}\nonumber\\&& -
\biggr(\frac{\zeta}{\sqrt{6}} U^C_{3a}+ \sqrt{\frac{2}{3}} i \eta
U^C_{1a}\biggr) V^H_{31} + \biggr( \frac {i\zeta }{\sqrt{2}}
U^C_{2a}-\frac {\rho}{3 \sqrt{2}} U^C_{3a} + \frac {i\bar\zeta
}{\sqrt{2}} U^C_{1a}\biggr) V^H_{51} \nonumber\\&& + \biggr(\frac
{\zeta }{\sqrt{6}} U^C_{2a} + \frac {\bar\zeta }{\sqrt{6}}
U^C_{1a}\biggr)V^H_{61} \biggr| ^2 F_{12}(m^Q, m^C_a, Q)\eea

\bea K_{T\bar E} & = &\sum_ {a =1}^{\mbox{d(T)}}\sum_ {a'
=1}^{\mbox{d(E)}} \biggr| \biggr(\gamma
   V^T_{5a} U^E_{1a'} -\gamma V^T_{3a} U^E_{4a'} -\bar\gamma V^T_{2a} U^E_{4a'} -\bar\gamma
   V^T_{5a} U^E_{2a'} -i\bar\gamma V^T_{4a} U^E_{3a'}
    +k V^T_{6a} U^E_{3a'} \nonumber \\ &&+ ik V^T_{5a} U^E_{6a'}-
   \frac{ik }{\sqrt{2}} V^T_{7a} U^E_{4a'}\biggr) V^H_{11} +\biggr( 2\sqrt{\frac{2}{3}}\eta V^T_{3a} U^E_{3a'} +
   2 \sqrt{3}\eta V^T_{5a} U^E_{1a'}+
   i\bar\gamma \sqrt{\frac{2}{3}} V^T_{1a} U^E_{3a'}\nonumber \\ &&+
   \frac {\bar\gamma}{\sqrt{3 } } V^T_{1a} U^E_{4a'} - \frac {\bar\zeta}{\sqrt{3 } }V^T_{6a} U^E_{3a'}  -i \sqrt{3}\bar\zeta V^T_{5a} U^E_{6a'} +
   \frac{2 i\bar\zeta }{\sqrt{3}} V^T_{7a} U^E_{3a'}-
   \frac{i\bar\zeta }{\sqrt{6}} V^T_{7a} U^E_{4a'} \nonumber \\ && +
   \sqrt{\frac{2}{3}}\bar\zeta V^T_{6a} U^E_{4a'}\biggr) V^H_{21} +\biggr( \frac{i\zeta}{\sqrt{6}} V^T_{7a} U^E_{4a'}-\frac {\zeta}{\sqrt{3 } } V^T_{6a}U^E_{3a'} -
   \sqrt{\frac{2}{3}}\zeta V^T_{6a} U^E_{4a'} +
   \frac{i}{\sqrt{3}} \zeta V^T_{5a} U^E_{6a'}\nonumber \\ &&
    + \sqrt{\frac{2}{3}}i\gamma V^T_{1a} U^E_{3a'}-
   \frac{\gamma}{\sqrt{3}} V^T_{1a} U^E_{4a'} -2 \sqrt{\frac{2}{3}}\eta V^T_{2a} U^E_{3a'} -
   \frac{2 i\eta}{\sqrt{3}} V^T_{4a} U^E_{3a'}  +
   \frac{2\eta}{\sqrt{3}} V^T_{5a} U^E_{2a'} \nonumber \\ &&-
   2  \sqrt{\frac{2}{3}}i\eta V^T_{4a} U^E_{4a'}\biggr) V^H_{31} +\biggr( 2 \eta i V^T_{3a} U^E_{2a'}- 2\eta i V^T_{2a} U^E_{1a'}
    -
   2 \sqrt{2}\eta V^T_{4a} U^E_{1a'} - \gamma V^T_{1a} U^E_{1a'} \nonumber \\ &&- \bar\gamma V^T_{1a} U^E_{2a'} -
    k V^T_{1a} U^E_{6a'}-
    \frac {\sqrt{2}\rho}{ 3} V^T_{6a} U^E_{6a'} -\frac{i\rho }{3\sqrt{2}} V^T_{7a}
 U^E_{6a'} + \sqrt{2}\zeta i V^T_{6a} U^E_{1a'} -\zeta
   V^T_{3a} U^E_{6a'} \nonumber \\ &&+
   \sqrt{2}i\bar\zeta V^T_{6a} U^E_{2a'} +\bar\zeta
   V^T_{2a} U^E_{6a'} -
   \sqrt{2} i\bar\zeta V^T_{4a} U^E_{6a'}
   - \sqrt{2}\bar\zeta V^T_{7a} U^E_{2a'}\biggr) V^H_{41} +\biggr( \zeta V^T_{5a} U^E_{1a'}\nonumber \\ && -k V^T_{1a} U^E_{4a'} - \frac{i\rho}{3 }V^T_{5a} U^E_{6a'}
   -\frac {i \rho }{3 \sqrt{2}}V^T_{7a} U^E_{4a'} + i \bar\zeta V^T_{4a} U^E_{3a'} + \bar\zeta V^T_{5a} U^E_{2a'}\biggr) V^H_{51} \nonumber \\ && +\biggr(\sqrt{\frac{2}{3}} i k V^T_{1a} U^E_{3a'} +  \sqrt{\frac{2}{3}}i\zeta V^T_{3a} U^E_{3a'} +
   \sqrt{3}i\zeta V^T_{5a} U^E_{1a'} -
    \frac {i \zeta}{\sqrt{3}}V^T_{3a} U^E_{4a'}
-\sqrt{\frac{2}{3}}i \bar\zeta V^T_{2a} U^E_{3a'}\nonumber \\ &&-
   \frac{i \bar\zeta }{\sqrt{3}}V^T_{5a} U^E_{2a'} +
   \sqrt{\frac{2}{3}} \bar\zeta V^T_{4a} U^E_{4a'} -
    \frac{ i \bar\zeta}{\sqrt{3}} V^T_{2a} U^E_{4a'}
   +\frac{ \bar\zeta }{ \sqrt{3}}V^T_{4a} U^E_{3a'} -\frac {i\rho}{3\sqrt{3}} V^T_{6a} U^E_{3a'}
   \nonumber \\ &&- \frac {\rho }{3\sqrt{3}} V^T_{7a} U^E_{3a'} - \frac {\rho }{3\sqrt{6}}V^T_{7a} U^E_{4a'}
   + \frac {2\rho }{3\sqrt{3}} V^T_{5a} U^E_{6a'}\biggr) V^H_{61} \biggr| ^2 F_{12} (m^T_a,
   m^E_{a'}, Q) \nonumber \\ && -2{g_{10}^2} \biggr|-U^{T*}_{1a}V^H_{11}+\frac{i}{\sqrt{3}}U^{T*}_{2a}V^H_{21}
  +\biggr(\frac{i}{\sqrt{3}}U^{T*}_{3a}+\sqrt{\frac{2}{3}}U^{T*}_{4a}\biggr)V^H_{31}-\frac{i}{\sqrt{2}}U^{T*}_{7a}V^H_{51}\nonumber \\ &&-\biggr(\frac{U^{T*}_{7a}}{\sqrt{6}}+i \sqrt{\frac{2}{3}}U^{T*}_{6a}\biggr) V^H_{61}\biggr|^2 F_{12}(m^T_a,
   m_{\lambda_{E}}, Q)\eea

\bea
  K_{Y\bar L} & = &\sum_{a=1}^{\mbox{d(L)}}
  \biggr|\biggr( k U^L_{2a}-i\gamma U^L_{1a} \biggr) V^H_{11} +\biggr(\frac {2 i\eta} {\sqrt{3}}  \ U^L_{1a} + \frac {\bar\zeta}
{\sqrt{3}} U^L_{2a}\biggr)
 V^H_{21} +\frac {\zeta }{\sqrt{3}}U^L_{2a}
  V^H_{31} +i \zeta U^L_{1a}
 V^H_{51} \nonumber \\ && +\biggr(\frac {\zeta }{\sqrt{3}}U^L_{1a} -
   \frac {i \rho}{3\sqrt{3}} U^L_{2a}\biggr) V^H_{61}\biggr|^2 F_{12}(m^Y, m^L _a,
   Q)\qquad
  \eea

\bea K_{W\bar B} &=& \biggr| \gamma V^H_{11} - \frac{2
\eta}{\sqrt{3}}
 V^H_{21} + \zeta V^H_{51} -
 \frac{i\zeta }{\sqrt{3}} V^H_{61} \biggr| ^2 F_{12}(m^W,m^B, Q)
\qquad\eea

\bea K_{C\bar Z} &=& \sum_ {a =
  1}^{\mbox{d(C)}} \biggr| \biggr( \bar\gamma V^C_{2a}-\gamma V^C_{1a} -
  ik V^C_{3a}\biggr) V^H_{11} + \biggr(\frac {2\eta}{ \sqrt{3}} V^C_{1a} - \frac
{i\bar\zeta }{ \sqrt{3}} V^C_{3a}\biggr) V^H_{21}\nonumber\\&& -
\biggr(\frac {i\zeta }{ \sqrt{3}}  V^C_{3a} + \frac {2 \eta}{
\sqrt{3}} V^C_{2a}\biggr) V^H_{3 1} + \biggr(\frac {i\rho }{ 3}
V^C_{3a}\nonumber\\&& - \zeta V^C_{1a} - \bar\zeta V^C_{2a}\biggr)
V^H_{51} + \biggr( \frac{i\zeta }{\sqrt{3}} V^C_{1a} +
  \frac{i\bar\zeta }{\sqrt{3}} V^C_{2a}\biggr) V^H_{61} \biggr| ^2 F_{12}(m^Z, m^C_a, Q) \eea

\bea
 K_{E\bar U} &= &\sum_ {a =
  1}^{\mbox{d(E)}} \biggr| \biggr(\frac{ i\gamma}{\sqrt{2}} V^E_{2a} - \frac{i\bar\gamma }{\sqrt{2}} V^E_{1a} +
  \frac{k}{\sqrt{2}} V^E_{6a}\biggr) V^H_{11} + \biggr(\sqrt{6} i\eta V^E_{2a} -
  \sqrt{\frac{3}{2}} \bar\zeta V^E_{6a}\biggr) V^H_{21}\nonumber\\&&+ \biggr(
  \sqrt{\frac{2}{3}} i \eta V^E_{1a} +\frac{\zeta}{\sqrt{6}} V^E_{6a} \biggr) V^H_{31}
  + \biggr( \sqrt{2} \lambda V^E_{4a} -2 \lambda  V^E_{3a}\biggr) V^H_{41} \nonumber\\&& +
  \biggr(\frac {\rho}{3 \sqrt{2}} V^E_{6a} - \frac
{i\zeta }{\sqrt{2}} V^E_{2a}- \frac{i\bar\zeta }{\sqrt{2}}
V^E_{1a}\biggr) V^H_{51} \nonumber\\&&+ \biggr(
\sqrt{\frac{3}{2}}\zeta V^E_{2a} - \frac{\bar\zeta}{\sqrt{6}}
V^E_{1a}+
   \frac{\sqrt{2}i \rho}{3\sqrt{3}} V^E_{6a}\biggr) V^H_{61}\biggr | ^2 F_{12} (m^U, m^E_a, Q)
  \nonumber\\&&-2{g_{10}^2} \biggr|\frac{-V^H_{41}}{\sqrt{2}}\biggr|^2 F_{12}(m^U, m_{\lambda_{E}}, Q) \eea

   \bea
  K_{U \bar D} &=& \sum_ {a =
  1}^{\mbox{d(D)}} \biggr| \biggr(\frac{i\gamma }{\sqrt{2}} U^D_{1a} - \frac{i \bar\gamma }{\sqrt{2}} U^D_{2a} +
  \frac{k}{\sqrt{2}} U^D_{3a}\biggr) V^H_{11} + \biggr( \frac{\bar\zeta}{\sqrt{6}} U^D
_{3a}- \sqrt{\frac{2}{3}}i\eta U^D_{1a} \biggr) V^H_{21}
\nonumber\\&&+ \biggr(-\sqrt{\frac{3}{2}} \zeta U^D_{3a} -i
\sqrt{6} \eta U^D_{2a}\biggr) V^H_{31} + \biggr(\frac{\rho}{ 3
\sqrt{2}} U^D _{3a} - \frac{i\zeta }{\sqrt{2}} U^D_{1a} -
\frac{i\bar\zeta }{\sqrt{2}} U^D_{2a}\biggr) V^H_{51}
\nonumber\\&&+ \biggr( \sqrt{\frac{3}{2}} \bar\zeta U^D_{2a}-\frac
{\zeta}{\sqrt{6}} U^D_{1a} -
   \frac{\sqrt{2} i\rho}{3 \sqrt{3}}U^D_{3a}\biggr) V^H_{61}\biggr|^2 F_{12} (m^U,
  m^D_a, Q)
\eea

\bea K_{HO} & = & \sum_{a = 2}^{\mbox{d(H)}} \biggr|
 \gamma V^H_{4a}  V^H_{11} + 2\sqrt{3}\eta V^H_{4a} V^H_{21} +\biggr(\gamma V^H_{1a} +
 2\sqrt{3}\eta  V^H_{2a} + \zeta  V^H_{5a}+ \sqrt{3} i \zeta  V^H_{6a}\biggr)  V^H_{41}
  \nonumber\\&&+ \zeta  V^H_{4a}  V^H_{51} + \sqrt{3} i \zeta  V^H_{4a}  V^H_{61} \biggr| ^ 2 F_{12} (m^O, m^H_a, Q)
  +\biggr| 2\gamma  V^H_{41}  V^H_{11} \nonumber\\&&+
  4\sqrt{3}\eta  V^H_{41}  V^H_{21} +2 \zeta  V^H_{51}  V^H_{41} +
  2  \sqrt{3} i \zeta  V^H_{61}  V^H_{41}\biggr| ^ 2 F_{11} (m^O, Q) \eea

\bea K_{\bar V \bar A} &=& \biggr|\sqrt{2} i \bar\gamma V^H_{11} +
  2 \sqrt{6} i \eta V^H_{31} - \sqrt{2} i \bar\zeta V^H_{51} - \sqrt{6} \bar\zeta V^H_{61} \biggr| ^2 F_{12}(m^V,
m^A, Q) \eea

\bea K_{K \bar X} &=& \sum_ {a = 1}^{\mbox{d(K)}} \sum_ {a' =
  1}^{\mbox{d(X)}}\biggr| \biggr(\sqrt{2} i\bar\gamma V^K_{1a} U^X_{1a'} +
  ik V^K_{2a} U^X_{2a'}\biggr) V^H_{11} \nonumber\\&&+ \biggr(
  \frac{i\bar\zeta }{\sqrt{3}} V^K_{2a} U^X_{2a'}-2\sqrt{\frac {2}{3}} i\bar\zeta V^K_{2a} U^X_{1a'} \biggr) V^H_{21} \nonumber\\&&+
  \biggr( 2 \sqrt{\frac{2}{3}} i\eta V^K_{1a} U^X_{1a'}- \frac{ i\zeta}{\sqrt{3} }
  V^K_{2a}
  U^X_{2a'}
  +
 \frac{ 4 i\eta}{\sqrt{3}} V^K_{1a} U^X_{2a'}\biggr) V^H_{31}\nonumber\\&& + \biggr( \frac{i\rho }{3} V^K_{2a} U^X_{2a'}-
  \sqrt{2}i \bar\zeta V^K_{1a} U^X_{1a'}\biggr) V^H_{51} + \biggr( \frac{\rho}{3} \sqrt{\frac{2}{3}}
  V^K_{2a}
  U^X_{1a'}
 \nonumber\\&&+ \frac{\rho}{3 \sqrt{3} } V^K_{2a} U^X_{2a'}-\frac{2
\bar\zeta}{\sqrt{3}} V^K_{1a} U^X_{2a'} -
  \sqrt{\frac{2}{3}} \bar\zeta V^K_{1a} U^X_{1a'} \biggr) V^H_{61} \biggr| ^2
F_{12} (m^K_a,
 m^X_{a'},
 Q)\nonumber\\&&-2{g_{10}^2} \biggr|-\frac{2 U^{K*}_{1a}}{\sqrt{3}}V^H_{31}+iU^{K*}_{2a}V^H_{51}
  +\frac{U^{K*}_{2a}}{\sqrt{3}}V^H_{61}\biggr|^2F_{12}(m^K_a,m_{\lambda_X},Q)
\eea

\bea K_{H\bar F} &= &\sum_ {a = 2}^{\mbox{d(H)}}\sum_ {a' =
  1}^{\mbox{d(F)}} \biggr| \biggr( \bar\gamma U^F_{2a'}  V^H_{2a} -
  i \bar\gamma U^F_{1a'} V^H_{4a} -\gamma U^F_{2a'}  V^H_{3a} + k U^F_{4a'} V^H_{4a} -
  i k U^F_{2a'}  V^H_{6a}\biggr)  V^H_{11}
 \nonumber\\&&+ \biggr( \bar\zeta U^F_{2a'}  V^H_{5a}- \sqrt{3} \bar\zeta U^F_{4a'}  V^H_{4a} +
  \frac{2i \bar\zeta }{ \sqrt{3} } U^F_{2a'}  V^H_{6a} -\frac{4 \eta }{ \sqrt{3} } U^F_{2a'}  V^H_{3a}-\bar\gamma U^F_{2a'}  V^H_{1a}\biggr)  V^H_{21}
\nonumber\\&& + \biggr(\frac{4 \eta }{ \sqrt{3} } U^F_{2a'}
V^H_{2a}-\sqrt{3} \zeta U^F_{4a'}  V^H_{4a} +\frac{2i \zeta }{
\sqrt{3} }  U^F_{2a'}  V^H_{6a} + \zeta U^F_{2a'} V^H_{5a} -2
\sqrt{3} \eta i U^F_{1a'} V^H_{4a}\nonumber\\&&+ \gamma U^F_{2a'}
V^H_{1a}\biggr) V^H_{31}
 + \biggr(2  \sqrt{3} \eta i U^F_{1a'}  V^H_{3a} + \frac{i \rho}{\sqrt {3} } U^F_{4a'}  V^H_{6a} -
  i \bar\zeta U^F_{1a'}  V^H_{5a} - \sqrt{3} \bar\zeta U^F_{1a'}  V^H_{6a}\nonumber\\&&+ i \bar\gamma U^F_{1a'}  V^H_{1a}+\sqrt{3} \zeta U^F_{4a'}  V^H_{3a} -k U^F_{4a'}  V^H_{1a}
 +
  \sqrt{3} \bar\zeta U^F_{4a'}  V^H_{2a} \biggr)  V^H_{41}\nonumber\\&&+
\biggr( i \bar\zeta U^F_{1a'}  V^H_{4a}-\bar\zeta U^F_{2a'}
V^H_{2a} - \zeta U^F_{2a'} V^H_{3a}
 +
 \frac{i \rho}{3} U^F_{2a'}  V^H_{6a} \biggr)  V^H_{51}\nonumber\\&&+\biggr( i k U^F_{2a'}  V^H_{1a}- \frac{2 i \bar\zeta }{ \sqrt{3} }  U^F_{2a'}  V^H_{2a}
- \frac{i \rho}{\sqrt {3} } U^F_{4a'}  V^H_{4a} -\frac{2 i \zeta
}{ \sqrt{3} } U^F_{2a'}  V^H_{3a} +\sqrt{3}\bar\zeta U^F_{1a'}
V^H_{4a}\nonumber\\&&-
  \frac{i \rho}{3} U^F_{2a'}  V^H_{5a}\biggr)  V^H_{61}
 | ^2 F_{12} (m^H _a,m^F_{a'}, Q)
 -\sum_ {a = 2}^{\mbox{d(H)}}2{g_{10}^2}
\biggr|i\biggr(U^{H*}_{1a}V^H_{11}\nonumber\\&&+U^{H*}_{2a}V^H_{21}+U^{H*}_{3a}V^H_{31}
  +U^{H*}_{5a}V^H_{51}+U^{H*}_{6a}V^H_{61}\biggr)\biggr|^2F_{12} (m^H _a, m_{\lambda_F}, Q)\nonumber\\&&-
2{g_{10}^2}
\biggr|i\biggr(U^{H*}_{11}V^H_{11}+U^{H*}_{21}V^H_{21}+U^{H*}_{31}V^H_{31}
  +U^{H*}_{51}V^H_{51}\nonumber\\&&+U^{H*}_{61}V^H_{61}\biggr)\biggr|^2F_{11} (m_{\lambda_F},
  Q)\qquad
\eea

\bea K_{N \bar Y} &=& \biggr| \sqrt{2} i \bar\gamma V^H_{11} -
  2 \sqrt{\frac{2}{3}}i \eta  V^H_{31} - \sqrt{2} i \bar\zeta V^H_{51} + \sqrt{\frac{2}{3}} \bar\zeta V^H_{61}\biggr | ^2 F_{12}(m^N,
 m^Y, Q)
\eea

\bea K_{Y\bar W} &=& \biggr| \bar\gamma V^H_{11} - \frac{2
\eta}{\sqrt{3}} V^H_{31} + \bar\zeta V^H_{51} +
 \frac{i \bar\zeta}{\sqrt{3}}  V^H_{61} \biggr| ^2 F_{12} (m^Y, m^W, Q) \eea

\bea K_{V\bar O} &=& \biggr| \bar\gamma V^H_{11} + 2 \sqrt{3} \eta
V^H_{31} + \bar\zeta V^H_{51} -
 \sqrt{3}i\bar\zeta V^H_{61} \biggr| ^2 F_{12}(m^V,
 m^O, Q)
\eea

\vspace{.2cm}

\bea
 K_{L\bar B} &=& \sum_ {a =
  1}^{\mbox{d(L)}} \biggr| \biggr( k V^L_{2a}-i \bar\gamma V^L_{1a} \biggr) V^H_{11} + \frac{\bar\zeta}{\sqrt{3} } V^L_{2a} V^H_{21}\nonumber\\&&
  + \biggr(\frac{ \zeta}{\sqrt{3}} V^L_{2a} + \frac
{2 i\eta }{\sqrt{3}} V^L_{1a}\biggr) V^H_{31} + i\bar\zeta
V^L_{1a} V^H_{51} \nonumber\\&&+ \biggr(
  \frac{i \rho}{3\sqrt{3}} V^L_{2a}-\frac {\bar\zeta }{\sqrt{3}} V^L_{1a} \biggr) V^H_{61}\biggr|^2 F_{12}(m^B, m^L_a, Q)) \eea

\vspace{.2cm}

\bea K_{S\bar H} & = &\sum_ {a =
  2}^{\mbox{d(H)}}\biggr | \biggr( \frac{i\bar\gamma }{\sqrt{2} } U^H_{2a} -\frac{i\gamma }{\sqrt{2} }  U^H_{3a} -
  \frac{k }{\sqrt{2}}  U^H_{6a}\biggr)  V^H_{11} - \biggr(2  \sqrt{\frac{2}{3}}i  \eta U^H_{3a} -
  \sqrt{\frac{2}{3}} \bar\zeta U^H_{6a} + \frac{i\bar\zeta }{\sqrt{2}} U^H_{5a} \nonumber\\&&+ \frac{ i\bar\gamma}{\sqrt{2} }  U^H_{1a}\biggr)
   V^H_{21} + \biggr(\sqrt{\frac{2}{3}} \zeta U^H_{6a} -  \frac {i\zeta  }{\sqrt{2} } U^H_{5a} + \frac{i\gamma }{\sqrt{2} } U^H_{1a} +
  2  \sqrt{\frac{2}{3}} i \eta U^H_{2a}\biggr)  V^H_{31}\nonumber\\&& - \sqrt{6 } i \lambda  U^H_{4a}  V^H_{41} -
  \biggr(\frac{\rho }{3 \sqrt{2}}  U^H_{6a} -
\frac{i\bar\zeta }{\sqrt{2} } U^H_{2a} - \frac{i \zeta }{\sqrt{2}
} U^H_{3a}\biggr)  V^H_{51} + \biggr(\frac{k}{\sqrt{2}} U^H_{1a} -
\sqrt{\frac{2}{3}} \bar\zeta U^H_{2a}\nonumber\\&& -
\sqrt{\frac{2}{3}} \zeta U^H_{3a} +
 \frac{\rho}{3\sqrt{2 }} U^H_{5a}\biggr)  V^H_{61} \biggr| ^2 F_{12} (m^S, m^H_a, Q)\nonumber\\&& +\biggr| \biggr(\frac{i \bar\gamma }{\sqrt{2}
}  U^H_{21} - \frac{i\gamma }{\sqrt{2} }  U^H_{31} -
\frac{k}{\sqrt{2}} U^H_{61}\biggr)  V^H_{11} - \biggr(2
\sqrt{\frac{2}{3}} i \eta U^H_{31}- \sqrt{\frac{2}{3}} \bar\zeta
U^H_{61} + \frac{i\bar\zeta }{\sqrt{2} } U^H_{51} \nonumber\\&&+
\frac{i \bar\gamma }{\sqrt{2} } U^H_{11}\biggr) V^H_{21} +
\biggr(\sqrt{\frac{2}{3}} \zeta U^H_{61} - \frac{i \zeta
}{\sqrt{2} } U^H_{51} + \frac{i\gamma }{\sqrt{2} } U^H_{11} +
  2  \sqrt{\frac{2}{3}} i \eta U^H_{21}\biggr)  V^H_{31} \nonumber\\&&- \sqrt{6}i  \lambda  U^H_{41} V^H_{41} -
  \biggr(\frac{\rho}{3 \sqrt{2}} U^H_{61} - \frac{i\bar\zeta }
{ \sqrt{2}}  U^H_{21}- \frac{i\zeta }{\sqrt{2}} U^H_{31}\biggr)
V^H_{51} \nonumber\\&&+ \biggr( \frac{k}{\sqrt{2} } U^H_{11} -
\sqrt{\frac{2}{3}} \bar\zeta U^H_{21} - \sqrt{\frac{2}{3}} \zeta
U^H_{31} +
  \frac{\rho}{3\sqrt{2 }} U^H_{51}\biggr)  V^H_{61} \biggr| ^2 F_{11} (m^S, Q) \eea

\bea K_{G\bar H} &=& \sum_ {a = 2}^{\mbox{d(H)}} \sum_ {a' =
  1}^{\mbox{d(G)}} \biggr|\biggr( \frac{\gamma } {\sqrt{2}} V^G_{3a'} U^H_{3a} -\gamma V^G_{2a'} U^H_{3a} - \sqrt{2} i\gamma
V^G_{4a'} U^H_{4a}  - \bar\gamma V^G_{2a'} U^H_{2a} -
  \frac{ \bar\gamma  } {\sqrt{2}} V^G_{3a'} U^H_{2a}\nonumber\\&&+  \frac{ ik  } {\sqrt{2}} V^G_{3a'}U^H_{6a} +
  k V^G_{1a'} U^H_{5a}\biggr) V^H_{11} + \biggr(
  2  \sqrt{\frac{2}{3}} \eta V^G_{3a'} U^H_{3a} - \frac{4 \eta  } {\sqrt{3}} V^G_{2a'} U^H_{3a} -
  2  \sqrt{6} i \eta V^G_{4a'} U^H_{4a} \nonumber\\&& - \sqrt{\frac{2}{3}} i\bar\zeta V^G_{3a'} U^H_{6a} +
  i \bar\zeta V^G_{1a'}  U^H_{6a} -
  \frac{\bar\zeta } {\sqrt{2}} V^G_{3a'} U^H_{5a}- \bar\gamma V^G_{2a'}  U^H_{1a} + \frac{ \bar\gamma  } {\sqrt{2}} V^G_{3a'}
U^H_{1a} \biggr)  V^H_{21} \nonumber\\&&- \biggr(
\sqrt{\frac{2}{3}} i\zeta V^G_{3a'} U^H_{6a} +
  i \zeta V^G_{1a'} U^H_{6a} +
 \frac{\zeta}{\sqrt{2} } V^G_{3a'}  U^H_{5a} +
  \gamma V^G_{2a'}  U^H_{1a} +
  \frac{\gamma}{\sqrt{2} } V^G_{3a'} U^H_{1a}\nonumber\\&& +
  \frac{4 \eta}{\sqrt{3} } V^G_{2a'} U^H_{2a} +
  2 \sqrt{\frac{2}{3}} \eta V^G_{3a'} U^H_{2a}\biggr)V^H_{31} + \biggr( \sqrt{6}  \lambda V^G_{3a'} U^H_{4a}-2 \sqrt{3} \lambda V^G_{2a'}
U^H_{4a} \nonumber\\&&- \sqrt{2} i \bar\zeta V^G_{5a'} U^H_{5a}
-\sqrt{6} \bar\zeta V^G_{5a'} U^H_{6a} + \sqrt{2}i \bar\gamma
V^G_{5a'} U^H_{1a} +
  2  \sqrt{6}i  \eta  V^G_{5a'} U^H_{3a}\biggr)  V^H_{41} \nonumber\\&&+ \biggr(
  k V^G_{1a'} U^H_{1a}-\frac{i\rho }{3 \sqrt{2}} V^G_{3a'} U^H_{6a} + \frac{\bar\zeta} {\sqrt{2} } V^G_{3a'}
U^H_{2a}
 + \frac{\zeta}{\sqrt{2} } V^G_{3a'} U^H_{3a} +
 \sqrt{2} i \zeta V^G_{4a'} U^H_{4a} \biggr)  V^H_{51} \nonumber\\&&
 + \biggr(\sqrt{\frac{2}{3}} \bar\zeta i V^G_{3a'} U^H_{2a} + i\bar\zeta  V^G_{1a'} U^H_{2a}-\frac{2  \rho}{3 \sqrt{3} } V^G_{2a'} U^H_{6a}
 - \frac{ ik}{\sqrt{2}}  V^G_{3a'} U^H_{1a}
  +
 \sqrt{\frac{2}{3}}i \zeta V^G_{3a'} U^H_{3a}  \nonumber\\&&-
  i \zeta V^G_{1a'} U^H_{3a} -
 \sqrt{6} \zeta V^G_{4a'} U^H_{4a} + \frac{i\rho }{3  \sqrt{2} }
  V^G_{3a'}  U^H_{5a}\biggr)  V^H_{61}\biggr| ^ 2 F_{12} (m^H_a, m^G_{a'}, Q) \nonumber\\&&
 -\sum_ {a = 2}^{\mbox{d(H)}} 2{g_{10}^2} \biggr|\frac{i}{\sqrt{5}}\biggr(V^{H*}_{1a}V^H_{11}+V^{H*}_{2a}V^H_{21}+V^{H*}_{3a}V^H_{31}-4 V^{H*}_{4a}V^H_{41}
  +V^{H*}_{5a}V^H_{51}\nonumber\\&&+V^{H*}_{6a}V^H_{61}\biggr)\biggr|^2F_{12} (m^H_a, m_{\lambda_G}, Q)
+\sum_ {a' =
 1}^{\mbox{d(G)}} \biggr|\biggr( \frac{\gamma } {\sqrt{2}} V^G_{3a'} U^H_{31} -\gamma V^G_{2a'} U^H_{31} - \sqrt{2} i\gamma
V^G_{4a'} U^H_{41} \nonumber\\&& - \bar\gamma V^G_{2a'} U^H_{21} -
  \frac{ \bar\gamma  } {\sqrt{2}} V^G_{3a'} U^H_{21}+  \frac{ i k  } {\sqrt{2}} V^G_{3a'}U^H_{61} +
  k V^G_{1a'} U^H_{51}\biggr) V^H_{11} + \biggr(
  2  \sqrt{\frac{2}{3}} \eta V^G_{3a'} U^H_{31} \nonumber\\&&- \frac{4 \eta  } {\sqrt{3}} V^G_{2a'} U^H_{31} -
  2  \sqrt{6} i \eta V^G_{4a'} U^H_{41} - \sqrt{\frac{2}{3}} i\bar\zeta V^G_{3a'} U^H_{61} +
  i \bar\zeta V^G_{1a'}  U^H_{61}-
  \frac{\bar\zeta } {\sqrt{2}} V^G_{3a'} U^H_{51} \nonumber\\&&- \bar\gamma V^G_{2a'}  U^H_{11} + \frac{ \bar\gamma  } {\sqrt{2}} V^G_{3a'}
U^H_{11} \biggr)  V^H_{21} - \biggr( \sqrt{\frac{2}{3}} i\zeta
V^G_{3a'} U^H_{61} +
  i \zeta V^G_{1a'} U^H_{61} +
 \frac{\zeta}{\sqrt{2} } V^G_{3a'}  U^H_{51} \nonumber\\&&+
  \gamma V^G_{2a'}  U^H_{11} +
  \frac{\gamma}{\sqrt{2} } V^G_{3a'} U^H_{11} +
  \frac{4 \eta}{\sqrt{3} } V^G_{2a'} U^H_{21} +
  2 \sqrt{\frac{2}{3}} \eta V^G_{3a'} U^H_{21}\biggr)V^H_{31} + \biggr( \sqrt{6}  \lambda V^G_{3a'} U^H_{41}\nonumber\\&&-2 \sqrt{3} \lambda V^G_{2a'}
U^H_{41} - \sqrt{2} i \bar\zeta V^G_{5a'} U^H_{51} -\sqrt{6}
\bar\zeta V^G_{5a'} U^H_{61} + \sqrt{2}i \bar\gamma V^G_{5a'}
U^H_{1a} \nonumber\\&&+
  2  \sqrt{6}i  \eta  V^G_{5a'} U^H_{31}\biggr)  V^H_{41} + \biggr(
  k V^G_{1a'} U^H_{11}-\frac{i\rho }{3 \sqrt{2}} V^G_{3a'} U^H_{61}  + \frac{\bar\zeta} {\sqrt{2} } V^G_{3a'}
U^H_{21}
 \nonumber\\&& + \frac{\zeta}{\sqrt{2} } V^G_{3a'} U^H_{31} +
 \sqrt{2} i \zeta V^G_{4a'} U^H_{41} \biggr)  V^H_{51}+
 \biggr(\sqrt{\frac{2}{3}} \bar\zeta i V^G_{3a'} U^H_{21}
 + i\bar\zeta  V^G_{1a'} U^H_{21}-\frac{2  \rho}{3 \sqrt{3} } V^G_{2a'} U^H_{61}\nonumber\eea \bea &&
- \frac{ ik}{\sqrt{2}}  V^G_{3a'} U^H_{11}
  +
 \sqrt{\frac{2}{3}}i \zeta V^G_{3a'} U^H_{31} -
  i \zeta V^G_{1a'} U^H_{31} -
 \sqrt{6} \zeta V^G_{4a'} U^H_{41}\nonumber\\&& + \frac{i\rho }{3  \sqrt{2} }
  V^G_{3a'}  U^H_{51}\biggr)  V^H_{61}\biggr| ^ 2 F_{11} (m^G_{a'}, Q)
  -2{g_{10}^2} \biggr|\frac{i}{\sqrt{5}}\biggr(V^{H*}_{11}V^H_{11}+V^{H*}_{21}V^H_{21}+V^{H*}_{31}V^H_{31}\nonumber\\&&-4 V^{H*}_{41}V^H_{41}
  +V^{H*}_{51}V^H_{51}+V^{H*}_{61}V^H_{61}\biggr)\biggr|^2F_{11} ( m_{\lambda_G}, Q)
\eea \vspace{.3cm}For the ${\overline H}[1,2,-1]$ line we have
 \bea (16 \pi^2){\cal
K}_{\overline H} &=&8K_{R C}+
 + 9K_{U\bar E}+ 9K_{P \bar X}+3K_{T\bar X}+ 6K_{M\bar B}
 3K_{D\bar J} + 9K_{E\bar P} + K_{\bar V\bar F}\nonumber\\&&
  +3K_{J\bar E}
  + 6K_{L\bar Y}
  + 8K_{Z\bar C}
   +
 3K_{E\bar T}+ 3K_{I\bar D}
  + 3K_{SH}
+
 24K_{Q C}
+ 9K_{D\bar U}\nonumber\\&&
 + 6K_{Y\bar N}
  + K_{F\bar H}
   + 3K_{X \bar K}
+ K_{ V A}
 + 6K_{B\bar L}
+ 18K_{B\bar W}
  + 3K_{\bar H\bar O}
 \nonumber\\&&
 + 18K_{W\bar Y}
 + 3K_{\bar V O}
+ K_{GH}\eea

\bea K_{RC} &=&\sum_ {a = 1}^{\mbox{d(R)}} \sum_ {a' =
  1}^{\mbox{d(C)}} \biggr|\biggr(\gamma V^R_ {1 a}V^C_ {1 a'} +
    \frac{ \gamma }{\sqrt{2}}V^R_ {2 a}V^C_ {1 a'} + \bar \gamma V^
   R_ {1 a}V^C_ {2 a'} -\frac{\bar \gamma }{\sqrt{2}}V^R_ {2 a} V^C_ {2 a'} + \frac{i k}{\sqrt {2}}V^
   R_ {2 a} V^C_ {3 a'}\biggr) U^
   H_ {11} \nonumber \\ && + \biggr( \frac{i\bar\zeta}{\sqrt {6}} V^R_ {2 a} V^
   C_ {3 a'} -\frac{2\eta}{\sqrt {3}} V^R_ {1 a} V^
   C_ {1 a'} -\sqrt {\frac{2}{3}} \eta V^R_ {2 a} V^
   C_ {1 a'}\biggr) U^
   H_ {21} \nonumber \\ &&+ \biggr(
   \sqrt {\frac{2}{3}}\eta V^R_ {2 a}V^C_ {2 a'} -\frac{2\eta}{\sqrt {3}} V^R_ {1 a}V^C_ {2 a'}  + \frac{i\zeta}{\sqrt {6}} V^
   R_ {2 a} V^C_ {3 a'}\biggr) U^
   H_ {31} \nonumber \\ &&+ \biggr(  \frac{\zeta}{\sqrt{2}}V^R_ {2 a}V^C_ {1 a'} +
    \frac{ \bar\zeta}{\sqrt{2}}V^R_ {2 a}V^C_ {2 a'}-\frac{i \rho}{3\sqrt {2}} V^R_ {2 a} V^
   C_ {3 a'}\biggr) U^
   H_ {51} \nonumber \\ &&- \biggr(\frac{\rho}{3\sqrt {3}} V^R_ {1 a} V^
    C_ {3 a'}  + \frac{i\zeta}{\sqrt {6}} V^R_ {2 a}V^C_ {1 a'}+
    \frac{i\bar\zeta}{\sqrt {6}} V^R_ {2 a} V^C_ {2 a'}\biggr) U^
   H_ {61}\biggr|^2 F_{12} (m^R_a, m^C_{a'},Q)
\eea

 \bea
  K_{U\bar E} &=&\sum_ {a = 1}^{\mbox{d(E)}}\biggr|\biggr( i\bar\gamma U^E_ {2 a}-i \gamma U^E_ {1 a}
  - k U^E_ {6 a} \biggr) \frac{U^H_ {11}}{\sqrt{2}} + \biggr( 2i\eta U^
   E_ {1 a}-\bar\zeta U^
   E_ {6 a} \biggr)\frac{U^H_ {21}}{\sqrt {6}}\nonumber \\ && + \biggr( 6i \eta U^
   E_{2 a} +3 \zeta U^E_ {6 a}\biggr) \frac{U^H_ {31}}{\sqrt {6}}  +
    \biggr(
   2 \lambda U^E_ {3 a}-\sqrt {2} \lambda U^E_ {4 a} \biggr) U^
   H_ {41} \nonumber \\ &&+ \biggr(i \zeta U^
   E_ {1 a}- \frac{\rho}{3} U^E_ {6 a} + i \bar\zeta U^E_ {2 a}\biggr) \frac{U^
   H_ {51}}{\sqrt {2}} + \biggr(\zeta U^E_{1 a}-3 \bar\zeta U^
   E_ {2 a} +\frac{2i\rho}{3} U^
  E_ {6 a}\biggr) \frac{U^H_ {61}}{\sqrt {6}}\biggr|^2 \nonumber \\ && \times F_{12} (m^U,m^E_a,Q)-2{g_{10}^2} \biggr|\frac{ U^H_{41}}{\sqrt{2}}
\biggr|^2F_{12}(m^U, m_{\lambda_{E}}, Q)
  \eea

\bea K_{P\bar X} &=&\sum_ {a = 1}^{\mbox{d(P)}} \sum_ {a' =
  1}^{\mbox{d(X)}} \biggr|\biggr(
 \frac{k}{\sqrt{2}}V^P_{2 a} U^X_{2 a'}-\gamma V^P_{1a}U^X_{1a'} \biggr) U^
  H_{11} + \biggr( \frac{2\sqrt{2}\eta}{\sqrt{3}} V^P_{1 a} U^X _{2 a'} -\frac{2\eta}{\sqrt{3}}
  V^P_{1 a} U^X_{1 a'} \nonumber \\ &&-
  \frac{\bar \zeta}{\sqrt{6}} V^P_{2 a} U^X_{2 a'}\biggr) U^H_{21} +
 \frac{\zeta}{ \sqrt{3}} V^
  P_{2 a} \biggr(2 U^X_{1 a'} + \frac{U^X_{2 a'}}{\sqrt {2}}\biggr) U^ H_{31}\nonumber \\ && - \biggr(\frac{\rho}{3\sqrt {2}} V^P_{2 a} U^X _{2 a'} +
 \zeta V^P_{1 a} U^X_{1 a'}\biggr) U^
  H _{51}  + \biggr( \frac{i\sqrt{2}\zeta}{\sqrt{3}}V^P_{1 a} U^X_{2 a'}-\frac{i\zeta}{\sqrt{3}}V^P_{1 a} U^X_{1 a'}\nonumber \\ && +
  \frac{i\rho}{3 \sqrt {3}}V^ P_{2 a} U^X _{1 a'}- \frac{i\rho}{3 \sqrt {6}}V^
  P_{2 a}U^X _{2 a'}\biggr ) U^H_{61}\biggr|^ 2 F_{12} (m^P_a, m^X_{a'},
  Q) \nonumber\\&&
  -2{g_{10}^2} \biggr|i \sqrt{\frac{2}{3}} U^{P*}_{1a}U^H_{21}+\frac{U^{P*}_{2a}}{\sqrt{2}}U^H_{51}
  +\frac{i}{\sqrt{6}}U^{P*}_{2a}U^H_{61}\biggr|^2 F_{12} (m^P_a, m_{\lambda_X}, Q) \eea

\bea K_{T\bar X} &=&\sum_ {a' = 1}^{\mbox{d(X)}} \sum_ {a =
1}^{\mbox{d(T)}}\biggr|\biggr(\gamma U^X_ {2 a'} V^T_ {3 a} -
i\bar \gamma U^X_ {1 a'} V^T_ {4 a} + \bar\gamma U^X_ {2 a'} V^T_
{2 a} -\frac{i k }{\sqrt {2}} U^X_ {2 a'} V^T_ {7 a} -k U^X_ {1
a'} V^T_ {6 a}\biggr) U^H_{11}\nonumber \\ && + \biggr( \frac{\bar
\zeta}{\sqrt {3}} U^X_ {1 a'} V^T_ {6 a} + \frac{2i\bar
\zeta}{\sqrt {3}} U^X_ {1 a'}V^T_ {7 a} - \frac{i\bar\zeta}{\sqrt
{6}} V^T_ {7 a}U^X_ {2 a'} - \sqrt{\frac{2}{3}}\bar\zeta U^X_ {2
a'}V^T_ {6 a}-\frac{\sqrt {2}i\bar \gamma}{\sqrt {3}} U^X_ {1
a'}V^T_ {1 a}\nonumber \\ && -\frac{\bar \gamma}{\sqrt {3}} U^X_
{2 a'} V^T_ {1 a} -\frac{4\eta}{\sqrt {6}}U^X_ {1 a'} V^T_ {3 a}
\biggr) U^ H_ {21} + \biggr( \frac{\gamma}{\sqrt {3}} U^X_ {2 a'}
V^T_ {1 a}-\frac{i\sqrt {2}\gamma}{\sqrt {3}} U^X_ {1 a'} V^T_ {1
a} + 2\sqrt{\frac{2}{3}}\eta U^X_ {1 a'}V^T_ {2 a} \nonumber \\
&&- 2\sqrt{\frac{2}{3}}i \eta U^X_ {2 a'}V^T_ {4 a} -
\frac{2i\eta}{\sqrt{3}} U^X_ {1 a'}V^T_ {4 a} + \frac{\zeta}{\sqrt
{3}}U^X_ {1 a'} V^T_ {6 a} + \frac{i\zeta}{\sqrt {6}} U^X_ {2 a'}
V^T_ {7 a} + \frac{\sqrt {2}\zeta}{\sqrt {3}}U^X_ {2 a'} V^T_ {6
a}\biggr) U^H_ {31} \nonumber \\ &&+ 2 i\lambda \biggr(U^X_ {2 a'}
+ \sqrt {2} U^X_ {1 a'}\biggr) V^T_ {5 a} U^H_ {41} + \biggr(k
U^X_ {2 a'}V^T_ {1 a} - \frac{i\rho}{3\sqrt {2}} U^ X_ {2 a'} V^T_
{7 a} + i \bar\zeta U^X_ {1 a'} V^T_ {4 a} \biggr) U^ H_ {51}
\nonumber \\ &&+ \biggr( \frac{i\sqrt {2}\bar\zeta}{\sqrt {3}}
U^X_ {1 a'}V^T_ {2 a} + \frac{\bar\zeta}{\sqrt {3}} U^X_ {1
a'}V^T_ {4 a} + \frac{\sqrt {2}\bar\zeta}{\sqrt {3}} U^ X_ {2 a'}
V^T_ {4 a}+ \frac{i\bar\zeta}{\sqrt {3}} U^ X_ {2 a'}V^T_ {2
a}-\sqrt{\frac{2}{3}}i k U^X_ {1 a'}V^T_ {1 a}\nonumber \\ && +
\frac{i \zeta}{\sqrt {3}} U^X_ {2 a'}V^ T_ {3 a} - \frac{\sqrt
{2}i \zeta }{\sqrt {3}}
 U^X_ {1 a'}V^ T_ {3 a}
-\frac{\rho}{3\sqrt {3}}U^X_ {1 a'} V^T_ {7 a} -
\frac{\rho}{3\sqrt {6}} U^X_ {2 a'} V^T_ {7 a}\nonumber \\ && +
\frac{i\rho}{3\sqrt {3}} U^X_ {1 a'} V^T_ {6 a} \biggr) U^H_
{61}\biggr|^2 F_{12} (m^X_{a'},m^T_a,Q)\nonumber \\ &&
 -2{g_{10}^2} \biggr|U^{T*}_{1a}U^H_{11}-\frac{i}{\sqrt{3}}U^{T*}_{2a}U^H_{21}+\biggr(\sqrt{\frac{2}{3}}U^{T*}_{4a}
 -\frac{i U^{T*}_{3a}}{\sqrt{3}}\biggr)U^H_{31}\nonumber\\&&- iU^{T*}_{5a}U^H_{41}-\frac{i}{\sqrt{2}}U^{T*}_{7a}U^H_{51}
 +\biggr(i\sqrt{\frac{2}{3}}U^{T*}_{6a}-\frac{U^{T*}_{7a}}{\sqrt{6}} \biggr)U^H_{61}\biggr|^2F_{12} (m_{\lambda_X}, m^T_a, Q) \eea

\bea
  K_{M\bar B} &=&\biggr|- \sqrt{2}i\bar \gamma U^H_ {11} + 2\sqrt{\frac{2}{3}}i\eta U^H_
  {31}+\sqrt{2} i\bar\zeta U^H_ {51} -\sqrt{\frac{2}{3}}\bar \zeta U^
  H_ {61}\biggr|^2 F_{12} (m^M,m^B,Q)
  ~~~~~~~~~\eea

\bea K_{D \bar J} &=&\sum_ {a = 1}^{\mbox{d(D)}} \sum_ {a' =
  1}^{\mbox{d(J)}} \biggr|\biggr( \frac{\bar \gamma}{\sqrt{2}} V^D_ {1 a}U^J_ {3 a'}- \bar \gamma V^D_ {1 a}U^
   J_ {2 a'} - \gamma V^D_ {2 a}U^J_ {2 a'} - \frac{\gamma }{\sqrt{2}}V^D_ {2 a}U^J_ {3a'} - \frac{i k}{\sqrt{2}} V^
   D_ {3 a}U^J_ {3a'} \biggr) U^
   H_ {11} \nonumber \\ &&+ \biggr( \sqrt{\frac{2}{3}}\eta V^
   D_ {2 a}U^J_ {3 a'}- \frac{2\eta}{\sqrt{3}}V^
   D_ {2 a}U^J_ {2 a'} - \frac{2i\bar\zeta}{\sqrt {3}} V^D_ {3 a}U^
   J_ {2 a'} - \frac{i\bar\zeta}{\sqrt {6}}V^D_ {3 a}U^J_ {3 a'}\biggr) U^
   H_ {21}\nonumber \\ && + \biggr(
   \sqrt {6}\eta V^ D_ {1 a} U^J_ {3 a'}-\frac{2\eta}{\sqrt {3}} V^ D_ {1 a}U^J_ {2 a'}-
    \frac{2i\zeta}{\sqrt {3}} V^ D_ {3 a} U^J_ {2 a'} + \sqrt {\frac{3}{2}} i\zeta V^ D_ {3 a}U^J_ {3 a'}\biggr) U^
   H_ {31} \nonumber \\ &&+ \biggr(\frac{i \rho}{3} V^D_ {3 a}U^J_ {5 a'}-4\eta V^D_ {1 a} U^
   J_ {1 a'} -
   2\zeta V^D_ {2 a}U^J_ {5 a'} + 2i \zeta V^
   D_ {3 a}U^J_ {1 a'}\biggr) U^
   H_ {41} \nonumber \\ &&+ \biggr(\frac{i\rho}{3} V^
   D_ {3 a} - \zeta V^D_ {2 a} - \bar\zeta V^
   D_ {1 a}\biggr) \frac{U^J_ {3 a'}}{\sqrt {2}} U^
   H_ {51} + \biggr( \frac{2i\zeta}{\sqrt {3}} V^D_ {2 a} U^J_ {2 a'}+
    \frac {i\zeta }{\sqrt {6}}V^D_ {2 a}U^J_ {3 a'}  -\sqrt {\frac{3}{2}}i\bar\zeta V^ D_ {1 a}U^J_ {3 a'} \nonumber \\ &&+\frac{2i\bar\zeta }{\sqrt {3}}
   V^ D_ {1 a} U^J_ {2 a'} -\frac{\rho}{3\sqrt {3}}V^D_ {3 a}U^
   J_ {2 a'} + \frac{\sqrt {2}\rho}{3\sqrt {3}} V^D_ {3 a} U^J_ {3 a'}\biggr) U^
  H_{61}|^2 F_{12} (m^D_a, m^J_{a'},Q)\nonumber \\ &&
  -2{g_{10}^2} \biggr|\frac{2i}{\sqrt{3}}\biggr(U^{D*}_{2a}U^H_{21}+U^{D*}_{1a}U^H_{31}
  +U^{D*}_{3a'}U^H_{61}\biggr)\biggr|^2 F_{12}(m^D_a, m_{\lambda_{J}},Q)
\eea

\bea K_{E\bar P} &=&\sum_ {a = 1}^{\mbox{d(E)}} \sum_ {a' =
1}^{\mbox{d(P)}}\biggr|\biggr( \frac{k}{\sqrt{2}} V^ E_ {4 a}
 U^P_ {2 a'}-\bar \gamma V^E_ {3 a}U^P_ {1 a'} \biggr) U^ H_ {11} +
\biggr(\frac{2\bar\zeta}{\sqrt {3}} V^E_ {3 a} U^ P_ {2 a'} +
\frac{\bar \zeta}{\sqrt {6}}V^E_ {4 a} U^ P_ {2 a'}\biggr) U^H_
{21} \nonumber \\ &&+ \biggr( \frac{2\sqrt{2} \eta}{\sqrt {3}}V^E_
{4 a} U^P_ {1 a'} -\frac{2 \eta}{\sqrt {3}}V^E_ {3 a}U^P_ {1 a'} -
\frac{ \zeta}{\sqrt {6}}V^E_ {4 a} U^P_ {2 a'} \biggr) U^ H_
{31}\nonumber \\ &&+ \biggr(2\sqrt {2}i \eta V^ E_ {2 a} U^P_ {1
a'} + \frac{\rho}{3\sqrt {2}}V^ E_ {6 a} U^P_ {2 a'}  + i\sqrt {2}
\bar\zeta V^ E_ {1 a} U^P_ {2 a'} +\sqrt {2} \bar\zeta V^E_ {6 a}
U^P_ {1 a'}\biggr) U^ H_ {41} \nonumber \\
&&- \biggr(\frac{\rho}{3\sqrt {2}} V^E_ {4 a} U^P_ {2 a'}+ \bar
\zeta V^E_ {3 a} U^P_ {1 a'}\biggr) U^ H_ {51} +
\biggr(\frac{i\bar \zeta}{\sqrt {3}} V^E_ {3 a}U^P_ {1
a'}-\frac{i\sqrt{2}\bar \zeta}{\sqrt {3}}V^ E_ {4 a}U^P_ {1 a'}
\nonumber \\
&&-\frac{i\rho}{ 3\sqrt{3}}V^E_ {3 a} U^ P_ {2 a'} +\frac{i\rho}{
3\sqrt{6}} V^E_ {4 a} U^ P_ {2 a'}\biggr) U^ H_ {61}\biggr|^2
F_{12} (m^E_a, m^P_{a'}, Q) \nonumber\\&&
  -2{g_{10}^2} \biggr|-i \sqrt{\frac{2}{3}}V^{P*}_{1a'}U^H_{31}+\frac{V^{P*}_{2a'}}{\sqrt{2}}U^H_{51}
  -\frac{i}{\sqrt{6}}V^{P*}_{2a'}U^H_{61}\biggr|^2 F_{12} ( m_{\lambda_E},m^P_{a'}, Q) \eea

\bea
 K_{\bar V \bar F} &=&\sum_ {a = 1}^{\mbox{d(F)}}\biggr|-\biggr(i \bar \gamma U^F_ {1 a} + k U^F_ {4
a}\biggr) U^H_ {11} + \sqrt {3}
   \bar\zeta U^F_ {4a}U^H_ {21} + \biggr( \sqrt{3} \zeta U^F_ {4a}-2 \sqrt{3}i\eta U^F_ {1 a} \biggr)
U^H_ {31}\nonumber \\ && +2 \sqrt {3} \lambda U^F_ {2a} U^H_ {41}
+
   i \bar\zeta U^F_ {1a}U^H_ {51} +\biggr(\frac{i \rho}{
  \sqrt {3}} U^F_ {4a} + \sqrt{3} \bar\zeta U^F_ {1a}\biggr) U^H_ {6
 1}\biggr|^2 F_{12}(m^V, m^F_{a}, Q)\nonumber \\ &&-2{g_{10}^2} \biggr|i U^H_{41}
\biggr|^2F_{12}(m^V, m_{\lambda_F}, Q) \quad \quad \eea

\bea K_{J \bar E} &=&\sum_ {a = 1}^{\mbox{d(J)}} \sum_ {a' =
  1}^{\mbox{d(E)}} \biggr|\biggr( \frac{\bar \gamma}{\sqrt {2}} V^J_ {3 a}U^E_ {2 a'}- i\sqrt {2}\bar \gamma U^E_ {3 a'} V^J_ {1 a} -\bar \gamma U^E_ {2 a'}V^J_ {2
a} - \gamma V^J_ {2 a} U^E_ {1 a'} - \frac{\gamma}{\sqrt {2}} V^J_
{3 a}U^E_ {1 a'}
  \nonumber \\ &&  -\frac{i k }{\sqrt {2}}V^ J_ {3 a}U^E_ {6 a'}
-i k V^J_ {5 a}U^E_ {4 a'}\biggr) U^H_ {11} + \biggr( \sqrt
{\frac{3}{2}}i\bar \zeta V^
 J_ {3 a} U^E_ {6 a'} + \frac{2i\bar \zeta}{\sqrt{3}} V^J_ {2 a} U^E_ {6 a'} +
 \frac{ 2\sqrt {2}i\bar \zeta}{\sqrt{3}}V^J_ {5 a}U^E_ {3 a'} \nonumber \\ &&- \frac{i\bar \zeta}{\sqrt{3}}
 V^J_ {5 a}U^E_ {4 a'}-\frac{2\eta}{\sqrt{3}} V^J_ {2 a} U^ E_ {1 a'}  - \sqrt {6}\eta
V^J_ {3 a} U^ E_ {1 a'}\biggr) U^H_ {21}+ \biggr( -
\frac{i\zeta}{\sqrt{6}} V^J_ {3 a} U^
 E_ {6 a'} +\frac{2i\zeta}{\sqrt{3}} V^J_ {2 a} U^E_ {6 a'} \nonumber \\ &&+ \frac{i\zeta}{\sqrt {3}} V^J_ {5 a}U^E_ {4 a'}
 - \frac{2\eta}{\sqrt {3}} V^J_ {2 a} U^
E_ {2 a'} -\sqrt {\frac{2}{3}}\eta V^J_ {3 a} U^ E_ {2 a'}-
\frac{4i \eta}{\sqrt{3}} V^J_ {1 a}U^E_ {4 a'}
 - \sqrt {\frac{8}{3}}i \eta V^J_ {1 a} U^E_ {3 a'}\biggr) U^
  H_ {31}\nonumber \\ &&+ \biggr(2\sqrt {2}i \lambda V^ J_ {2 a}U^E_ {3 a'} - \sqrt{2}i \lambda V^J_ {3 a}U^E_ {4
a'}- 2i \lambda V^J_ {3 a}U^E_ {3 a'}\biggr) U^H_ {41} \nonumber
\\ &&+ \biggr( \frac{i \rho}{3\sqrt {2}}V^J_ {3 a} U^ E_ {6 a'}
-\frac{i\rho}{3} V^J_ {5 a} U^E_ {4 a'} - \frac{\bar
\zeta}{\sqrt{2}} V^J_ {3 a}U^E_ {2 a'} + \sqrt{2}i\bar \zeta U^
 E_ {3 a'} V^J_ {1 a} -\frac{\zeta}{\sqrt{2}} V^J_ {3 a} U^E_ {1 a'}\biggr) U^
 H_ {51} \nonumber \\ && + \biggr(\frac{2}{\sqrt {3}}\bar \zeta V^J_{1 a}U^E_ {4 a'}+ \sqrt{\frac{2}{3}}\bar \zeta  V^J_{1 a}U^E_ {3 a'}
 + \frac{ i \bar \zeta}{\sqrt{6}} V^J_ {3 a} U^E_ {2 a'} -
 \frac{2i}{\sqrt {3}} \bar \zeta  V^J_ {2 a} U^E_ {2 a'}- \sqrt{\frac{3}{2}}i\zeta V^J_ {3 a}U^
  E_ {1 a'}\nonumber \\ &&-\frac{2i}{\sqrt {3}}\zeta V^J_ {2 a} U^
  E_ {1 a'} - \frac{\rho}{3 \sqrt{3}} V^J_ {5 a}U^E_ {4 a'} -\frac{\sqrt{2}\rho}{3 \sqrt{3}} V^J_ {5 a} U^E_ {3 a'} -
   \frac{\sqrt {2}\rho}{3 \sqrt{3}} V^J_ {3 a} U^E_ {6 a'}\nonumber \\ && -
  \frac{\rho}{3 \sqrt{3}} V^J_ {2 a}U^E_ {6 a'}\biggr) U^H_{61}\biggr|^2 F_{12 }(m^J_a,m^E_{a'},Q)\nonumber \\ &&
   -2{g_{10}^2} \biggr|-\frac{2i}{\sqrt{3}}V^{E*}_{1a'}U^H_{21}-\frac{2i}{\sqrt{3}}
   V^{E*}_{2a'}U^H_{31}-\sqrt{2} V^{E*}_{3a'}U^H_{41}-\frac{2i}{\sqrt{3}}V^{E*}_{6a'}U^H_{61}\biggr|^2
   F_{12}(m_{\lambda_J}, m^E _{a'}, Q) \nonumber \\ &&
   -2{g_{10}^2} \biggr|\frac{2}{\sqrt{3}}U^{J*}_{1a}U^H_{31}-iU^{J*}_{2a}U^H_{41}
   +\frac{i}{\sqrt{2}}U^{J*}_{3a}U^H_{41}-i U^{J*}_{5a}U^H_{51}\nonumber \\ &&-\frac{1}{\sqrt{3}}U^{J*}_{5a}U^H_{61}\biggr|^2
   F_{12} (m^J_a,m_{\lambda_E},
   Q)~~~~~~~~~~
 \eea

\bea
  K_{L\bar Y} &=&\sum_ {a = 1}^{\mbox{d(L)}} \biggr|-\biggr(i \bar \gamma V^L_ {1 a} + k V^L_ {2 a}\biggr) U^H_ {11}-
   \frac{\bar \zeta}{\sqrt{3}} V^L_ {2 a}U^H_ {21} + \biggr(2 i \eta V^L_ {1 a} - \zeta V^L_ {2 a}\biggr) \frac{U^
   H_ {31}}{\sqrt {3}} +
    i\bar \zeta V^
   L_ {1 a}U^H_ {51}\nonumber \\ && - \biggr(\frac{i\rho}{3\sqrt {3}} V^L_ {2 a} +
  \frac{\bar \zeta}{\sqrt {3}} V^L_ {1 a}\biggr) U^
  H_ {61}\biggr|^ 2 F_{12}(m^L_a,m_Y,Q)
\qquad\quad\eea

 \bea K_{Z \bar C} &=&\sum_ {a = 1}^{\mbox{d(C)}}\biggr| \biggr(i
k U^C_ {3 a} + \gamma U^C_ {2 a} - \bar\gamma U^C_ {1 a}\biggr) U^
   H_ {11} +\biggr( i \bar\zeta U^C_ {3 a}-2 \eta U^
   C_ {2 a} \biggr) \frac{U^H_ {21}}{\sqrt {3}} + \biggr( i\zeta U^C_ {3 a} +
   2 \eta U^C_ {1 a}\biggr) \frac{U^H_ {31}}{\sqrt {3}} \nonumber \\ &&+
   \biggr( \zeta U^C_ {2 a}-\frac{i
   \rho}{
   3} U^C_ {3 a} +
    \bar\zeta U^C_ {1 a}\biggr) U^
   H_ {51} -\frac{i}{\sqrt {3}} \biggr(\zeta U^C_ {2 a} +
  \bar\zeta U^C_ {1 a}\biggr) U^
  H_ {61}\biggr|^ 2 F_{12} (m^Z,m^C_a,Q)
  \eea

\bea
 K_{E\bar T} &=& \sum_ {a'
= 1}^{\mbox{d(T)}}\sum_ {a = 1}^{\mbox{d(E)}}\biggr|\biggr(\gamma
U^T_ {3 a'} V^E_ {4 a} - \gamma U^T_ {5 a'} V^E_ {2 a} - i\gamma
U^T_ {4 a'} V^E_ {3 a} + \bar \gamma U^T_ {2 a'} V^E_ {4 a} +\bar
\gamma U^T_ {5 a'} V^E_ {1 a} -k U^T_ {6 a'} V^E_ {3 a} \nonumber
\\ &&-i k U^T_ {5 a'} V^E_ {6 a} -\frac{i k}{\sqrt {2}}
U^T_ {7 a'} V^E_ {4 a}\biggr) U^ H_ {11} + \biggr(\frac{\bar
\gamma}{\sqrt {3}} U^ T_ {1 a'} V^E_ {4 a} -\frac{\sqrt {2}i\bar
\gamma}{\sqrt {3}} U^ T_ {1 a'} V^E_ {3a}
+\frac{2\sqrt {2}\eta}{\sqrt {3}} U^T_ {3 a'} V^E_ {3 a}\nonumber \\
&& - \frac{2i\eta}{\sqrt {3}} U^T_ {4 a'} V^E_ {3 a}
+\frac{2\eta}{\sqrt {3}} U^T_ {5 a'} V^E_ {2 a} - \frac{2\sqrt
{2}i\eta}{\sqrt {3}} U^ T_ {4 a'} V^E_ {4 a} + \frac{\bar
\zeta}{\sqrt {3}} U^T_ {6 a'} V^ E_ {3 a} + \frac{\sqrt {2}\bar
\zeta}{\sqrt {3}} U^T_ {6 a'} V^E_ {4 a}\nonumber \\ && +
\frac{i\bar\zeta}{\sqrt {6}} U^T_ {7 a'} V^E_ {4 a} -
\frac{i\bar\zeta}{\sqrt{3}} U^T_ {5 a'} V^E_ {6 a}\biggr) U^ H_
{21} + \biggr( 2\sqrt{3}\eta U^T_{5 a'} V^E_{1 a} -\frac{2\sqrt
{2}\eta}{\sqrt {3}} U^T_{2 a'} V^E_{3 a} - \frac{ \sqrt
{2}i\gamma}{\sqrt {3}}U^ T_ {1 a'} V^E_ {3 a}\nonumber \\ &&
-\frac{\gamma}{\sqrt {3}}U^T_{1 a'}V^E_ {4 a}  +
 \frac{ \zeta}{\sqrt {3}} U^T_ {6 a'}V^ E_ {3 a} -\frac{ \sqrt {2} \zeta}{\sqrt {3}} U^T_ {6 a'} V^E_
{4 a}+i\zeta\sqrt{3} U^T_ {5 a'} V^ E_ {6 a}
+\frac{2i\zeta}{\sqrt{3}} U^T_ {7 a'} V^E_ {3 a}\nonumber \\
&& - \frac{i\zeta}{\sqrt {6}}U^T_ {7 a'} V^E_ {4 a}\biggr) U^H_
{31}+ \biggr( 2 i \eta U^T_ {2 a'} V^E_ {2 a}-2 i \eta U^ T_ {3
a'} V^E_ {1 a} + 2\sqrt {2} \eta U^T_ {4 a'} V^ E_ {1 a} - \bar
\gamma U^T_ {1 a'}V^E_ {1 a} \nonumber \\
&&- \gamma U^T_ {1 a'} V^ E_ {2 a}+ k U^T_ {1 a'}V^E_ {6 a} +
 \frac{\sqrt {2}\rho}{3} U^T_ {6 a'}V^E_ {6 a} - \frac{i\rho}{3\sqrt {2}} U^T_ {7 a'} V^E_ {6 a}
+ \sqrt {2} i\zeta U^T_ {6 a'} V^E_ {2 a} \nonumber \\ &&-\zeta
U^T_ {3 a'} V^E_ {6 a} - \sqrt {2} i\zeta U^T_ {4 a'} V^E_ {6 a}+
\sqrt {2}\zeta U^T_ {7 a'} V^E_ {2 a} + \sqrt {2} i\bar\zeta U^T_
{6 a'} V^ E_ {1 a} +\bar\zeta U^T_ {2 a'} V^E_ {6 a}\biggr) U^H_
{41} \nonumber \\
&& + \biggr(k U^T_ {1 a'} V^E_ {4 a} - \frac{i\rho}{3\sqrt{2}}
U^T_ {7 a'}V^E_ {4 a} +\frac{i\rho}{3} U^T_ {5 a'} V^E_ {6 a}-
\zeta U^T_ {5 a'}V^E_ {2 a}+i\zeta U^T_ {4 a'} V^E_ {3 a} -
\bar\zeta U^T_ {5 a'}V^E_ {1 a}\biggr) U^ H_ {51}\nonumber \\ &&+
\biggr(ik \sqrt {\frac{2}{3}} U^T_ {1 a'} V^E_ {3 a} +
\frac{i\zeta}{\sqrt {3}} U^ T_ {5 a'}V^E_ {2 a} - \frac{\sqrt
{2}\zeta}{\sqrt {3}} U^T_ {4 a'} V^E_ {4 a} - \frac{i\zeta}{\sqrt
{3}} U^T_ {3 a'} V^E_ {4 a} - \frac{\zeta}{\sqrt {3}}U^T_ {4 a'}
V^E_ {3 a} \nonumber \\ &&- \frac{ \sqrt {2}i\zeta}{\sqrt {3}}
U^T_ {3 a'} V^E_ {3 a} +
 \sqrt {\frac{2}{3}}i\bar \zeta U^ T_ {2 a'} V^E_ {3 a} - \sqrt
{3} i\bar \zeta U^T_ {5 a'}V^ E_ {1 a} - \frac{i\bar \zeta }{\sqrt
{3}}U^T_ {2 a'} V^E_ {4 a}\nonumber \\ &&+ \frac{\rho}{3\sqrt {3}}
(U^T_ {7 a'}V^E_ {3 a} + U^T_ {7 a'}\frac{V^E_ {4 a}}{\sqrt {2}} +
2 U^T_ {5 a'}V^E_ {6 a} - i U^T_ {6 a'} V^ E_ {3 a})\biggr) U^H_
{61}\biggr|^2 F_{12} ( m^T_{a'},m^E_a, Q)\nonumber \\ &&
-2{g_{10}^2}
\biggr|V^{T*}_{1a'}U^H_{11}+\biggr(\frac{i}{\sqrt{3}}V^{T*}_{2a'}-\sqrt{\frac{2}{3}}V^{T*}_{4a'}\biggr)U^H_{21}
+\frac{i}{\sqrt{3}}V^{T*}_{3a'}U^H_{31}-\frac{i}{\sqrt{2}}V^{T*}_{7a'}U^H_{51}\nonumber
\\ &&+\biggr(\frac{V^{T*}_{7a'}}{\sqrt{6}}-i
\sqrt{\frac{2}{3}}V^{T*}_{6a'} \biggr)U^H_{61}\biggr|^2 F_{12} (
m^T_{a'},m_{\lambda_E}, Q) \eea

\vspace{.2cm} \bea
  K_{I \bar D} &=&\sum_ {a = 1}^{\mbox{d(D)}}\biggr|\biggr( \bar\gamma U^D_ {2 a}-i k U^D_ {3 a} - \gamma U^D_ {1 a}
  \biggr) U^H_ {11} + \sqrt {3}\biggr(i \bar\zeta U^D_ {3 a} -
   2 \eta U^D_ {1 a}\biggr) U^H_ {21}\nonumber \\ && +
   \biggr( - i \zeta U^D_ {3 a}-2 \eta U^D_ {2 a}\biggr) \frac{U^H_ {31}}{\sqrt {3}}  + \biggr(\frac{i\rho U^D_ {3 a}}{3} -
   \zeta U^D_ {1 a} - \bar\zeta U^D_ {2 a}\biggr) U^
    H_ {51}\nonumber \\ && - \biggr(\sqrt {3} i\zeta U^D_ {1 a} -\frac{i\bar\zeta U^D_ {2 a}}{\sqrt {3}}
   + \frac{2 \rho U^D_ {3 a}}{3 \sqrt{3}} \biggr) U^
  H_ {61}\biggr|^2 F_{12} (m^I,m^D_a,Q)
 \qquad \qquad ~~~~\eea

\bea
  K_{S H} &=&\sum_ {a = 2}^{\mbox{d(H)}}\biggr |\biggr(i \gamma V^H_ {3 a} - i\bar\gamma V^H_ {2 a} +
   k V^H_ {6 a}\biggr) \frac{U^H_ {11}}{\sqrt {2}} + \biggr(2\sqrt
   {\frac{2}{
   3}}i \eta V^H_ {3 a} - \sqrt {\frac{2}{3}} \bar\zeta V^
   H_ {6 a} + \frac{i\bar\zeta}{\sqrt {2}} V^
   H_ {5 a}\nonumber \\ && +\frac{i\bar\gamma}{\sqrt {2}} V^H_ {1 a}\biggr) U^
   H_ {21} + \biggr(
   \frac{i\zeta}{\sqrt {2}} V^
   H_ {5 a}-\sqrt {\frac{2}{3}} \zeta V^H_ {6 a} - \frac{i\gamma}{\sqrt {2}} V^H_ {1 a} -
   2 \sqrt {\frac{2}{3}}i \eta V^H_ {2 a}\biggr) U^
   H_ {31}\nonumber \\ && - \sqrt {6}i \lambda V^H_ {4 a} U^
   H_ {41} + \biggr(\frac{\rho}{3} V^
   H_ {6 a} - i\zeta V^H_ {3 a}-i
   \bar\zeta V^H_ {2 a}\biggr) \frac{U^
   H_ {51}}{\sqrt {2}} \nonumber \\ && + \biggr( \sqrt {\frac{2}{3}} \bar\zeta V^
   H_ {2 a}-\frac{k}{\sqrt {2}}V^
   H_ {1 a} + \sqrt {\frac{2}{3}} \zeta V^H_ {3 a} -
  \frac{\rho}{3\sqrt {2}} V^H_ {5 a}\biggr) U^
  H_ {61}\biggr|^2 F_{12} (m^H_a,m^S,Q)\nonumber\\&&+
   \biggr|\biggr(i \gamma V^H_ {31} - i\bar\gamma V^H_ {21} +
   k V^H_ {61}\biggr) \frac{U^H_ {11}}{\sqrt {2}} + \biggr(2\sqrt
   {\frac{2}{
   3}}i \eta V^H_ {31} - \sqrt {\frac{2}{3}} \bar\zeta V^
   H_ {61} + \frac{i\bar\zeta}{\sqrt {2}} V^
   H_ {51}\nonumber \\ && +\frac{i\bar\gamma}{\sqrt {2}} V^H_ {11}\biggr) U^
   H_ {21} + \biggr(
   \frac{i\zeta}{\sqrt {2}} V^
   H_ {51}-\sqrt {\frac{2}{3}} \zeta V^H_ {61} - \frac{i \gamma}{\sqrt {2}} V^H_ {11} -
   2 \sqrt {\frac{2}{3}}i \eta V^H_ {21}\biggr) U^
   H_ {31}\nonumber \\ && - \sqrt {6}i \lambda V^H_ {41} U^
   H_ {41} + \biggr(\frac{\rho}{3} V^
   H_ {61} - i\zeta V^H_ {31}-i
   \bar\zeta V^H_ {21}\biggr) \frac{U^
   H_ {51}}{\sqrt {2}}\nonumber \\ && + \biggr( \sqrt {\frac{2}{3}} \bar\zeta V^
   H_ {21} -\frac{k}{\sqrt {2}}V^
   H_ {11} + \sqrt {\frac{2}{3}} \zeta V^H_ {31} -
  \frac{\rho}{3\sqrt {2}} V^H_ {51}\biggr) U^
  H_ {61}\biggr|^2 F_{11} (m^S,Q)
  \eea
\bea
 K_{QC} &=&\sum_ {a = 1}^{\mbox{d(C)}}\biggr|\biggr(i \gamma V^C_ {1 a} + k V^C_ {3 a} - i \bar\gamma V^
   C_ {2 a}\biggr) \frac{U^H_ {11}}{\sqrt {2}} - \biggr(2i \eta V^C_ {1 a} +
   \bar\zeta V^C_ {3 a}\biggr) \frac{U^H_ {21}}{\sqrt {6}}\nonumber \\ && +
    \biggr(\zeta V^C_ {3 a} + 2i \eta V^
   C_ {2 a}\biggr) \frac{U^H_ {31}}{\sqrt {6}} + \biggr(\frac{\rho}{3} V^C_ {3 a} - i\zeta V^
   C_ {1 a} - i \bar\zeta V^C_ {2 a}\biggr) \frac{U^
  H_ {51}}{\sqrt {2}}\nonumber \\ &&- \biggr(\zeta V^C_ {1 a} +
  \bar\zeta V^C_ {2 a}\biggr) \frac{U^H_ {61}}{\sqrt {6}}\biggr|^2 F_{12}(m^Q,m^C_a,Q)
  \eea

\bea
  K_{D\bar U } & = &\sum_ {a =
  1}^{\mbox {d(D)}}\biggr | \biggr( i\bar\gamma V^D_ {1 a}-i\gamma V^D_ {2 a} - k V^
  D_ {3 a} \biggr)\frac {U^
 H_ {11}} {\sqrt {2}} + \biggr(\sqrt{\frac{3}{2}}\bar\zeta V^D_ {3 a}
 -\sqrt{6} i\eta V^D_ {2 a}\biggr)U^
 H_ {21}\nonumber \\ && - \biggr(\frac{\zeta}{\sqrt{6}} V^
  D_ {3 a}+\sqrt{\frac{2}{3}}i\eta V^D_ {1 a} \biggr)U^H_ {31} + \biggr( i\zeta V^D_ {2 a}-\frac{\rho}{3} V^D_ {3 a} +
 i\bar\zeta V^D_ {1 a}\biggr)\frac {U^H_ {51}} {\sqrt {2}} \nonumber \\ &&+ \biggr( \bar\zeta V^
  D_ {1 a} -3\zeta V^
  D_ {2 a} -\frac {2 i\rho} {3} V^
  D_ {3 a}\biggr) \frac {U^H_ {61}} {\sqrt {6}} \biggr|^2 F_ {12} (m^U, m^D_a, Q)
  \eea
\vspace{.2cm} \bea
  K_{Y\bar N} &=&\biggr|- \sqrt{2}i \gamma U^H_ {11} + 2\sqrt{\frac{2}{3}}i\eta U^H_
  {21}+\sqrt{2} i\zeta U^H_ {51} +\sqrt{\frac{2}{3}}\zeta U^
  H_ {61}\biggr|^2 F_{12} (m^M,m^B,Q)\qquad \eea

  \bea
  K_{F \bar H} &=&\sum_ {a = 1}^{\mbox{d(F)}} \sum_ {a'= 2}^{\mbox{d(H)}}
 \biggr|\biggr(i \gamma V^F_ {1 a} U^H_ {4 a'} + \gamma U^H_ {3 a'} V^
   F_ {2 a} - \bar \gamma V^F_ {2 a} U^H_ {2 a'} +
   k V^F_ {4 a} U^H_ {4 a'} + i k V^
   F_ {2 a}U^H_ {6 a'}\biggr) U^H_ {11} \nonumber \\ &&+ \biggr(\frac {4\eta}{\sqrt{3}} V^F_ {2 a} U^H_ {3 a'}
   +2\sqrt {3}i \eta V^F_ {1 a}U^H_ {4 a'} - \sqrt {3}\bar \zeta V^F_ {4 a} U^
   H_ {4 a'} - \frac{2i\bar \zeta}{\sqrt {3}} V^F_ {2 a}U^
   H_ {6 a'} - \bar \zeta V^F_ {2 a}U^H_ {5 a'} \nonumber \\ &&+ \bar \gamma V^
   F_ {2 a} U^H_ {1 a'}\biggr) U^H_ {21} -\biggr(\frac {4}{\sqrt{3}} \eta V^
   F_ {2 a}U^H_ {2 a'}  + \frac{2 i \zeta }{\sqrt {3}}V^F_ {2 a}U^H_ {6 a'} +\zeta
   V^F_ {2 a} U^H_ {5 a'} + \sqrt {3}\zeta V^F_ {4 a} U^
   H_ {4 a'} \nonumber \\ &&+ \gamma V^F_ {2 a} U^
   H_ {1 a'}\biggr) U^H_ {31} + \biggr( i \zeta V^F_ {1 a}U^H_ {5 a'} -\sqrt {3}\zeta
   V^F_ {1 a}U^H_ {6 a'} -\frac{i \rho}{\sqrt{3}} V^
   F_ {4 a} U^H_ {6 a'} \nonumber \\ &&- 2\sqrt {3}i \eta V^
   F_ {1 a} U^H_ {2 a'} - i\gamma V^
   F_ {1 a} U^H_ {1 a'} -
   k V^F_ {4 a} U^H_ {1 a'} + \sqrt {3}\bar\zeta V^
   F_ {4 a} U^
   H_ {2 a'} +\sqrt {3} \zeta V^
   F_ {4 a}U^H_ {3 a'}\biggr) U^H_ {41}\nonumber \\ && + \biggr( \bar\zeta V^F_ {2 a} U^
   H_ {2 a'}-\frac{i\rho}{3} V^F_ {2 a} U^
   H_ {6 a'} - i\zeta V^F_ {1 a} U^
   H_ {4 a'} + \zeta V^F_ {2 a} U^H_ {3 a'}\biggr) U^
   H_ {51} \nonumber \\ && + \biggr(\sqrt {\frac{4}{3}}i \zeta V^
   F_ {2 a} U^H_ {3 a'}+\sqrt {\frac{4}{3}} i \bar \zeta V^
   F_ {2 a}U^H_ {2 a'}+ \sqrt {3}\zeta  V^
   F_ {1 a}U^H_ {4 a'} +\frac{i\rho}{\sqrt{3}} V^F_ {4 a} U^
   H_ {4 a'}\nonumber\\&& + \frac{i\rho}{3} V^F_ {2 a} U^
  H_ {5 a'} - i k V^F_ {2 a} U^H_ {1 a'}\biggr) U^
   H_ {61}\biggr|^2 F_{12} (m^F_a, m^H_{a'},Q)\nonumber\\&&
- \sum_ {a'= 2}^{\mbox{d(H)}}2{g_{10}^2}
\biggr|i\biggr(V^{H*}_{1a'}U^H_{11}+V^{H*}_{2a'}U^H_{21}+V^{H*}_{3a'}U^H_{31}
  +V^{H*}_{5a'}U^H_{51}\nonumber \\ &&+V^{H*}_{6a'}U^H_{61}\biggr)\biggr|^2F_{12} (m^H_{a'}, m_{\lambda_F}, Q)-
  2{g_{10}^2} \biggr|i\biggr(V^{H*}_{11}U^H_{11}+V^{H*}_{21}U^H_{21}\nonumber \\ &&+V^{H*}_{31}U^H_{31}
+V^{H*}_{51}U^H_{51}+V^{H*}_{61}U^H_{61}\biggr)\biggr|^2F_{11} (
m_{\lambda_F}, Q) \eea

\bea
  K_{X\bar K} &=&\sum_ {a = 1}^{\mbox{d(X)}} \sum_ {a' =
  1}^{\mbox{d(K)}} \biggr| -\biggr(i \gamma \sqrt{2}V^X_ {1 a} U^K_{1 a'} +i
   k V^X_ {2 a}U^K_{2 a'}\biggr) U^
  H_ {11} - \biggr( \frac{2 \sqrt{2}i\eta}{\sqrt{3}} V^X_ {1 a}U^K_{1 a'} +\frac{4i\eta}{\sqrt{3}}
   V^X_ {2 a} U^K_{1 a'} \nonumber \\ && -
  \frac{i\bar \zeta}{\sqrt {3}} V^X_ {2 a}U^K_{2 a'} \biggr) U^
  H_ {21} + \frac{i\zeta}{\sqrt {3}} \biggr(2\sqrt{2} V^X_ {1 a} -
   V^X_ {2 a}\biggr) U^K_{2 a'} U^
  H_ {31} + \biggr(
  \sqrt{2} i \zeta V^X_ {1 a} U^K_{1 a'}\nonumber \\ &&-\frac{i\rho}{3} V^X_ {2 a} U^K_{2 a'}\biggr) U^
  H_ {51}  + \biggr(\frac{\rho}{3\sqrt {3}}V^X_ {2 a}U^ K_{2 a'} +
   \frac{\sqrt {2}\rho}{3\sqrt {3}}V^X_ {1 a} U^ K_{2 a'}- \frac{\sqrt{2}\zeta}{\sqrt{3}}V^X_ {1 a} U^
   K_{1 a'} \nonumber\\&&- \frac{2\zeta}{\sqrt{3}} V^X_ {2 a}U^
   K_{1 a'} \biggr) U^
  H_ {61}\biggr|^2 F_{12} (m^X _{a},m^K _{a'}, Q)\nonumber\\&&-2{g_{10}^2} \biggr|\frac{-2}{\sqrt{3}}V^{K*}_{1a'}U^H_{21}
  -iV^{K*}_{2a'}U^H_{51}
  +\frac{V^{K*}_{2a'}}{\sqrt{3}}U^H_{61}\biggr|^2 F_{12} (m_{\lambda_X},m^K _{a'}, Q)
  \eea

\bea
  K_{VA} &=&\biggr|- \sqrt{2}i\gamma U^H_ {11} - 2\sqrt{6}i\eta U^H_
  {21}+\sqrt{2}i \zeta U^H_ {51} -\sqrt{6} \zeta U^
  H_ {61}\biggr|^2 F_{12} (m^V,m^A,Q)\qquad
  \eea

\bea
  K_{B\bar L} &=&\sum_ {a = 1}^{\mbox{d(L)}}\biggr|-\biggr(i \gamma U^L_ {1 a} + k U^L_ {2 a}\biggr)U^H_ {11} + \biggr( 2i\eta U^
   L_ {1 a}-\bar\zeta U^
   L_ {2 a}\biggr)\frac{U^H_ {21}}{\sqrt {3}} - \zeta U^L_ {2 a} \frac{U^H_ {31}}{\sqrt {3}}\nonumber \\ && +
   i \zeta U^
   L_ {1 a}U^H_ {51}  + \biggr(\zeta U^
   L_ {1 a} + \frac{i\rho}{3} U^
  L_ {2 a}\biggr) \frac{U^
  H_ {61}}{\sqrt{3}}\biggr|^2 F_{12} (m^B,m^L_a,Q)
  \eea

\bea
  K_{B\bar W} &=&\biggr|-\bar \gamma U^H_ {11} + \frac{2\eta}{\sqrt{3}} U^H_
  {31}-\bar \zeta U^H_ {51} -\frac{i\bar \zeta}{\sqrt{3}} U^
  H_ {61}\biggr|^2 F_{12} (m^B,m^W,Q)
  \eea

  \bea
  K_{\bar O\bar H} &=&\sum_ {a = 2}^{\mbox{d(H)}} \biggr|\bar \gamma U^H_ {4 a} U^H_ {11} + 2\sqrt {3} \eta U^
   H_{4 a} U^H_{31} +\biggr(2\sqrt {3} \eta U^H_{3 a} +
  \bar\zeta U^H_ {5 a} - \sqrt {3}i\bar\zeta U^H_ {6 a}+\bar\gamma U^H_ {1 a}\biggr) U^
  H_ {41}\nonumber \\ && +
   \bar\zeta U^H_ {4 a} U^H_ {51} - \sqrt {3}i \bar\zeta U^
  H_ {4 a} U^
  H_ {61}\biggr|^2 F_{12}(m^H_a, m^O,Q)+\biggr|2\bar \gamma U^H_ {41} U^H_ {11}\nonumber \\ && + 4\sqrt {3} \eta U^
   H_{41} U^H_{31}+2
   \bar\zeta \biggr(U^H_ {41} U^H_ {51} - \sqrt {3}i U^
  H_ {41} U^
  H_ {61}\biggr)\biggr|^2 F_{11}( m^O,Q)
  \eea

\bea
  K_{W\bar Y} &=&\biggr|- \gamma U^H_ {11} + \frac{2\eta}{ \sqrt{3}} U^H_
  {21}- \zeta U^H_ {51} +\frac{i\zeta}{\sqrt{3}} U^
  H_ {61}\biggr|^2 F_{12} (m^W,m^Y,Q)
  \eea

\bea
  K_{\bar V O} &=&\biggr|- \gamma U^H_ {11} -2\sqrt{3}\eta U^H_
  {21}- \zeta U^H_ {51} -\sqrt{3}i\zeta U^
  H_ {61}\biggr|^2 F_{12} (m^V,m^O,Q)
  \eea

  \bea
  K_{G H} &=&\sum_ {a = 1}^{\mbox{d(G)}} \sum_ {a'= 2}^{\mbox{d(H)}}
  \biggr|\biggr(\gamma V^G_ {2 a}V^H_ {3 a'} +\frac{ \gamma}{\sqrt {2}} V^G_ {3 a}V^H_ {3 a'}-
    \sqrt {2}i \bar\gamma V^G_ {5 a}V^H_ {4 a'} +
   \bar\gamma V^G_ {2 a}V^H_ {2 a'} - \frac{\bar\gamma}{\sqrt {2}} V^G_ {3 a}V^H_ {2 a'}
   \nonumber \\ &&-k V^G_ {1 a} V^H_ {5 a'} +
   \frac{ik}{\sqrt {2}} V^G_ {3 a} V^H_ {6 a'}\biggr) U^
   H_ {11} + \biggr(2 \sqrt {\frac{2}{3}} \eta V^
   G_ {3 a}V^H_ {3 a'} + \frac{4\eta}{\sqrt{3}} V^G_ {2 a} V^H_ {3 a'}
   -\sqrt {\frac{2}{3}} i \bar\zeta V^G_ {3 a}V^H_ {6 a'}\nonumber \\ &&  - \frac{\bar\zeta}{\sqrt {2}}V^G_ {3 a}V^H_ {5 a'}
    -
   \bar\zeta i V^G_ {1 a}V^H_ {6 a'} +
   \bar\gamma V^G_ {2 a}V^H_ {1 a'} + \frac{\bar\gamma}{\sqrt {2}} V^G_ {3 a}V^H_ {1 a'}
    \biggr ) U^
   H_ {21} + \biggr( \frac{4\eta}{\sqrt {3}} V^G_ {2 a} V^
   H_ {2 a'}\nonumber \\ && - \frac{4\eta}{\sqrt {6}} V^G_ {3 a} V^
   H_ {2 a'}-2\sqrt {6} \eta i V^G_ {5 a} V^
   H_ {4 a'} -\sqrt {\frac{2}{3}}i \zeta V^G_ {3 a}V^H_ {6 a'} +i \zeta
   V^G_ {1 a} V^H_ {6 a'} - \frac{\zeta}{\sqrt {2}}V^G_ {3 a} V^
   H_ {5 a'} \nonumber \\ &&+
   \gamma V^G_ {2 a}V^H_ {1 a'} - \frac{\gamma}{\sqrt {2}} V^G_ {3 a} V^H_ {1 a'}
    \biggr) U^
   H_ {31} + \biggr(2\sqrt {6} i \eta V^
   G_ {4 a}V^H_ {2 a'} -\sqrt {2}i\zeta V^G_ {4 a} V^H_ {5 a'} \nonumber \\ && +
    \sqrt {6}\zeta V^G_ {4 a} V^H_ {6 a'} +
   \sqrt {2}\gamma i V^G_ {4 a}V^H_ {1 a'} +
   2\sqrt {3}\lambda V^G_ {2 a}V^H_ {4 a'} - \sqrt {6}\lambda V^G_ {3 a}V^H_ {4 a'}
   \biggr) U^
   H_ {41} \nonumber \\ &&+ \biggr( \frac{\zeta}{\sqrt {2}} V^G_ {3 a} V^H_ {3 a'} - k V^G_ {1 a} V^H_ {1 a'}-\frac{i\rho}{3\sqrt {2}} V^
   G_ {3 a}V^H_ {6 a'} +
   \sqrt {2} i \bar\zeta V^
   G_ {5 a}V^H_ {4 a'} +
    \frac{\bar\zeta}{\sqrt {2}} V^G_ {3 a}V^H_ {2 a'} \biggr) U^
   H_ {51} \nonumber \eea

   \bea &&+ \biggr( \sqrt {\frac{2}{3}}i\zeta V^G_ {3 a}V^H_ {3 a'} +i\zeta
   V^G_ {1 a} V^H_ {3 a'} +
   \sqrt {\frac{2}{3}} i \bar\zeta V^G_ {3 a}V^
   H_ {2 a'} - i \bar\zeta V^G_ {1 a} V^
   H_ {2 a'} + \frac{2 \rho}{3\sqrt {3}} V^H_ {6 a'} V^
   G_ {2 a} \nonumber \\ &&+ \frac{i\rho}{3\sqrt {2}} V^H_ {5 a'} V^G_ {3 a} -
   \frac{ik}{\sqrt {2}} V^H_ {1 a'} V^
   G_ {3 a} + \sqrt {6} \bar\zeta V^H_ {4 a'} V^G_ {5 a}\biggr) U^
  H_ {61}\biggr|^2 F_{12} (m^G_a, m^H_{a'},Q)\nonumber\\&&
   -\sum_ {a'= 2}^{\mbox{d(H)}}2{g_{10}^2} \biggr|\frac{-i}{\sqrt{5}}\biggr(U^{H*}_{1 a'}U^H_{11}+U^{H*}_{2a'}U^H_{21}+U^{H*}_{3a'}U^H_{31}-4 U^{H*}_{4a'}U^H_{41}
  +U^{H*}_{5a'}U^H_{11}\nonumber\\&&+U^{H*}_{6 a'}U^H_{61}\biggr)\biggr|^2F_{12}(m^H_{a'}, m_{\lambda_G},Q)\nonumber\\&&
  + \sum_ {a = 1}^{\mbox{d(G)}}
  \biggr|\biggr(\gamma V^G_ {2 a}V^H_ {3 1} + \frac{\gamma}{\sqrt {2}} V^G_ {3 a}V^H_ {31}-
    \sqrt {2} i\bar\gamma V^G_ {5 a}V^H_ {41} +
   \bar\gamma V^G_ {2 a} V^H_ {21} - \frac{\bar\gamma}{\sqrt {2}}V^G_ {3 a}V^H_ {21}
   -
   k V^G_ {1 a} V^H_ {51}\nonumber \\ && +
   \frac{ik}{\sqrt {2}} V^G_ {3 a} V^H_ {61}\biggr) U^
   H_ {11} + \biggr(2\sqrt {\frac{2}{3}} \eta V^
   G_ {3 a}V^H_ {31} + \frac{4\eta}{\sqrt{3}} V^G_ {2 a} V^H_ {31}
   -\sqrt {\frac{2}{3}} i \bar\zeta V^G_ {3 a}V^H_ {61} - \frac{\bar\zeta}{\sqrt {2}}V^G_ {3 a}V^H_ {51}
  \nonumber \\ &&  -
    i \bar\zeta V^G_ {1 a}V^H_ {61}  +
   \bar\gamma V^G_ {2 a} V^H_ {11} +
   \frac{\bar\gamma }{\sqrt {2}}V^G_ {3 a}V^H_ {11}\biggr) U^
   H_ {21} + \biggr(i \zeta
   V^G_ {1 a} V^H_ {61} -\sqrt {\frac{2}{3}}i \zeta V^G_ {3 a}V^H_ {61}\nonumber \\ && - \frac{\zeta}{\sqrt {2}}V^G_ {3 a} V^
   H_ {51}-2\sqrt {6}i \eta V^G_ {5 a} V^
   H_ {41}  +
   \gamma V^G_ {2 a}V^H_ {11} -
   \frac{\gamma }{\sqrt {2}}V^G_ {3 a}V^H_ {11} + \frac{4\eta}{\sqrt {3}} V^G_ {2 a} V^
   H_ {21} - \frac{4\eta}{\sqrt {6}}V^G_ {3 a} V^
   H_ {2 1}\biggr) U^
   H_ {31}\nonumber \\ && + \biggr(2\sqrt {6} i \eta V^
   G_ {4 a}V^H_ {21} -\sqrt {2}i\zeta V^G_ {4 a}V^H_ {51}  +
    \sqrt {6}\zeta  V^G_ {4 a} V^H_ {61}+
   \sqrt {2}i\gamma V^G_ {4 a} V^H_ {11} +
   2\sqrt {3}\lambda V^G_ {2 a}V^H_ {41}\nonumber \\ && - \sqrt {6}\lambda
   V^G_ {3 a}V^H_ {41}\biggr) U^
   H_ {41} + \biggr(-\frac{i\rho}{3\sqrt {2}} V^
   G_ {3 a}V^H_ {61} - k V^G_ {1 a} V^H_ {1 a'} +
   \sqrt {2} i \bar\zeta V^
   G_ {5 a} V^H_ {41} + \frac{\zeta }{\sqrt {2}}V^G_ {3 a}V^H_ {31} \nonumber \\ &&+
   \frac{ \bar\zeta}{\sqrt {2}} V^G_ {3 a} V^H_ {21} \biggr) U^
   H_ {51} + \biggr( \sqrt {\frac{2}{3}}i\zeta V^G_ {3 a}V^H_ {31} +i\zeta
   V^G_ {1 a} V^H_ {31} +
    \sqrt {\frac{2}{3}}i \bar\zeta V^G_ {3 a} V^
   H_ {2 1}\nonumber \\ &&- i \bar\zeta V^G_ {1 a} V^
   H_ {2 1} + \frac{2 \rho}{3\sqrt {3}} V^
   G_ {2 a} V^H_ {61} + \frac{i\rho}{3\sqrt {2}} V^G_ {3 a}V^H_ {5 1} -
   \frac{ik}{\sqrt {2}} V^
   G_ {3 a}V^H_ {1 1}+ \sqrt {6} \bar\zeta V^G_ {5 a}V^H_ {41}\biggr) U^
  H_ {61}\biggr|^2 F_{11} (m^G_a,Q)\nonumber\\&&
  -2{g_{10}^2} \biggr|\frac{-i}{\sqrt{5}}\biggr(U^{H*}_{11}U^H_{11}+U^{H*}_{21}U^H_{21}+U^{H*}_{31}U^H_{31}-4 U^{H*}_{41}U^H_{41}
  +U^{H*}_{51}U^H_{51}\nonumber\\&&+U^{H*}_{61}U^H_{61}\biggr)\biggr|^2F_{11}( m_{\lambda_G},Q)
  \eea
\chapter{Loop Corrected Susy Spectra} \label{loop}
\section{Introduction}
The discovery that the Higgs mass is, at around 126 GeV, almost
28$\% $ larger than the tree level upper limit in the MSSM
\cite{Aad:2012tfa,Chatrchyan:2012ufa}, has emphasized the crucial
role of loop corrections in the MSSM \cite{Zwirner}. For the
precise estimation of Susy particle masses, it is essential to
consider the loop effects. Susy threshold corrections to the SM
Yukawas are already incorporated in the NMSGUT \cite{nmsgut}. The
NMSGUT requires GUT scale \cite{aulakhgargkhosa} and Susy
threshold corrections \cite{nmsgut} to suppress fast B-decay rates
and for fermion fitting respectively. In the previous chapter tree
level Susy spectrum is presented (however the Higgs mass is
one-loop corrected). The next step regarding inclusion of quantum
effects is to calculate one loop corrections to the Susy spectrum.
NMSGUT fits \cite{nmsgut,aulakhgargkhosa} prefer mini-split Susy
spectrum: gaugino/higgsino masses $\sim$ $10^2$ TeV, heavy third
s-generation $\sim$ 10 TeV, $\mu, A_0, M_{H,\bar H}$ $\sim$ $10^2$
TeV and first and second generation lie in between
neutralino/chargino
 and third generation sparticle masses.  In this chapter we present
a one-loop corrected effective MSSM spectrum in the context of
NMSGUT. We will investigate the significant corrections received
from the NMSGUT kind of low energy spectrum. Formulae for the one
loop corrections to the entire Susy spectrum are well known
\cite{deltar,higgs,gluino1,gluino2,gluino3,piercebagger}.
Particularly Ref. \cite{piercebagger} is a pedagogical manual for
one loop corrections to fermion, sparticle and gauge boson
spectra. The modified dimensional reduction
($\overline{\mbox{DR}}$) renormalization scheme \cite{DRbar1} is
convenient to use in Susy theories. The poles of the loop
corrected propagators determine the physical masses. Thus the
physical mass ($m_{phys}$) of a boson is given by the solution of
\be M^2=\hat M^2(Q)-  {\cal R}e \, \Pi(M^2)\ee Here $\hat M^2(Q)$
is tree level mass parameter at scale $Q$ and $\Pi(M^2)$ is self
energy contribution ( the sensitivity  to $Q$ of $m_{phys}$ should
decrease as one increases the order of perturbation since it
should be a RG invariant). To avoid negative pole mass self energy
is iteratively calculated. One needs to calculate self energy of
$W$ and $Z$ boson to calculate $\overline{\mbox{DR}}$ EW symmetry
breaking VEV. Fermion masses are also calculated from the pole of
corresponding propagator. One-loop self energy of Susy particles
affects :
\begin{itemize}
\item Gluino, Chargino and Neutralino Masses \item Higgs sector :-
heavy Higgs boson masses ($M_A$, $M_H$, $M_{H^+}$) and light Higgs
boson ($M_h$) \item Squarks and Sleptons \end{itemize} Complete
formulae of these self energy calculations are presented in
\cite{piercebagger}. In the next section we will discuss dominant
corrections with explicit expressions.
\section{Dominant Corrections}\vspace{.2cm}
FORTRAN subroutines implementing formulae of \cite{piercebagger}
are also available \cite{porod}. We have interfaced these
subroutines with our FORTRAN code.
 We calculated the loop corrections
 to the tree level spectrum presented in the previous chapter (see Tables \ref{table1}-\ref{table5}) and  \cite{aulakhgargkhosa}. Including loop corrections (some of) the first and second generation squark and slepton
  masses can turn negative as shown in the example Tables \ref{com} and \ref{tablesol2}. We identified the
   dominant dangerous corrections- i.e. those which can drive some sparticle masses to negative values- using the solutions presented in the previous chapters. The significant corrections are discussed below :-
\subsection{Gluino Mass}
 Tree level gluino mass is \be M_{\tilde g}\ =\ M_3(Q) \ee
 The physical gluino mass is given by the solution of
\begin{equation}
M_{\tilde g}\ =\ M_3(Q)\ -\ {\cal R}e\,\Sigma_{\tilde g}(M_{\tilde
g}^2)
\end{equation}
where $\Sigma_{\tilde g}(M_{\tilde g}^2)$ is gluino self-energy :
\begin{eqnarray}
\Sigma_{\tilde g}(p^2) &=& {g_3^2\over16\pi^2}\Biggl\{ -\
M_{\tilde g} \left(15 + 9\ln{Q^2\over M^2_{\tilde g}}\right)\ -\
\sum_q\sum_{i=1}^2 M_{\tilde g}B_1(p,m_q,m_{\tilde q_i})\nonumber\\
&& \qquad +\ \sum_q m_q \sin {2\theta_q} \biggl[ B_0(p,m_q,
m_{\tilde q_1})- B_0(p,m_q,m_{\tilde q_2})\biggr]\Biggr\}
\label{siglu}
\end{eqnarray}
where $Q$ is the renormalization scale (which we take to be
$M_Z$), $g_3$ and $\theta_q$ are SU(3) gauge coupling constant and
squark mixing angle respectively.
 One loop correction to gluino mass comes from gluon/gluino and quark($m_q$)/squark($m_{\tilde{q}}$) loops \cite{piercebagger}.   The loop-function $B_0$ has the form
\begin{equation}
 B_0(p, m_1, m_2) \ =\ {1\over\hat\epsilon} - \ln\left(p^2\over
Q^2\right) - f_B(x_+) - f_B(x_-)
\end{equation}
where \bea
 x_{\pm}\ =\ {s \pm \sqrt{s^2 - 4p^2(m_1^2-i\varepsilon)}\over2p^2} \quad ;
\quad s=p^2-m_2^2+m_1^2 \nnu f_B(x) \ =\ \ln(1-x) -
x\ln(1-x^{-1})-1 \eea and $1/\hat\epsilon$ represents infinite
contribution. The function $B_1$ is defined in terms of
loop-functions $B_0$ and $A_0$ :
\begin{equation}
A_0(m)\ =\ m^2\left({1\over\hat\epsilon} + 1 - \ln{m^2\over
Q^2}\right)\label{A}
\end{equation}
\begin{equation}
 B_1(p, m_1,m_2) \ =\ {1\over2p^2}\biggl[ A_0(m_2) - A_0(m_1) + (p^2
+m_1^2 -m_2^2) B_0(p, m_1, m_2)\biggr]
\end{equation}
First term of Eq.\eqref{siglu} represents the gluon/gluino
contribution. Quark/squark contribution is given by second and
third terms. Out of these the third term has negligible
contribution since it is proportional to quark masses.  First and
second term have opposite contribution. Second term involves the
loop function $A_0$ which is proportional to $m^2_{\tilde q}$ ,
therefore it provides the dominant contribution. Gluino mass gets
approximately 30$\%$ corrections (see Tables \ref{com} and
\ref{tablesol2}).
\subsection{Neutralino and Chargino Masses}
MSSM gauginos and Higgsinos  form chargino and neutralino
eigenstates. Neutralino mass matrix including radiative
corrections is given by:
\begin{equation}
 \mathcal M_{\tilde
\chi}^{{\tiny{\mbox{1-loop}}}}={\cal M}_{\tilde\chi^0} +
{1\over2}\left(\delta {\cal M}_{\tilde\chi^0}(p^2) + \delta{\cal
M}_{\tilde\chi^0}^T(p^2)\right) \label{chi0-1}
\end{equation}
where
 \begin{equation}
 \delta{\cal M}_{\tilde\chi^0}(p^2) \ =
\ -\ \Sigma_R^0(p^2){\cal M}_{\tilde\chi^0} \ -\ {\cal
M}_{\tilde\chi^0}\Sigma_L^0(p^2) \ -\ \Sigma_S^0(p^2)
\end{equation}
Here ${\cal M}_{\tilde\chi^{0}}$ is the tree-level neutralino mass
matrix (Eq. \ref{neutmass}), and the factors
$\Sigma_{L,R,S}^{0}(p^2)$ are {\em matrix} corrections. Self
energy has contributions from (quark/squark, lepton/slepton,
neutrino/sneutrino),
 chargino/W-boson, neutralino/Z-boson and gaugino/Higgs loops. Dominant
  contribution comes from- quark/squark, lepton/slepton, neutrino/sneutrino loops. Third generation quark/squark provide maximum corrections :
\be (\Sigma_{L}^{0}(p^2))^{ q/\tilde{q}}_{ij} \simeq
\sum_{k=1}^{2} a^*_{\tilde\chi_i^0 q\tilde q_k}\,
a_{\tilde\chi_j^0 q\tilde q_k} \ {\cal R}e\,B_1(p,m_q,m_{\tilde
q_k}) \ee \be (\Sigma_S^{0}(p^2))^{q/\tilde{q}}_{ij} \simeq
\sum_{k=1}^{2} b^*_{\tilde\chi_i^0 q \tilde
q_k}\,a_{\tilde\chi_j^0q \tilde q_k} m_t \ {\cal
R}e\,B_0(p,m_q,m_{\tilde q_k}) \ee Here $q$ represents top/bottom
quark and $\tilde{q}$ refer to the corresponding scalar.
$a_{\tilde\chi_j^0q\tilde q_k}$ are neutralino-fermion-sfermion
couplings and sum over $k$ includes contribution of  scalar
partners of both left and right handed quarks. Similarly, one-loop
chargino mass matrix is as follows :
\begin{equation}
\mathcal M_{\tilde \chi^+}^{{\tiny{\mbox{1-loop}}}}={\cal
M}_{\tilde\chi^+} -\Sigma_R^+(p^2){\cal M}_{\tilde\chi^+}  - {\cal
M}_{\tilde\chi^+}\Sigma_L^+(p^2) - \Sigma_S^+(p^2)~
\end{equation}
where ${\cal M}_{\tilde\chi^{+}}$ is the tree-level chargino mass
matrix (Eq \ref{chargmass}). Quark/squark corrections have form :
\be (\Sigma_L^{+}(p^2))^{q/\tilde{q}}_{ij} \approx \sum_{k=1}^2
a^*_{\tilde\psi_i^+q\tilde {q'}_k}\,a_{\tilde\psi_j^+ q \tilde
{q'}_k} \ {\cal R}e\,B_1(p,m_q,m_{\tilde {q'}_k}) \ee

\be (\Sigma_S^{+}(p^2))^{q/\tilde{q}}_{ij} \approx \sum_{k=1}^2
b^*_{\tilde\psi_i^+ q \tilde {q'}_k}\, a_{\tilde\psi_j^+ q\tilde
{q'}_k}\,m_q \ {\cal R}e\,B_0(p,m_q ,m_{\tilde {q'}})\ee Here $q'$
denotes bottom (top) when $q$ is top (bottom). ${\Sigma_R^{0,+}}$
can be obtained from ${\Sigma_L^{0,+}}$ by replacing the
couplings. Neutralino/chargino corrections are approximately
7$\%$.
\subsection{Higgs Mass}
The MSSM has two Higgs doublets $H_1$ and $H_2$ whose VEVs
generate fermion masses, so to start with we have 8 degrees of
freedom. $W^{\pm}$ and Z bosons eat 3 degrees of freedom and
become massive so we are left with 5 degrees of freedom i.e. the
neutral Higgs ($h$, $H$), charged Higgs $H^{\pm}$, CP odd Higgs
$A$.
 Higgs soft mass parameters $m_{H_{1,2}}^2$ at $M_Z$ determine $\mu$ and $B$ :
\bea \mu^2 &=& {1\over2}\Biggl[\, \tan2\beta\biggl(
m_{H_2}^2\tan\beta- m_{H_1}^2\cot\beta\biggr) \ -\ M_Z^2\
 \Biggr] \nonumber\\  && B=\frac{1}{2}\Biggl[\,\tan2\beta \biggl(
m_{H_2}^2- m_{H_1}^2\biggl)- M_Z^2 \sin2\beta\Biggr]
\label{symbreak}\eea $M_A$ $(B(\tan\beta+\cot\beta))$ is computed
from $B$ and then tree level Higgs masses are calculated. Tadpoles
need to be calculated to take into account the one loop radiative
corrections to the Higgs masses. Vanishing of tadpoles provides :
\begin{eqnarray}
\mu^2 &=& {1\over2}\Biggl[\, \tan2\beta\biggl(\overline
m_{H_2}^2\tan\beta-\overline m_{H_1}^2\cot\beta\biggr)  - M_Z^2\ -
\ {\cal R}\!e\,\Pi_{ZZ}^T(M_Z^2)\, \Biggr]\ \nonumber\\ M_A^2 &=&
{1\over c_{2\beta}}\biggl(\overline m_{H_2}^2-\overline
m_{H_1}^2\biggr)  - M_Z^2  - {\cal R}e\,\Pi_{ZZ}^T(M_Z^2)  - {\cal
R}e\,\Pi_{AA}(m_A^2)  +\ b_A
\end{eqnarray}

\be \bar{m}_{H_1}^2 = m_{H_1}^2 - \frac{t_1}{v_1}  \qquad
\bar{m}_{H_2}^2 = m_{H_2}^2 - \frac{t_2}{v_2} \ee

  \be  b_A = s_\beta^2\, \frac{t_1}{v_1} +
c_\beta^2\,\frac{t_2}{v_2} \ee Here $\Pi_{ZZ}$ and $\Pi_{AA}$ are
self energies of Z-boson and CP odd pseudo-scalar $A$. Charged
Higgs mass including loop corrections is given by:
\begin{equation}
M^2_{H^+}\ =\ M_A^2 + M_W^2 + {\cal R}e\biggl[\Pi_{AA}(M^2_A) -
\Pi_{H^+H^-}(m_{H^+}^2)+\Pi^T_{WW}(M^2_W)\biggr]~
\label{chargedHiggs}\end{equation} Remaining two CP-even Higgs
mass are determined from the pole of matrix :\vspace{.2cm}
\[
{\cal M}^2_s(p^2) \ =\ \left(\begin{array}{cc} \hat M_Z^2c_\beta^2
+ \hat M_A^2s_\beta^2 -\Pi_{s_1s_1}(p^2) + \frac{t_1}{v_1} &
-(\hat
M_Z^2 + \hat M_A^2)s_\beta c_\beta -\Pi_{s_1s_2}(p^2)\\[2mm]
-(\hat M_Z^2 + \hat M_A^2)s_\beta c_\beta -\Pi_{s_2s_1}(p^2) &
\hat M_Z^2s_\beta^2 + \hat M_A^2c_\beta^2 -\Pi_{s_2s_2}(p^2) +
\frac{t_2}{v_2}
\end{array}\right)\ .
\vspace{.3cm}\] Here $\hat M_Z^2$ and $\hat M_A^2$ are
$\overline{\mbox{DR}}$ masses. Explicit form of self-energies
($\Pi_{ss'}$) can be found in \cite{piercebagger}.

\begin{table}
 $$
 \begin{array}{ccc}
 \hline \hline&\vspace{-.3 cm}\\{\mbox {Parameter } }&\mbox { Tree level Masses (GeV)}& \mbox{ Loop Corrected Masses (GeV)} \vspace{0.1 cm}\\
 \hline \hline
                M_{\tilde{g}}&       1000.14     & 1297.10 \\
               M_{\chi^{\pm}}&       569.81,         109858.12       &  628.52,         123993.68  \\
       M_{\chi^{0}}&     210.1,            569.8,         109858.1    ,         109858.1      &  215.9,            628.5,         123993.7   ,         123993.7 \\
              M_{\tilde{\nu}}^2&         \{2.34, 2.33, 4.55\} \times 10^8 & \{2.0, 2.02, 8.42\} \times 10^9 \\
                M_{\tilde{e}}^2&         \{3.10, 234.34\}\times 10^6, 45071.29   &\{-3.50, 1.99, -3.46, 2.01\}\times 10^9 \\
                &\{2.33, 4.29, 4.57\}\times 10^8 &  \{ 8.42, 9.38\}\times 10^9 \\
                M_{\tilde{u}}^2&    \{1.27, 2.09, 1.27, 2.09\} \times 10^8       &   \{-.45, 2.56,-.45, 2.56\}\times 10^9 \\   & \{6.06, 16.22\} \times 10^8  & \{2.95, 26.14 \}\times 10^9\\

                M_{\tilde{d}}^2&   \{0.71, 1.27, .71, 1.27\}   \times 10^8      & \{-1.1, -.45, -1.1, -.45\}\times 10^9  \\ & \{16.22, 26.88\}   \times 10^8&\{26.1, 52.9\}\times 10^9\\

                        M_{A}&     517662.74       & 40556.46 \\
                  M_{H^{\pm}}&    517662.75       & 40590.03 \\
                    M_{H}&      517662.74     & 40624.33 \\
                    M_{h}&         89.09   &   493.46 \\
 \hline \hline
 \end{array}
 $$
 \caption{\small{Tree level and loop corrected Susy spectra
 corresponding to the soft paramter presented in the previous
 chapter. The loop corrected squark and slepton masses turn
 negative.
 \label{com} }}\end{table}
 \vspace{.4cm}
 \begin{table}
  $$
 \begin{array}{ccc}
 \hline\hline &\vspace{-.3 cm}\\{\mbox{Parameter}}&\mbox{Tree level Masses (GeV)}& \mbox{Loop Corrected Masses (GeV)} \vspace{0.1 cm}\\
 \hline\hline
 M_{\tilde{g}}&  1200.01    &  1542.55 \\
 M_{\chi^{\pm}}&   590.18, 155715.46     &  646.67,  160020.65  \\
 M_{\chi^{0}}&    246.4,            590.2, 155715.4 , 155715.4       &  255.8, 646.7, 160020.6,  160020.6 \\
 M_{\tilde{\nu}}^2& \{{2.35, 2.35, 9.08}\} \times 10^8 & \{25.43, 25.71, 115.44\} \times 10^8 \\
 M_{\tilde{e}}^2&         \{1.43, 2.35, 1.43, 2.35\}\times 10^8   &\{-4.42, 2.52, -4.35, 2.55\}\times 10^9 \\
 & \{9.08, 14.87\}\times 10^8& \{14.22, 19.38\}\times 10^9 \\
 M_{\tilde{u}}^2&    \{1.64, 1.81, 1.64, 1.81\} \times 10^8       &   \{-.57, 3.21, -.57, 3.21\}\times 10^9 \\
& \{23.26, 24.01\} \times 10^8  & \{ 5.02, 24.69\}\times 10^9  \\
 M_{\tilde{d}}^2 &   \{1.26, 1.81, 1.26, 1.81\}   \times 10^8      & -\{1.39, .57, 1.39, .57\}\times 10^9  \\
& \{23.88, 24.42\}   \times 10^8&\{ 24.69, 47.34\}\times 10^9   \\
 M_{A}&   584560.76 &  25390.88 \\
 M_{H^{\pm}}&     584560.76    & 25398.06 \\
 M_{H}&    584560.76     & 25409.02 \\
 M_{h}&     89.35     &  411.44  \\
 \hline\hline
 \end{array}
 $$\caption{\small{Tree level and loop corrected Susy spectra
 corresponding to solution 2 of \cite{aulakhgargkhosa}.
 The loop corrected squark and slepton masses turn
 negative.\label{tablesol2}}}
 \end{table}
\vspace{.1cm}
\subsection{Squark and Slepton Masses}\vspace{.3cm}
Tree level squark and slepton masses are calculated from $6 \times
6$ mass matrices. It is convenient to breakup self energy
corrections as $3 \times 3$ blocks $(\Pi_{\tilde f_L\tilde f_L},$
$\Pi_{\tilde f_L\tilde f_R}$, $\Pi_{\tilde f_R\tilde f_R})$:
\vspace{.3cm}
\begin{equation}
{\cal M}_{\tilde f}^2(p^2)\ =\ \left(\begin{array}{cc} M_{\tilde
f_L\tilde f_L}^2-\Pi_{\tilde f_L\tilde f_L}(p^2) & \qquad
M_{\tilde f_L\tilde f_R}^2-\Pi_{\tilde f_L\tilde f_R}(p^2)
\\[2mm] M_{\tilde f_R\tilde f_L}^2-\Pi_{\tilde f_R\tilde
f_L}(p^2) & \qquad M_{\tilde f_R\tilde f_R}^2-\Pi_{\tilde
f_R\tilde f_R}(p^2)
\end{array}\right)~.
\vspace{.4cm}\end{equation}Self energies are calculated at the
tree mass scale for each particle. Sfermion masses are fixed as
poles of propagator \be Det[p_i^2-M_{\tilde f}^2(p_i^2)]=0  \ee
Squark and slepton receive corrections from electroweak gauge
bosons, neutralinos, charginos, sfermion quartic interactions (up
squark, down squark, charged slepton, sneutrinos), pseudoscalar
Higgs, neutral Higgs and charged Higgs. Out of these Higgs sector
corrections are dominant because of decoupled Susy spectra. For up
type squarks, these are :- \bea \Pi_{\tilde u_{L}\tilde
u_{L}}^{\mbox{{\tiny{Higgs-sector}}}}(p^2) &\approx&
{1\over2}\sum_{n=1}^4 \biggl(Y_{u}^2\,D_{nu} - {g^2 g_{u_L}\over2
\cos^2{\theta_W}} C_n\biggr) A_0(m_{H^0_n}) \nonumber\\ & +&
\sum_{n=3}^4 \biggl(Y_{d}^2D_{nu}+g^2\left(
{g_{u_L}\over2\cos^2{\theta_W}}-I_3^u\right)C_n\biggr)
A_0(m_{H^+_{n-2}}) \nonumber\\ &+& \sum_{n=1}^4\sum_{i=1}^2
(\lambda_{H^0_n{\tilde {u}}_L{\tilde {u}}_i})^2
B_0(p,m_{H^0_n},m_{{\tilde {u}}_i}) \nonumber\\ & +&
\sum_{i,n=1}^2 (\lambda_{H^+_n{\tilde {u}}_L{\tilde {d}}_i})^2
B_0(p,m_{{\tilde d}_i},m_{H^+_n}) \vspace{.35cm}\eea Here $H_n^0$
denotes $H$, $h$, $G^0$ and $A^0$, $H^+_{1}(H^+_{2})$ refer to
$H^+(G^+)$, $Y_u/Y_d$ is the Yukawa coupling of up/down quarks,
the factors $D_{nu}$ and $C_n$ are sine/cosine functions of Higgs
mixing angles ($\alpha,\beta$) \cite{piercebagger}. The parameters
$I_3^u$, $g$ and $g_{f}$ ($I_3^{f}-Q_f \sin^2 \theta_W$) represent
SU(2) quantum number, SU(2) gauge coupling and weak neutral
current couping. Self energy $\Pi_{\tilde f_R\tilde
f_R}^{\mbox{\tiny{Higgs-sector}}}(p^2)$ can be obtained by
replacing $g_{u_L}$ by $g_{u_R}$ and $\lambda_{H^0_n{\tilde
t}_L{\tilde t}_i}/\lambda_{H^+_n{\tilde t}_L{\tilde b}_i} $ by
$\lambda_{H^0_n{\tilde t}_R{\tilde t}_i}/\lambda_{H^+_n{\tilde
t}_R{\tilde b}_i}$ in $\Pi_{\tilde f_L\tilde
f_L}^{\mbox{\tiny{Higgs-sector}}}(p^2)$. \bea \Pi_{\tilde
u_{L}\tilde u_{R}}^{\mbox{{\tiny{Higgs-sector}}}}(p^2) &\approx &
\sum_{n=1}^4\sum_{i=1}^2 \lambda_{H^0_n\tilde {u}_L\tilde {u}_i}
\lambda_{H^0_n\tilde {u}_R\tilde {u}_i} B_0(p,m_{H^0_n},m_{{\tilde
{u}}_i}) \nonumber \\&& + \sum_{i,n=1}^2 \lambda_{H^+_n\tilde
{u}_L\tilde {d}_i} \lambda_{H^+_n\tilde {u}_R\tilde {d}_i}
B_0(p,m_{\tilde {d}_i},m_{H^+_n}) \vspace{.35cm}\eea Corrections
due to quartic Higgs  couplings involve loop function $A_0$ which
is proportional to $m^2$, so these terms dominate for the heavy
$M_A$, as predicted by the NMSGUT. If we switch off the terms
containing $A_0$ in the Higgs contribution then we get positive
loop corrected masses. But if we ignore all the Higgs corrections
then third generation masses become negative as shown in Tables
\ref{tablecom2} and \ref{tablecom22}. Chargino/neutralino
contribution is appreciable for third generation :\vspace{.2cm}
\bea \Pi_{\tilde q_L\tilde q_L}^{\mbox{{\tiny{C/N}}}}(p^2)
&&\approx\sum_{i=1}^4 \biggl[ f_{iq\tilde q_{LL}}
G(p,m_{\tilde\chi_i^0},m_q) - 2\,g_{iq\tilde q_{LL}}
m_{\tilde\chi_i^0}m_t B_0(p,m_{\tilde\chi_i^0},m_q) \biggr] \nonumber\\
&&+ \sum_{i=1}^2 \biggl[ f_{iq'\tilde q_{LL}}
G(p,m_{\tilde\chi_i^+},m_{q'}) - 2\,g_{iq'\tilde q_{LL}}
m_{\tilde\chi_i^+}m_{q'} B_0(p,m_{\tilde\chi_i^+},m_{q'})\biggr]
\eea

\be \Pi_{\tilde q_L\tilde q_R}^{\mbox{{\tiny{C/N}}}}(p^2) \approx
\sum_{i=1}^2 \biggl[\, f_{iq'\tilde q_{LR}}
G(p,m_{\tilde\chi_i^+},m_{q'}) - 2\,g_{iq'\tilde q_{LR}}
m_{\tilde\chi_i^+}m_{q'} B_0(p,m_{\tilde\chi_i^+},m_{q'}) \ee
Couplings $f$ and $g$ are defined as
\bea f_{if\bar{f}_j}&=&|a_{\tilde{\chi}_i f \tilde{f}_j}|^2+ |b_{\tilde{\chi}_i f \tilde{f}_j}|^2 \nonumber\vspace{.3cm}\\
g_{if\bar{f}_j}&=&2 {\cal R}e\, (b_{\tilde{\chi}_i f
\tilde{f}_j}^* a_{\tilde{\chi}_i f \tilde{f}_j}) \eea Loop
function $G$ is defined in terms of functions $B_0$ and $A_0$.
\begin{eqnarray}
G(p,m_1,m_2) &=&
(p^2-m_1^2-m_2^2)B_0(p,m_1,m_2)-A_0(m_1)-A_0(m_2)\
\end{eqnarray}
Chargino/neutalino and Higgs corrections have opposite
contribution. One needs to decrease
 the magnitude of Higgs corrections to get positive first and second generation sparticle
  masses. The chargino and neutralino masses are proportional to $\mu$ parameter. We have
   softened this contribution via penalty on $\mu$ and $M_A$ which we will discuss in the next section. In the
    tree level example spectra presented in the previous chapter $M_A$ $\sim$ $3 \mu$.
\begin{table}
 $$
 \begin{array}{cc}
 \hline\hline&\vspace{-.3 cm}\\ {\mbox {Parameter } }&\mbox {Loop Corrected Masses (GeV)}\vspace{0.1 cm} \\
 \hline\hline
 M_{\tilde{e}}^2&         \{2.61, 22.35, 2., 22.05\}\times 10^7, -4.41\times 10^8,305993.39   \}\\

                M_{\tilde{u}}^2&    \{ 1.36, 1.96, 1.36, 1.96, -21.61, -12.93 \} \times 10^8   \\

                M_{\tilde{d}}^2&   \{0.81, 1.36, 0.81, 1.36, -12.93, -4.46\}   \times 10^8     \\
\hline\hline
 \end{array}
 $$
 \caption{\small{Charged slepton and squark masses ignoring Higgs sector corrections for the solution presented in the previous
 chapter. Third generation squark and slepton masses turn negative.
  }\label{tablecom2}}
\vspace{.8cm}
 $$
 \begin{array}{cc}
 \hline \hline&\vspace{-.3 cm}\\{\mbox {Parameter } }&\mbox {Loop Corrected Masses(GeV)}\vspace{0.1 cm} \\
 \hline\hline
 M_{\tilde{e}}^2&         \{1.488, 2.33, 1.431, 2.301, -1.777, 0.82\}\times 10^8\\

                M_{\tilde{u}}^2&    \{ 1.64, 1.87, 1.64, 1.87, -24.52, -20.98 \} \times 10^8   \\

                M_{\tilde{d}}^2&   \{1.32, 1.87, 1.31, 1.87, -20.10, -18.07\}   \times 10^8     \\
\hline\hline
 \end{array}
 $$
 \caption{\small{Charged slepton and squark masses ignoring Higgs sector corrections corresponding to solution 2 of \cite{aulakhgargkhosa}.
  Third generation squark and slepton masses turn negative.}\label{tablecom22}}\vspace{.3cm}\end{table}
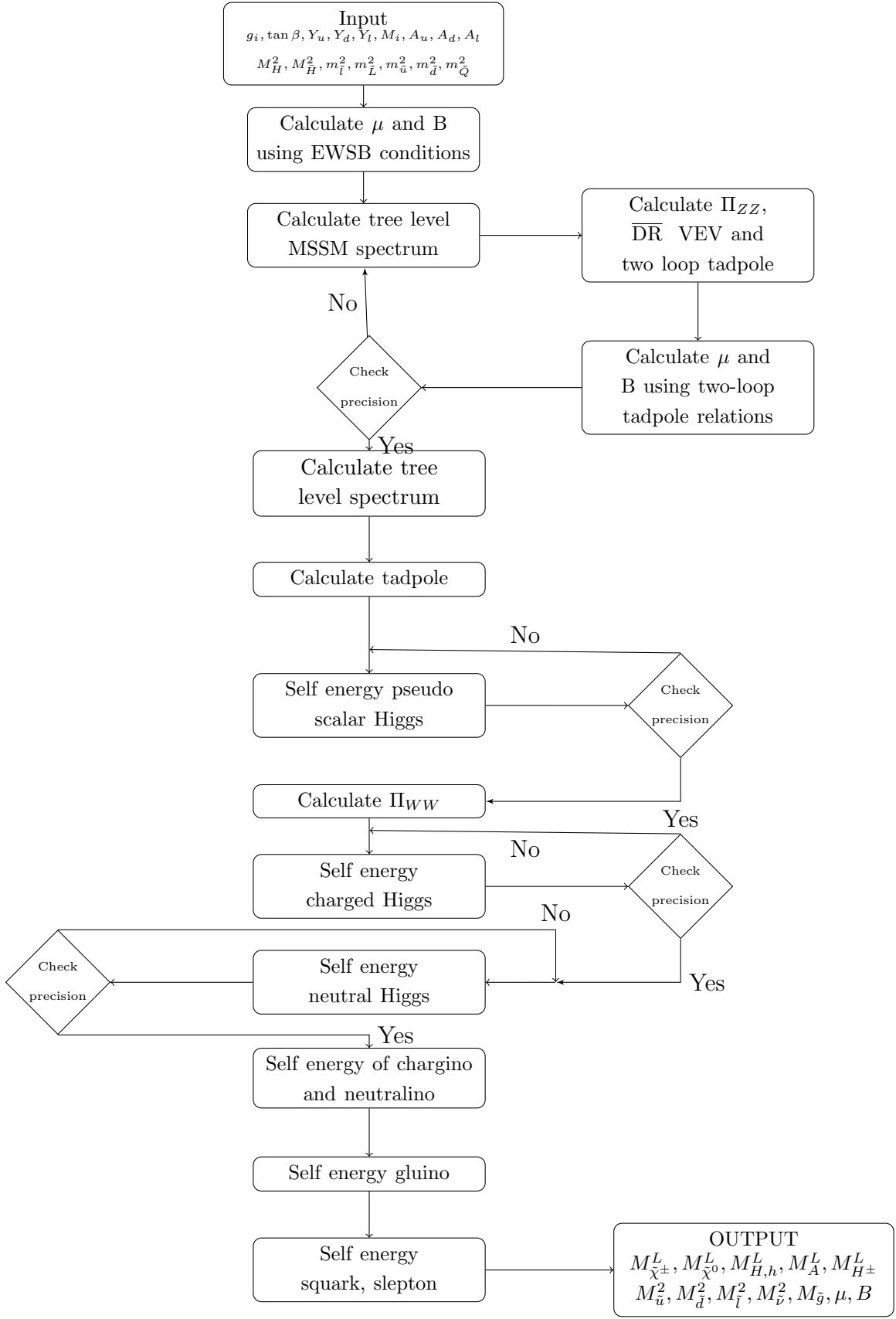
\begin{figure}
\vspace{-.4cm}
 \begin{center}
  \begin{tikzpicture}[node distance=1.65cm,scale=.8]
\node[block4] (a) {{\footnotesize{Input\\{\tiny{$g_i,\tan\beta,
Y_u,Y_d,Y_l,M_i,A_u,A_d,A_l$\\$M_H^2,M^2_{\bar
H},m_{\tilde{l}}^2,m_{\tilde{L}}^2,m_{\tilde u}^2,m_{\tilde
d}^2,m_{\tilde {Q}}^2$}}}}};
\node[block2] (b)  [below of=a] {{\footnotesize{Calculate $\mu$
and B using EWSB conditions}}}; \node[block3] (c)  [below of=b]
{{\footnotesize{Calculate tree level MSSM spectrum}}};
\node[block3] (d)  [right=1.75cm of c] {{\footnotesize{Calculate
$\Pi_{ZZ}$, $\rm{\overline{\mbox{DR}}~~ VEV}$ and two loop
tadpole}}};
\node[block3] (f)  [below=1.00cm of d] {{\footnotesize{Calculate
$\mu$ and B using two-loop tadpole relations}}}; \node[decision2]
(ff)  [left=2.75cm of f] {{\tiny{Check precision}}}; \node[block2]
(g)  [below of=ff] {{\small{Calculate tree level spectrum}}};
\node[block2] (h) [below of=g] {{\footnotesize{Calculate
tadpole}}}; \node[inner sep=0,minimum size=0,below=.90cm of h]
(kkk) {}; \node[block2] (i)  [below=.40cm of kkk]
{{\footnotesize{Self energy pseudo scalar Higgs}}};
\node[decision2] (ii)  [right=2.45cm of i]   {{\tiny{Check
precision}}};

\node[block2] (j)  [below of=i] {{\footnotesize{Calculate
$\Pi_{WW}$}}}; \node[inner sep=0,minimum size=0,below=.20cm of j]
(jjj) {};

\node[block2] (k)  [below=.40cm of jjj] {\footnotesize{{Self
energy charged Higgs}}}; \node[decision2] (jj)  [right=2.45cm of
k]   {{\tiny{Check precision}}}; \node[block2] (l)  [below of=k]
{\footnotesize{{Self energy neutral Higgs}}}; \node[decision2]
(ll)  [left=2.45cm of l]   {{\tiny{Check precision}}}; \node[inner
sep=0,minimum size=0,right=1.20cm of l] (lll) {}; \node[block2]
(m)  [below of=l]                       {\footnotesize{{Self
energy of chargino and neutralino}}};

\node[block2] (n)  [below of=m] {\footnotesize{{Self energy
gluino}}};

\node[block2] (o)  [below of=n] {\footnotesize{{Self energy
squark, slepton}}};
\node[block4] (q)  [right=2.20cm of o]                       {\footnotesize{{OUTPUT\\$M_{\tilde{\chi}^\pm}^L,M_{\tilde{\chi}^0}^L,M_{H,h}^L,M^L_A,M^L_{H^\pm}$}\\
$M_{\tilde u}^2,M_{\tilde d}^2,M_{\tilde
l}^2,M_{\tilde\nu}^2,M_{\tilde{g}},\mu,B$}}; \draw[->] (a) -- (b);
   \draw[->] (b) -- (c);
\draw[->] (c) -- (d);
   \draw[->] (d) -- (f);
 \draw[->] (f) -- (ff);
  \draw[->] (ff) --node[right] {Yes} (g);
 \path [line] (ff) -- node[left] {No}(c);
   \draw[->] (g) -- (h);
\draw[->] (h) -- (i); \draw[->] (i) -- (ii); \draw[->] (ii.north)
--node[above] {No} (kkk);
  \path [line] (ii) |-node[below] {Yes} (j);
  \draw[->] (jj.north) --node[below] {No} (jjj);
 \draw[->] (j) -- (k);
   \draw[->] (k) -- (jj);
   \path [line] (jj) |-node[right] {Yes} (lll);
   \draw[->] (lll) -- (l);
   \draw[->] (ll.north) -| node[above] {No}(lll);
   \draw[->] (l) -- (ll);
\draw[->] (ll.south) -|node[right] {Yes} (m);

   \draw[->] (m) -- (n);
 \draw[->] (n) -- (o);
   \draw[->] (o) -- (q);

\end{tikzpicture}
  \caption{Flowchart of SPheno subroutine \emph{LoopMassesMSSM}.}
  \label{flowchart}
 \end{center}
\end{figure}
\section{Numerical Procedure}
Our fitting criteria are the same as discussed in the previous
chapters. To calculate the one loop Susy spectrum, we have
interfaced the \emph{LoopmassesMSSM} subroutine of SPheno
\cite{porod} with our low scale calculation subroutine FUNKTUNE
(see flowchart \ref{flowc1}). We will discuss only the extra
penalties imposed and structure of the additional subroutine.
Algorithm of \emph{LoopmassesMSSM} is represented by a flowchart
\ref{flowchart}. It calculates one-loop radiative
\cite{piercebagger} corrections to the Susy particles but for the
Higgs masses two-loop corrections
\cite{higgstwoloop1,higgstwoloop2,higgstwoloop} are also included.
We run the MSSM RGEs from $M_{GUT}$ to $M_Z$ to get hard and soft
parameters at $M_Z$ (considered the renormalization matching
scale), which are the inputs to \emph{LoopmassesMSSM}. First of
all the subroutine calculates $\mu$ and $B$ parameter using
electroweak symmetry breaking conditions (EWSB) (Eq.
\ref{symbreak}). Using calculated $\mu$ and $B$ tree-level Susy
spectra is calculated. Z-boson self energy ($\Pi_{ZZ}$) is used to
compute
 $\overline{\mbox{DR}}$ VEV  \vspace{.3cm}\be v^2(Q)=4\frac{M_Z^2+{\cal R}e\, \Pi_{ZZ}(M_Z^2)}{g_1^2(Q)+g_2^2(Q)}\vspace{.3cm} \ee
This VEV is used to calculate one-loop and two-loop tadpoles
further in the spectrum calculations. $\mu$ and $B$ are calculated
using  two-loop effective potential. This process is repeated
until consistent values are achieved. Then the tree level
sparticle spectrum is calculated using updated $\mu$ and $B$
value. \vspace{.3cm}

As the Higgs sector masses are parameterized in terms of CP- odd
pseudo-scalar Higgs mass $(M_A)$ and $\tan\beta$, so first of all
two loop corrections to
  pseudo-scalar Higgs are calculated and the process is repeated to get the convergent value. To calculate one-loop corrected charged
   Higgs masses (Eq. \ref{chargedHiggs}),
   self energy of charged Higgs and W-boson is calculated. Then neutral Higgs masses are calculated taking into account the two loop corrections \cite{higgstwoloop1,higgstwoloop2,higgstwoloop,2loopHiggs}. After this one loop correction to
    gaugino-higgsino sector are calculated. At the end loop corrections to the squarks and sleptons are
     calculated. Subroutine \emph{LoopmassesMSSM} calls several subroutines from SPheno modules - LoopMasses, Couplings and TwoLoopHiggs to compute loop corrections to Susy particles. Each
      subroutine has \emph{in/out} argument \emph{kont} which traces the occurrence of negative mass square parameter. We
       have modified the subroutine to accumulate \emph{kont} from all the called subroutines to the end so that program moves
        forward using absolute value of the negative quantity. Output \emph{kont} of \emph{LoopmassesMSSM} is added to $\chi_{Z}^2$. In
         other words  penalties are imposed to get positive
         masses.
\vspace{.3cm}

\begin{table}
\vspace{-1.0cm}
 $$
 \begin{array}{cccc}
 \hline\hline&&&\vspace{-.4 cm}\\
 {\rm Parameter }&{\rm Value} &{\rm  Field }& {\rm Masses }\\
 &&\hspace{-10mm}{\rm}[SU(3),SU(2),Y]&({\rm Units\,\,of 10^{16} GeV})\vspace{0.1 cm}\\ \hline\hline
       \chi_{X}&  0.0794           &A[1,1,4]&      1.53 \\ \chi_{Z}&
  .0234
                &B[6,2,{5/3}]&            0.1096\\
           h_{11}/10^{-6}&  5.4571         &C[8,2,1]&{      0.96,      2.56,      5.81 }\\
           h_{22}/10^{-4}&  4.6430    &D[3,2,{7/ 3}]&{      0.06,      3.80,      6.83 }\\
                   h_{33}&  0.0225     &E[3,2,{1/3}]&{      0.14,      0.78,      2.05 }\\
 f_{11}/10^{-6}&
 -0.0149+  0.2047i
                      &&{     2.045,      3.14,      6.04 }\\
 f_{12}/10^{-6}&
 -2.3914-  2.7940i
          &F[1,1,2]&      0.20,      0.65
 \\f_{13}/10^{-5}&
 -0.2130-  0.2161i
                  &&      0.65,      3.99  \\
 f_{22}/10^{-5}&
  9.8088+  8.2589i
              &G[1,1,0]&{     0.023,      0.21,      0.70 }\\
 f_{23}/10^{-4}&
 -0.5695+  3.2942i
                      &&{     0.702,      0.76,      0.81 }\\
 f_{33}/10^{-3}&
 -1.2891+  0.6842i
              &h[1,2,1]&{     0.357,      2.69,      3.83 }\\
 g_{12}/10^{-4}&
  0.2339+  0.1782i
                 &&{      5.63,     24.55 }\\
 g_{13}/10^{-5}&
-10.6053-  0.3780i
     &I[3,1,{10/3}]&      0.34\\
 g_{23}/10^{-4}&
 -1.4787-  0.6802i
          &J[3,1,{4/3}]&{     0.300,      0.58,      1.33 }\\
 \lambda/10^{-2}&
 -5.6612+  0.3859i
                 &&{      1.33,      4.29 }\\
 \eta&
 -0.2656+  0.0830i
   &K[3,1, {8/ 3}]&{      2.00,      4.25 }\\
 \rho&
  0.6135-  0.3515i
    &L[6,1,{2/ 3}]&{      1.80,      2.92 }\\
 k&
  0.5779-  0.0493i
     &M[6,1,{8/ 3}]&      1.92\\
 \zeta&
  0.9346+  0.7021i
     &N[6,1,{4/ 3}]&      1.82\\
 \bar\zeta &
  0.3302+  0.8020i
          &O[1,3,2]&      2.57\\
       m/10^{16} \mbox{GeV}&  0.0130    &P[3,3,{2/ 3}]&{      0.71,      3.99 }\\
     m_\Theta/10^{16}\mbox{GeV}&  -2.873e^{-iArg(\lambda)}     &Q[8,3,0]&     0.272\\
             \gamma& -0.1121        &R[8,1, 0]&{      0.12,      0.38 }\\
              \bar\gamma& -3.5300     &S[1,3,0]&    0.4397\\
 x&
  0.7947+  0.6059i
         &t[3,1,{2/ 3}]&{ \hspace{-8mm}     0.26,      0.53,      0.88,      2.76        }\\\Delta_X&      0.52 &&{      4.40,      4.73,     27.10 }\\
                                \Delta_{G}^{\footnotesize{\mbox{Tot}}},\Delta_{G}^{{\scriptsize{\mbox{GUT}}}}&-19.0057,-21.7990           &U[3,3,{4/3}]&     0.368\\
      \{\Delta\alpha_{3}^{\footnotesize{\mbox{Tot}}},\Delta\alpha_{3}^{{\scriptsize{\mbox{GUT}}}}\}{\footnotesize{(M_{Z})}}& -0.0126, -0.0024               &V[1,2,3]&     0.288\\
    M^{\nu^c}/10^{12}\mbox{GeV}&{0.001043,    2.68,   81.22    }&W[6,3,{2/ 3}]&              1.68  \\
 M^{\nu}_{ II}/10^{-11}\mbox{eV}&\hspace{-8mm} 15.9, 40763.9,       1237293.6               &X[3,2,{5/ 3}]&  \hspace{-8mm}   0.098,     2.259,     2.259\\
                  M_\nu(\mbox{meV})&{1.351328,    7.17,   40.36    }&Y[6,2, {1/3}]&              0.13  \\
  \{\rm{Evals[f]}\}/ 10^{-6}&\hspace{-8mm} {0.01977,   50.71, 1538.58         }&Z[8,1,2]&              0.38  \\
 \hline\hline
 \mbox{Soft parameters}&{\rm m_{\frac{1}{2}}}=
           304.546
 &{\rm m_{0}}=
         12345.27
 &{\rm A_{0}}=
         -1.22 \times 10^{   4}
 \\
 \mbox{at $M_{X}$}&\mu=
          8.077 \times 10^{   4}
 &{\rm B}=
         -3.49 \times 10^{   8}
  &{\rm tan{\beta}}=           51.00\\
 &{\rm M^2_{\bar H}}=
         -5.6 \times 10^{   9}
 &{\rm M^2_{  H} }=
         -8.4\times 10^{   9}
 &
 {\rm R_{\frac{b\tau}{s\mu}}}=
  8.3873
  \\
\hspace{-6mm}\rm{ Max(|L_{ABCD}|,|R_{ABCD}|)}&
          4.8 \times 10^{ -22}
  {\,\rm{GeV^{-1}}}&& \\
 \hline\hline
 \mbox{Susy contribution to}&M_{{\tiny{\mbox{Susy}}}}=2.89~{\small{\mbox{TeV}}}&&
 \\
 {\rm \Delta_{X,G,3}}&{\rm \Delta_X^{{\tiny{\mbox{Susy}}}}}=
            -0.242
 &{\rm \Delta_G^{{\tiny{\mbox{Susy}}}}}=
             2.793
 &{\rm \Delta\alpha_3^{{\tiny{\mbox{Susy}}}}}=
            -0.010
 \\
 \hline\hline\end{array}
 $$
  \caption{\small{Fit 1 : Values of the NMSGUT-SUGRY-NUHM  parameters at $M_X$
  derived from an  accurate fit to all 18 fermion data and compatible with RG constraints.
 Unification parameters and mass spectrum of superheavy and superlight fields are  also given.
 \label{table41} }}\end{table}

NMSGUT solutions \cite{nmsgut,aulakhgargkhosa} have  $A_0$, $\mu$
$|M_{H,\bar H}|$ parameter of order of $10^5$ GeV. Large value of
$M_A$ gives huge corrections to sleptons and squarks masses and
can turn these negative. To overcome these obstacles one way is to
decouple each Susy particle at its mass threshold \cite{XTata}.
But implementation of this procedure is not trivial in multiple
iteration search code. Other possibility is to limit the size of
loop correction factor which is controlled by $\mu$ and $M_A$.
While finding the solution we restricted the
(\emph{LoopmassesMSSM} output) parameters such that
$.3<\big(\frac{\mu}{M_A^{Loop}}\big)^2<2.7$ by imposing the
penalty in
the program. Lower limit is decided from the tree level fits which have $M_A \approx 3 \mu$. 
Two successful solution sets are given in Tables
\ref{table41}-\ref{table-end} which have $\mu> M_A$. We also give
the tree level spectrum for both the solutions in Table
\ref{tree1} and \ref{tree2}. Spectrum calculated using one loop
EWSB conditions and including only one-loop corrections to CP-odd
Higgs boson and neutral Higgs, is given in Tables
\ref{table-sol1-oneloop} and \ref{table-sol2-oneloop}. Since the
two-loop corrections are small the spectrum is slightly modified.
The values of $\mu$ and $B$ calculated using one-loop EWSB
conditions are also provided. Results of nucleon decay rate
calculations corresponding to the solution found are given in
Table \ref{bdecayloop}. These fits have $A_0$, $\mu$, $M_H^2$ and
$M_{\bar H}^2$ parameter lesser by a factor of 10 as compared to
the tree level fits. Parameter $B$ is decreased by a factor of
100. Loop corrected spectrum have $M_A^{Tree}$ slightly greater
than $\mu$ parameter but earlier (tree level fits) we had $M_A
\approx 3 \mu$ where as now $M_A^{{\scriptsize{\mbox{loop}}}}
\approx .5 \mu$.
 \begin{table}
 $$
\begin{array}{ccccc}
 \hline\hline
 &&&&\vspace{-.3 cm}\\
 {\rm  Parameter }&{\rm Target} =\bar O_i &{\rm Uncert.= \delta_i }   &{\rm Achieved= O_i}
 & {\rm Pull =\frac{(O_i-\bar O_i)}{\delta_i }}\vspace{0.1 cm}\\
 \hline\hline
    y_u/10^{-6}&  2.092493&  0.799332&  2.092761&  0.000336\\
    y_c/10^{-3}&  1.019953&  0.168292&  1.019585& -0.002187\\
            y_t&  0.349569&  0.013983&  0.349591&  0.001589\\
    y_d/10^{-5}&  9.205462&  5.366784&  9.268855&  0.011812\\
    y_s/10^{-3}&  3.289656&  1.552718&  3.324815&  0.022643\\
            y_b&  0.282268&  0.146497&  0.290937&  0.059176\\
\hline\hline
 \end{array}
 $$
 \label{table b} \caption{\small{Fit 1 with $\chi_X=\sqrt{ \sum\limits_{i=1}^{17}
 \frac{(O_i-\bar O_i)^2}{\delta_i^2}}=
    0.0794
 $. Target values,  at $M_X$ of the fermion Yukawa
 couplings and mixing parameters, together with the estimated uncertainties, achieved values and pulls.
 The eigenvalues of the wavefunction renormalization for fermion  and
 Higgs lines are given with Higgs fractions $\alpha_i,{\bar{\alpha_i}}$ which control the MSSM fermion
  Yukawa couplings.
 }}
 \end{table}

 \begin{table}
 $$
 \begin{array}{cccc}
 \hline \hline &&&\vspace{-.3 cm}\\ {\rm  Parameter }& {\rm SM(M_Z)} &{\rm m^{GUT}(M_Z)} & {\rm m^{MSSM}=(m+\Delta m)^{GUT}(M_Z)}\vspace{0.1 cm} \\
 \hline \hline
    m_d/10^{-3}&   2.90000&   0.96198&   2.90145\\
    m_s/10^{-3}&  55.00000&  34.51146&  55.27487\\
            m_b&   2.90000&   2.46374&   2.95452\\
    m_e/10^{-3}&   0.48657&   0.48374&   0.48402\\
         m_\mu &   0.10272&   0.10222&   0.10228\\
         m_\tau&   1.74624&   1.72797&   1.72682\\
    m_u/10^{-3}&   1.27000&   1.12553&   1.27181\\
            m_c&   0.61900&   0.54834&   0.61960\\
            m_t& 172.50000& 147.29259& 172.36306\\
 \hline \hline
 \end{array}
 $$
 \label{table c}
 \caption{\small{Fit 1 : Values of the SM
 fermion masses in GeV at $M_Z$ compared with the masses obtained from
 values of GUT derived  Yukawa couplings  run down from $M_X$ to
 $M_Z$  both before and after threshold corrections.
  Fit with $\chi_Z=\sqrt{ \sum\limits_{i=1}^{9} \frac{(m_i^{{\tiny{\mbox{MSSM}}}}- m_i^{{\tiny{\mbox{SM}}}})^2}{ (m_i^{{\tiny{\mbox{MSSM}}}})^2} } =
.0234$.}}
 \end{table}
 \begin{table}
 $$
 \begin{array}{cccc}
 \hline \hline&&&\vspace{-.3 cm}\\
 {\rm  Parameter}  &{\rm Value}&  {\rm  Parameter}& {\rm Value }\vspace{0.1 cm}\\
 \hline \hline
                       M_{1}&            132.41&   m_{{\tilde {\bar {u}}_1}}&           5433.34\\
                       M_{2}&            260.08&   m_{{\tilde {\bar {u}}_2}}&           5435.12\\
                       M_{3}&            900.00&   m_{{\tilde {\bar {u}}_3}}&          38053.13\\
     m_{{\tilde {\bar l}_1}}&          17466.95&               A^{0(l)}_{11}&          -8987.41\\
     m_{{\tilde {\bar l}_2}}&          17545.48&               A^{0(l)}_{22}&          -8977.63\\
     m_{{\tilde {\bar l}_3}}&          33074.48&               A^{0(l)}_{33}&          -6202.67\\
        m_{{\tilde {L}_{1}}}&           9270.57&               A^{0(u)}_{11}&          -9976.33\\
        m_{{\tilde {L}_{2}}}&           9344.03&               A^{0(u)}_{22}&          -9976.12\\
        m_{{\tilde {L}_{3}}}&          21841.66&               A^{0(u)}_{33}&          -5835.50\\
     m_{{\tilde {\bar d}_1}}&          13413.69&               A^{0(d)}_{11}&          -9998.92\\
     m_{{\tilde {\bar d}_2}}&          13418.47&               A^{0(d)}_{22}&          -9998.43\\
     m_{{\tilde {\bar d}_3}}&          28449.38&               A^{0(d)}_{33}&          -6484.53\\
          m_{{\tilde {Q}_1}}&          12775.38&                   \tan\beta&             51.00\\
          m_{{\tilde {Q}_2}}&          12778.24&                    \mu(M_Z)&          76197.57\\
          m_{{\tilde {Q}_3}}&          34335.50&                      B(M_Z)&
          1.8027 \times 10^{   8}
 \\
 M_{\bar {H}}^2&
         -4.2990 \times 10^{   9}
 &M_{H}^2&
         -6.1500 \times 10^{   9}
 \\
 \hline \hline
 \end{array}
 $$
  \caption{\small{Fit 1 : Values (in GeV) of the soft Susy parameters  at $M_Z$
 (evolved from the soft SUGRY-NUHM parameters at $M_X$) which
 determine the Susy threshold corrections to the fermion Yukawas $(M_{{\tiny{\mbox{Susy}}}}=2.89~{\small{\mbox{TeV}}})$.\label{soft1}}}
 \end{table}

\begin{table}
 $$
 \begin{array}{cc}
 \hline \hline &\vspace{-.3 cm}\\{\mbox {Field } }&{\rm Mass(GeV)}\vspace{0.1 cm}\\
 \hline \hline
                M_{\tilde{g}}&            900.00\\
               M_{\chi^{\pm}}&            260.08,          76197.65\\
       M_{\chi^{0}}&            132.41,            260.08,          76197.62    ,          76197.62\\
              M_{\tilde{\nu}}&          9270.348,          9343.811,         21841.570\\
                M_{\tilde{e}}&           9270.69,          17467.00,           9344.11   ,          17545.56,          21840.17,          33075.53  \\
                M_{\tilde{u}}&           5433.23,          12775.27,           5435.00   ,          12778.13,          34335.69,          38053.43  \\
                M_{\tilde{d}}&          12775.51,          13413.72,          12778.34   ,          13418.53,          28445.36,          34338.89  \\
                        M_{A}&          95903.61\\
                  M_{H^{\pm}}&          95903.65\\
                    M_{H}&          95903.61\\
                    M_{h}&             90.40\\
 \hline \hline
 \end{array}
 $$
 \caption{\small{Fit 1 : Tree level spectra of supersymmetric partners calculated ignoring generation mixing effects.
  }\label{tree1}}\end{table}

\begin{table}
 $$
 \begin{array}{cc}
 \hline \hline&\vspace{-.3 cm}\\ {\mbox {Field } }&{\rm Mass(GeV)}\vspace{0.1 cm}\\
 \hline \hline
                M_{\tilde{g}}&           1174.85\\
               M_{\chi^{\pm}}&            297.42,          82465.86\\
       M_{\chi^{0}}&            141.08,            297.27,          82465.83    ,          82465.83\\
              M_{\tilde{\nu}}&         11506.049,         11573.595,         24714.071\\
                M_{\tilde{e}}&          11506.29,          14622.41,          11573.76   ,          14725.34,          24712.40,          33944.89  \\
                M_{\tilde{u}}&           9584.69,          12249.73,           9585.36   ,          12253.39,          34217.01,          34917.59  \\
                M_{\tilde{d}}&          12248.89,          12251.97,          12250.38   ,          12260.36,          34536.37,          34896.71  \\
                        M_{A}&          46913.36\\
                  M_{H^{\pm}}&          46913.38\\
                    M_{H}&          46913.56\\
                    M_{h}&            127.00\\
 \hline \hline
 \end{array}
 $$
 \label{table e}\caption{\small{Fit 1 : Loop corrected spectra of supersymmetric partners calculated ignoring generation mixing effects.}}\end{table}

\begin{table}
 $$
 \begin{array}{cc}
 \hline \hline&\vspace{-.3 cm}\\ {\mbox {Field } }&{\rm Mass(GeV)}\vspace{0.1 cm}\\
 \hline \hline
                M_{\tilde{g}}&           1174.85\\
               M_{\chi^{\pm}}&            297.55,          81878.03\\
       M_{\chi^{0}}&            141.11,            297.40,          81878.00    ,          81878.00\\
              M_{\tilde{\nu}}&         11443.477,         11510.892,         24614.190\\
                M_{\tilde{e}}&          11443.72,          14720.30,          11511.06   ,          14821.88,          24612.54,          33887.74  \\
                M_{\tilde{u}}&           9484.66,          12272.04,           9485.34   ,          12275.64,          34327.99,          34878.81  \\
                M_{\tilde{d}}&          12271.74,          12292.45,          12274.80   ,          12299.65,          34375.94,          34845.54  \\
                        M_{A}&          44572.62\\
                  M_{H^{\pm}}&          44570.38\\
                    M_{H}&          44570.54\\
                    M_{h}&            124.90\\
 \hline \hline
 \end{array}
 $$
 \caption{\small{Loop corrected spectra of supersymmetric partners calculated ignoring
  two loop tadpoles and two loop corrections to CP-odd pseudo scalar and neutral Higgs. $\mu$=75664.64 and $B$=1.7528 $\times$ $10^{8}$.
 \label{table-sol1-oneloop} }}\vspace{.7cm}\end{table}

\begin{table}
\vspace{-1.0cm}
 $$
 \begin{array}{cccc}
 \hline \hline&&&\vspace{-.4 cm}\\
 {\rm Parameter }&{\rm Value} &{\rm  Field }& {\rm Masses }\\
 &&\hspace{-10mm}{\rm}[SU(3),SU(2),Y]&({\rm Units\,\,of 10^{16} GeV})\vspace{0.1 cm}\\ \hline \hline
       \chi_{X}&  0.0986           &A[1,1,4]&      1.53 \\ \chi_{Z}&
  .0206
                &B[6,2,{5/3}]&            0.0762\\
           h_{11}/10^{-6}&  5.5774         &C[8,2,1]&{      0.99,      2.47,      5.06 }\\
           h_{22}/10^{-4}&  4.6169    &D[3,2,{7/ 3}]&{      0.11,      3.48,      5.93 }\\
                   h_{33}&  0.0240     &E[3,2,{1/3}]&{      0.10,      0.72,      1.90 }\\
\hline \hline \end{array}
 $$
 \label{table a} \caption{\small{Fit 2 : Values of the NMSGUT-SUGRY-NUHM  parameters at $M_X$
  derived from an  accurate fit to all 18 fermion data and compatible with RG constraints.
 Unification parameters and mass spectrum of superheavy and superlight fields are  also given.
  }}\end{table}

\begin{table}
 $$
 \begin{array}{ccccc}
 \hline \hline
 &&&&\vspace{-.3 cm}\\
 {\rm  Parameter }&{\rm Target} =\bar O_i &{\rm Uncert.= \delta_i }   &{\rm Achieved= O_i}
 & {\rm Pull =\frac{(O_i-\bar O_i)}{\delta_i }}\vspace{0.1 cm}\\
 \hline  \hline
    y_u/10^{-6}&  2.105633&  0.804352&  2.106028&  0.000491\\
    y_c/10^{-3}&  1.026362&  0.169350&  1.023794& -0.015164\\
            y_t&  0.353072&  0.014123&  0.353092&  0.001457\\
    y_d/10^{-5}& 10.797324&  6.294840& 10.836219&  0.006179\\
    y_s/10^{-3}&  3.338681&  1.575858&  3.354115&  0.009794\\
            y_b&  0.295397&  0.153311&  0.302034&  0.043295\\
\hline \hline
 \end{array}
 $$
 \label{table b} \caption{\small{Fit 2 with $\chi_X=\sqrt{ \sum\limits_{i=1}^{17}
 \frac{(O_i-\bar O_i)^2}{\delta_i^2}}=
    0.0986$. Target values,  at $M_X$ of the fermion Yukawa
 couplings and mixing parameters, together with the estimated uncertainties, achieved values and pulls.
 The eigenvalues of the wavefunction renormalization for fermion  and
 Higgs lines are given with Higgs fractions $\alpha_i,{\bar{\alpha_i}}$ which control the MSSM fermion
  Yukawa couplings.
 }}
 \end{table}
\begin{table}
 $$
 \begin{array}{cccc}
 \hline \hline &&&\vspace{-.3 cm}\\ {\rm  Parameter }&{\rm SM(M_Z)} & {\rm m^{GUT}(M_Z)} & {\rm m^{MSSM}=(m+\Delta m)^{GUT}(M_Z)}\vspace{0.1 cm} \\
 \hline  \hline
    m_d/10^{-3}&   2.90000&   1.09359&   2.90288\\
    m_s/10^{-3}&  55.00000&  33.84880&  55.12840\\
            m_b&   2.90000&   2.46581&   2.94708\\
    m_e/10^{-3}&   0.48657&   0.48450&   0.48488\\
         m_\mu &   0.10272&   0.10320&   0.10327\\
         m_\tau&   1.74624&   1.72938&   1.72822\\
    m_u/10^{-3}&   1.27000&   1.13124&   1.27398\\
            m_c&   0.61900&   0.54992&   0.61931\\
            m_t& 172.50000& 147.89942& 172.73656\\
 \hline \hline
 \end{array}
 $$
 \label{table c}
 \caption{\small{Fit 2 : Values of the SM
 fermion masses in GeV at $M_Z$ compared with the masses obtained from
 values of GUT derived  Yukawa couplings  run down from $M_X$ to
 $M_Z$  both before and after threshold corrections.
  Fit with $\chi_Z=\sqrt{ \sum\limits_{i=1}^{9} \frac{(m_i^{{\tiny{\mbox{MSSM}}}}- m_i^{{\tiny{\mbox{SM}}}})^2}{ (m_i^{{\tiny{\mbox{MSSM}}}})^2} } =
.0206$.}}
 \end{table}
 \begin{table}
 $$
 \begin{array}{cccc}
 \hline  \hline&&&\vspace{-.3 cm}\\
 {\rm  Parameter}  &{\rm Value }&  {\rm  Parameter}&{\rm  Value }\vspace{0.1 cm}\\
 \hline  \hline
                       M_{1}&            117.91&   m_{{\tilde {\bar {u}}_1}}&           3840.82\\
                       M_{2}&            231.79&   m_{{\tilde {\bar {u}}_2}}&           3842.97\\
                       M_{3}&            800.05&   m_{{\tilde {\bar {u}}_3}}&          35037.40\\
     m_{{\tilde {\bar l}_1}}&          15569.94&               A^{0(l)}_{11}&          -7954.28\\
     m_{{\tilde {\bar l}_2}}&          15650.58&               A^{0(l)}_{22}&          -7944.94\\
     m_{{\tilde {\bar l}_3}}&          30854.82&               A^{0(l)}_{33}&          -5342.45\\
        m_{{\tilde {L}_{1}}}&           7850.18&               A^{0(u)}_{11}&          -8933.50\\
        m_{{\tilde {L}_{2}}}&           7929.53&               A^{0(u)}_{22}&          -8933.32\\
        m_{{\tilde {L}_{3}}}&          20331.96&               A^{0(u)}_{33}&          -5164.69\\
     m_{{\tilde {\bar d}_1}}&          11787.72&               A^{0(d)}_{11}&          -8852.66\\
     m_{{\tilde {\bar d}_2}}&          11792.39&               A^{0(d)}_{22}&          -8852.22\\
     m_{{\tilde {\bar d}_3}}&          26506.85&               A^{0(d)}_{33}&          -5605.50\\
          m_{{\tilde {Q}_1}}&          11197.93&                   \tan\beta&             52.00\\
          m_{{\tilde {Q}_2}}&          11200.73&                    \mu(M_Z)&          69679.92\\
          m_{{\tilde {Q}_3}}&          31725.86&                      B(M_Z)&
          1.4655 \times 10^{   8}
 \\
 M_{\bar {H}}^2&
         -3.6171 \times 10^{   9}
 &M_{H}^2&
         -5.1774 \times 10^{   9}
 \\
 \hline  \hline
 \end{array}
 $$
 \caption{\small{Fit 2 : Values (in GeV) of the soft Susy parameters  at $M_Z$
 (evolved from the soft SUGRY-NUHM parameters at $M_X$) which
 determine the Susy threshold corrections to the fermion Yukawas $(M_{{\tiny{\mbox{Susy}}}}=2.92~{\small{\mbox{TeV}}})$.
 \label{soft2}  }}
 \end{table}

 \begin{table}
 $$
 \begin{array}{cc}
 \hline \hline &\vspace{-.3 cm}\\{\mbox {Field } }&{\rm Mass(GeV)}\vspace{0.1 cm}\\
 \hline  \hline
                M_{\tilde{g}}&            800.05\\
               M_{\chi^{\pm}}&            231.78,          69680.01\\
       M_{\chi^{0}}&            117.91,            231.78,          69679.97    ,          69679.98\\
              M_{\tilde{\nu}}&          7849.919,          7929.269,         20331.856\\
                M_{\tilde{e}}&           7850.32,          15570.00,           7929.62   ,          15650.67,          20330.35,          30855.95  \\
                M_{\tilde{u}}&           3840.66,          11197.80,           3842.81   ,          11200.61,          31726.08,          35037.73  \\
                M_{\tilde{d}}&          11198.08,          11787.75,          11200.84   ,          11792.46,          26502.28,          31729.75  \\
                        M_{A}&          87313.43\\
                  M_{H^{\pm}}&          87313.47\\
                    M_{H}&          87313.43\\
                    M_{h}&             90.47\\
 \hline  \hline
 \end{array}
 $$
 \caption{\small{Fit 2 : Tree level spectra of supersymmetric partners calculated
  ignoring generation mixing effects.
  }\label{tree2}}\end{table}

 \begin{table}
 $$
 \begin{array}{cc}
 \hline \hline &\vspace{-.3 cm}\\{\mbox {Field } }& {\rm Mass(GeV)} \vspace{0.1 cm}\\
 \hline  \hline
                M_{\tilde{g}}&           1046.78\\
               M_{\chi^{\pm}}&            264.60,          75524.08\\
       M_{\chi^{0}}&            125.28,            264.44,          75524.04    ,          75524.04\\
              M_{\tilde{\nu}}&         9970.212,         10040.719,         22929.231\\
                M_{\tilde{e}}&            9970.50,          12962.67,          10040.90   ,          13067.75,          22927.44,          31698.40   \\
                M_{\tilde{u}}&           8060.81,          10699.37,           8061.49   ,          10702.88,          31650.65,          32370.72  \\
                M_{\tilde{d}}&          10698.71,          10701.36,          10699.79   ,          10709.91,          32068.70,          32371.52  \\
                        M_{A}&          42900.74\\
                  M_{H^{\pm}}&          42900.76\\
                    M_{H}&          42900.90\\
                    M_{h}&            127.00\\
 \hline  \hline
 \end{array}
 $$
\caption{\small{Fit 2 : Loop corrected spectra of supersymmetric
partners calculated ignoring
  generation mixing effects.
  } \label{table-end}}\end{table}

\begin{table}
 $$
 \begin{array}{cc}
 \hline\hline &\vspace{-.3 cm}\\ {\mbox {Field } }& {\rm Mass(GeV)} \vspace{0.1 cm}\\
 \hline\hline
                M_{\tilde{g}}&           1046.78\\
               M_{\chi^{\pm}}&            264.73,          75039.61\\
       M_{\chi^{0}}&            125.32,            264.56,          75039.57    ,          75039.58\\
              M_{\tilde{\nu}}&         9903.666,         9974.010,         22818.584\\
                M_{\tilde{e}}&           9903.95,          13064.36,           9974.20   ,          13167.78,          22816.81,          31624.16  \\
                M_{\tilde{u}}&           7951.10,          10722.79,           7951.80   ,          10726.22,          31729.56,          32294.13  \\
                M_{\tilde{d}}&          10722.63,          10744.40,          10725.47   ,          10751.32,          31874.24,          32275.46  \\
                        M_{A}&          40376.80\\
                  M_{H^{\pm}}&          40373.97\\
                    M_{H}&          40374.08\\
                    M_{h}&            124.85\\
 \hline\hline
 \end{array}
 $$
\caption{\small{Loop corrected spectra of supersymmetric partners
calculated ignoring
  two loop tadpoles and two loop corrections to CP-odd pseudo scalar and neutral Higgs. $\mu$=69242.22 and $B$=1.4198 $ \times $ $10^{8}$.
 \label{table-sol2-oneloop} }}\end{table}

\begin{table}
  $$
  \begin{array}{ccc}
  \hline \hline &&\vspace{-.3 cm}\\{\rm Parameter } &\multicolumn{2}{ c }{\rm Value}\\
  & {\rm Fit-1} & {\rm Fit-2}\vspace{0.1 cm} \\
  \hline \hline
   \rm{ \tau_p(M^+ \bar\nu)}&
                 7.28 \times 10^{  34} & 1.39 \times 10^{  35}

 \\\hline
    \rm{ \Gamma(p\rightarrow \pi^+ \bar\nu)}&
                 1.71 \times 10^{ -36}& 6.71 \times 10^{ -37}

 \\
             \rm{BR( p\rightarrow\pi^+\bar \nu_{e,\mu,\tau})}&
 \{
                 5.15 \times 10^{  -3}
 ,
                 0.007
 ,
                 0.987
 \}&
 \{
                 .013
 ,
                 0.006
 ,
                 0.981
 \}

\\
     \rm{\Gamma(p\rightarrow K^+ \bar\nu)}&
                 1.20 \times 10^{ -35}
&  6.544 \times 10^{ -36}\\
        \rm{ BR( p\rightarrow K^+ \bar\nu_{e,\mu,\tau})} &
 \{
                 6.36 \times 10^{  -3}
 ,
                 0.049
 ,
                 0.944
 \}&
\{
                 8.93 \times 10^{  -3}
 ,
                 0.031
 ,
                 0.96
 \}

 \\
\hline  \hline
 \end{array}
 $$
\caption{\small{Table of d=5 operator mediated proton
and neutron lifetimes $\tau_{p,n}$(yrs), decay rates
$\Gamma$(yr$^{-1}$)  and branching ratios in the
different channels.\label{bdecayloop}}}
 \end{table}
 \vspace{1.2cm}
\section{Discussion}
We have found the NMSGUT superpotential couplings and mSUGRY-NUHM
parameter sets which provide realistic fermion mass-mixing data,
B-decay rates respecting experimental limits and consistent loop
corrected Susy
   spectra as shown in Table \ref{table41}-\ref{table-end}. Computation of sfermion self eneries require special treatment as heavy CP-pseudo scalar $A$ and chargino/neutralino give huge corrections. To control these corrections the $(\frac{\mu}{M_A})^2$ ratio is
   kept
    within the range : 0.3-2.7 . Consequently the heavy chargino/neutralino masses and heavy Higgs masses ($M_A,H,H^\pm$) are smaller (O(80) TeV, 40-50 TeV in our example fits).
 Loop corrected spectra retain the distinctive feature (of NMSGUT): normal s-hierarchy.  The reason is the preference for huge negative soft Higgs masses
($M^2_{H,\bar H}$) $\approx$ -$10^{9}$ GeV$^2$ in the fits. LSP is
pure Bino as before. We shall see (in Chapter 6) that the NMSGUT
provides a natural reason why the soft Higgs masses become
negative already at the GUT scale. Loop
      corrected spectrum have right handed up squarks as the lightest sfermion instead of light smuon. The proton decay rates are still suppressed as
explained in Chapter 3.

As discussed this ratio $\frac{\mu}{M_A}$ is crucial for the Higgs
sector and Higgsino loop correction
    to the scalars. Values of this ratio found
in the fits approach the upper limit applied. We have not yet
found light smuon solution after including loop
    corrections which require parameter space scan with different ranges of $\frac{\mu}{M_A}$. As a check we calculated
     the one-loop spectrum using package SuSpect \cite{suspect} by providing it soft masses (at $M_Z$, given
      in Tables \ref{soft1} and \ref{soft2}) along with $\mu$ and $M_A$. We found acceptable agreement between the two packages except for sfermion
masses which are not  at all  accurate in SuSpect  when $M_A$ is
very large since no contribution at all from $M_A$  is included on
the grounds that large $M_A$($>>$ 500 GeV)  corresponds to high
fine tuning (as noted on the SuSpect  webpage \cite{suspectweb}:
comment on bug in Version 2.43). However the actual corrections
\cite{piercebagger} are proportional to $M_A^2$ and  can be as
large as  100\%!  Fortunately the subroutines from SPheno used by
us include the complete corrections of \cite{piercebagger}.

We are still searching for fits with light smuons so as to keep
open the possibility of muon g-2 anomaly resolution and as DM
co-annihilation channel.

\newpage\thispagestyle{empty}\mbox{}\newpage
\chapter{Lepton Flavour
Violation} \label{ch:method}
\section{Introduction}
The SM renormalizable Lagrangian respects four global U(1)
symmetries (individual family lepton numbers $L_e$, $L_{\mu}$,
$L_{\tau}$ and B). Right handed neutrinos are absent by choice in
the SM. Without including non-renormalizable operators, neutrinos
remain massless in the SM. However it is clear from the neutrino
oscillation data that neutrinos are massive and their flavour
states mix with each other. Therefore  it is confirmed that lepton
flavour is violated in nature. Most mechanisms for generating
neutrino masses inevitably lead to lepton number violation and the
simplest one is the so called `seesaw mechanism'. SO(10) GUTs can
naturally accommodate seesaw and Susy together since they embed
the minimal Susy LR models which have high scale breaking of B-L
symmetry \cite{LR1,LR2,rparso10.1,rparso10.2,LR3}. In Susy GUTs
lepton flavour violation (LFV) in the neutrino sector generates
LFV in the charged lepton sector.

Besides explaining smallness of neutrino masses, another
attractive feature of the seesaw mechanism is that it provides a
natural mechanism for generating the observed baryon asymmetry of
the universe through leptogenesis.  Once the MSSM is extended with
three right handed neutrinos, the lepton number asymmetry can
arise via a lepton number and CP violating out of equilibrium
decay of the right handed neutrino. This can be followed by
processing of the net lepton number into baryon asymmetry via
sphaleron processes at scales near to but higher than $M_Z$
\cite{lepto1}. The E821 experiment \cite{BNL} at Brookhaven
National Laboratory  observed a significant deviation of the muon
g-2 from its SM prediction. This observation provides hint for
physics beyond SM . In Susy GUTs extra contribution to muon g-2
than SM will come from loop diagrams involving Susy particles. In
this chapter we discuss rare muon decay in extended SM and lepton
flavour violation in Susy-GUTs and in particular NMSGUT
predictions for mentioned lepton sector observables.
\section{Standard Model Muon Decay}
Muon decay $(\mu \rightarrow e \nu_e \bar\nu_{\mu}  )$ in the SM
is mediated by $W^{\pm}$ gauge bosons. Due to suppression of heavy
gauge boson propagation this decay is represented as point like
4-fermion interaction governed by the interaction Lagrangian \be
\mathcal {L}_{int}=-\frac{G_F}{\sqrt{2}}[\bar e
\gamma^{\mu}(1-\gamma_5)\nu_{e} \bar\nu_{\mu}
\gamma_{\mu}(1-\gamma_5)\mu ]\ee Here $G_F$(=$ \frac{g^2 }
{4\sqrt{2}M_W^2 }$) is Fermi constant . \be \Gamma( \mu
\rightarrow e \nu_e \bar\nu_{\mu} )=\frac{ G_F^2 m_{\mu}^5} {192
\pi^3 }\ee
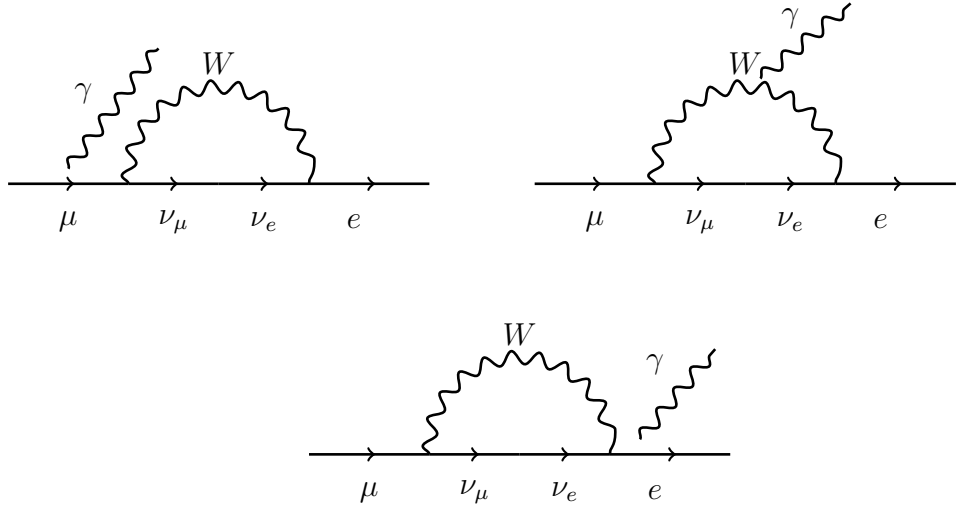
\begin{figure}[tbh]
\vspace{1.0cm} \centering
\begin{tikzpicture}[line width=1.0 pt, scale=2.0]
    \draw[fermion] (0,0)--(.8,0);
     \node at (.4,-.25) {$\mu$};
     \node at (2.3,-.25) {$e$};
     \node at (.5,.6) {$\gamma$};
     \node at (1.1,-.25) {$\nu_\mu$};
     \node at (1.7,-.25) {$\nu_e$};
     \node at (1.4,.8) {$W$};
    \draw[vector] (.8,0) arc (180:0:.60);
   \draw[fermion] (.8,0) -- (1.4,0);
   \draw[fermion] (1.4,0) -- (2.0,0);
    \draw[fermion] (2.0,0) --(2.8,0);
     \draw[vector] (0.4,.1) -- (1.0,.9);
\draw[fermion] (3.5,0)--(4.3,0);
 \node at (3.9,-.25) {$\mu$};
     \node at (5.8,-.25) {$e$};
    \node at (5.2,1.1) {$\gamma$};
     \node at (4.6,-.25) {$\nu_\mu$};
      \node at (5.2,-.25) {$\nu_e$};
    \node at (4.9,.8) {$W$};
    \draw[vector] (4.3,0) arc (180:0:.6);
    \draw[fermion] (4.3,0) -- (4.9,0);
\draw[fermion] (4.9,0) -- (5.5,0);
    \draw[fermion] (5.5,0) --(6.3,0);
     \draw[vector] (5.0,.7) -- (5.6,1.2);

\draw[fermion] (2,-1.8)--(2.8,-1.8);
     \node at (2.4,-2.05) {$\mu$};
     \node at (4.3,-2.05) {$e$};
     \node at (4.3,-1.2) {$\gamma$};
     \node at (3.1,-2.05) {$\nu_\mu$};
     \node at (3.7,-2.05) {$\nu_e$};
     \node at (3.4,-1.0) {$W$};
    \draw[vector] (2.8,-1.8) arc (180:0:.60);
   \draw[fermion] (2.8,-1.8) -- (3.4,-1.8);
   \draw[fermion] (3.4,-1.8) -- (4.0,-1.8);
    \draw[fermion] (4.0,-1.8) --(4.8,-1.8);
     \draw[vector] (4.2,-1.7) -- (4.7,-1.1);

\end{tikzpicture}
\vspace{.5cm} \caption{Feynman diagrams representing SM $\mu
\rightarrow e \gamma$ decay\label{SMLFV}} \vspace{1.0cm}
\end{figure}
The $\mu \rightarrow e \nu_e \bar\nu_{\mu}  $ channel accounts for
nearly 100 $\%$ of the muon decay width. However many other
channels for muon decay width are conceivable in the SM and
beyond. For example the diagram \ref{SMLFV} accounts for the
lepton flavour violating decay $\mu \rightarrow e \gamma$ in the
SM supplemented by Majorana mass term for neutrino mass (coded in
the Weinberg operator $((M^{\bar\nu})^{-1}_{ij} L_i H L_j H$).
Decay rate and branching ratio (BR) are given by : \be \Gamma(\mu
\rightarrow e \gamma )=\frac{ G_F^2 m_{\mu}^5} {192 \pi^3 }
\frac{3 \alpha } { 32 \pi} \bigg[\sum_{i} U^*_i U_i
\frac{m_{\nu_i}^2 } {M_W^2 }\bigg]^2\ee
 \begin{table}
 \vspace{.8cm}
 $$
\begin{array}{cc}
 \hline\hline &\vspace{-.3 cm}\\{\mbox {Process } }& \mbox{Present Upper Bound}\vspace{0.1 cm}\\
 \hline \hline
    \mu \rightarrow e \gamma
 &
          5.7 \times 10^{ -13}
 \\
 \tau \rightarrow \mu \gamma
 &
          4.4 \times 10^{  -8}
 \\
\tau \rightarrow e \gamma
 &
          3.3 \times 10^{  -8}
 \\
\mu \rightarrow e e e
 &
          1.0 \times 10^{ -12}
 \\
 \tau \rightarrow e e e
 &
          2.7 \times 10^{  -8}
 \\
 \tau \rightarrow \mu \mu \mu
 &
          2.1 \times 10^{-8}
 \\
\hline \hline
 \end{array}
 $$
 \label{table e}\caption{\small{Experimental upper bound \cite{MEG,BaBar,SINDRUM} for the BR of LFV processes.
 \label{BR} }}\end{table} \be B(\mu \rightarrow e
\gamma)=\frac{\Gamma(\mu \rightarrow e \gamma )  }{\Gamma(\mu
\rightarrow e \nu_e \bar\nu_{\mu} ) }\simeq
 \frac{\alpha  }{2\pi  } \bigg(\frac{\Delta m^2 } {M_W^2 } \bigg)^2 \simeq 10^{-55}\ee
Here $U$ is the neutrino mixing matrix, $\Delta m^2$ is neutrino
mass square difference parameter and $M_W$ is mass of W-boson. SM
predicts  unmeasurable BR for $\mu \rightarrow e \gamma  $ even
after the introduction of neutrino masses and mixing. MEG
experiment \cite{MEG}(and several others \cite{BaBar,SINDRUM})
have provided an improved upper bound on BR for rare LFV decay
processes (see Table \ref{BR}). One needs to include the
contribution of new physics (beyond SM) to get BR comparable to
the experimental limit.
\section{LFV in Supersymmetric GUTs}
The SM extensions predict lepton flavour violation. In the unified
scenario LFV has been studied by Barbieri \emph{et al.}
considering large top Yukawa couplings
\cite{Barbieri:1,Barbieri:2}. The main consequence of the seesaw
mechanism in supersymmetric theories is the violation of lepton
number in the scalar sector leading to rare lepton flavour
violating decays. Many groups have discussed this issue
considering different frameworks
\cite{hisano,Hisano2,lfv1,lfv2,lfv3,lfv4,lfv5}. However NMSGUT
also predicts neutrino Yukawa coupling like the Yukawa couplings
of other SM fermions. LFV ultimately originates from the neutrino
Yukawa coupling, because neutrino Yukawa couplings in general are
not diagonal in the basis in which charged lepton Yukawa coupling
and right-handed neutrino mass matrix are diagonal. The
off-diagonal neutrino Yukawa couplings will give rise to LFV in
the soft slepton masses through the RGEs \cite{hisano,Hisano2}.
Lepton flavour violation will be generated through the RGE even if
the supersymmetry breaking mechanism at the high scale conserves
flavour as it does in mSUGRY. Leptonic superpotential of the MSSM
+ three right handed neutrino is given by :- \vspace{.15cm}\be
W_{lep}=e^c_{i} (Y_e)_{ij}L_jH_d+{\nu}^c_i
(Y_\nu)_{ij}L_jH_u-\frac{1}{2}{\nu}^c_{i}M_{ij}^{\bar{\nu}}{\nu}^c_{j}
\vspace{.15cm}\ee Here $H_u$ and $H_d$ are MSSM Higgs doublets
whose VEVs generate mass of the up-type and down-type fermions,
($Y_e$, $Y_{\nu}$) are charged lepton and neutrino Yukawa
couplings and $M^{\bar{\nu}} $ is right handed neutrino Majorana
mass. In Susy-GUTs, right-handed neutrino Yukawa couplings
generate lepton flavour violation in the soft Susy breaking
lagrangian \vspace{.15cm}\bea -
{\cal{L}}_{{\scriptsize{\mbox{soft}}}}^{lep}&=&(m_{\tilde{L}}^2)_{ij}
\tilde{L}^*_i\tilde{L}_j +
(m_{\tilde{\nu}}^2)_{ij}\tilde{\nu}^*_{Ri}\tilde{\nu}_{Rj}
\nonumber\\&&+ (A_{ij}^e\tilde{e}_{Ri} H_d \tilde{L}_{Rj}+
A_{ij}^{\nu }\tilde{\nu}_{Ri} H_u \tilde{L}_{Rj} + h.c.
)\vspace{.15cm}\eea where the $\tilde{L}$, $ \tilde{e}_{R} $ and $
\tilde{\nu}_{R} $ are the leptons Susy partners (sleptons). The
parameters ($m_{\tilde{L}}^2$, $ m_{\tilde{\nu}}^2 $) and ($A^e$,
$A^\nu$) are slepton's soft mass matrices squared and trilinear
couplings. Additional contribution by neutrino Yukawa couplings to
the one loop RGE of the slepton masses $(m^2_{\tilde{L}})$ is
given by \cite{hisano} \be \frac{1}{16
\pi^2}[(m^2_{\tilde{L}}Y_{\nu}^\dag Y_{\nu}+ Y_{\nu}^\dag
Y_{\nu}m^2_{\tilde{L}})_{ij}+2(Y_{\nu}^\dag
m_{\tilde{\nu}}^2Y_{\nu}+M_{H}^2 Y_{\nu}^\dag Y_{\nu}+A_{\nu}^\dag
A_{\nu})_{ij}]\ee The leading contribution to the off-diagonal
entries of the slepton mass squared matrix can be estimated as
\cite{hisano}
 \be \vspace{.3cm} \Delta
(m_{\tilde{L}}^2)_{ij}\simeq \frac{2m_0^2+M_{H}^2+A_0^2}{8 \pi^2}
\sum_{k}(Y_\nu^*)_{ik}(Y_\nu)_{jk}
\log\bigg(\frac{M_X}{M^{\bar{\nu}}_k} \bigg)\ee The flavour
violation in the slepton sector contributes through the one loop
diagrams to charged flavour violating processes such as rare muon
and tau decay $(\mu\rightarrow e \gamma, \tau \rightarrow \mu
\gamma, \tau \rightarrow e, e, e\,\,\, \mbox{etc.})$ and yields BR
: \be \mbox{BR}(l_i \rightarrow l_j \gamma) \thickapprox
\frac{\alpha^3}{G_F^2} \frac{((m_{\tilde L}^2)_{ij})^2}{M_S^8} \ee
Where $\alpha$ is fine structure constant and $M_S$ is Susy
particles mass scale. \vspace{1.0cm}
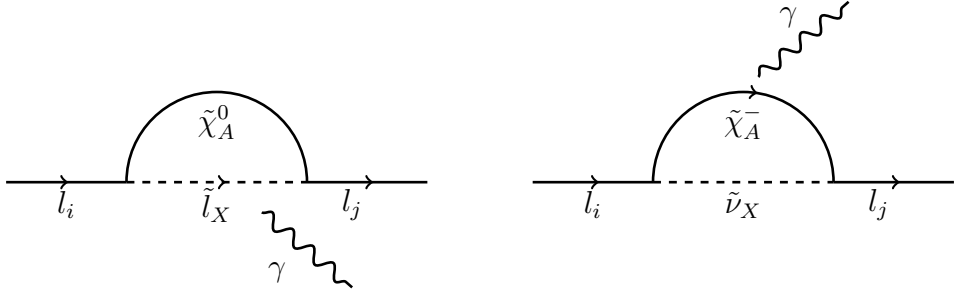
\begin{figure}[tbh]
\centering
\begin{tikzpicture}[line width=1.0 pt, scale=2.0]
    \draw[fermionbar] (.8,0)--(0,0);
     \node at (.4,-.15) {$l_i$};
     \node at (2.3,-.15) {$l_j$};
     \node at (1.8,-.6) {$\gamma$};
     \node at (1.4,-.15) {$\tilde{l}_X$};
     \node at (1.4,.4) {$\tilde{\chi}_A^0$};
    \draw[fermion1] (.8,0) arc (180:0:.6);
    \draw[scalar] (.8,0) -- (2.0,0);
    \draw[fermion] (2.0,0) --(2.8,0);
     \draw[vector] (1.7,-.2) -- (2.3,-.7);
\draw[fermion] (3.5,0)--(4.3,0);
 \node at (3.9,-.15) {$l_i$};
     \node at (5.8,-.15) {$l_j$};
     \node at (5.2,1.1) {$\gamma$};
     \node at (4.9,-.15) {$\tilde{\nu}_X$};
     \node at (4.9,.4) {$\tilde{\chi}_A^-$};
    \draw[fermion] (4.3,0) arc (180:0:.6);
    \draw[scalar1] (4.3,0) -- (5.5,0);
    \draw[fermion] (5.5,0) --(6.3,0);
     \draw[vector] (5.0,.7) -- (5.6,1.2);
\end{tikzpicture}
\vspace{.5cm} \caption{Feynman diagrams representing the
contribution of neutralino $-$ slepton and chargino $-$ sneutrino
loop for the process $l_i \rightarrow l_j \gamma$. The photon line
is to be attached on each of the charged particle
lines.\label{lfvbsm}}
\end{figure}
\vspace{1.0cm} Feynman diagrams contributing to these processes
are shown in Fig. \ref{lfvbsm}. The upper bounds on BR of these
processes may constrain very strictly the elements of soft mass
matrices. It is important to estimate these LFV  rates from NMSGUT
on the basis of fits of the fermion data. To estimate LFV, one
needs to consider the effect of right handed neutrino evolution
upto their mass scale which we will discuss in the following
section.
\section{NMSGUT LFV Predictions}
In NMSGUT using the formulae of Eq. \eqref{yukawa} Yukawa
couplings of SM fermion and neutrino are calculated. Then GUT
scale threshold corrections
 (discussed in Chapter \ref{chgut}) are applied to the tree level Yukawas (see Eq. \ref{Ynutilde}). At GUT scale right handed neutrino are present
  in the theory (along with the other fermions) which need to be integrated out below their mass scale. We shift to the right handed neutrino and charged lepton diagonal basis by redefining the fields :
\be  {Y'}_{\nu}=U^T_{\bar\nu}\tilde{Y_\nu}  \quad ; \quad {Y'}_e=
V^T_e \tilde{Y_e} U_e \quad ;\quad
M_{\bar\nu}^D=U^T_{\bar\nu}M_{\bar\nu} U_{\bar\nu} \ee Here
$\tilde{Y}_f$ ($f=\nu$, $e$) is threshold corrected Yukawa
coupling, redefined in a basis where kinetic terms have canonical
form.
\subsection{Right Handed Neutrino Thresholds}
We incorporated the three thresholds associated with the heavy
right handed neutrino masses in the theory. For this  neutrino
Dirac coupling is progressively introduced \cite{nuflow} on
passing these thresholds when going higher in energy and right
handed neutrino are decoupled one by one going the other way.
\[ \overbrace{\underbrace{MSSM}_\text{EFT1}}^\text{ $\kappa^{(1)} $} \overset{M_{\bar\nu_1}}{\longrightarrow}
\overbrace{\underbrace{ MSSM+\bar\nu_1}_\text{EFT2}}^\text{
$\kappa^{(2)} $, $Y^{(2)}$ and $M^{(2)}$}\overset{M_{\bar\nu_2}}
{\longrightarrow} \overbrace{\underbrace{MSSM+\bar\nu_1,
\bar\nu_2}_\text{EFT3}}^\text{ $\kappa^{(3)}$, $Y^{(3)}$ and
$M^{(3)}$} \overset{M_{\bar\nu_3}}{\longrightarrow}\overbrace{
MSSM+\bar\nu_1, \bar\nu_2, \bar\nu_3}^\text{Full Theory $Y^{\nu}$,
$M^{\bar\nu}$ } \]
 To do this one needs
to consider different effective theories (EFTs) in each of the
energy ranges defined by right handed neutrino masses. We use the
RGEs of the MSSM parameters extended with three right handed
neutrino superfields \cite{rhn} upto the energy scale of the
lightest right handed neutrino by using the matching condition at
the thresholds of right handed neutrino as discussed below. Along
with these parameters we also consider the running
\cite{Babu:1993qv} of dimension five (Weinberg \cite{weinbergd5})
operators which provide effective neutrino mass matrices below
successive heavy mass thresholds. In the full theory there is no
$\kappa$ (coefficient of dimension five operator). Once third
generation right handed neutrino is eliminated, a dimension five
Weinberg operator will appear in the theory. Leptonic
superpotential in the above discussed basis is: \be W_{lep}={\bar
e}^T_{A} Y^e_{AB}e_B+{\bar \nu}^T_{A}
Y^\nu_{AB}\nu_B-\frac{1}{2}{\bar
\nu}^T_{A}M^{\bar{\nu}}_{AA}{\bar\nu}_{A} \label{sp}\ee here
$A,B={1,2,3}$ are generation indices. Split A, B into a, 3 and b,
3 where a, b run from 1 to 2. At scale $M_3^{\bar\nu}$ we
diagonalize the mass matrix of right handed neutrinos. Writing
only last 2 terms of the superpotential (Eq. \ref{sp}) with
indices a, b and 3 in right handed diagonal basis : \bea \nonumber
W_{lep}^{{\tiny{\mbox{FT}}}}=&{\bar \nu}^T_{a}
Y^\nu_{ab}\nu_b+{\bar \nu}^T_{3} Y^\nu_{33}\nu_3 +{\bar \nu}^T_{3}
Y^\nu_{3a}\nu_a+{\bar\nu}^T_{a} Y^\nu_{a3}\nu_3\\
&-\frac{1}{2}{\bar\nu}^T_{a}M_{aa}^{\bar{\nu}}{\bar\nu}_{a}
-\frac{1}{2}{\bar\nu}^T_{3}M_{33}^{\bar{\nu}}{\bar\nu}_{3}\label{effp}
\eea By solving the superpotential equation of motion for the
heavy singlet ${\bar\nu}_3$ to leading order in
$(M^{\bar\nu})^{(-1)}$ we get \be
 {\bar\nu}_{3}=\frac{Y^\nu_{33}\nu_3+Y^\nu_{3a}\nu_a}{M_{33}^{\bar{\nu}}}=\frac{Y^\nu_{3A}\nu_A}{M_{33}^{\bar{\nu}}} \ee
Substituting back in \eqref{effp} the effective superpotential is
given
 by
 \be W_{eff}={\bar\nu}^T_{a}
Y^\nu_{aB}\nu_B-\frac{1}{2}{\bar\nu}^T_{a}M_{aa}{\bar{\nu}}_{a}
+\nu_A^T\biggr(\frac{1}{2}\frac{Y^\nu_{3A} Y^\nu_{3B}
 }{M_{33}^{\bar{\nu}}}\biggr)\nu_B
\ee  From third term we identify the coupling in EFT3
$\kappa^{(3)} $: \be \kappa^{(3)}_{AB}=\frac{1}{2} (Y^{
\nu})^T_{A3}M_{33}^{-1}(Y^{ \nu})_{3B}  \ee
This condition should be imposed at $\mu=M^{\bar{\nu} }_3 $
(largest eigenvalue of $M^{\bar\nu}$). Now the Yukawa matrix
$Y_{aB}^{(3)}$ is $2\times 3$ and $M^{(3)}_{ab}$ is $2 \times 2$.
\be W_{eff}^{{\tiny{\mbox{EFT3}}}}=
{\bar\nu}^T_{a}Y^{(3)}_{aB}\nu_B-\frac{1}{2}{\nu}^c_{a}M_{ab}^{(3)}{\bar{\nu}}^c_{b}
+\nu_A^T \kappa^{(3)}_{AB} \nu_B \label{eff2} \ee In the full
theory light neutrino mass matrix is given by\vspace{.2cm} \be
m_\nu^{{\tiny{\mbox{FT}}}}=\frac{v^2}{2}Y_\nu^T
M^{-1}_{\bar\nu}Y_\nu  \ee After integrating out $\bar\nu_3$ \be
m_\nu^{{\tiny{\mbox{EFT3}}}}=v^2
\kappa^{(3)}+\frac{v^2}{2}(Y_\nu^{(3)})^T
(M^{(3)})^{-1}_{\bar\nu}Y_\nu^{(3)} \ee Another way of calculating
$\kappa^{(3)}$ is to compare $m_\nu$ in both the theories at
$\mu=M^3_{\bar\nu}$ because light neutrino mass matrix in both the
theory should match
($m_\nu^{{\tiny{\mbox{FT}}}}=m_\nu^{{\tiny{\mbox{EFT3}}}}$) at
right handed neutrino threshold. By evolving the RGEs down to the
scale
 $M^{\bar{\nu}}_2$, one has to repeat the same procedure. After
 integrating out ${\bar\nu}_2$ at $\mu=M^{\bar{\nu}}_2$, the Yukawa matrix
  and the effective neutrino mass operator
 are further modified. At Scale $M^{\bar{\nu}}_2$
superpotential is given by Eq. \eqref{eff2}. Again using the
superpotential equation of motion for ${\bar\nu}_2$ effective
superpotential is given by
 \bea  W_{eff}^{{\tiny{\mbox{EFT2}}}}=&{\bar\nu}^T_{1}
Y^{(3)}_{1A}\nu_A-\frac{1}{2}{\bar\nu}^T_{1}M_{11}^{(3)}{\bar\nu}_{1}+
\nu_A \biggr(\frac{(Y^{(3)})^T_{A2}Y^{(3)}_{2B}}{2 M_{22}^{(3)}}
 +\kappa^{(3)}_{AB} \biggr) \nu_B \eea \be
\kappa^{(2)}_{AB}=\frac{(Y^{(3)})^T_{A2} Y^{(3)}_{2B}}{2
M_{22}^{(3)}}
   +\kappa^{(3)}_{AB} \ee
  \be m_\nu^{{\tiny{\mbox{EFT2}}}}=v^2 \kappa^{(2)}+\frac{v^2}{2}(Y_\nu^{(2)})^T (M^{(2)})^{-1}_{\bar\nu}Y_\nu^{(2)} \ee
  Again $\kappa^{(2)}$ can be obtained from matching of $m_\nu^{{\tiny{\mbox{EFT3}}}}$ and $m_\nu^{{\tiny{\mbox{EFT2}}}}$ at $\mu=M_2$.
 After a further RGE running to $\mu=M^{\bar{\nu}}_1$,
the above steps are repeated for ${\bar\nu}_1$, so that
$\kappa^{(1)}$ is given by \be
\kappa^{(1)}_{AB}=\frac{(Y^{(2)})^T_{A1}Y^{(2)}_{1B}}{2
M_{11}^{(2)} }
   +\kappa^{(2)}_{AB}   \ee
\be  W_{eff}^{{\tiny{\mbox{EFT1}}}}=\nu^T_A \kappa^{(1)}_{AB}
\nu_B\ee After this the effective theory is the MSSM
\cite{MartinVaughn} with Weinberg operators giving all three light
neutrino a mass so we use the RGEs of MSSM and
$\kappa=\kappa^{(1)}$ to run the parameters from $M^{\bar\nu}_1$
to $M_Z$. Left handed neutrino masses at electroweak scale are
given by
\[ m_\nu=v^2\kappa (M_Z) \]
\subsubsection*{Soft Sector}
In the soft sector we have terms involving right handed sneutrino
in addition to the ${\cal
L}_{{\scriptsize{\mbox{soft}}}}^{{\tiny{\mbox{MSSM}}}}$. \be -
{\cal L}_{{\scriptsize{\mbox{soft}}}}=-{\cal
L}_{{\scriptsize{\mbox{soft}}}}^{{\tiny{\mbox{MSSM}}}}+\tilde{\bar\nu}^*_A
(m^2_{\bar\nu})_{AB} \tilde{\bar\nu}_B+ (\tilde{\bar\nu}^T_A
A^{\nu}_{AB} \tilde{L}_B H+\frac{1}{2}\tilde{\bar\nu}^T_A
(b_M)_{AB}\tilde{\bar\nu}_B+h.c) \ee We treat heavy right handed
sneutrino in the same way as their fermion partner. At each
threshold we move to diagonal basis of the right handed neutrino.
We apply the same rotation to the right handed sneutrino, because
sparticle also follow superpotential equation of motion. In order
to integrate out right handed sneutrino, we simply remove the last
row and column from soft mass matrix in right handed neutrino
diagonal basis. Similarly our trilinear coupling $A_{\nu}$ is
modified at each threshold. These thresholds have very small
effect on gauge and Yukawa unification as shown in Table
\ref{effect}. But these thresholds are important for lepton
flavour violation phenomenology which is our next task.
\begin{table}
$$
\begin{array}{ccc }
\hline \hline&&\vspace{-.3 cm}\\ {\rm Parameter}&\multicolumn{2}{ c }{\rm Value ( at\,\, M_X)}\vspace{0.1 cm}\\
 & {\mbox{ Without RHN}} & {\mbox{ With RHN}}\\
\hline \hline
{\rm Y_u}& \{2.082 \times 10^{-6},1.014 \times 10^{-3},0.348\} &\{2.076\times 10^{-6},1.002\times 10^{-3},0.346\} \\
{\rm Y_d}& \{9.269\times 10^{-5},3.325\times 10^{-3},0.291\} &\{8.303\times 10^{-5},3.516\times 10^{-3},0.286\}\\

{\rm Y_l}& \{1.101 \times 10^{-4},2.328 \times 10^{-2},0.456 \}
 &  \{1.225\times 10^{-4},2.294\times 10^{-2},0.452 \}\\
\hline
{ g_1,g_2,g_3}& \{0.728,0.734,0.728\} &\{0.728,0.734,0.728\}\\
\hline \hline
\end{array}
$$\caption{ Effect of right handed neutrino(RHN) thresholds on gauge and Yukawa couplings for the fit presented in Tables \ref{compsol1}-\ref{compsoll11}.\label{effect}}
\end{table}
\subsection{LFV Decay Rate}
The decay rates for these processes are calculated using the
amplitudes depicted by Fig. \ref{lfvbsm} and take the form \be
\Gamma( l_i \rightarrow l_j \gamma  )= \frac{e^2}{16 \pi}
m_{l_i}^5
(|A^L|^2+|A^R|^2)\,\,\,\,;\,\,\,\,A^{L,R}=A^{(n)L,R}+A^{(c)L,R}\ee
where $A^{(n)L,R}$ and $A^{(c)L,R}$ are the contributions from the
neutralino and chargino loops respectively. In order to calculate
these loop contributions one must write the interaction Lagrangian
(fermion-sfermion-neutralino, fermion-sfermion-chargino) in the
mass diagonal basis. To do this one needs to diagonalize the
slepton mass matrices and to consider the mixing in the neutralino
and chargino sectors. Fermion-sfermion-gaugino/higgsino
interaction Lagrangian relevant to the $\mu \rightarrow e \gamma$,
$\tau \rightarrow e \gamma$ and $\tau \rightarrow \mu \gamma$
processes is \vspace{1.0cm}
\begin{figure}
\begin{tikzpicture}[line width=1.0 pt, scale=1.2]
\hspace{1.2cm}
        \node at (1.9,1.9) {$l_i$};
     \node at (5.90,1.9) {$l_j$};
     \node at (1.9,-.65) {$l_j$};
     \node at (5.9,-.65) {$l_j$};
     \node at (4.2,2.4) {$\tilde{\nu}$};
     \node at (4.5,.1) {$\gamma,Z$};
     \node at (4.9,1.2) {$\tilde{\chi}^-$};
        \draw[fermion] (2,2.3) -- (3,2);
        \draw[scalar1] (3,2) -- (5,2);
        \draw[fermion] (5,2) -- (6,2.3);
        \draw[fermion] (3,2) -- (4,.5);
        \draw[fermion] (4,.5) -- (5,2);
        \draw[vector] (4,.5) -- (4,-.5);
        \draw[fermion] (6,-1) -- (4,-.5);
        \draw[fermion] (4,-.5) -- (2,-1);

\draw[fermion] (9,2.3) -- (10,2); \node at (8.9,1.9) {$l_i$};
     \node at (12.90,1.9) {$l_j$};
     \node at (8.9,-.65) {$l_j$};
     \node at (12.9,-.65) {$l_j$};
     \node at (11.2,2.4) {$\tilde{l}$};
     \node at (11.5,.1) {$\gamma,Z$};
     \node at (11.9,1.2) {$\tilde{\chi}^0$};

        \draw[scalar] (10,2) -- (12,2);
        \draw[fermion] (12,2) -- (13,2.3);
        \draw[fermion1] (10,2) -- (11,.5);
        \draw[fermion1] (11,.5) -- (12,2);
        \draw[vector] (11,.5) -- (11,-.5);
        \draw[fermion] (13,-1) -- (11,-.5);
        \draw[fermion] (11,-.5) -- (9,-1);
\end{tikzpicture}
\vspace{.5cm} \caption{{\small{Penguin type diagrams for the
process $l_i$ $\rightarrow$ 3$l_j$}}\label{penguin}}
\vspace{1.15cm}
\begin{tikzpicture}[line width=1.0 pt, scale=1.0]
\hspace{2.2cm} \node at (-.7,1.9) {$l_i$}; \node at (-.7,0.2)
{$l_j$};
     \node at (2.7,1.9) {$l_j$};
     \node at (2.7,0.2) {$l_j$};
 \node at (2.3,1.0) {$\tilde{l}$};
 \node at (.3,1.0) {$\tilde{l}$};
 \node at (1.2,2.3) {$\tilde{\chi}^0$};
  \node at (1.2,-0.3) {$\tilde{\chi}^0$};
        \draw[fermion] (-1,2.3) -- (0,2);
        \draw[fermion1] (0,2) -- (2,2);
        \draw[fermion] (2,2) -- (3,2.3);
        \draw[fermionbar] (-1,-.3) -- (0,0);
        \draw[fermion1] (0,0) -- (2,0);
        \draw[fermionbar] (2,0) -- (3,-.3);
        \draw[scalar] (0,2) -- (0,0);
        \draw[scalar] (2,0) -- (2,2);

\node at (5.3,1.9) {$l_i$}; \node at (5.3,0.2) {$l_j$};
     \node at (8.7,1.9) {$l_j$};
     \node at (8.7,0.2) {$l_j$};
 \node at (8.3,1.0) {$\tilde{\nu}$};
 \node at (6.3,1.0) {$\tilde{\nu}$};
 \node at (7.2,2.3) {$\tilde{\chi}^-$};
  \node at (7.2,-0.3) {$\tilde{\chi}^-$};
\draw[fermion] (5,2.3) -- (6,2);
        \draw[fermion] (6,2) -- (8,2);
        \draw[fermion] (8,2) -- (9,2.3);
        \draw[fermionbar] (5,-.3) -- (6,0);
        \draw[fermion] (6,0) -- (8,0);
        \draw[fermionbar] (8,0) -- (9,-.3);
        \draw[scalar1] (6,2) -- (6,0);
        \draw[scalar1] (8,2) -- (8,0);

\node at (2.3,-1.7) {$l_i$}; \node at (2.3,-3.4) {$l_j$};
     \node at (5.7,-1.6) {$l_j$};
     \node at (5.7,-3.3) {$l_j$};
 \node at (5.3,-2.5) {$\tilde{l}$};
 \node at (3.3,-2.5) {$\tilde{l}$};
 \node at (3.6,-1.6) {$\tilde{\chi}^0$};
  \node at (3.6,-3.4) {$\tilde{\chi}^0$};
\draw[fermion] (2,-1.4) -- (3,-1.5);
        \draw[fermion1] (3,-1.5) -- (5,-3.5);
        \draw[fermion1] (3,-3.5) -- (5,-1.5);

        \draw[fermion] (5,-1.5) -- (6,-1.2);
        \draw[fermionbar] (2,-3.8) -- (3,-3.5);
        \draw[fermionbar] (5,-3.5) -- (6,-3.8);
        \draw[scalar] (3,-1.5) -- (3,-3.5);
        \draw[scalar] (5,-3.5) -- (5,-1.5);

     \end{tikzpicture}
     \vspace{.5cm}
\caption{{\small{Box diagrams for the process $l_i$ $\rightarrow$
3$l_j$}}\label{box}} \vspace{1.0cm}
\end{figure}
\bea \mathcal{L}^{int}&=&\bar l_i (N^{R
}_{iAX}P_R+N^{L}_{iAX}P_L)\tilde{\chi}^0_A \tilde l_X \nonumber
\\&& + \bar l_i (C^{R}_{iAX}P_R+C^{L}_{iAX}P_L)\tilde{\chi}^-_A
\tilde \nu_X +h.c.\eea One obtains \cite{hisano} \bea
A^{(n)L}&=&\frac{1}{32 \pi^2}
\frac{1}{m_{\tilde{l}_X}^2}\big[N^{L}_{jAX} N^{L*}_{i
AX}\frac{(1-6x+3 x^2+2 x^3-6 x^2 \ln x )}{6(1-x )^4}\nonumber\\&&
+N^{L}_{j AX} N^{R*}_{i AX}\frac{ M_{\tilde{\chi}_A^0 }} {m_{\mu}
}\frac{(1-x^2+2 x \ln x)}{(1-x)^3}\big ] \label{anl}\eea \bea
A^{(c)L}&=&-\frac{1}{32 \pi^2}
\frac{1}{m_{\tilde{\nu}_X}^2}\big[C^{L}_{j AX} C^{L*}_{i
AX}\frac{(2+3y-6 y^2+ y^3+6 y \ln y )}{6(1-y )^4}\nonumber\\&&
+C^{L}_{j AX} C^{R*}_{i AX}\frac{ M_{\tilde{\chi}_A ^-}} {m_{\mu}
}\frac{(-3+4y-y^2-2  \ln y)}{(1-y)^3}\big]\eea \be
A^{(n)R}=A^{(n)L}|_{L \leftrightarrow R} \quad ; \quad
A^{(c)R}=A^{(c)L}|_{L \leftrightarrow R}\label{acr} \ee Here
$x=M_{{\tilde{\chi}^{0}_A}}^2 /m_{\tilde{l}_X}^2$,
$y=M^2_{\tilde{\chi}^{-}_A} ,m_{\tilde{\nu}_X}^2 $ are ratios of
neutralino mass squared to the charged slepton mass square and
chargino mass squared to the sneutrino mass square respectively.
The neutralino-slepton vertices used in the Eqns.
\ref{anl}-\ref{acr} have form \be
N^{R}_{iAX}=-\frac{g_2}{\sqrt{2}}{\{( -O^N_{A2}-O^N_{A1}\tan
\theta_W )(U^*)_{X,i}^l+\frac{m_{l_i} }{M_W \cos\beta
}O^N_{A3}(U^*)_{X,i+3}^l }\}\ee \be
N^{L}_{iAX}=-\frac{g_2}{\sqrt{2}}\{ \frac{m_{l_i} }{M_W \cos\beta
}O^N_{A3}(U^*)^l_{X,i}+2O^N_{A1}\tan\theta_W (U^*)_{X,i+3}^l \}\ee
Similarly chargino-sneutralino vertices are :-

\be C^{R}_{i AX}=-g_2 O^C_{A1} (U^*)^{\nu}_{X,i}\ee

\be C^{L}_{i AX}=-g_2 \frac{m_{l_i} }{\sqrt{2} M_W \cos\beta
}O^C_{A2} (U^*)^{\nu}_{X,i}\ee The matrices $O^c$, $U^l$, $U^\nu$
and $O^N$ are the unitary matrices which diagonalize chargino,
slepton and neutralino mass matrices respectively. Notice before
diagonalization slepton mass matrices are rotated to the diagonal
fermion basis. The processes $l_i$ $\rightarrow$ 3$l_j$ also
involve same vertices. These have contributions from both
Penguin-type and box diagram given in Fig. \ref{penguin} and
\ref{box}. Explicit expression for the decay rate can be found in
\cite{hisano}. \vspace{.25cm}
\section{Numerical Fit}\vspace{.2cm}
We used the solution sets presented in the previous chapter to
calculate LFV. If we directly use these
 fits (which are found integrating out all the heavy right handed neutrinos at GUT scale) and run down
  hard and soft parameter using neutrino thresholds then soft Susy parameters at $M_Z$ are slightly changed as shown
   in Tables \ref{rhn1} and \ref{rhn2}, but this give rise to different sparticle LR mixing. Low scale fermion masses
    for both the fits are given in Tables \ref{lowf1} and \ref{lowf2}. Susy threshold corrections are sensitive
     to LR mixing, so down and strange quark masses (Susy threshold corrections modify them by factor of 3) do not match with MSSM data. Therefore we rerun multiple iteration search program using
      these solutions to get appropriate Susy threshold corrections to down type quarks. As shown RHN thresholds do not change
       Yukawa unification, so while running experimental data to get a target for GUT scale fitting we used the MSSM RGEs. We
        assume the normal hierarchy for the left handed neutrino. After some iteration, program found
         a reasonable fermion fitting at the low scale. The fitting criteria is same as discussed in previous chapters. We need
          off-diagonal running from GUT scale to EW scale to calculate BR for LFV processes. Since the loop corrections
           including generation mixing are not available, we use tree level spectrum to calculate BR for LFV processes. We calculate
            the loop corrected Susy spectrum using diagonal running corresponding to the same solutions to avoid possibility of tachyons after
             inclusion of one loop corrections. Loop corrected Susy spectrum for previous chapter fits taking into account the effect
              of heavy neutrino thresholds is presented in Tables \ref{specrhn1} and \ref{specrhn2}. BR for the LFV processes for these fits
               is given in Table \ref{brf1f2}. Complete solution with acceptable down type quark masses is presented
                in Table \ref{compsol1}-\ref{compsoll11} and LFV BR is given in Table \ref{brbest}. For fitting
                 purpose we have calculated the PMNS mixing angles at GUT scale which is permissible only if we integrate
                  out all the heavy right handed neutrino at that scale. Since with the fixed hierarchy of left handed neutrino mixing angles do not change dramatically with RG evolution.
\begin{table}
 $$
\begin{array}{cccc}
 \hline \hline&&&\vspace{-.3 cm}\\
 {\rm  Parameter}  &{\rm Value }&  {\rm  Parameter}&{\rm  Value }\vspace{0.1 cm}\\
 \hline  \hline
                       M_{1}&            132.46&   m_{{\tilde {\bar {u}}_1}}&           5448.42\\
                       M_{2}&            260.51&   m_{{\tilde {\bar {u}}_2}}&           5450.16\\
                       M_{3}&            899.60&   m_{{\tilde {\bar {u}}_3}}&          37816.78\\
     m_{{\tilde {\bar l}_1}}&          17463.31&               A^{0(l)}_{11}&          -8989.22\\
     m_{{\tilde {\bar l}_2}}&          17542.19&               A^{0(l)}_{22}&          -8978.32\\
     m_{{\tilde {\bar l}_3}}&          32843.01&               A^{0(l)}_{33}&          -6123.23\\
        m_{{\tilde {L}_{1}}}&           9267.41&               A^{0(u)}_{11}&          -9898.21\\
        m_{{\tilde {L}_{2}}}&           9378.74&               A^{0(u)}_{22}&          -9898.01\\
        m_{{\tilde {L}_{3}}}&          23412.11&               A^{0(u)}_{33}&          -5795.95\\
     m_{{\tilde {\bar d}_1}}&          13416.10&               A^{0(d)}_{11}&         -10000.39\\
     m_{{\tilde {\bar d}_2}}&          13420.89&               A^{0(d)}_{22}&          -9999.90\\
     m_{{\tilde {\bar d}_3}}&          28511.31&               A^{0(d)}_{33}&          -6488.84\\
          m_{{\tilde {Q}_1}}&          12774.49&                   \tan\beta&             51.00\\
          m_{{\tilde {Q}_2}}&          12777.36&                    \mu(M_Z)&          75869.67\\
          m_{{\tilde {Q}_3}}&          34229.55&                      B(M_Z)&
          1.7771 \times 10^{   8}
 \\
 M_{\bar {H}}^2&
         -4.3012 \times 10^{   9}
 &M_{H}^2&
         -6.0983 \times 10^{   9}
 \\
 \hline  \hline
 \end{array}
 $$
 \caption{ \small {Values (in GeV) of the soft Susy parameters  at $M_Z$
 (evolved from the soft SUGRY-NUHM parameters at $M_X$) which
 determine the Susy threshold corrections to the fermion Yukawas for the Fit 1 of Chapter \ref{loop} including heavy neutrino thresholds.
 \label{rhn1}}}
 \end{table}
 \begin{table}
 $$
 \begin{array}{cc}
 \hline \hline &\vspace{-.3 cm}\\{\mbox {Field } }& {\rm Mass(GeV)}\vspace{0.1 cm}\\
 \hline \hline
                M_{\tilde{g}}&           1174.36\\
               M_{\chi^{\pm}}&            298.03,          82096.67\\
       M_{\chi^{0}}&            141.16,            297.89,          82096.64    ,          82096.64\\
              M_{\tilde{\nu}}&         11481.787,         11580.152,         26082.412\\
                M_{\tilde{e}}&          11482.03,          14652.15,          11580.32   ,          14755.14,          26080.54,          33699.23  \\
                M_{\tilde{u}}&           9559.21,          12256.54,           9559.87   ,          12260.19,          34000.41,          34793.04  \\
                M_{\tilde{d}}&          12256.10,          12267.83,          12258.83   ,          12275.49,          34484.99,          34785.22  \\
                        M_{A}&          46231.20\\
                  M_{H^{\pm}}&          46231.22\\
                    M_{H}&          46231.39\\
                    M_{h}&            126.76\\
 \hline \hline
 \end{array}
 $$
 \caption{\small{Loop corrected spectra of supersymmetric partners calculated ignoring
  generation mixing effects for the Fit 1 of Chapter \ref{loop} including heavy neutrino thresholds.
 \label{specrhn1} }}\end{table}

\begin{table}
 $$
 \begin{array}{cccc}
 \hline \hline &&&\vspace{-.3 cm}\\ {\rm  Parameter }&{\rm SM(M_Z)} &{\rm m^{GUT}(M_Z) }& {\rm m^{MSSM}=(m+\Delta m)^{GUT}(M_Z)}\vspace{0.1 cm} \\
 \hline \hline
    m_d/10^{-3}&   2.90000&   0.96362&   1.48676\\
    m_s/10^{-3}&  55.00000&  34.56512&  53.08815\\
            m_b&   2.90000&   2.46805&   2.96674\\
    m_e/10^{-3}&   0.48657&   0.48450&   0.48478\\
         m_\mu &   0.10272&   0.10240&   0.10245\\
         m_\tau&   1.74624&   1.72406&   1.72292\\
    m_u/10^{-3}&   1.27000&   1.12199&   1.26776\\
            m_c&   0.61900&   0.54662&   0.61764\\
            m_t& 172.50000& 147.05640& 172.08579\\
 \hline \hline
 \end{array}
 $$

 \caption{\small{Values of the SM
 fermion masses in GeV at $M_Z$ compared with the masses obtained from
 values of GUT derived  Yukawa couplings  run down from $M_X$ to
 $M_Z$ (for the Fit 1 of Chapter \ref{loop} including heavy neutrino thresholds) both before and after threshold corrections.
  Fit with $\chi_Z=\sqrt{ \sum\limits_{i=1}^{9} \frac{(m_i^{{\tiny{\mbox{MSSM}}}}- m_i^{{\tiny{\mbox{SM}}}})^2}{ (m_i^{{\tiny{\mbox{MSSM}}}})^2} } =
0.9516$.\label{lowf1}}}
 \end{table}

\begin{table}
 $$
 \begin{array}{cccc}
 \hline \hline&&&\vspace{-.3 cm}\\
 {\rm  Parameter}  &{\rm Value }&  {\rm  Parameter}&{\rm  Value }\vspace{0.1 cm}\\
 \hline \hline
                       M_{1}&            117.93&   m_{{\tilde {\bar {u}}_1}}&           3864.14\\
                       M_{2}&            232.11&   m_{{\tilde {\bar {u}}_2}}&           3866.24\\
                       M_{3}&            799.50&   m_{{\tilde {\bar {u}}_3}}&          34774.44\\
     m_{{\tilde {\bar l}_1}}&          15566.41&               A^{0(l)}_{11}&          -7964.82\\
     m_{{\tilde {\bar l}_2}}&          15647.11&               A^{0(l)}_{22}&          -7954.65\\
     m_{{\tilde {\bar l}_3}}&          30584.36&               A^{0(l)}_{33}&          -5286.11\\
        m_{{\tilde {L}_{1}}}&           7847.40&               A^{0(u)}_{11}&          -8865.38\\
        m_{{\tilde {L}_{2}}}&           7959.05&               A^{0(u)}_{22}&          -8865.20\\
        m_{{\tilde {L}_{3}}}&          21805.42&               A^{0(u)}_{33}&          -5140.46\\
     m_{{\tilde {\bar d}_1}}&          11791.57&               A^{0(d)}_{11}&          -8862.73\\
     m_{{\tilde {\bar d}_2}}&          11796.24&               A^{0(d)}_{22}&          -8862.29\\
     m_{{\tilde {\bar d}_3}}&          26543.70&               A^{0(d)}_{33}&          -5625.41\\
          m_{{\tilde {Q}_1}}&          11198.87&                   \tan\beta&             52.00\\
          m_{{\tilde {Q}_2}}&          11201.66&                    \mu(M_Z)&          69411.39\\
          m_{{\tilde {Q}_3}}&          31595.82&                      B(M_Z)&
          1.4439 \times 10^{   8}
 \\
 M_{\bar {H}}^2&
         -3.6225 \times 10^{   9}
 &M_{H}^2&
         -5.1348 \times 10^{   9}
 \\
 \hline \hline
 \end{array}
 $$
  \caption{ \small {Values (in GeV)  of the soft Susy parameters  at $M_Z$
 (evolved from the soft SUGRY-NUHM parameters at $M_X$) which
 determine the Susy threshold corrections to the fermion Yukawas.
 The matching of run down fermion Yukawas in the MSSM to the SM  parameters
 determines  soft SUGRY parameters at $M_X$ for the Fit 2 of Chapter \ref{loop} including heavy neutrino thresholds.
\label{rhn2}  }}
 \end{table}
 \begin{table}
 $$
 \begin{array}{cc}
 \hline \hline &\vspace{-.3 cm}\\{\mbox {Field } }& {\rm Mass(GeV)}\vspace{0.1 cm}\\
 \hline \hline
                M_{\tilde{g}}&           1046.16\\
               M_{\chi^{\pm}}&            265.08,          75198.58\\
       M_{\chi^{0}}&            125.32,            264.92,          75198.54    ,          75198.55\\
              M_{\tilde{\nu}}&         9947.132,         10043.320,         24208.033\\
                M_{\tilde{e}}&           9947.42,          12990.49,          10043.51   ,          13095.20,          24206.02,          31400.95  \\
                M_{\tilde{u}}&           8038.40,          10707.75,           8039.08   ,          10711.23,          31389.36,          32198.29  \\
                M_{\tilde{d}}&          10706.71,          10717.49,          10709.12   ,          10724.95,          31979.96,          32224.47  \\
                        M_{A}&          42237.34\\
                  M_{H^{\pm}}&          42237.36\\
                    M_{H}&          42237.49\\
                    M_{h}&            126.56\\
 \hline \hline
 \end{array}
 $$
 \caption{\small{Loop corrected spectra of supersymmetric partners calculated ignoring
  generation mixing effects for the Fit 2 of Chapter \ref{loop} including heavy neutrino thresholds.
 \label{specrhn2}}}\end{table}

\begin{table}
 $$
 \begin{array}{cccc}
 \hline \hline &&&\vspace{-.3 cm}\\ {\rm  Parameter }&{\rm SM(M_Z)} &{\rm m^{GUT}(M_Z)} & {\rm m^{MSSM}=(m+\Delta m)^{GUT}(M_Z)}\vspace{0.1 cm} \\
 \hline \hline
    m_d/10^{-3}&   2.90000&   1.09275&   1.69684\\
    m_s/10^{-3}&  55.00000&  33.82299&  52.55803\\
            m_b&   2.90000&   2.46505&   2.92709\\
    m_e/10^{-3}&   0.48657&   0.48414&   0.48450\\
         m_\mu &   0.10272&   0.10312&   0.10319\\
         m_\tau&   1.74624&   1.72120&   1.72004\\
    m_u/10^{-3}&   1.27000&   1.12569&   1.26768\\
            m_c&   0.61900&   0.54722&   0.61625\\
            m_t& 172.50000& 147.39488& 172.15940\\
 \hline \hline
 \end{array}
 $$

 \caption{\small{Values of the SM
 fermion masses in GeV at $M_Z$ compared with the masses obtained from
 values of GUT derived  Yukawa couplings  run down from $M_X$ to
 $M_Z$ (for the Fit 2 of Chapter \ref{loop} including heavy neutrino thresholds) both before and after threshold corrections.
  Fit with $\chi_Z=\sqrt{ \sum\limits_{i=1}^{9} \frac{(m_i^{{\tiny{\mbox{MSSM}}}}- m_i^{{\tiny{\mbox{SM}}}})^2}{ (m_i^{{\tiny{\mbox{MSSM}}}})^2} } =
0.7108$.\label{lowf2}}}
 \end{table}
\begin{table}
 $$
 \begin{array}{ccc}

 \hline \hline &&\vspace{-.3 cm}\\{\mbox {Process } } &\multicolumn{2}{ c }{\mbox{Branching Ratio from NMSGUT}}\\
  & {\rm Fit-1} & {\rm Fit-2}\vspace{0.1 cm} \\
  \hline \hline
 \mu \rightarrow e \gamma &
         1.1906 \times 10^{ -18}
 &
         1.3544 \times 10^{ -18}
 \\
 \tau \rightarrow \mu \gamma &
          1.2475 \times 10^{ -17}
 &
          9.1312 \times 10^{ -18}
 \\
 \tau \rightarrow e \gamma &
          1.8641 \times 10^{ -19}
 &
          2.1977 \times 10^{ -19}
 \\

  \mu \rightarrow e e e &
         1.8303 \times 10^{ -18}
 &
         3.2487 \times 10^{ -18}
 \\
 \tau \rightarrow \mu \mu \mu &
          3.5397 \times 10^{ -19}
 &
          2.6097 \times 10^{ -19}
 \\
 \tau \rightarrow e ee  &
          6.8617 \times 10^{ -21}
 &
          8.1293 \times 10^{ -21}
 \\

 \hline  \hline
 \end{array}
 $$
 \caption{\small{BR of LFV processes.
  \label{brf1f2}}}\end{table}

\begin{table}
\vspace{-1.0cm}
 $$
 \begin{array}{cccc}
 \hline \hline&&&\vspace{-.3 cm}\\
 {\rm Parameter }&{\rm Value} &{\rm  Field }& {\rm Masses }\\
 &&\hspace{-10mm}{\rm}[SU(3),SU(2),Y]&({\rm Units\,\,of 10^{16} GeV})\vspace{0.1 cm}\\ \hline \hline
       \chi_{X}&  0.3907           &A[1,1,4]&      1.48 \\ \chi_{Z}&
  0.1495
                &B[6,2,{5/3}]&            0.1074\\
           h_{11}/10^{-6}&  6.7245         &C[8,2,1]&{      0.92,      2.49,      5.58 }\\
           h_{22}/10^{-4}&  5.2504    &D[3,2,{7/ 3}]&{      0.06,      3.64,      6.61 }\\
                   h_{33}&  0.0254     &E[3,2,{1/3}]&{      0.14,      0.75,      2.01 }\\
\hline\hline\end{array}
 $$
  \caption{\small{Values of the NMSGUT-SUGRY-NUHM  parameters at $M_X$
  derived from an  accurate fit to all 18 fermion data and compatible with RG constraints.
 Unification parameters and mass spectrum of superheavy and superlight fields are  also given.
 \label{compsol1}}}\end{table}
 \begin{table}
 $$
 \begin{array}{ccccc}
 \hline \hline
 &&&&\vspace{-.3 cm}\\
 {\rm  Parameter }&{\rm Target} =\bar O_i &{\rm Uncert.= \delta_i }   &{\rm Achieved= O_i}
 & {\rm Pull =\frac{(O_i-\bar O_i)}{\delta_i }}\vspace{0.1 cm}\\
 \hline \hline
    y_u/10^{-6}&  2.070378&  0.790884&  2.075721&  0.006756\\
    y_c/10^{-3}&  1.009175&  0.166514&  1.001722& -0.044756\\
            y_t&  0.345391&  0.013816&  0.345835&  0.032130\\
    y_d/10^{-5}&  8.392364&  4.892748&  8.303114& -0.018241\\
    y_s/10^{-3}&  3.492497&  1.648459&  3.515603&  0.014017\\
            y_b&  0.281969&  0.146342&  0.286114&  0.028327\\
\hline \hline
 \end{array}
 $$
 \label{table b} \caption{\small{Fit   with $\chi_X=\sqrt{ \sum\limits_{i=1}^{17}
 \frac{(O_i-\bar O_i)^2}{\delta_i^2}}=
    0.3907
 $. Target values,  at $M_X$ of the fermion Yukawa
 couplings and mixing parameters, together with the estimated uncertainties, achieved values and pulls.
 The eigenvalues of the wavefunction renormalization for fermion  and
 Higgs lines are given with Higgs fractions $\alpha_i,{\bar{\alpha_i}}$ which control the MSSM fermion
  Yukawa couplings.
 }}
 \end{table}
 \begin{table}
 $$
 \begin{array}{cccc}
 \hline  \hline &&&\vspace{-.3 cm}\\ {\rm  Parameter }&{\rm SM(M_Z)} &{\rm m^{GUT}(M_Z)} & {\rm m^{MSSM}=(m+\Delta m)^{GUT}(M_Z)}\vspace{0.1 cm} \\
 \hline \hline
    m_d/10^{-3}&   2.90000&   0.86783&   3.39332\\
    m_s/10^{-3}&  55.00000&  36.74367&  55.05208\\
            m_b&   2.90000&   2.45010&   2.92919\\
    m_e/10^{-3}&   0.48657&   0.49774&   0.49787\\
         m_\mu &   0.10272&   0.10140&   0.10143\\
         m_\tau&   1.74624&   1.71479&   1.71386\\
    m_u/10^{-3}&   1.27000&   1.11735&   1.27186\\
            m_c&   0.61900&   0.53922&   0.61378\\
            m_t& 172.50000& 146.60676& 172.52755\\
 \hline \hline
 \end{array}
 $$
 \label{table c}
 \caption{\small{Values of the SM
 fermion masses in GeV at $M_Z$ compared with the masses obtained from
 values of GUT derived  Yukawa couplings  run down from $M_X$ to
 $M_Z$  both before and after threshold corrections.
  Fit with $\chi_Z=\sqrt{ \sum\limits_{i=1}^{9} \frac{(m_i^{{\tiny{\mbox{MSSM}}}}- m_i^{{\tiny{\mbox{SM}}}})^2}{ (m_i^{{\tiny{\mbox{MSSM}}}})^2} } =
0.1495$.}}
 \end{table}
 \begin{table}
 $$
 \begin{array}{cccc}
 \hline \hline&&&\vspace{-.3 cm}\\
 {\rm  Parameter}  &{\rm Value }&  {\rm  Parameter}&{\rm  Value }\vspace{0.1 cm}\\
 \hline \hline
                       M_{1}&            139.08&   m_{{\tilde {\bar {u}}_1}}&           9611.03\\
                       M_{2}&            273.94&   m_{{\tilde {\bar {u}}_2}}&           9612.67\\
                       M_{3}&            941.89&   m_{{\tilde {\bar {u}}_3}}&          49603.05\\
     m_{{\tilde {\bar l}_1}}&          23415.65&               A^{0(l)}_{11}&         -10267.66\\
     m_{{\tilde {\bar l}_2}}&          23515.87&               A^{0(l)}_{22}&         -10255.53\\
     m_{{\tilde {\bar l}_3}}&          43323.72&               A^{0(l)}_{33}&          -7063.33\\
        m_{{\tilde {L}_{1}}}&          13753.64&               A^{0(u)}_{11}&         -11188.30\\
        m_{{\tilde {L}_{2}}}&          13882.30&               A^{0(u)}_{22}&         -11188.05\\
        m_{{\tilde {L}_{3}}}&          31213.03&               A^{0(u)}_{33}&          -6589.55\\
     m_{{\tilde {\bar d}_1}}&          18412.68&               A^{0(d)}_{11}&         -11323.83\\
     m_{{\tilde {\bar d}_2}}&          18419.55&               A^{0(d)}_{22}&         -11323.20\\
     m_{{\tilde {\bar d}_3}}&          37754.82&               A^{0(d)}_{33}&          -7403.99\\
          m_{{\tilde {Q}_1}}&          17659.29&                   \tan\beta&             51.00\\
          m_{{\tilde {Q}_2}}&          17663.29&                    \mu(M_Z)&          99537.13\\
          m_{{\tilde {Q}_3}}&          45002.01&                      B(M_Z)&
          2.7908 \times 10^{   8}
 \\
 M_{\bar {H}}^2&
         -7.5398 \times 10^{   9}
 &M_{H}^2&
         -1.0487 \times 10^{  10}
 \\
 \hline \hline
 \end{array}
 $$
 \label{table d} \caption{ \small {Values (in GeV) of the soft Susy parameters  at $M_Z$
 (evolved from the soft SUGRY-NUHM parameters at $M_X$) which
 determine the Susy threshold corrections to the fermion Yukawas.
 The matching of run down fermion Yukawas in the MSSM to the SM   parameters
 determines  soft SUGRY parameters at $M_X$ ($M_{{\tiny{\mbox{Susy}}}}=3.25 {\small{\mbox{TeV}}}$).
  }}
 \end{table}
 \begin{table}
 $$
 \begin{array}{cc}
 \hline\hline&\vspace{-.3 cm}\\ {\mbox {Field } }&{\rm Mass(GeV)}\vspace{0.1 cm}\\
 \hline\hline
                M_{\tilde{g}}&            941.89\\
               M_{\chi^{\pm}}&            273.93,          99537.20\\
       M_{\chi^{0}}&            139.08,            273.93,          99537.17    ,          99537.18\\
              M_{\tilde{\nu}}&         13753.492,         13882.149,         31212.961\\
                M_{\tilde{e}}&          13753.72,          23415.69,          13882.35   ,          23515.93,          31211.82,          43324.63  \\
                M_{\tilde{u}}&           9610.96,          17659.21,           9612.61   ,          17663.21,          45002.16,          49603.27  \\
                M_{\tilde{d}}&          17659.39,          18412.70,          17663.35   ,          18419.60,          37751.68,          45004.69  \\
                        M_{A}&         119324.64\\
                  M_{H^{\pm}}&         119324.67\\
                    M_{H}&         119324.64\\
                    M_{h}&             90.31\\
 \hline\hline
 \end{array}
 $$
 \label{table e}\caption{\small{Tree level spectra of supersymmetric partners calculated
  ignoring generation mixing effects.
  }}\end{table}

 \begin{table}
 $$
 \begin{array}{cc}
 \hline \hline&\vspace{-.3 cm}\\ {\mbox {Field } }& {\rm Mass(GeV)}\vspace{0.1 cm}\\
 \hline \hline
                M_{\tilde{g}}&           1259.36\\
               M_{\chi^{\pm}}&            314.27,         107937.67\\
       M_{\chi^{0}}&            148.50,            314.14,         107937.64    ,         107937.64\\
              M_{\tilde{\nu}}&         16243.435,         16359.851,         34250.256\\
                M_{\tilde{e}}&          16243.61,          20059.00,          16359.96   ,          20183.25,          34248.81,          44010.02  \\
                M_{\tilde{u}}&          13963.35,          17100.49,          13964.06   ,          17105.21,          44115.72,          44720.70  \\
                M_{\tilde{d}}&          17100.52,          17105.69,          17103.78   ,          17114.48,          44482.14,          44736.79  \\
                        M_{A}&          59582.86\\
                  M_{H^{\pm}}&          59582.88\\
                    M_{H}&          59583.05\\
                    M_{h}&            127.17\\
 \hline \hline
 \end{array}
 $$
 \caption{\small{Loop corrected spectra of supersymmetric partners calculated ignoring
  generation mixing effects.
 \label{compsoll11} }}\end{table}
\begin{table}
 $$
 \begin{array}{cc}
 \hline \hline&\vspace{-.3 cm}\\ {\mbox {Process } }&\mbox{Branching Ratio from NMSGUT}\vspace{0.1 cm}\\
 \hline \hline
    \mu \rightarrow e \gamma &
         1.2740 \times 10^{ -19}
 \\
 \tau \rightarrow \mu \gamma &
           1.9050 \times 10^{ -18}
 \\
 \tau \rightarrow e \gamma &
          4.7212 \times 10^{ -20}
 \\

  \mu \rightarrow e e e &
         6.8500 \times 10^{ -19}
 \\
 \tau \rightarrow \mu \mu \mu &
           7.0904 \times 10^{ -20}
 \\
 \tau \rightarrow e e e &
          2.1471 \times 10^{ -21}
 \\

 \hline \hline
 \end{array}
 $$
 \caption{\small{BR of LFV processes.
 \label{brbest} }}\end{table}
\section{Anomalous Magnetic Moment of Muon }
Dirac magnetic moment of the muon corresponding to the tree level
Feynman diagram is equal to 2. Difference between the classical
results and observed value is called anomalous magnetic moment
denoted by $a_\mu$. So, anomalous magnetic moment of muon in the
MSSM is a contribution of loops involving sparticles to the
magnetic moment of the muon. The magnetic moment $ \vec{\mu} $ of
a muon is related to gyromagnetic ratio as \be \vec{\mu}= g
\bigg(\frac{e}{2 m_{\mu}} \bigg) \vec{S}=
2(1+a_{\mu})\bigg(\frac{e}{2 m_{\mu}} \bigg)
\vec{S}\,\,\,\,;\,\,\,\,a_\mu=\frac{1}{2} (g_\mu-2) \,\,\,\, \ee
Its SM prediction consists of three type of contribution from QED,
hadronic loops and weak interactions \be
a_\mu(\mbox{SM})=a_\mu(\mbox{QED})+a_\mu(\mbox{Had})+a_\mu(\mbox{Weak})
\ee The precisely measured \cite{BNL} magnetic moment of muon has
a significant deviation \cite{PDG} from the theoretical prediction
: \be \Delta
a_\mu=a_\mu\mbox{(exp)}-a_\mu\mbox{(SM)}=287(63)(49)\times
10^{-11}\ee The $\Delta a_\mu$ may thus represent the contribution
of new physics beyond SM. Supersymmetry is one of the leading
candidate for new physics. The deviation $a_\mu$ may be due to
heavy sparticle contributions \cite{Stockinger}.
\subsection{Analytic Formulae}
\begin{figure}[tbh]
\vspace{1.0cm} \centering
\begin{tikzpicture}[line width=1.0 pt, scale=2.0]
    \draw[fermion1] (0,0)--(.8,0);
     \node at (.4,-.15) {$\mu$};
     \node at (2.3,-.15) {$\mu$};
     \node at (1.8,-.6) {$\gamma$};
     \node at (1.4,-.15) {$\tilde{\mu}$};
     \node at (1.4,.4) {$\tilde{\chi}^0$};
    \draw[fermion1] (.8,0) arc (180:0:.6);
    \draw[scalar] (.8,0) -- (2.0,0);
    \draw[fermion1] (2.0,0) --(2.8,0);
     \draw[vector] (1.7,-.2) -- (2.3,-.7);
\draw[fermion1] (3.5,0)--(4.3,0);
 \node at (3.9,-.15) {$\mu$};
     \node at (5.8,-.15) {$\mu$};
     \node at (5.2,1.1) {$\gamma$};
     \node at (4.9,-.15) {$\tilde{\nu}$};
     \node at (4.9,.4) {$\tilde{\chi}^+$};
    \draw[fermion] (4.3,0) arc (180:0:.6);
    \draw[scalar1] (4.3,0) -- (5.5,0);
    \draw[fermion1] (5.5,0) --(6.3,0);
     \draw[vector] (5.0,.7) -- (5.6,1.2);
\end{tikzpicture}
\vspace{.5cm} \caption{Feynman diagrams for the lowest-order
supersymmetric contribution to $a_{\mu}$\label{loworder}}
\vspace{1.0cm}
\end{figure}
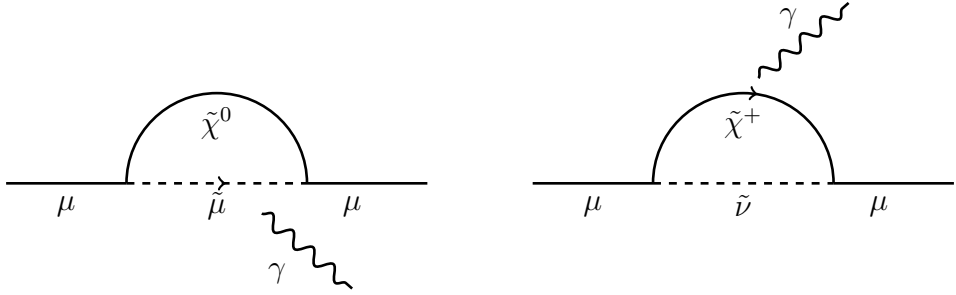
The lowest order supersymmetric contribution
($a_{\mu}^{\tiny{\mbox{Susy}}}$) to $a_{\mu}$ from
sneutrino-chargino and smuon-neutralino loops
\cite{hisano,Stockinger} is shown in Figure \ref{loworder}
\[a_{\mu}^{\tiny{\mbox{Susy,1L}}}=
a_{\mu}^{\tilde{\chi}^{\pm}}+ a_{\mu}^{\tilde{\chi}^0}\] where
$a_{\mu}^{\tilde{\chi}^{\pm}} $ and $a_{\mu}^{\tilde{\chi}^0} $
denote chargino and neutralino contribution, given as\be
 a_{\mu}^{\tilde{\chi}^{\pm}}= \frac{m_\mu^2}{48 \pi^2 m_{\tilde{\nu}_X}^2}(|C_{2AX}^L|^2+|C_{2AX}^R|^2)F^c_1(x_{AX}) +
 \frac{ m_\mu m_{{\tilde{\chi}^{\pm}_k}}}{8 \pi^2 m_{\tilde{\nu}_X}^2}Re[C_{2AX}^L C_{2AX}^{R*} ]F^c_2(x_{AX})  \label{charg} \ee
\be a_{\mu}^{\tilde{\chi}^0}= - \frac{m_\mu^2}{48 \pi^2
m_{\tilde{\mu}_X}^2}(|N_{2AX}^L|^2+|N_{2AX}^R|^2)F^n_1(x_{AX}) -
 \frac{ m_\mu m_{{\tilde{\chi}^{0}_A}}}{8 \pi^2 m_{\tilde{\mu}_X}^2}Re[N_{2AX}^L N_{2AX}^{R*} ]F^n_2(x_{AX}) \label{neut}
\ee Here A=1...4 (1, 2) and X=1, 2 denote the neutralino(chargino)
and smuon indices respectively. Variables of loop functions are
defined as ratios of mass squares
$x_{AX}=m_{{\tilde{\chi}^{-}_A}}^2/m_{\tilde{\mu}_m}^2 $ $(
m^2_{\tilde{\chi}^{0}_A}/m_{\tilde{\nu}_X}^2 )$ for chargino
(neutralino) contribution and loop functions have form : \be
F^c_1(x) =\frac{2}{(1-x)^4}(2+3 x-6 x^2+x^3+6x \log x)\ee \be
F^c_2(x)=\frac{3}{(1-x)^3}(-3+4 x- x^2+-2 \log x)\ee \be
F^n_1(x)=\frac{2}{(1-x)^4}(1-6 x+3 x^2+2 x^3-6x^2 \log x)\ee \be
F^n_2(x)==\frac{3}{(1-x)^3}(1-x^2+2 x \log x )\ee It is clear from
Eq. \eqref{neut} that $a_{\mu}^{\tiny{\mbox{Susy}}}$ is sensitive
to smuon mass and $ \tan \beta $ ( terms linear in
$m_{{\tilde{\chi}^{0,\pm}}} $ in Eqns. \eqref{charg} and
\eqref{neut} are proportional to it). Susy contribution to g-2 is
completely independent of color sparticles at leading order. It
requires sleptons as light as $O(10^2)$ GeV and $(g-2)_\mu$
motivated LHC Susy searches have been discussed in
\cite{mkst,Iwamoto:2013mya}. To understand the behavior of
$a_{\mu}^{\tiny{\mbox{Susy}}}$ one needs to investigate
approximate relations for the different diagram contributions.
Mass insertion method is used to calculate different
  diagrams \cite{tm,Iwamoto:2013mya,Stockinger,mkst,detail}. MSSM prediction of $\Delta a_\mu$ depend upon left and right
   handed smuon masses ($m_{\tilde \mu_L}$, $m_{\tilde \mu_R}$), gaugino
    mass parameters ($M_1$, $M_2$), higgsino mass ($\mu$) and ratio of VEVs ($\tan\beta$).
Fits presented have heavy smuon so we use exact formulae of Eq.
\eqref{charg} and \eqref{neut} for Susy contribution to $a_\mu$.
There are $\tan\beta$ enhanced two-loop diagrams also
\cite{Stockinger} which can modify the leading order results by 10
$\%$ but we calculated only one loop diagrams. NMSGUT fits
prediction for $\Delta a_\mu$ is given in Table \ref{amu}. We see
that except for the light smuon solution (presented in Chapter 3),
the $\Delta a_{\mu}$ values are too small to resolve the muon g-2
anomaly.
\begin{table}
 $$
 \begin{array}{cc}
 \hline \hline &\vspace{-.3 cm}\\{\mbox {Fit } }&\Delta a_{\mu}/10^{-10}\vspace{0.1 cm}\\
 \hline \hline
             {\rm Fit 1 (4^{th} Chapter)}&      0.0084     \\
             {\rm Fit 2 (4^{th} Chapter)} &     0.0111       \\
             {\rm Fit 1 (5^{th} Chapter)} &     0.0033        \\
              {\rm Fit 1 (3^{rd} Chapter)} &     10.6        \\
   \hline \hline
 \end{array}
 $$
 \caption{\small{Susy contribution to muon g-2.
 \label{amu} }}\end{table}
\section{Leptogenesis}\vspace{.2cm}
The baryon asymmetry of the universe is defined as : \be
\eta_B=\frac{ n_B-n_{\bar{B}}}{n_\gamma }=6.1 \pm 0.3 \times
10^{-10} \ee This can be generated via :- B violation, C and CP
violation and departure from thermal equlibrium as pointed out by
Sakharov \cite{Sakharov}. Leptogenesis \cite{lepto1,Luty:1992un}
is most promising mechanism to explain observed baryon asymmetry
of the universe as CP violating decay of right handed neutrino (
shown in Fig. \ref{leptofig}) can fulfill these conditions.
Leptonic CP asymmetry can be converted to baron asymmetry through
sphaleron processes. Non-thermal leptogenesis involves right
handed neutrinos generation through inflaton decay whose further
decay can  generate lepton and ( after sphaleron processes) baryon
asymmetry.
\begin{figure}
\vspace{1.0cm}
\begin{tikzpicture}[line width=1.0 pt, scale=1.2]
 \node at (3.2,1.6) {$N_j$};
     \node at (5.6,2.9) {$H$};
     \node at (5.6,.9) {$l$};
        \draw[fermion] (3.2,2) -- (5,2);
        \draw[scalar] (5,2) -- (6,3);
        \draw[fermion] (5,2) -- (6,1.0);

\node at (0.2,-1.4) {$N_j$};
     \node at (4.0,-0.4) {$H$};
     \node at (4.0,-1.6) {$l$};
\node at (2.3,-0.4) {$l$};
     \node at (2.3,-1.6) {$H$};
\node at (3.4,-1.0) {$N_j$};

\draw[fermion] (0.2,-1) -- (2,-1);
        \draw[scalar] (2,-1) -- (3.1,-.1);
        \draw[fermion] (2,-1) -- (3.1,-1.9);
        \draw[fermion] (3.1,-0.1) -- (3.1,-1.9);
       \draw[scalar] (3.1,-0.1) -- (3.1,-1.9);
\draw[scalar] (3.1,-0.1) -- (4.6,-0.1); \draw[fermion] (3.1,-1.9)
-- (4.6,-1.9);

\node at (6.5,-1.4) {$N_j$}; \node at (9.3,-1.4) {$N_j$};
     \node at (11,-0.4) {$H$};
     \node at (11,-1.6) {$l$};

\node at (8.1,0) {$l$};

\draw[fermion] (5.8,-1) -- (7.4,-1);
\draw[fermion] (7.4,-1) arc (180:0:.7);
    \draw[scalar] (8.8,-1) arc (0:-180:.7);
\draw[fermion] (8.8,-1) -- (10,-1);
        \draw[scalar] (10,-1) -- (11,0);
        \draw[fermion] (10,-1) -- (11,-2);
\end{tikzpicture}
\vspace{.5cm} \caption{CP violating decay of right handed
neutrino\label{leptofig}} \vspace{1.0cm}\end{figure}
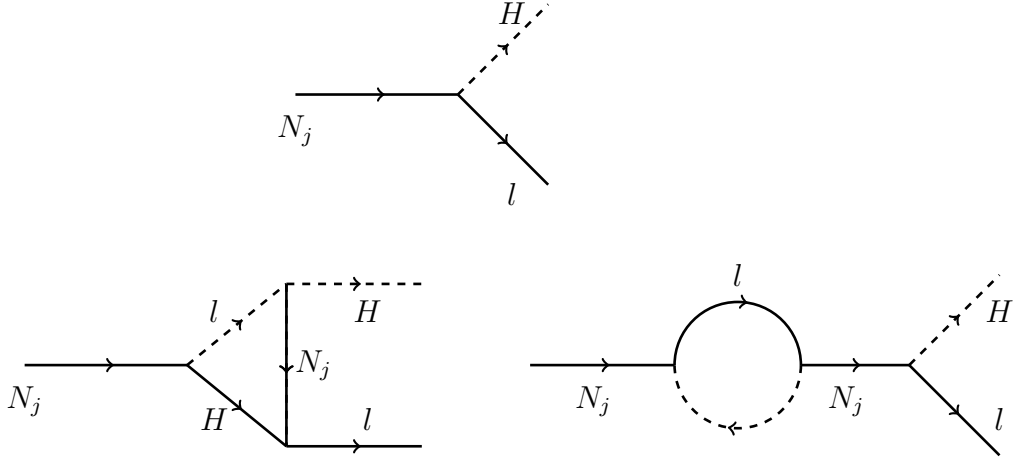 CP asymmetry
parameter relevant for leptogenesis is given by \be
\epsilon{\tiny{\mbox{CP}}}= \frac{\Gamma(N \rightarrow l_L + H ) -
\Gamma(N \rightarrow\bar{l}_L + \bar{H} ) }{\Gamma(N \rightarrow
l_L + H ) + \Gamma(N \rightarrow\bar{l}_L + \bar{H} )} \ee  In
case of hierarchical spectrum: \be \epsilon_{{\tiny{\mbox{CP}}}}
\simeq -\frac{3}{8\pi}\frac{Im\{[Y_\nu^\dag Y_\nu ]_{12}^2
\}}{[Y_\nu^\dag Y_\nu ]_{11}}
\frac{M^{\bar\nu}_1}{M^{\bar\nu}_2}\ee Model independent upper
bound on CP asymmetry \cite{Davidson} :- \be
|\epsilon_{\tiny{\mbox{CP}}}| \leqslant \frac{3}{ 8 \pi}
\frac{M_1^{\bar\nu}(m_3-m_1)}{\langle H^0_u \rangle}\ee Here
$m_{1,3} $ are left handed neutrino masses. The
$\epsilon_{\tiny{\mbox{CP}}}$  parameter for the three discussed
fits is given in Table \ref{eps}. The desirable range of values
for successful leptogenesis \cite{Davidson} is
$\epsilon_{\tiny{\mbox{CP}}}\sim 10^{-7}$ . Although the generic
values we obtain are somewhat small, it should be noted that we
have not optimized our fits to improve
$\epsilon_{\tiny{\mbox{CP}}}$ yet we are not too far off. Since it
depends sensitively on the off diagonal structure of
$Y_\nu^\dagger Y_\nu$ and linearly  on
$M^{\bar\nu}_1/M^{\bar\nu}_2$ , optimization could easily yields
more  satisfactory values. We will return to these questions in
future research.
 \begin{table}
 $$
 \begin{array}{cc}
 \hline \hline&\vspace{-.3 cm}\\ {\mbox {Fit } }&\epsilon_{{\tiny{\mbox{CP}}}}/10^{-8}\vspace{0.1 cm}\\
 \hline \hline
             {\rm Fit 1 (4^{th} Chapter)}&    0.18        \\
             {\rm Fit 2 (4^{th} Chapter)} &       0.10     \\
             {\rm Fit 1 (5^{th} Chapter)} &     0.24       \\
   \hline \hline
 \end{array}
 $$
 \caption{\small{Lepton sector parameter.
 \label{eps} }}\end{table}
\section{Conclusion}
Neutrino Dirac Yukawa coupling is crucial for LFV predictions
which require inclusion of heavy right
 handed neutrino thresholds. We found a reasonable fit implementing these thresholds and calculated
  BR for $l_i$ $\rightarrow$ $l_j$$\gamma$ and $l_i$ $\rightarrow$ 3$l_j$ LFV processes. NMSGUT predict
   BR for the these processes several order of magnitude smaller than the upper bound from
 experiments because the fits have negative soft Higgs masses ($M_H^2$) and heavy Susy spectrum. Negative soft Higgs masses provide
  cancellation in the off-diagonal entries
 of slepton mass matrices \cite{LFVSO10} which estimate the LFV. Solution presented has all the NMSGUT superpotential and soft parameters in the range
  corresponding to the previous chapter solutions. The smuon is heavy in all the solutions
   with loop corrected sfermion masses that we have so far found. Hence Susy contribution to muon g-2 is very small $\sim$ $10^{-12}$. $\Delta a_\mu$ is not sensitive to neutrino Yukawa coupling but depend upon the soft Susy spectrum which further depend upon mSUGRY parameters at GUT scale. The value of the leptogenesis CP violation parameter is roughly in the desired ball park even without optimization.

This study shows that NMSGUT fits found are compatible with LFV
constraints but further searches are needed to explore whether
light smuon and adequate $\epsilon_{{\tiny{\mbox{CP}}}}$ are
achievable.

\newpage\thispagestyle{empty}\mbox{}\newpage
\chapter{Renormalization Group Evolution Equations of the NMSGUT} \label{ch:method}
\section{Introduction}
Renormalization group equations (RGEs) are used to evolve the
gauge couplings, superpotential parameters and soft terms from UV
scales into physically meaningful quantities that describe physics
near the electroweak scale. The scale of generation of soft Susy
breaking parameters in mSUGRY and string motivated mechanisms is
above the GUT scale, typically $M_{P}$. Thus a complete RG study
of a Susy GUT requires evolution of GUT parameters  from $M_{P}$
to GUT scale and then of the effective MSSM to electroweak/Susy
breaking scale. Evolution between $M_{P}$ and $M_{GUT}$ can be
(and is) very important for both the hard and soft Susy breaking
parameters of the NMSGUT due to the large RG $\beta$ functions in
SO(10). In this chapter we give formulae for NMSGUT $\beta$
function upto two loops for the first time and examine their
effect. The form of the RGEs for supersymmetric theories is
governed by the supersymmetric non-renormalization theorem
\cite{nonrenorth}. According to this theorem the logarithmically
divergent contributions to a particular process arise only from
wave-function renormalisation, without any superpotential coupling
renormalization. Variation of parameter $X$ with energy scale is
given by \vspace{.2cm} \be { \frac{d}{dt}X=\frac{ 1} {16 \pi^2 }
\beta_X^{(1)}+ \frac{ 1} {(16 \pi^2)^2 } \beta_X^{(2)} }  \ee
where $t=log(Q/Q_0)$, Q and $Q_0$ are the renormalization  and
reference scale respectively. $\beta_X^{(1)}$ and $\beta_X^{(2)}$
are one-loop and two-loop $\beta$ functions. The 2-loop RGEs for
the MSSM (effective theory of Susy-GUTs) and the soft Susy
breaking parameters are well known \cite{MartinVaughn}, and are
useful from low scale to GUT scale. However, to consider the
effect of renormalization from Planck scale to GUT scale, one
needs the explicit form of GUT dependent RGEs. We have computed
the two-loop RGEs for gauge coupling constant, superpotential  and
soft Susy breaking parameters for the  New Minimal Supersymmetric
SO(10) Grand Unified Theory. General formulae for the evolution
functions for any softly broken Susy simple gauge group theory are
available \cite{MartinVaughn} to compute SO(10) two-loop $\beta$
function. A Mathematica package immediately gets stuck
\cite{Susyno} on combinatorial complexity while performing the
sums over irrep indices required to obtain RGE coefficients for
SO(10) irreps. However special tricks using the properties of the
model and SO(10) irrep index contraction make the sums over the
components of the large irreps (${\bf{210}}$, ${\bf{126}}$
,${\boot}$ and ${\bf{120}}$) used in the NMSGUT tractable. We got
explicit results for all RGEs upto second order using gauge
invariance as a guiding principle.  This work is done in
collaboration with Prof. C.S. Aulakh and Ila
Garg\cite{csaigckkRG}.
\section{Formalism}
  The one-loop $\beta$-function for the gauge coupling \cite{MartinVaughn} is :  \be
{\beta_g^{(1)}=g^3[S(R)-3 C(G)]}\ee where $S(R)$ and $C(G)$ are
Dynkin index (including contribution of all superfields) and
Casimir invariant respectively. Note that C(R) d(R)=S(R) d(G).
Since the values are S(45,10,16,120,126,210)=(8,1,2,28,35,56) and
C(R)=45(8/45,1/10,2/16,28/120,\\35/126,56/210) we get one-loop
$\beta$ function for the SO(10) gauge coupling to be :
\vspace{.2cm}\be \beta_{g_{10}}^{(1)}=137 g_{10}^3 \ee Notice that
this implies very rapid change of $g_{10}$ and hence require great
care to avoid nonsensical results. The generic form of one-loop
$\beta$-function for the superpotential parameters is
$(W=\frac{1}{6}{Y^{ijk} \phi_i\phi_j\phi_k+.. })$ : \be{
[\beta_Y^{(1)}]^{ijk}=Y^{ijp} \gamma_p^{(1)k}+(k\leftrightarrow i
)+(k \leftrightarrow j ) }\ee where i, j, k are the indices
running over all the chiral fields in the theory and
$\gamma^{(1)}$ is the one loop anomalous dimension matrix. SO(10)
gauge invariance implies that $\gamma^i_j$ must be fieldwise and
componentwise diagonal : thus simplifies their computation
enormously. The NMSGUT has $(\lambda, k, \rho, \gamma, \bar\gamma,
\eta, \zeta, \bar\zeta)$ superpotential couplings representing the
following interactions : \bea
 \lambda &:& {\bf{210}}^{\bf{3}} \qquad ; \qquad  \eta   :
 {\bf{210\cdot 126\cdot{\overline {126}}}}
 ;\qquad\qquad \rho :{\bf{120\cdot 120 \cdot{  { 210}}}}
\nnu k &:& {\bf{ 10\cdot 120\cdot{ {210}}}} \qquad;\qquad\gamma
\oplus {\bar\gamma}  : {\bf{10 \cdot 210}\cdot(126
\oplus{\overline {126}}}) \nnu && \zeta \oplus {\bar\zeta}  :
{\bf{120 \cdot 210}\cdot(126 \oplus {\overline {126}}})
  \eea
  and mass parameters :
 \bea
 \mu_{\Phi}: {\bf{210}}^{\bf{2}} \quad ;\quad  \mu_{\Sigma} : {\bf{126\cdot{\overline {126}}}}
 ;\qquad \mu_H : {\bf{10}}^{\bf{2}};\qquad \mu_\Theta :{\bf{120}}^{\bf{2}}
\eea The generic one-loop anomalous dimension parameters
associated with superfields carry the crucial structure governing
the NMSGUT RGEs and are given by \be { \gamma_i^{(1)j}=\frac{1}{2}
Y_{ipq}Y^{jpq}-2 g^2 \delta_i^j C(i) }\ee Now we discuss the
contribution of superpotential invariant $\rho \Phi_{ijkl}
\Theta_{ijm} \Theta_{klm} $ to $\gamma_\Phi^{(1)}$. Let us focus
on what couples to a given component of the $\bf{210}$
\[ \frac{\rho}{4!}  \Phi_{ijkl} \Theta_{ijm} \Theta_{klm} =
\sum\limits_{m} \frac{\rho}{4!} 4.2.\Phi_{1234} (\Theta_{12m}
\Theta_{34m}-\Theta_{13m} \Theta_{24m}+\Theta_{14m} \Theta_{23m} )
\] Here m runs over remaining 6 values since the $\bf{120}$ plet
is totally antisymmetric. We see that since the SO(10) symmetry
will require $\gamma^i_j$ diagonality within an irrep one can
 obtain the $\gamma^i_j$ by counting the possibilities for any representative irrep element
  on the external lines and SO(10) allowed index combinations on the summed indices. In this example we can have 18 possible combinations that couple to $\Phi_{1234}$. Therefore
  \be  \frac{1}{2}\bigg|Y^{\{\Phi_{1234}.\Theta.\Theta\}}\bigg|^2=\frac{18|\rho|^2}{9}=2 |\rho|^2 \ee
Similarly \be \frac{\gamma}{4!} \Phi_{ijkl}H_m \Sigma_{ijklm} =
\gamma \Phi_{1234} (H_5 \Sigma_{12345}+H_6 \Sigma_{12346}+....)
\ee The six allowed index values for $H$ (i.e. 5-10) give \be
\sum_{H,\Sigma} Y_{\{\Phi_{1234}.H.\Sigma \}}
Y^{\{\Phi_{1234}.H.\Sigma\}}=6|\gamma|^2 \ee The invariant $k H_i
\Theta_{jkl} \Phi_{ijkl}$ will contribute to $\gamma_\Phi^{(1)}$
\be \frac{k}{3!} H_i \Theta_{jkl} \Phi_{ijkl} = k \Phi_{1234} (H_1
\Theta_{234}-H_2 \Theta_{134}+H_3 \Theta_{124}-H_4 \Theta_{312}
)+.... \ee \be \sum_{H,\Theta} Y_{\{\Phi_{1234}.H.\Theta \}}
Y^{\{\Phi_{1234}.H.\Theta\}}=4|k|^2 \ee Thus the anomalous
dimension matrix reduces to a common anomalous dimension for each
independent component of each field.
 \be \gamma_\Phi^{(1)}=240 |\eta |^2+4 |k
|^2+180 |\lambda|^2+2 |\rho |^2+6 (|\gamma|^2+|\bar{\gamma}|^2)+60
(|\zeta |^2+|\bar{\zeta}|^2)-24 g_{10}^2 \ee \be
\gamma_\Sigma^{(1)}= 200 |\eta |^2+10 |\gamma |^2+100 |\zeta
|^2-25 g_{10}^2 \ee \be  \gamma_{\bar\Sigma}^{(1)}=  200 |\eta
|^2+10 |\bar{\gamma}|^2+100 |\bar{\zeta}|^2+32
\text{Tr}[f^{\dag}.f]-25 g_{10}^2 \ee \be \gamma_H^{(1)}= 84 |k
|^2+126 (|\gamma|^2+|\bar{\gamma}|^2)+8 \text{Tr}[h^{\dag}.h] -9
g_{10}^2 \ee \be \gamma_\Theta^{(1)}= 7 (|k |^2+|\rho |^2)+105
(|\zeta |^2+|\bar{\zeta}|^2)+8 \text{Tr}[g^{\dag}.g]-21 g_{10}^2
\ee \be (\gamma_\Psi^{(1)})_A^B= (\gamma_\Psi^{(1)})_{AB} =252
f^{\dag}.f+120 g^{\dag}.g+10 h^{\dag}.h-\frac{45 g_{10}^2}{4} \ee
Here $h$, $f$ and $g$ are Yukawa couplings of $\bf{10}$, $\boot$
and $\bf{120}$ Higgs irreps. Using  the $\gamma$'s one can compute
RGEs for all the superpotential parameters. For example one loop
$\beta$ function for $\lambda$ is : \be   \quad \quad
\beta_\lambda^{(1)}=  3 \gamma_\Phi^{(1)}  \lambda \ee In addition
to superpotential parameters and gauge coupling, we need to
compute ${\mathcal L}_{{\scriptsize{\mbox{soft}}}}$ parameters
RGEs: \be {\mathcal L}_{{\scriptsize{\mbox{soft}}}} = -{1\over 6}
h^{ijk} \phi_i\phi_j\phi_k - {1\over 2} b^{ij} \phi_i \phi_j -
{1\over 2}(m^2)^i_j \phi^{*i} \phi_j - {1\over 2}M \lambda \lambda
+ {\rm h.c.} \ee In total we have $\{ \tilde\lambda, \tilde{k},
\tilde{\rho}, \tilde{\gamma}, \tilde{\bar\gamma}, \tilde{\eta},
\tilde{\zeta}, \tilde{\bar\zeta}$, $\tilde{h}$, $\tilde{f}$,
$\tilde{g}$ $\}$, $\{b_{\Phi}, b_{\Sigma}, b_H, b_{\Theta}\}$ and
$\{{m}^2_{\Phi}, {m}^2_{\Sigma}, {m}^2_{\bar\Sigma},
{m}^2_{\Theta},$ $ {m}^2_{H}, {m}^2_{\tilde\Psi} \}$ parameters in
the NMSGUT soft Lagrangian. One loop $\beta$- function for
$h^{ijl}$ is \cite{MartinVaughn} : \bea [\beta^{(1)}_h ]^{ijk}  &
=&
  {1\over 2} h^{ijl} Y_{lmn} Y^{mnk}
+ Y^{ijl} Y_{lmn} h^{mnk} - 2  (h^{ijk} - 2 M Y^{ijk} ) g^2  C(k)
\nonumber\\&&+ (k \leftrightarrow i) + (k \leftrightarrow j)  \eea
We define : \be \bar\gamma_i^{(1)j}=\frac{1}{2} Y_{ipq}Y^{jpq}
\quad ; \quad \tilde\gamma_i^{(1)j}=\frac{1}{2} Y_{ipq}h^{jpq}
\quad ; \quad \hat\gamma_i^{(1)j}=\frac{1}{2} h_{ipq}h^{jpq} \ee
Then arguments similar to those given above yield : \be  \bar
\gamma_\Phi^{(1)}=240 |\eta |^2+4 |k |^2+180 |\lambda|^2+2 |\rho
|^2+6 (|\gamma|^2+|\bar{\gamma}|^2)+60 (|\zeta
|^2+|\bar{\zeta}|^2)         \ee

\be   \tilde{ \gamma}_\Phi^{(1)}=  240
   \tilde{\eta } \eta ^{*}+ 4 \tilde{k} k^*+180
   \tilde{\lambda} \lambda ^{*}+2
\tilde{\rho } \rho ^{*}+6
   (\tilde{\gamma}  \gamma
   ^{*}+\tilde{\bar{\gamma}}
   \bar\gamma^{*})+60
   (\tilde{\zeta}  \zeta
   ^{*}+\tilde{\bar{\zeta}}
   \bar \zeta ^{*})         \ee

\be
\hat{\gamma}_{\Phi}^{(1)}=240|\tilde{\eta}|^2+4|\tilde{\kappa}|^2+180|\tilde{\lambda}|^2+2|\tilde{\rho}|^2+
6(|\tilde{\gamma}|^2+|\tilde{\bar{\gamma}}|^2)+60(|\tilde{\zeta}|^2+|\tilde{\bar{\zeta}}|^2)\ee

\be \beta_{\tilde{\lambda}}^{(1)}=3 \tilde{\lambda}
\bar{\gamma}_{\Phi}^{(1)}+6 \lambda \tilde{\gamma}_{\Phi}^{(1)}
-72g_{10}^2(\tilde{\lambda}-2 {M} \lambda)\ee

\bea [\beta^{(1)}_b  ]^{ij} & =&
 {1\over 2} b^{il} Y_{lmn} Y^{mnj} +{1\over 2}Y^{ijl} Y_{lmn} b^{mn}
+ \mu^{il} Y_{lmn} h^{mnj} \nonumber\\&&- 2  (b^{ij} - 2 M
\mu^{ij}  )g^2 C(i)+ (i \leftrightarrow j)  \eea

\be \beta_{b_{\Phi}}^{(1)}=2 b_{\Phi}
\bar{\gamma}_{\Phi}^{(1)}+4\mu_{\Phi}
\tilde{\gamma}_{\Phi}^{(1)}-48 g_{10}^2(b_{\Phi}-2 {M}
\mu_{\Phi})\ee

\bea [\beta^{(1)}_{m^2} ]_i^j &= &
 {1\over 2} Y_{ipq} Y^{pqn} {(m^2)}_n^j
+ {1\over 2} Y^{jpq} Y_{pqn} {(m^2)}_i^n + 2 Y_{ipq} Y^{jpr}
{(m^2)}_r^q \nonumber\\&&+ h_{ipq} h^{jpq}- 8\delta_i^j M
M^\dagger g^2 C(i) +
 2g^2{\bf t}^{Aj}_i {\rm Tr}[{\bf t}^A m^2 ]  \eea

\bea \beta_{{m}^2_{\Phi}}^{(1)}&=&2
\bar{\gamma}_{\Phi}^{(1)}{m}^2_{\Phi}+720 {m}^2_{\Phi}|\lambda|^2+
{m}^2_{H}(12|\gamma|^2+12|\bar{\gamma}|^2+8|k|^2)\nonumber\\&&
+{m}^2_{\Theta}(8|\rho|^2+120(|\zeta|^2+|\bar{\zeta}|^2)+8|k|^2)
+{m}^2_{\Sigma}(480|\eta|^2+12|\gamma|^2+120|\zeta|^2)\nonumber\\&&+{m}^2_{\bar{\Sigma}}(480|\eta|^2
+12|\bar{\gamma}|^2+120|\bar{\zeta}|^2)
+2\hat{\gamma}_{\Phi}^{(1)}-96|{M}|^2 g_{10}^2\eea The two loop
anomalous dimensions $\gamma^{(2)}$ are the main building blocks
of two loop $\beta$ functions, having generic form : \bea
\gamma_i^{(2)j}&=&-\frac{1}{2}Y_{imn}Y^{npq}Y_{pqr}Y^{mrj}+g^2_{10}Y_{ipq}Y^{jpq}[2C(p)-C(i)]\nonumber\\&&+
2 \delta_i^j g^4_{(10)}[C(i)S(R)+2 C(i)^2-3 C(G)C(i)] \eea Again
they are fieldwise and independent component wise diagonal. Only
the first term require attention. The intermediate sums over $n,r$
can be broken field wise and thereafter using diagonality of the
one loop anomalous dimensions (with respect to independent irrep
components) already computed: \be
Y_{imn}Y^{npq}Y_{pqr}Y^{mrj}=Y_{im n_H}\bar\gamma^{(1)}_H Y^{m n_H
j}+Y_{i m n_{\Theta}} \bar\gamma^{(1)}_{\Theta} Y^{m
n_{\Theta}j}+... \ee As discussed for the one loop, SO(10) gauge
invariance provide $\gamma^{(2)}$ and constraint n=r (in the first
term), so we have (as already the sum is over independent field
components) \be Y_{imn}Y^{npq}Y_{pqr}Y^{mrj}= Y_{imn}
(\bar\gamma)^n_n Y^{mnj} \ee One needs to examine the
superpotential invariant involving  two fields carrying field
component indices mn. Thus the total contribution can be written
with the help of one loop anomalous dimension parameters. For
example :\bea \gamma_\Phi^{(2)}&=&-(240 |\eta|^2
(\bar{\gamma}_\Sigma^{(1)}+ \bar{\gamma}_{\bar{\Sigma}}^{(1)}  )+4
|k|^2(\bar{\gamma}_H^{(1)}+ \bar{\gamma}_\Theta^{(1)} )+6
|\gamma|^2(\bar{\gamma}_H^{(1)}+ \bar{\gamma}_\Sigma^{(1)} )
\nonumber\\&&+360|\lambda|^2\bar{\gamma}_\Phi^{(1)} +
4|\rho|^2\bar{\gamma}_\Theta^{(1)} +6
|\bar{\gamma}|^2(\bar{\gamma}_H^{(1)}+
\bar{\gamma}_{\bar\Sigma}^{(1)} )+60
|\zeta|^2(\bar{\gamma}_\Theta^{(1)}+ \bar{\gamma}_\Sigma^{(1)}
)\nonumber\\&&+60 |\bar{\zeta}|^2(\bar{\gamma}_\Theta^{(1)}+
\bar{\gamma}_{\bar\Sigma}^{(1)} ))+g^2_{10}(6240|\eta|^2+ 24|k|^2
+4320|\lambda|^2+36|\rho|^2\nonumber\\&&+60|\gamma|^2+60|\bar\gamma|^2+1320
|\zeta|^2+1320 |\bar{\zeta}|^2)+3864 g^4_{10}\eea Two-loop $\beta$
functions for other superpotential parameters are given in the
Appendix and for the soft couplings can be found in
\cite{csaigckkRG,ilathesis}. We will use these RGEs to estimate
the variation of the soft parameters between $M_{P}$ and $M_X^0$.
NMSGUT fits discussed in the previous chapters have large negative
soft Higgs masses $(M_{H,\bar H}^2)$. SO(10) RGEs can explain
origin of these kind of couplings. \vspace{.5cm}
\section{Numerical Analysis}
\begin{figure}[tbh]
\centering
\includegraphics[scale=1.2]{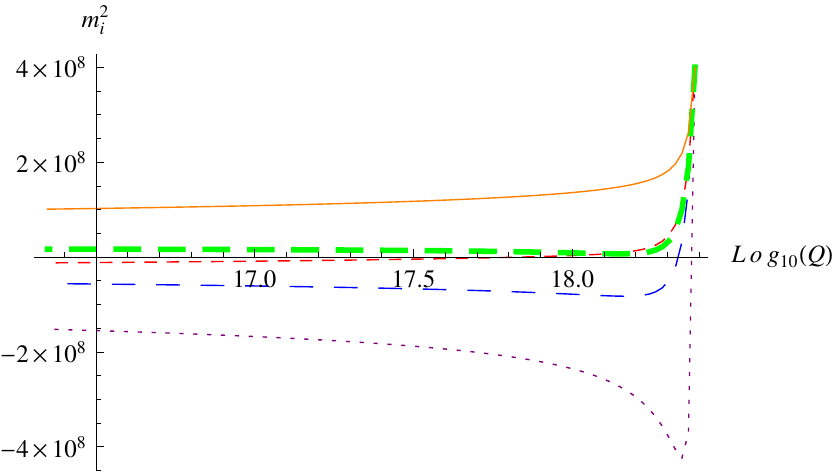}
\caption{Evolution of soft masses from Planck scale to GUT scale.
Dashed (red), dotted (purple), medium dashed (blue), thick dashed
(green) and solid (orange) lines represent
$\tilde{m}^2_{\bar{\Phi}}$, $\tilde{m}^2_{H}$,
$\tilde{m}^2_{\Theta}$, $\tilde{m}^2_{\Sigma}$ and
$\tilde{m}^2_{\bar\Sigma}$
respectively.\label{softmassplot}}\vspace{.5cm}
\end{figure}

We throw SO(10) gauge and Yukawa couplings and soft parameters
randomly in the perturbative range. Along with this we choose soft
breaking parameters according to SUGRY soft term form (assuming
canonical soft terms). Susy breaking i.e. with all gaugino masses
zero, all soft scalar masses equal, $A_0$=2$m_{3/2}$,
$b_i$=($A_0$-$m_{3/2}$)$\mu_i$ at the Planck scale. We chose
$m_{3/2}$=20 TeV and renormalize them from $M_{P}$ and $M_X^0$.
Large coefficient of the trilinear couplings in the anomalous
dimension make these RGEs to evolve fast between $M_{P}$ to
$M_X^0$. The values of hard and soft parameters at two scales
($M_P$ and $M_X^0$) are given in Tables \ref{t1} and \ref{t2}
respectively. This shows that the evolution can be very
significant and in particular the soft masses change rapidly. The
large value of $\beta_g$ makes a UV fixed point impracticable.
Soft mass evolution is shown in Fig. \ref{softmassplot}.
\begin{table}
 $$
 \begin{array}{ccc}
 \hline \hline&&\vspace{-.3 cm}\\ {\rm  Parameter }&{\mbox {Value at $M_P$} }& {\mbox {Value at $ M_{X}^0(10^{16.33}$ GeV}) }\vspace{0.1 cm} \\
 \hline\hline
    \lambda &  -0.0434 + 0.0078 i & -0.0098 + 0.0018 i  \\
    \eta&  -0.3127 +0.0798 i& -0.0969 + 0.0247 i \\
            \rho &   0.9544 - 0.2698  i&0.141 - 0.0399 i \\
   k &   0.0273 + 0.0991 i& 0.0015 + 0.0053 i  \\
         \gamma  &   0.4711& 0.0318 \\
         \bar\gamma &   -3.2719&  -0.2922 \\
    \zeta&   1.0091 +
 0.8305 i & 0.1876 + 0.1544 i  \\
            \bar \zeta&   0.3596 + 0.5898 i & 0.0885 + 0.1452 i  \\
            \hline\hline
   h_{11}/10^{-6} & 4.4602   & 1.0843  \\
      h_{22}/10^{-4} &  4.1031  & 0.9971  \\
       h_{33} &  0.0244  &  .0059 \\

 h_{12}/10^{-12} & 0.0  & -2.9819+4.8131i  \\
      h_{13}/10^{-11} &  0.0  & -2.3318+4.0693i  \\
       h_{23}/10^{-9} &  0.0  &  -6.1280+11.4938i \\

     f_{11}/10^{-6} &  -.0044+.16207 & -0.0049+.1811 i\\
      f_{22}/10^{-5} & 6.675+4.8457i   & 7.4587+ 5.4144i\\
       f_{33}/10^{-4} & -9.264+2.7876i   &  -10.3507+3.1146i \\
       f_{12}/10^{-6} &  -.84951-1.7825  &  -0.9492-1.9917i \\
      f_{13} /10^{-6}&  .54964+1.1479 i  &  0.6141+1.2826i \\
       f_{23} /10^{-4} &  -.4266+2.231i  &  -0.4767+2.4927i \\
       g_{12}/10^{-5} &  1.4552+1.599i  &  0.9755+1.0718i \\
      g_{13}/10^{-5} & -1.1784+.49613i   & -7.8988+3.3255i  \\
       g_{23}/10^{-4} &  -1.6648-1.18436i  &  -1.1159-0.7939i \\
           \hline\hline
     \mu_\Phi & 10^{15}~\mbox{GeV}& 37.10 \times 10^{13}~\mbox{GeV}  \\
    \mu_H &10^{15} ~\mbox{GeV}&  3.19 \times 10^{13}~\mbox{GeV} \\
            \mu_\Sigma &  10^{15} ~\mbox{GeV}  & 50.87 \times 10^{13}~\mbox{GeV}  \\
  \mu_{\Theta} &  10^{15}~\mbox{GeV} & 24.25 \times 10^{13}~\mbox{GeV} \\
         \hline
          g &2.2519& 0.3445 \\\hline\hline
       \end{array}
 $$
 \label{table c}
 \caption{\small{Values of NMSGUT parameters at two different scales evolved by using one-loop SO(10) RGEs.\label{t1}}}
 \end{table}

\begin{table}
 $$
 \begin{array}{ccc}
 \hline\hline &&\vspace{-.3 cm}\\  {\rm  Parameter }&\mbox {Value at $M_P$}& \mbox {Value at $ M_{X}^0(10^{16.33})$ GeV}\vspace{0.1 cm} \\
\hline\hline&&\vspace{-.4 cm}\\
 \tilde\lambda &-43.4231 + 7.7508i &   -0.3413 + 0.0613 i\\
    \tilde\eta& -312.69 +79.7879 i&  -450.6968 + 115.0163 i \\
            \tilde\rho & 954.387 -269.837 i &-556.0236 + 157.1827 i \\
  \tilde{ k} & 27.3142 + 99.0676 i & -12.3372 -44.7845 i \\
         \tilde\gamma  & 471.122  &-185.8855 \\
         \tilde{\bar\gamma} & -3271.91  & 707.8617  \\
    \tilde{\zeta}& 1009.13 + 830.517 i & -255.3635 - 210.1669i  \\
            \tilde{\bar \zeta}& 359.587 + 589.788 i  &183.1857 + 300.4532 i  \\
           \hline  \hline&&\vspace{-.4 cm}\\
            \tilde h_{11} & 0.1784  & .0219  \\
     \tilde h_{22} &16.4052 & 2.014 \\
     \tilde  h_{33} &976.7480 & 119.8772 \\
     \tilde h_{12}/10^{-7} & 0.0  & -2.1449  \\
     \tilde h_{13}/10^{-6} &0.0 & -1.3053 \\
     \tilde  h_{23}/10^{-4} &0.0 & -5.1246 \\

     \tilde f_{11}/10^{-3} & -0.1744+6.483i &-0.1496+5.56i \\
     \tilde  f_{22} &2.6701+1.9383i  & 2.2899+1.6623i\\
     \tilde   f_{33} & -37.0558+11.1505 &-31.7754+9.5616i \\
    \tilde   f_{12}/10^{-2} & -3.3980-7.1301i &-2.9143-6.1149i \\
    \tilde  f_{13}/10^{-2} & 2.1986+4.5916i & 1.8853+3.937i\\
    \tilde   f_{23} & -1.7064+8.9336 & -1.4633+7.6523i \\


     \tilde  g_{12} & 0.5821+0.6396i & 0.2408+0.2646i \\
    \tilde  g_{13} & -4.7136+1.9845 i& -1.9500+0.821i \\
     \tilde  g_{23} & -6.6592-4.7374i & -2.7549-1.9599i \vspace{0.1 cm}\\
         \hline   \hline&&\vspace{-.4 cm}\\
           {M} & 0 & 0 \\\hline\hline&&\vspace{-.4 cm}\\
    {b_\Phi} & 2.0 \times 10^{19} \mbox{GeV}^2 & -2.465 \times 10^{18} \mbox{GeV}^2  \\
    {b}_H & 2.0 \times 10^{19}  \mbox{GeV}^2& -6.2536 \times 10^{17} \mbox{GeV}^2  \\
            {b}_\Sigma & 2.0 \times 10^{19} \mbox{GeV}^2 & -1.0311 \times 10^{18} \mbox{GeV}^2 \\
  {b}_{\Theta} & 2.0 \times 10^{19} \mbox{GeV}^2 & -2.576 \times 10^{18} \mbox{GeV}^2 \\
       \hline  \hline&&\vspace{-.4 cm}\\
        {m}^2_{\Phi} &  m_{3/2}^2&-12577864.9856 \mbox{GeV}^2  \\

 {m}^2_{H} & m_{3/2}^2 & -151814083.220 \mbox{GeV}^2 \\
         {m}^2_{\Theta}& m_{3/2}^2 &-55926753.8102 \mbox{GeV}^2 \\
{m}^2_{\Sigma}&  m_{3/2}^2 & 16687663.4739 \mbox{GeV}^2\\
         {m}^2_{\bar{\Sigma}}&  m_{3/2}^2 &100530987.171 \mbox{GeV}^2 \vspace{0.1 cm}\\
          {\rm Eval~{m}^2_{\tilde\Psi}}& \{m_{3/2}^2,m_{3/2}^2,m_{3/2}^2 \}&\{3.9999,3.9999,3.9993\} \times 10^8 \mbox{GeV}^2 \vspace{0.1 cm}\\
     \hline\hline
 \end{array}
 $$
 \label{table c}
 \caption{\small{Values of NMSGUT soft parameters at two different scales evolved by using one-loop
  SO(10) RGEs. $\{ \tilde\lambda, \tilde{k}, \tilde{\rho}, \tilde{\gamma},$ $\tilde{\bar\gamma},
   \tilde{\eta},$ $\tilde{\zeta}, \tilde{\bar\zeta},$ $\tilde{h},\tilde{f},\tilde{g}\}$=$A_0(\lambda, k, \rho,$ $\gamma, \bar\gamma, \eta, \zeta,$
   $\bar\zeta,h,f,g)$, $A_0=40$ TeV, $m_{3/2}=20$ TeV.\label{t2}}}
 \end{table}
 \vspace{.5cm}
\section{Discussion and Outlook}
We derived the NMSGUT RG equations to determine the RG evolution
of couplings between
  $M_P$ and $M_X^0$ (the matching scale between GUT and effective theory) assuming pure supergravity canonical scenario for the starting parameter ansatz.
  Evaluating the effects of the evolution on randomly chosen sets of parameter values we see that the RG
 evolution has dramatic effects on the soft susy breaking parameters. Firstly most of the soft Susy squared masses of
  the SO(10) Higgs irreps become negative. It provides a potentially robust justification of the negative values
   of $M_{H,\bar H}^2$  the NMSGUT fits already needed. Note that the distinctive normal s-hierarchy at low scale
    is strongly correlated with the large negative $M_{H,\bar H}^2$ we use in the fits. Gaugino masses will be
     generated by two loop RG evolution, however even $M$=0 at the GUT scale yielded adequate gaugino
      masses at the electroweak scale. The other dramatic effect is the intermediate scale $(O({m_{3/2}M_X}))$ values of the soft
       parameters $b_{\Phi,\Sigma,\Theta,H}$ required by the canonical SUGRY ansatz and induced by the dependence $\frac{db}{dt}$ $\sim$ $\mu m_{3/2}$. Actually the running  values of  $ b_{\Phi,H,\Sigma,\Theta}  $ turn   negative after starting positive
so it is possible that they run to smaller   and more acceptable
values $O( m^2_{ 3/2})$   for a suitable set of starting
parameters. We are currently studying the detailed implications.
\eject
\section*{Appendix}
\subsection*{One-loop RGEs\label{oneloop}}
One-loop beta functions for the SO(10) superpotential parameters
and  Yukawa couplings of $\bf{10}$, $\bf{\overline{126}}$ and
$\bf{120}$ :

\be  \beta_\lambda^{(1)}=  3 \gamma_\Phi^{(1)}  \lambda  \quad ;
\quad \beta_\eta^{(1)}= \eta (\gamma_\Sigma^{(1)} +\gamma_{\bar
\Sigma}^{(1)} +\gamma_\Phi^{(1)} )   \label{one}     \ee

\be  \beta_\gamma^{(1)}= \gamma
(\gamma_H^{(1)}+\gamma_\Sigma^{(1)} +\gamma_\Phi^{(1)} )     \quad
; \quad  \beta_{\bar\gamma}^{(1)}= \bar\gamma
(\gamma_H^{(1)}+\gamma_{\bar\Sigma}^{(1)} +\gamma_\Phi^{(1)} ) \ee

\be  \beta_k^{(1)}= k (\gamma_H^{(1)}+\gamma_\Theta^{(1)}
+\gamma_\Phi^{(1)} )     \quad ; \quad  \beta_\zeta^{(1)}=
\zeta(\gamma_\Theta^{(1)} +\gamma_\Sigma^{(1)} +\gamma_\Phi^{(1)}
) \ee

\be  \beta_{\bar\zeta}^{(1)}= \bar\zeta(\gamma_\Theta^{(1)}
+\gamma_{\bar\Sigma}^{(1)} +\gamma_\Phi^{(1)} ) \quad ; \quad
\beta_\rho^{(1)}= \rho  (\gamma_\Phi^{(1)}+2 \gamma_\Theta^{(1)} )
\ee

\be  \beta_h^{(1)}= h
\gamma_H^{(1)}+(\gamma_\Psi^{(1)})^{T}.h+h.\gamma_\Psi^{(1)} \quad
; \quad  \beta_f^{(1)}= f
\gamma_{\bar\Sigma}^{(1)}+(\gamma_\Psi^{(1)})^{T}.f+f.\gamma_\Psi^{(1)}
\ee

\be  \beta_g^{(1)}= g
\gamma_\Theta^{(1)}-(\gamma_\Psi^{(1)})^{T}.g+g.\gamma_\Psi^{(1)}
\ee

\be \beta_{\mu_{\Phi}}^{(1)}=2  \gamma_{\Phi}^{(1)}\mu_{\Phi}
\quad ; \quad \beta_{\mu_H}^{(1)}=2 \gamma_{H}^{(1)} \mu_H\ee

\be
\beta_{\mu_{\Sigma}}^{(1)}=(\gamma_{\Sigma}^{(1)}+\gamma_{\bar{\Sigma}}^{(1)})
\mu_{\Sigma}\quad ; \quad  \beta_{\mu_{\Theta}}^{(1)}=2
\gamma_{\Theta}^{(1)} \mu_{\Theta} \label{last} \ee

\subsection*{Two-loop RGEs}
\be   \bar\gamma_\Sigma^{(1)}=  200 |\eta |^2 +10 |\gamma |^2+100
|\zeta |^2\ee

\be    \bar\gamma_{\bar\Sigma}^{(1)}= 200 |\eta |^2+ 10
|\bar\gamma |^2+100 |\bar \zeta |^2+32 \text{Tr}[f^{\dag }.f] \ee

\be  \bar \gamma_H^{(1)}= 84 |k |^2+126
(|\gamma|^2+|\bar{\gamma}|^2)+8 \text{Tr}[h^{\dag }.h] \ee

\be   \bar\gamma_\Theta^{(1)}= 7(|k|^2+|\rho |^2)+ 105 (|\zeta
|^2+|\bar\zeta |^2)+8\text{Tr}[g^{\dag }.g]         \ee

\be \bar\gamma_\Psi^{(1) }=252 f^{\dag}.f+120 g^{\dag}.g+10
h^{\dag}.h \ee

\bea  \beta_{g_{10}}^{(2)}&=&  9709
g^5_{10}-\frac{g^3_{10}}{45}(9\bar\gamma_{H}^{(1)}+21\bar\gamma_{\Theta}^{(1)}+
25\bar\gamma_{\Sigma}^{(1)}+25\bar\gamma_{\bar\Sigma}^{(1)}
+24\bar\gamma_{\Phi}^{(1)}\nonumber\\&&+\frac{135}{4}(8\text
Tr[h^\dag.h]+ 8\text Tr[g^\dag.g]+32\text Tr[f^\dag.f])) \eea

\bea  \gamma_\Sigma^{(2)}&=&-(200 |\eta|^2
(\bar{\gamma}_\Phi^{(1)}+ \bar{\gamma}_{\bar{\Sigma}}^{(1)}  ) +
10 |\gamma|^2(\bar{\gamma}_H^{(1)}+ \bar{\gamma}_\Phi^{(1)} ) +
100 |\zeta|^2(\bar{\gamma}_\Theta^{(1)}+ \bar{\gamma}_\Phi^{(1)}
))\nonumber\\&&+g^2_{10}( 4800|\eta|^2+80  |\gamma|^2
+2000|\zeta|^2) + 4050 g^4_{10} \eea

\bea  \gamma_{\bar{\Sigma}}^{(2)}&=&-(200 |\eta|^2
(\bar{\gamma}_\Phi^{(1)}+ \bar{\gamma}_{\Sigma}^{(1)}  ) +  10
|\bar{\gamma}|^2(\bar{\gamma}_H^{(1)}+ \bar{\gamma}_\Phi^{(1)} ) +
100 |\bar{\zeta}|^2(\bar{\gamma}_\Theta^{(1)}+
\bar{\gamma}_\Phi^{(1)} )\nonumber\\&&+\text
Tr[f^{\dag}.\bar\gamma_{\Psi}^{(1)}.f] )+g^2_{10}( 4800|\eta|^2+80
|\bar\gamma|^2 +2000|\bar\zeta|^2\nonumber\\&&-80\text
Tr[f^{\dag}.f]) + 4050 g^4_{10} \eea

\bea \gamma_H^{(2)}&=&-( 84 |k|^2(\bar{\gamma}_\Phi^{(1)}+
\bar{\gamma}_\Theta^{(1)} )
+126|\gamma|^2(\bar{\gamma}_\Phi^{(1)}+ \bar{\gamma}_\Sigma^{(1)}
)+126 |\bar{\gamma}|^2(\bar{\gamma}_\Phi^{(1)}+
\bar{\gamma}_{\bar\Sigma}^{(1)} )\nonumber\\&& +\text
Tr[h^{\dag}.\bar\gamma_{\Psi}^{(1)}.h])+g^2_{10}(3024|k|^2+5040|\gamma|^2+5040|\bar{\gamma}|^2
\nonumber\\&&+108\text Tr[h^{\dag}.h])+1314 g^4_{10} \eea

\bea  \gamma_{\Theta}^{(2)}&=&-(7 |k|^2 (\bar{\gamma}_\Phi^{(1)}+
\bar{\gamma}_H^{(1)}  ) +7 |\rho|^2 (\bar{\gamma}_\Theta^{(1)}+
\bar{\gamma}_\Phi^{(1)} ) +105
|\zeta|^2(\bar{\gamma}_\Sigma^{(1)}+ \bar{\gamma}_\Phi^{(1)}
)\nonumber\\&&+\text Tr[ g^{\dag}.\bar\gamma_{\Psi}^{(1)}.g] + 105
|\bar{\zeta}|^2(\bar{\gamma}_{\bar\Sigma}^{(1)}+
\bar{\gamma}_\Phi^{(1)} ))+g^2_{10} (
84|k|^2+168|\rho|^2\nonumber\\&&+2940|\zeta|^2+2940|\bar\zeta|^2+12
Tr[g^{\dag}.g])+3318 g^4_{10} \eea

\bea \gamma_{\Psi}^{(2)}&=&
-(h^\dag.\bar{\gamma}_{\Psi}^{(1)T}.h-g^\dag.\bar{\gamma}_{\Psi}^{(1)T}.g+f^\dag.\bar{\gamma}_{\Psi}^{(1)T}.f+h^\dag.h
\gamma_H^{(1)}- g^\dag.g \gamma_\Theta^{(1)}+f^\dag.f \gamma_{\bar
\Sigma}^{(1)})\nonumber\\&&+g_{10}^2(90 h^\dag h+2520 g^\dag
g+6300 f^\dag f)+ \frac{26685 g_{10}^4}{16}.I \eea $\beta^{(2)}$
can be obtained from Eqns. \ref{one}-\ref{last} replacing
$\gamma^{(1)}$ by $\gamma^{(2)}$.

\chapter{Dynamical Yukawa Couplings} \label{ch:flavuni}
\section{Introduction}
The SM fermion mass-mixing data poses several questions. Fermion
masses vary from milli-eV from neutrino to between 0.5 MeV to 174
GeV for charged fermions. Leptonic mixing is large as compared to
quark sector mixing. Why do we have three fermion generations? Do
they follow some flavour symmetry ? The mass hierarchy is
different for up type quarks, down type quarks and for leptons.
All these questions  constitute the flavour puzzle posed by the SM
and neutrino oscillations data. To understand the origin of
observed flavour structure of the SM data is most basic problem of
flavour physics. Introduction of family symmetry and generation of
flavour structure by Yukawa couplings arising as VEVs of
``spurion" fields offers an attractive alternative prospect for
understanding flavour structure \cite{spurion}. Model builders
have considered various possibilities like discrete (tetrahedral
group $A_4$ and permutation group $S_3$), abelian/non-abelian
(global or local) symmetries. The establishment of the lepton
mixing pattern triggered great interest in the discrete family
symmetry approach (for reviews see
\cite{Altarelli:2010gt,King:2013eh,King:2014nza}). Mostly SU(5)
GUT and discrete family symmetry combination is considered. In the
so called Yukawa-on models \cite{koide} different symmetry is
considered for each type of fermion. The dimension-1 Yukawa-on
field ($\mathcal{Y}$) makes the Higgs vertex non-renormalizable
(${\cal{L}}= f^c \mathcal{Y} f H /\Lambda_\mathcal{Y} +...$) and
Yukawa-on dynamics is controlled by a high-scale
$\Lambda_\mathcal{Y}$.

In our view the strongest motivation and hint for the flavour
symmetry comes from third generation Yukawa unification in Susy
SO(10) GUTs at large $\tan\beta$. It indicates that the GUT gauge
symmetry breaking may generate the fermion hierarchy. Combining
this hint with the successful fitting of the fermion data in the
NMSGUT motivated us to extend the minimal renormalizable
supersymmetric SO(10) GUT with $O(N_g)$ family group
\cite{Aulakhkhosa}. In minimal Susy SO(10)GUT
\cite{aulmoh,ckn,abmsv,nmsgut}, MSSM Higgs pair emerges from the
large number of MSSM type doublets of UV theory and fermion
hierarchy is generated by the SO(10) matter Yukawa couplings. In
this extended scenario Higgs multiplets of SO(10) also carry
family index (``Yukawons") and their VEVs generate Yukawa
couplings of SM fermion and neutrino. In our study Yukawons also
carry representation of the gauge (SM/GUT) dynamics. As explained
in the previous chapters MSGUT completed with ${\bf{120}}$-plet
called NMSGUT \cite{nmsgut}, can generate realistic fermion mass
mixing data and experiment compatible B-decay rates after the
inclusion of superheavy thresholds. Therefore, from our viewpoint
of combined family and GUT unification, it is the logical base for
a dynamical theory of flavour. To start with, we study extension
of MSGUT based upon the $\bf{10 \oplus 210 \oplus \oot\oplus
 126} $ Higgs irreps. We will comment on the minor changes required to include the $\bf{120}$-plet: which may ultimately be necessary.
\section{Yukawon Ultra Minimal GUTs}
Yukawon Ultra Minimal GUTs are an extension of minimal
supersymmetric SO(10) model by $O(N_g)$ family gauge group. The
$\bf{10(H)}$ $\bf{\oplus}$ $\bf{210(\Phi) \oplus}$
$\bf{\oot(\bar\Sigma)}$ $\bf{\oplus
 126(\Sigma)}$ Higgs irreps  become symmetric representations of $O(N_g)$ family group. Matter
  fermions are present in the form of three copies of $\bf{16}(\Psi)$-plet. Superpotential of the model
  has same form as of MSGUT (with sum over flavour indices):
 \bea W_{GUT}&=&\mathrm{Tr}( m \Phi^2 + \lambda \Phi^3 + M \Sigb .\s +\eta \Phi .\Sigb.
  \s+\Phi.H.(\gamma \Sigma +\bar\gamma  . \Sigb) + { M_H}  H.H)\nnu
  W_{F}&=& \Psi_A .(h H_{AB} + f \s_{AB} + g\Theta_{AB} )\Psi_B \label{WF}\eea
Here A and B are the family indices. Now SO(10) Yukawa couplings
$h$, $f$, $g$ are complex number because flavour indices are
carried by MSGUT Higgs irreps themselves. Here we have included
$\bf{120}$-plet in $W_F$ but for simplicity we study only MSGUTs.
However addition of $\bf{120}$-plet does not effect GUT SSB since
it does not contain any MSSM singlet. Notice that $\bf{120}$-plet
carry antisymmetric representation of family group. \emph{In this
scenario matter fermion Yukawa couplings are  reduced from 15(21)
to just 3(5) parameters in MSGUT(NMSGUT) with 3 generations so we
call it \textbf{Yukawon Ultra Minimal Grand Unified Theory
(YUMGUTs)}}. Each Higgs irrep contains one MSSM Higgs type
multiplet ([$1,2,\pm 1$]). Mass matrix is given as

{\small{ \bee \vspace{.5cm}\cal {H} &=&\begin{pmatrix} -M_H
&\bar\gamma\sqrt{3} \Omega(\omega-a)&-\gamma\sqrt{3}
\Omega(\omega+a) &-\bar\gamma \Omega(\bar\sigma)\cr
 \gamma\sqrt{3} \Omega(\omega-a)&-(2 M  + 4 \eta \Omega(a-\omega))
 & \varnothing_d &  -2 \eta \sqrt{3} \Omega(\bar\sigma) \cr
-\bar\gamma\sqrt{3} \Omega(\omega+a)&\varnothing_d &-(2 M + 4 \eta
\Omega(\omega+a))&\varnothing_d &\cr
  - \gamma \Omega( \sigma)  &-2 \eta \sqrt{3} \Omega( \sigma)&\varnothing_d &
   6 \lambda \Omega(\omega-a)-2 m\cr
\end{pmatrix}\label{calH}\vspace{.5cm} \eee}}
The rows  are labelled by the $N_g(N_g+1)/2$-tuples (ordered and
normalized, for a symmetric $\phi_{AB},
  A,B=1..N_g$,  as $\{\phi_{11},\phi_{22},...\phi_{N_gN_g},
 \sqrt{2}\phi_{12},\sqrt{2}\phi_{13},.....,\sqrt{2}\phi_{N_g-1,N_g}\}$)\hfil\break
  containing MSSM type ${\overline{H}}[1,2,-1]$ doublets from
  $\mathbf{10,126,\oot,210}$. The columns represent ${ {H}}[1,2, 1]$ doublets
  in the order $\mathbf{10,\oot,126,210}$. $\varnothing_d$ is the d dimensional null square matrix.
The matrix function $\Omega$ ($\frac{N_g(N_g+1)}{2}$ dimensional)
is introduced to write $\cal {H}$ in compact notations and its
form is determined by symmetric invariant
$\phi_{AB}\phi_{BC}\phi_{CA}$ (here one field can have VEV and
other two should contain $H$ and $\bar{H}$).
  For
 $N_g=2$ it is \vspace{.2cm} \bea \Omega[V]= \begin{pmatrix} V_{11} &0&V_{12}/\sqrt{2}  \cr
 0 &V_{22}&V_{12}/\sqrt{2} \cr V_{12}/\sqrt{2} & V_{12}/\sqrt{2}
 &(V_{11}+V_{22})/2
  \cr \end{pmatrix} \vspace{.3cm}\eea
with labels $\{{\overline H}_{11},{\overline
H}_{22},\sqrt{2}{\overline H}_{12}\}\oplus \{{ H}_{11},{
H}_{22},\sqrt{2}{  H}_{12}\}$. Higgs mass matrix is now $2
N_g(N_g+1)$ dimensional which would become $N_g(3 N_g+1)$
dimensional if we include $\bf{120}$-plet.
 MSSM being a effective theory requires one light Higgs pair out of these large number of Higgs multiplets. Consistency condition of the light Higgs pair assumption (fine tuning $Det \cal{H}$=0) ensures this.
From left ($\hat W$) and right ($\hat V$) null eigenvectors we can
determine MSSM Yukawa couplings. For $N_g$=2 Yukawas of up and
down type quarks we get\vspace{-.2cm} \[ Y_u =
\begin{pmatrix} \hat h \hat V_1 + \hat f \hat V_4 & (\hat h \hat
V_3 + \hat f \hat V_6)/\sqrt{2} \cr (\hat h \hat V_3 + \hat f \hat
V_6)/\sqrt{2} & \hat h \hat V_2 + \hat f \hat V_5 \cr
\end{pmatrix} \quad ; \quad  \hat h = 2\sqrt{2} h \] \bea
 Y_d &=&\begin{pmatrix} \hat h \hat W_1 + \hat f \hat W_7 & (\hat h
\hat W_3 + \hat f \hat W_9)/\sqrt{2} \cr (\hat h \hat W_3 + \hat f
\hat W_9)/\sqrt{2} & \hat h \hat W_2 + \hat f \hat W_8 \cr
\end{pmatrix}   \qquad ;\qquad \hat
f=-4 i\sqrt{\frac{2}{3}} f
 \label{yukawas}\eea
By replacing $\hat f \rightarrow -3 \hat f$ in $Y_u,Y_d$ we can
get $Y_{\nu},Y_l$. Clearly for $f\sim h$ one can get $Y_{\nu},Y_l$
different from $Y_u,Y_d$ as $f\ll h$ implies $Y_u \approx Y_\nu$
and $Y_d \approx Y_l$. Higgs mass matrix consequently $\hat V,
\hat W$ are determined in terms of symmetry breaking VEVs
($p,a,\omega,\sigma,\bar\sigma$). Next step is to calculate these
VEVs.
\section{Spontaneous Symmetry Breaking}
The multiplets   $\bf{210}$, $\bf{\oot}$ $\bf{126}$ break the GUT
and flavour symmetry to MSSM. YUMGUT superpotential written in
terms of VEVs $(p,a,\omega,\sigma,\bar\sigma)$ of SM singlets:
 \bea W&=&
\mathrm{Tr }[m(p^2+3 a^2+6 \omega^2)+ 2 \lambda (a^3+3 p \omega^2
+6 a \omega^2)]\nonumber\\&&+\mathrm{Tr}[M \bar\sigma \sigma
 + \eta (p+3a-6 \omega)\frac{(\bar \sigma \sigma  +  \sigma \bar \sigma)}{2}]
  \eea
Susy vacuum is determined by the vanishing of F and D terms. The
F-term vanishing equations can  be written as: \be 2 m (p-a) -2
\lambda a^2 +2\lambda \omega^2 =0 \label{pa}  \ee \be
  2 m(p+ \omega ) +  \lambda (p+2 a + 3 \omega)\omega
  +  \lambda \omega (p+2 a + 3 \omega) =0  \label{pw}
 \ee \be
  M \sigma+\eta (\chi \sigma +\sigma \chi)/2 =0\label{homosig} \ee \be
 M \bar\sigma+\eta (\chi \bar\sigma +\bar \sigma \chi)/2  =0 \label{homosigbar}\ee \be
   \bar\sigma\sigma  + \sigma \bar\sigma  = -{\frac{4}{\eta}}(m p +3 \lambda
 \omega^2) \equiv   F  \label{sgbsg}
  \ee
where $\chi\equiv (p+3 a -6 \omega) $. D-terms include SO(10) and
family D-terms. SO(10) has only one non-trivial D-term : $D_{B-L}$
: \be \vspace{.3cm}\sigma_{AB}-\bar\sigma_{AB}=0 \vspace{.25cm}
\ee
    The set of homogenous Eqns.
(\ref{homosig},\ref{homosigbar})  can be written in a more
transparent form  as \vspace{.35cm}\be  \Xi \cdot\hat\Sigma =\Xi
\cdot \hat\os =0 \vspace{.35cm} \ee where ${\hat\Sigma,\hat\os} $
are $(N_g(N_g+1)/2)$-plet of $\sigma$, $\bar\sigma$ VEVs.
Nontrivial solutions of  Eqns. (\ref{homosig},\ref{homosigbar})
for $\sigma,\bar\sigma $ exist only if   $Det[\mathbf{\Xi}]=0$.
  In the MSGUT($N_g$=1) the linear condition ($\chi=-M/\eta$)
supplements the Eqns. (\ref{pa},\ref{pw})  and allows
determination of $p,a,\omega$ via a cubic equation for $\omega$.
After solving F-term conditions (actual procedure will be
discussed in the next section), the D-term conditions ($D_{B-L}=0$
from SO(10) and $D^A=0$ from $O(N_g)$) need to be solved. In
$N_g$=1 case \be \vspace{.25cm}
D_{B-L}=|\sigma|^2-|\bar\sigma|^2=0 \vspace{.25cm} \ee Since
$\mbox{Arg}[\sigma]-\mbox{Arg}[\bar\sigma]$ can be removed  by
$U(1)_{B-L}$ transformations, we choose $\sigma=\bar\sigma$. Here
also we only consider the cases corresponding to
$\sigma_{AB}=\bar{\sigma}_{AB}$, so that $D_{B-L}$ is
automatically zero.

The D-terms of the family group vanish automatically only for
trivial solutions of the F-terms conditions. We are interested in
non-trivial solutions because only these can generate generation
mixing. One needs to introduce additional fields to cancel GUT
sector contribution to the family D-terms. F-terms corresponding
to extra fields should not interfere with the GUT F terms so as
not to disturb the MSGUT SSB. The best possible choice is to
locate these fields in the hidden sector. In \cite{BMsugry}  it
has been shown that Bajc-Melfo (BM) two field superpotential is an
appropriate candidate. In the next section we will discuss how BM
superpotential enables YUMGUTs. \vspace{-.2cm}
\section{Bajc-Melfo Superpotential}
Two field $(S_s,\phi_s)$ BM superpotential reads:
 \be W_H=S_s
(\mu_B \phi_s  + \lambda_B \phi_s ^2) \ee It has a Susy preserving
global minima at $S_s=\phi_s=0$ and $S_s=0,
 \phi_s=-\frac{\mu_B}{\lambda_B}$ and Susy breaking local minima at
$ \langle \phi_s \rangle=-\frac{\mu_B}{2 \lambda_B}$ where $S_s$
remains undetermined with a condition $|\langle S_s \rangle |\geq
|\frac{\langle \phi_s \rangle }{\sqrt{2}}|$. $\langle S \rangle $
can be fixed either by radiative corrections \cite{BM} or by
couplings to N=1 \emph{supergravity }\cite{ovrab,hilo}. In
\cite{BMsugry} $\langle S \rangle$ is determined by coupling
$S,\phi$ fields to  N=1 \emph{supergravity } as reviewed below.
\subsection{Coupling to Supergravity}
Supergravity potential
\cite{cremmer,Ohta:1982wn,nath,halllykkwein} for the scalar field
$Z_I$ is \bea V&=& E (|F^I + {Z^*}^I \kappa^2 W|^2 - 3 \kappa^2
|W|^2) \qquad ;\qquad E \equiv e^{\kappa^2 \sum_i |Z_I|^2}\eea
Visible sector VEVs $(z_i)$ and $\phi_s$ preserve global Susy
$(\frac{\partial W}{\partial Z_I}=0=D(Z^I))$ \bea V(S_s)&=&
e^{\kappa^2 |S_s|^2 + \delta } \{(\delta \kappa^2 |\widehat{W}_0+
S_s \theta|^2 + |\theta + \kappa^2 S_s^*( \widehat{W}_0+ S_s
\theta)|^2 \nonumber\\&&- 3 \kappa^2 |\widehat{W}_0+ S_s \theta|^2
\}\eea Where $\delta =\kappa^2(|\bar{\phi}_s|^2+
\sum\limits_i|\bar z_i|^2)$ denotes Susy preserving VEVs
contribution and the contribution of $S_s$ has been separated.
 The potential written in terms of dimensionless variables \be x
=\kappa {\frac{\widehat{W}_0}{\theta}} \quad ;  \quad  y=\kappa
S_s \quad  ; \quad  \varphi_x =\mbox{Arg}[x] \quad  ;  \quad
\varphi_y=\mbox{Arg}[y]\ee \bea {\widetilde
V}&\equiv&{\frac{V}{|\theta|^2}} = {\big\{}(|x|^2
+|y|^2)(\delta-3) + (1+|y|^2)^2 + \nnu && |x|^2 |y|^2 + 2
\cos(\varphi_y-\varphi_x)|x||y|(|y|^2 +\delta-2) {\big\}}\eea
Minimum is achieved if  \bea V=V_{|y|}=V_{ \varphi_y}= 0=V_{|y|,
\varphi_y}\quad ; \quad V_{|y|,|y|},V_{\varphi_y \varphi_y}
>0 \eea The solution is \bea   \varphi_y &=& \varphi_x   \quad
;\qquad x= 2 - \sqrt{3 -\delta} \nnu  y&= &y_0= \sqrt{3-\delta} -1
{\overline{\partial_{|y|} \partial_{|y|} \tilde{V} }} = 4
\sqrt{3-\delta} \nnu {\overline{\partial_{\varphi_y }
\partial_{\varphi_y }\tilde{V} }} & =& 4 \delta \sqrt{3-\delta } -16
\delta -32 \sqrt{3-\delta }+56\quad \label{hidVEV}\eea provided
\bea \delta < 3- \bigg(1+
\sqrt{\bigg|\frac{\kappa^2\theta}{2\lambda_B}\bigg|}\bigg)^2
\simeq 2 \eea The globally undetermined VEV $\langle S_s \rangle
\sim M_p$ is now fixed. In gravity mediated scenario, gravitino
mass is given by \bea m_{\frac{3}{2}} &=& \kappa^2 |
\sqrt{\bar{E}}({\ovl{\widehat W}_0} + {\ovl{W_H}}) | \eea Typical
range of gravitino mass require cancellation among $\widehat W$
and $\ovl W_{GUT}$ such that \bea
   | { \ovl{\widehat W}}_0  + {\ovl{W_H}} |=M_p|\theta|   < 10^{39}-10^{41}
 \rm{GeV^3}\eea BM superpotential
parameters determine Susy breaking scale :\bea
\sqrt{|F^S|}&=&\sqrt{|\theta|}=\biggl|\frac{\mu_B}{
2\sqrt{\lambda_B}}\biggl|\sim
 10^{10.5}-10^{11.5} \rm{ GeV} \eea
\subsection{Gauged $O(N_g)$ with Hidden Sector Superpotential}
Considering BM superpotential as a hidden sector of supergravity,
total Superpotential is given by \be W=W_H+W_{GUT}(z_i) \ee where
$ W_H(S_s,\phi_s)= W_0+ S_s(\mu_B \phi_s + \lambda_B \phi_s^2)$,
$z_i$ represents GUT chiral and symmetric multiplets of $O(N_g)$
whose VEVs determine supersymmetric vacuum from vanishing of F and
D terms \bea {\ovl{F^i}}= {\ovl{{\frac{\partial W}{\partial
z_i}}}}=0 \nonumber \eea \be
{\ovl{D_{GUT}^\alpha}}(\bar{z}_i,\bar{z}_i^*)=0 \label{GUTmin}\ee
Supergravity potential representing D-term contribution is
 \be \vspace{.3cm} V_D= {\frac{g^2_\alpha}{4}}
{\bigg{[}}\bigg({z^*}^i + {\frac {F^i}{\kappa^2 W}}\bigg)
(T^{\alpha  })_i^j z_j + h.c.){\bigg{]}} \vspace{.3cm} \ee Since F
terms vanish for all $z_I$ except $S_s$ (which is gauge singlet)
we are left with just global supersymmetric D-terms. The $O(N_g)$
D terms
  are given as :
 \[ D^a_{O(N_g)}= \mathrm{Tr}( \hat\phi^\dagger[T^a,\hat\phi]+   \hat{S}^\dagger[T^a,
 \hat{S}] ) +{\ovl D^a_X} \] \be
{ {{\bar{D}}^a_X}}=\sum_i\bar z_i^\dagger {{\cal{T }}^a} \bar z_i
\ee
 where  ${\ovl D^a_X}$ is the visible sector contribution,
  and $T^a,{\cal T}^a$ are $O(N_g)$ generators in
  the fundamental and generic representations.
We can consider $S,\phi$ as traceful multiplets of $O(N_g)$,
traceless part still remains undetermined, to fix family D-terms :
\bea W_H&=& \mathrm{Tr} S(\mu_B \phi+ \sqrt{N_g} \lambda_B \phi^2)
\eea \bea S&=& {\hat{S}} +{\frac{1}{\sqrt{N_g}}} S_s {{\cal
I}_{N_g}} \qquad ;\qquad \mathrm{Tr} {\hat {S}}=0\eea here ${{\cal
I}_{N_g}}$ is the unit matrix of order $N_g$. All $O(N_g)$
non-singlet fields : $\hat{S}_{ab}$,  $\hat{\phi}_{ab}$ and
visible sector Higgs fields enter into the family D-terms. As
discussed in the previous section flavour singlet VEV $(S_s)$ is
detemined by supergravity effects and non-singlet part is  fixed
by the $O(N_g)$ D-terms and supergravity soft mass terms.
\subsection{$\langle S\rangle$ fixation}
\subsubsection{\textbf{A. $N_g=2$}} Using $O(2)\approx U(1)$ isomorphism,
S (similarly $\phi$) is defines as \bea S &=& {\frac{1}{2}}
\begin{pmatrix} \sqrt{2} S_s + S_{+} + S_{-}  &    i(S_{-}- S_{+})\cr
 i(S_{-}- S_{+}) & \sqrt{2} S_s -(  S_{+}+ S_{-})\cr \end {pmatrix}  \eea
 where $S_\pm,S_s$ are
properly normalized fields so that \[ \mathrm{Tr} S^\dagger S =
S_+^\dagger S_+ + S_{-}^\dagger S_{-} + S_s^\dagger S_s \] and \[
 \mathrm{Tr} S \phi= S_s \phi_s + S_{(+}\phi_{-)}\] The superpotential becomes
\bea W_H&=&
  \mu_B (S_s  \phi_s  + S_{(+}\phi_{-)})+ \lambda_B
  ( S_s \phi_s^2 + 2 S_s\phi_+\phi_{-} + 2 \phi_s S_{(+}\phi_{-)}) \label{WHNeq2}  \eea
$F_s$ is non zero($\theta$) and all other F-term vanish for
$\bar{\phi}_s=-\frac{\mu_B}{2\lambda_B},$ $\bar{\phi}_{\pm }=0$
and $S_{s,\pm}$ remain undetermined. Using the values of fields :
 \bea V(S_+,S_{-})&=& m_{3/2}^2\bigg(|S_+|^2 + |S_{-}|^2\bigg) +
{\frac{g_f^2}{2}}  \bigg(|S_{+}|^2 -|S_{-}|^2 +\bar{D}_X\bigg)^2
\eea here $\bar{D}_X=\sum q_i |\bar{z}_i|^2 $, $q_i$ is the family
symmetry charge of the  visible sector VEV $Z_i$. The minimum will
occur when \bea
  S_{-x} &=& \sqrt{|\bar{D}_X| - x {\frac{m_{3/2}^2}{g^2}}} \qquad ;
 \qquad   \bar{S}_{x}=0 \qquad (x=Sign[D_X])  \eea
 Detailed mass spectrum of $S$, $\phi$ fields can be found in \cite{BMsugry}.
\subsubsection{\textbf{B. $N_g=3$}}
$S_{ab}$ ($\phi_{ab}$) in terms of $T_3$ eigenfields
:\vspace{.2cm} {\small{\bea \begin{pmatrix}
 \frac {1} {6}  (\sqrt {6} {S_ 0} + 3 i{S_ {-2}} - 3 i {S_ {+2}} +
 2 \sqrt {3} {S_s} ) & \frac {1} {2}  ({S_{+2}} + {S_ {-2}}) & \frac{1}{2} ({{S_ {+}} + {S_{-}}})   \cr
\frac {1} {2}   ({S_{-2}} + {S_{+2}}) & \hspace{-.8mm}\frac {1}{6}
(\sqrt {6} {S_ 0} - 3 i{S_ {-2}} + 3 i{S_{+2}} + 2 \sqrt {3} {S_s}
) & \frac {1} {2} i ({S_ {+}} - {S_{-}})\cr \frac{1}{2}( {{S_ {+}}
+ {S_{-}}})   & \frac {1} {2} i ({S_{+}} - {S_{-}}) &
\frac{1}{\sqrt {3}} ({{S_s} - \sqrt {2} {S_ 0}})
    \end{pmatrix}
   \nonumber\eea}}
where  $S_{s,0,\pm,\pm 2} $ are properly normalized fields so that
\bea \mathrm{Tr} S^\dagger S =S_{+2 }^\dagger S_{+2} +
S_{-2}^\dagger S_{-2} + S_+^\dagger S_+ + S_{-}^\dagger S_{-} +
S_s^\dagger S_s\nnu  \mathrm{Tr} S \phi= S_s \phi_s +
S_{(+}\phi_{-)}  + S_{(+2}\phi_{-2)}\eea Now D-terms form a $O(3)$
vector. It is convenient to use a basis where D-terms point in the
third direction -${\bar{D'}}_X^a=\delta^a_3 { {|\bar{D_X}|}} $.
This can be achieved by performing the following rotations :
\vspace{-.2cm}
 \[
O = R_{23}[\theta_X] . R_{12}\bigg[{\frac{\pi}{2}}-\varphi_X\bigg]
\] \be \theta_X=ArcTan\bigg[\sqrt{\frac{V_1^2 +
V_2^2}{V_3^2}}\bigg]\quad ;\qquad \varphi_X  = ArcTan\bigg[ \frac
{V_2}{V_1}\bigg] \ee where $V^a= \bar{D}_X^a$. The potential for
the flat directions from $\hat S'$ is now\bea V[\hat S'] &=&
m_{3/2}^2(|S_0'|^2 +|S_{+}'|^2 +|S_{-1}'|^2 + |S_{+2}'|^2
+|S_{-2}'|^2)\nnu &+& {\frac{g_f^2 }{2}}\{(|S_{+}'|^2
+2|S_{+2}'|^2 -|S_{-}'|^2 -2|S'_{-2}|^2 +(\bar{D}_X^3)')^2  \nnu
&& +  2
 \mathrm{Tr}({S'}^\dagger [T_+,S']) \mathrm{Tr}({S'}^\dagger [T_{-},S'])  \}\eea Solution found is  \be
 |\bar{S}_{-2 }| = \sqrt{ {\frac{|{\vec D_X}|}{2}} - {\frac{m_{3/2}^2
}{4 g_f^2}}}  \qquad ; \qquad  |\bar{S}_{-,+,+2}| = 0\ee In
\cite{BMsugry} it is shown that BM type hidden sector necessarily
imply a number of light SM singlet scalars $(O(m_{3/2}))$ and even
lighter fermions that get mass only from radiative effects. These
modes are reminiscent of the light moduli in string theory. Note
that these light modes supplement the singlet (G[1,1,0] sector)
pseudo-Goldstones from the visible sector yielding a very rich set
of possible DM candidates. Light modes of the BM superpotential
may provide light DM candidates ($<50$) GeV as indicated by the
DAMA/LIBRA \cite{damalibra} experiments.
\section{Analytical and Numerical Analysis}
\vspace{.2cm} Writing VEVs $
\{p,\omega,a,\sigma,\bar\sigma,\chi\}$
=$\frac{m}{\lambda}\{P,W,A,\tilde \sigma,
\tilde{\bar\sigma},\tilde\chi\}$,
$\tilde\chi_A=\tilde\chi_{AA}+\xi$
 in units of $m/\lambda $, we can eliminate all the parameters in F-term equations (Eq. \ref{pa}-\ref{sgbsg}) except two ratios
  $\xi=\frac{\lambda
M}{\eta m}$ and  $\frac{\lambda}{\eta}$ : \be
2\left(\frac{m}{\lambda}\right)^2(P-A-A^2+W^2)=0 \label{dl6} \ee
\be \left(\frac{m}{\lambda}\right)^2(P+W+(P+2A+3W)W+W(P+2A+3W))=0
 \label{dl7}\ee \be \left(\frac{m}{\lambda}\right)^2(\xi
\tilde\sigma+(\tilde\chi \tilde\sigma+\tilde\sigma
\tilde\chi)/2)=0  \label{dl8} \ee \be
\left(\frac{m}{\lambda}\right)^2(\xi \tilde
{\bar\sigma}+(\tilde\chi \tilde{\bar\sigma}+\tilde{\bar\sigma}
\tilde\chi)/2)=0 \label{dl9}\ee  \be
\tilde{\bar\sigma}\tilde\sigma+\tilde\sigma\tilde{\bar\sigma}=-\frac{4
\lambda}{\eta}(P+3 W^2)=\frac{\lambda^2 F}{m^2}
=\tilde{F}\label{dl10} \vspace{.3cm} \ee It is convenient to use
dimensionless form of equations for SSB analysis because
 we can get most of the VEVs independent of model parameters. Before analyzing realistic  SM case ($N_g$=3) we will study the
simplest toy model ($N_g=2)$. \vspace{.3cm}
\subsection{Toy Model ($N_g=2$)}
\vspace{.3cm} For $N_g=2$, $\hat\Sigma
=\{\tilde\sigma_{11},\tilde\sigma_{22},\tilde\sigma_{12}\}$, the
matrix $\mathbf{\Xi}$ involves the combinations
$\tilde\chi_A=\tilde\chi_{AA} +\xi $: \bea \mathbf{\Xi}=
\begin{pmatrix}   \tilde\chi_1& 0&\tilde\chi_{12} \cr 0&\tilde\chi_{2}&\tilde\chi_{12} \cr
\tilde\chi_{12} &\tilde\chi_{12}& \tilde\chi_1+\tilde\chi_2
\end{pmatrix} \eea
\bea && Det[\mathbf{\Xi}] =
(\tilde\chi_1+\tilde\chi_2)(\tilde\chi_{12}^2-
\tilde\chi_1\tilde\chi_2)=0 \nnu &&\Rightarrow \quad
\tilde\chi_1=-\tilde\chi_2 \quad  {\rm{or}} \quad \tilde\chi_{12}
= \pm\sqrt{ \tilde\chi_1\tilde\chi_2}\eea $Det[\mathbf{\Xi}]$
should vanish for non-trivial solutions. Null $2 \times 2$ minors
provide $ \tilde\chi_1=\tilde\chi_2=\tilde\chi_{12}=0 $
 $(\mathbf{\Xi}\equiv 0)$.  Thus $Rank[\mathbf{\Xi}]<2  $ implies
  $Rank[\mathbf{\Xi}]=0$ so that all the six $\sigma,\bar\sigma$ remain undetermined. However we find
   that $Rank[\mathbf{\Xi}]=0$
is a degenerate case implying large colored and charged
pseudo-Goldstone multiplets so we consider only
$Rank[\mathbf{\Xi}]=2$ case. For a non-trivial solution,  one out
of two factors $(\tilde\chi_1+\tilde\chi_2)$ and
$(\tilde\chi_{12}^2-\tilde\chi_1 \tilde\chi_2)$ of
$Det[\mathbf{\Xi}] $ should vanish. One can calculate
$\tilde\sigma_{11}$ and $\tilde\sigma_{22}$ in terms of
$\tilde\sigma_{12}$ from ${ \Xi \cdot\hat\Sigma =0} $ as\be
\tilde\sigma_{11}=-\frac{\tilde\chi_{12}}{\tilde\chi_1}
\tilde\sigma_{12} \quad ; \quad
\tilde\sigma_{22}=-\frac{\tilde\chi_{12}}{\tilde\chi_2}
\tilde\sigma_{12} \ee \be
Det[\tilde\sigma]=\frac{(\tilde\chi_{12}^2-\tilde\chi_1
\tilde\chi_2)}{\tilde\chi_1 \tilde\chi_2} \tilde\sigma_{12}^2 \ee
$Det[\tilde\sigma]$ has  a factor $(\tilde\chi_{12}^2-\tilde\chi_1
\tilde\chi_2)$ in common with $Det[\mathbf{\Xi}] $ which will
cause $Det[\tilde\sigma]$ to also vanish if we choose this factor
to be zero to make $Det[\mathbf{\Xi}] $ vanish. In MSGUTs (also in
YUMGUTs), Majorana mass of the right handed neutrinos is
determined by $\langle \overline\Sigma
\rangle$=$\tilde{\bar\sigma}$, which requires invertible VEV for
Type I seesaw contribution. Therefore we analyze only  the branch
$(\tilde\chi_1+\tilde\chi_2)=0$ for vanishing $Det[\mathbf{\Xi}]$.
Eq. \eqref{dl10} then implies \be \tilde\sigma_{11}^2=
\frac{\tilde F_{11} \tilde\chi_{12}^2 }{2(\tilde\chi_{12}^2
+\tilde\chi_1^2)}\quad ; \quad \tilde F_{11} =
 \tilde F_{22}\qquad ;\qquad \tilde F_{12}=0\label{Fcnstrt}\ee
 \begin{table}
$$
\begin{array}{cccc }
\hline\hline &&&\vspace{-.3 cm}\\{\rm Parameter}&\multicolumn{3}{
c }{\rm Value}\vspace{0.1 cm}\\\hline
\multirow{2}{*}{\rm $M_H$} & \rm I  & \rm II & \rm III \\
&0.049 + 0.190 i &0.599 + 0.791 i & 1.39 + 0.80 i\\ \hline\hline
{\rm Y_u}&0.1537, 0.0080 &0.1293, 0.0118 & 0.0685, 0.0214\\
{\rm Y_d}&0.0537,0.0043&0.0562,0.0051&0.0359, 0.0052\\

{\rm Y_l}&0.0424, 0.0027 & 0.0712, 0.0065&0.0147,0.0063\\

{\rm Y_\nu}&0.2515,0.0233&0.0576,0.0053&0.0911, 0.0028\\
\hline
{\rm \theta_{CKM}(deg.)}&5.15&2.27\times 10^{-6}&7.41\\
{\rm \theta_{PMNS}(deg.)}&14.5&2.32 \times 10^{-5}& 33.7\\
\hline\hline
{\rm m_\nu(meV)} &0.0255,0.2791 &0.0013, 0.0144 &0.0011,  0.0121\\
{\rm \Delta m_\nu^2(eV^2)}&7.73 \times 10^{-8}&2.06 \times 10^{-10}&1.45 \times 10^{-10}\\
\hline\hline
\end{array}
$$\caption{\small{Yukawa eigenvalues and mixing angles for $N_g$=2, $f$=-0.13 . $\tilde{M}_{\nu^c}\equiv
   \lambda  {M}_{\nu^c}/m$=$\{0.6969, 0.0636\}$. $m/\lambda $ is taken to be $10^{16}$ GeV
  to estimate $\Delta m_\nu^2 $. $\lambda$= -0.038 + .005 i, $\eta$ = 0.4, $\gamma$= 0.32, $\bar\gamma$= -1.6, $h = .34$, $\xi$=0.8719+.5474i.\label{table:ng=2} }}
\vspace{1.0cm}
$$
\begin{array}{cccc }
\hline\hline &&&\vspace{-.3 cm}\\{\rm Parameter}&\multicolumn{3}{
c }{\rm Value}\vspace{0.1 cm}\\\hline
\multirow{2}{*}{\rm $M_H$} & \rm I  & \rm II & \rm III \\
&0.049 + 0.190 i &0.599 + 0.791 i & 1.39 + 0.80 i\\ \hline \hline
{\rm Y_u}& 0.1761, 0.0131& 0.1108,0.0101&0.0721, 0.0140\\
{\rm Y_d}&0.0507,0.0038&0.0569,0.0052&0.0283,0.00552\\

{\rm Y_l}& 0.0507, 0.0038&0.0569,0.0052&0.0283, 0.0055\\

{\rm Y_\nu}&0.1762,0.0131&0.1108, 0.0101&0.0721,0.0140\\
\hline
{\rm \theta_{CKM}(deg.)}& 0.00486& 2.47 \times 10^{-9}&0.00767\\
{\rm \theta_{PMNS}(deg.)}&8.7& 2.79 \times 10^{-6}& 26.8\\ \hline
\hline
{\rm m_\nu(meV)} & 10.05,  110.19 & 4.86, 53.29 & 4.39,  48.12\\
{\rm \Delta m_\nu^2(eV^2)}&0.01204 & 0.00282 & 0.00230\\
\hline \hline
\end{array}
$$\caption{\small{Effect of reducing $f$ :
  Yukawa eigenvalues and mixing angles for $N_g$=2, $f$=-0.00013 and other parameters same as in Table
  \ref{table:ng=2}. Notice that the   light neutrino masses are in an  acceptable range
  but  $Y_u$=$Y_\nu$, $Y_d$=$Y_l$ and the   quark mixing is negligible. $\tilde{M}_{\nu^c}\equiv
   \lambda  {M}_{\nu^c}/m$= $\{0.000697, 0.0000636\}$.\label{table:ng=2fsmall}}}
\end{table}
We solve Eq. \ref{dl7} (linear in P) for all the components of P.
Using calculated P values, solve  $\tilde\chi_1=-\tilde\chi_2$,
$\tilde F_{12}=0$ and $\tilde F_{11}-\tilde F_{22}=0$ for
$A_{11}$, $A_{12}$ and $A_{22}$. The remaining equations (Eq.
\ref{dl6} )  can be completely expressed in terms of $W$ and
$\xi$. We used a minimization method for a numerical solution of
$W$ for a convenient $\xi$. Using these numerical values of
$P,A,W,\tilde \sigma$ (given in Appendix A) and randomly chosen
YUMGUT parameters($\lambda,\eta,\gamma,\bar\gamma,h,f$), we find
$M_H$ values from $Det[\cal H]$=0. Then  Yukawas corresponding to
all allowed value of $M_H$ are determined. Yukawa eigenvalues,
mixing angles and neutrino masses are presented in Table
\ref{table:ng=2} for $f\sim h$ and in Table \ref{table:ng=2fsmall}
when $f$ is smaller by a factor of $10^{-3}$. In $f\sim h$ case we
have acceptable fermion hierarchy and mixing but too small
neutrino masses which is the main failure of MSGUT. One can boost
Type I seesaw contribution by suppressing $f$ which implies
$Y_u=Y_\nu$ and $Y_d=Y_l$ (see Table \ref{table:ng=2fsmall}). Type
II seesaw contribution is generated by VEV of O[1,3,-2] multiplet
\cite{ag1,ag2,blmdm,nmsgut} which is now family group triplet. In
the MSGUT(NMSGUT) Type I dominates over the Type II. Type II
contribution needs to be re-examined in the YUMGUT. Complete
superheavy spectra (in units of $m/\lambda$) for the solution
found is presented in Table \ref{Ngeq2PGoldies} in Appendix B.
Only the SM singlet sector $G[1,1,0]$ has pseudo-Goldstones (which
can act as DM candidates).
\subsection{Realistic Case ($N_g=3$)}\vspace{.3cm}
Symmetry breaking equations in this case are more complex and
offer a number of phenomenologically interesting possibilities
like light sterile neutrino and novel DM candidate from MSSM
singlet sector G[1,1,0]. Like $N_g=2$ case, $\tilde\sigma$
equations can be written as \vspace{.2cm}\be { \Xi \cdot\hat\Sig
=0} \label{sig3} \ee Now \bea \mathbf{\Xi}=
\begin{pmatrix}   \tilde\chi_1& 0 & 0&\tilde\chi_{12} &\tilde\chi_{13}& 0 \cr 0&\tilde\chi_{2}&0&\tilde\chi_{12}&0 & \tilde\chi_{23}\cr
0 & 0& \tilde\chi_{3}&0 &\tilde\chi_{13}& \tilde\chi_{23}\cr
\tilde\chi_{12} & \tilde\chi_{12}& 0&\tilde\chi_{1}+\tilde\chi_{2}
&\tilde\chi_{23}& \tilde\chi_{13}\cr \tilde\chi_{13} & 0&
\tilde\chi_{13}&\tilde\chi_{23} &\tilde\chi_{1}+\tilde\chi_{3}&
\tilde\chi_{12}\cr 0 & \tilde\chi_{23}&\tilde\chi_{23}
&\tilde\chi_{13}&\tilde\chi_{12}&
\tilde\chi_{2}+\tilde\chi_{3}\end{pmatrix} \eea and
$\hat\Sigma=\{\tilde\sigma_{11},\tilde\sigma_{22},
\tilde\sigma_{33}, \tilde\sigma_{12}, \tilde\sigma_{13},
\tilde\sigma_{23}\}$\vspace{.2cm} \bea Det[\mathbf{\Xi}]&=&
(\tilde\chi_1 \tilde\chi_{2} \tilde\chi_{3}-\tilde\chi_1
   \tilde\chi_{23}^2-\tilde\chi_{12}^2 \tilde\chi_3+2
   \tilde\chi_{12} \tilde\chi_{13} \tilde\chi_{23}-\tilde\chi_{13}^2 \tilde\chi_{2})\nonumber\\&& (\tilde\chi_{1}^2 \tilde\chi_2+
   \tilde\chi_{1}^2 \tilde\chi_{3}-\tilde\chi_{1} \tilde\chi_{12}^2-\tilde\chi_{1} \tilde\chi_{13}^2+\tilde\chi_{1}
   \tilde\chi_{2}^2+2 \tilde\chi_{1} \tilde\chi_{2} \tilde\chi_{3}+\tilde\chi_{1} \tilde\chi_{3}^2\nonumber\\&&-\tilde\chi_{12}^2
   \tilde\chi_{2}-2 \tilde\chi_{12} \tilde\chi_{13} \tilde\chi_{23}-\tilde\chi_{13}^2 \tilde\chi_{3}+\tilde\chi_{2}^2
   \tilde\chi_{3}-\tilde\chi_{2} \tilde\chi_{23}^2+\tilde\chi_{2}
   \tilde\chi_{3}^2-\tilde\chi_{23}^2 \tilde\chi_{3})\label{detXi} \eea
   $Det[\mathbf{\Xi}]$ should vanish for the non-trivial solution.
Order of mass matrices will be double that of $N_g=2$ case and can
be written using $\Omega_3[V]$ matrix function defined as: \bea
\Omega_3[V]\equiv
\begin{pmatrix} V_{11} & 0 & 0 & \frac{V_{12}}{\sqrt{2}} & \frac{V_{13}}{\sqrt{2}} & 0 \cr
0 & V_{22} & 0 & \frac{V_{12}}{\sqrt{2}} & 0 &
\frac{V_{23}}{\sqrt{2}}\cr 0& 0 & V_{33} & 0
&\frac{V_{13}}{\sqrt{2}} & \frac{V_{23}}{\sqrt{2}}\cr
\frac{V_{12}}{\sqrt{2}}&\frac{V_{12}}{\sqrt{2}}&0
&\frac{V_{11}+V_{22}}{2}& \frac{V_{23}}{2} &\frac{V_{13}}{2} \cr
\frac{V_{13}}{\sqrt{2}} & 0 & \frac{V_{13}}{\sqrt{2}}&
\frac{V_{23}}{2}&\frac{V_{11}+V_{33}}{2}&\frac{V_{12}}{2} \cr
 0 &\frac{V_{23}}{\sqrt{2}} & \frac{V_{23}}{\sqrt{2}}& \frac{V_{13}}{2}&\frac{V_{12}}{2}&\frac{V_{22}+V_{33}}{2}
\end{pmatrix} \eea
Matter Yukawas can be written by the same procedure as in the
$N_g$=2 case : \[ Y_u =
\begin{pmatrix} \hat h \hat V_1 + \hat f \hat V_7 & (\hat h \hat
V_4 + \hat f \hat V_{10})/\sqrt{2} & (\hat h \hat V_5 + \hat f
\hat V_{11})/\sqrt{2}  \cr (\hat h \hat V_4 + \hat f \hat
V_{10})/\sqrt{2} & \hat h \hat V_2 + \hat f \hat V_8 & (\hat h
\hat V_6 + \hat f \hat V_{12})/\sqrt{2}\cr (\hat h \hat V_5 + \hat
f \hat V_{11})/\sqrt{2} & (\hat h \hat V_6 + \hat f \hat
V_{12})/\sqrt{2} &  \hat h \hat V_3 + \hat f \hat V_9
\end{pmatrix} \] \be
 Y_d =\begin{pmatrix} \hat h \hat W_1 + \hat f \hat W_{13} & (\hat h
\hat W_4 + \hat f \hat W_{16})/\sqrt{2}  & (\hat h \hat W_5 + \hat
f \hat W_{17})/\sqrt{2} \cr (\hat h \hat W_4 + \hat f \hat
W_{16})/\sqrt{2} & \hat h \hat W_2 + \hat f \hat W_{14} & (\hat h
\hat W_6 + \hat f \hat W_{18})/\sqrt{2} \cr  (\hat h \hat W_5 +
\hat f \hat W_{17})/\sqrt{2} & (\hat h \hat W_6 + \hat f \hat
W_{18})/\sqrt{2} & \hat h \hat W_3 + \hat f \hat W_{15}
\end{pmatrix}
 \label{yukawas3d}\ee
To avoid pseudo-Goldstones, we start with the non-degenerate case
$Rank[\mathbf{\Xi}]= 5$.
\subsubsection{\textbf{A. $Rank[\mathbf{\Xi}]= 5$} }
Using Cramer's rule, we can solve  ${ \Xi \cdot\hat\Sigma =0} $
for five $\tilde\sigma$ variables in terms of undermined one (say
$\tilde\sigma_{23}$) \be
\begin{pmatrix} \Xi_5 & v \cr v^T & \tilde\chi_{2}+\tilde\chi_{3}
\end{pmatrix} \begin{pmatrix} \hat{\sigma} \cr \tilde\sigma_{23}
\end{pmatrix} = 0 \qquad  \Rightarrow \qquad
\hat{\sigma}=-(\Xi_5^{-1}v) \tilde\sigma_{23}\ee Here
$\hat{\sigma}$$=$ $(\tilde\sigma_{11}, \tilde\sigma_{22},
\tilde\sigma_{33}, \tilde\sigma_{12}, \tilde\sigma_{13})$, $\Xi_5$
and $v $ are upper left $5 \times 5$ block and $6^{th}$ column
(deleting the last element)  of $\mathbf{\Xi}$ respectively.
 We can
construct $\tilde\sigma$ from $\hat{v}=-(\Xi_5^{-1}v)$
\vspace{.2cm}\be \tilde\sigma= \begin{pmatrix} {\hat{v}}_1 &
{\hat{v}}_4 & {\hat{v}}_5 \cr {\hat{v}}_4 & {\hat{v}}_2 & 1 \cr
 {\hat{v}}_5 & 1 & {\hat{v}}_3\end{pmatrix} \tilde\sigma_{23} \label{sigexp}\ee
 Then
 \[ Det[\tilde\sigma]=  \frac{Det[\mathbf{\Xi}] N_5(\tilde\chi)}
 {D_5(\tilde\chi)}\tilde\sigma_{23}^3\]where \bea
N_5(\tilde\chi)=&& (\tilde\chi_{13}^2 \tilde\chi_2 - 2
\tilde\chi_{12} \tilde\chi_{13} \tilde\chi_{23} + \tilde\chi_1
\tilde\chi_{23}^2 + \tilde\chi_{12}^2 \tilde\chi_3 -
     \tilde\chi_1 \tilde\chi_2 \tilde\chi_3) (\tilde\chi_{12} \tilde\chi_{13}^2 + \tilde\chi_{13} \tilde\chi_2 \tilde\chi_{23}\nonumber\\&& - \tilde\chi_1 \tilde\chi_{12} \tilde\chi_3 -
     \tilde\chi_{12} \tilde\chi_2 \tilde\chi_3) (-\tilde\chi_{12}^2 \tilde\chi_{13} + \tilde\chi_1 \tilde\chi_{13} \tilde\chi_2 + \tilde\chi_{13} \tilde\chi_2 \tilde\chi_3 -
     \tilde\chi_{12} \tilde\chi_{23} \tilde\chi_3) \nonumber \eea \bea
D_5(\tilde\chi)=&&  (-\tilde\chi_1 \tilde\chi_{12}^2
\tilde\chi_{13}^2 + \tilde\chi_1^2 \tilde\chi_{13}^2 \tilde\chi_2
- \tilde\chi_{12}^2 \tilde\chi_{13}^2 \tilde\chi_2 +
    \tilde\chi_1 \tilde\chi_{13}^2 \tilde\chi_2^2 + \tilde\chi_1^2 \tilde\chi_{12}^2 \tilde\chi_3 \nonumber\\&& - \tilde\chi_{12}^2 \tilde\chi_{13}^2 \tilde\chi_3 - \tilde\chi_1^3 \tilde\chi_2 \tilde\chi_3 +
   \tilde\chi_1 \tilde\chi_{12}^2 \tilde\chi_2 \tilde\chi_3 + \tilde\chi_1 \tilde\chi_{13}^2 \tilde\chi_2 \tilde\chi_3 - \tilde\chi_1^2 \tilde\chi_2^2 \tilde\chi_3 + \tilde\chi_{13}^2 \tilde\chi_2^2 \tilde\chi_3 \nonumber\\&&-
   2 \tilde\chi_{12} \tilde\chi_{13} \tilde\chi_2 \tilde\chi_{23} \tilde\chi_3 + \tilde\chi_1 \tilde\chi_2 \tilde\chi_{23}^2 \tilde\chi_3 + \tilde\chi_1 \tilde\chi_{12}^2 \tilde\chi_3^2 -
   \tilde\chi_1^2 \tilde\chi_2 \tilde\chi_3^2 + \tilde\chi_{12}^2 \tilde\chi_2 \tilde\chi_3^2\nonumber\\&& - \tilde\chi_1 \tilde\chi_2^2 \tilde\chi_3^2  )^3  \eea
Thus \bea  \quad Det[\tilde\sigma] \sim Det[\mathbf{\Xi}]
\Rightarrow Det[\tilde\sigma]=0=Det[M_{\nu^c}] \eea It implies the
existence of one or more light sterile neutrino depending upon the
zero eigenvalues of $\tilde\sigma$ VEV. We proceed by solving the
$ Det[\mathbf{\Xi}]=0$ condition for $\tilde\chi_{1}$ : \be
\vspace{.2cm} \tilde\chi_1 = \frac{(\tilde\chi_{13}^2 \tilde\chi_2
-
  2 \tilde\chi_{12} \tilde\chi_{13} \tilde\chi_{23} + \tilde\chi_{12}^2 \tilde\chi_3)}{(\tilde\chi_2
\tilde\chi_3)-\tilde\chi_{23}^2  } \vspace{.2cm} \ee Like $N_g$=2
case, we solve for the undetermined variable $\tilde\sigma_{23}$
from one of  the equations of Eq. (\ref{dl10}) and P using Eq.
(\ref{dl7}). In the search program the remaining equations are
used to solve for  A and W. Notice the factor
\[(\tilde\chi_1 \tilde\chi_{2} \tilde\chi_{3}-\tilde\chi_1
   \tilde\chi_{23}^2-\tilde\chi_{12}^2 \tilde\chi_3+2
   \tilde\chi_{12} \tilde\chi_{13} \tilde\chi_{23}-\tilde\chi_{13}^2 \tilde\chi_{2})\]
   of $ Det[\mathbf{\Xi}]$ occurs twice in $Det[\tilde\sigma]$. We used this factor to acheive vanishing $ Det[\mathbf{\Xi}]$ in our numerical
   search program. So $\tilde\sigma$ VEVs (see Appendix A) have two zero eigenvalues. We therefore need to
    integrate out only one heavy right handed neutrino. Leptonic superpotential is : \be
W_{lep}={\bar \nu}^T_{A} Y^\nu_{AB}\nu_B+\frac{1}{2}{\bar
\nu}^T_{A}M^{\bar{\nu}}_{AB}{\bar\nu}_{B} \ee Using superpotential
equation of motion :
\begin{table}
 $$
 \begin{array}{cccc}
 \hline  \hline &&&\vspace{-.3 cm}\\
   {\rm  S.No.  }&  {\rm M_H}& {\rm  Y_u}&  {\rm Y_d} \vspace{0.1 cm}\\
  \hline  \hline
1.&2.55 + 0.13 i &0.007, 0.019, 0.368 &0.007, 0.014, 0.306  \\
2.&1.44 - 0.61 i &0.027, 0.13, 0.409 &0.009, 0.083, 0.242  \\
3.&1.28 + 0.75 i &0.063, 0.228, 0.424 &0.019, 0.083, 0.186  \\
4.&1.16 + 0.67 i & 0.062, 0.193, 0.439 & 0.02, 0.099, 0.188  \\
5.&1.06 - 0.73 i & 0.009, 0.076, 0.458 &0.008, 0.078, 0.321   \\
6.&0.02 - 0.03 i &0.022, 0.254, 0.604 & 0.009, 0.104, 0.289 \\
 \hline \hline &&&\vspace{-.3 cm}\\
   {\rm  S.No.  }&{\rm Y_l}&{\rm Y_{\nu}}&{\rm \{\theta_{13}, \theta_{12}, \theta_{23}\}^Q }\\&&&(deg.)  \vspace{0.1 cm}\\
 \hline  \hline
1.&.007, 0.026, 0.421 &0.014, 0.032, 0.533 &0.56, 13.18, 1.58  \\
2.&.023, 0.094, 0.314 &0.018, 0.213, 0.566&3.42, 8.71, 3.87  \\
3.&.031, 0.103, 0.212 &0.015, 0.187, 0.624 &6.65, 6.59, 1.11  \\
4.&.029, 0.094, 0.259 &0.018, 0.283, 0.42 &2.6, 5.18, 1.96  \\
5.&.009, 0.073, 0.4 &0.008, 0.214, 0.558 &1.51, 11.19, 4.61 \\
6.&.007, 0.148, 0.338 &0.01, 0.159, 0.608 & 1.04, 1.57, 6.03  \\
\hline \hline &&&\vspace{-.3 cm}\\
 {\rm  S.No.  }&\multicolumn{2}{c} {{\rm m_\nu(eV)}}&{\rm {M}_{\nu^c}} \vspace{0.1 cm}\\
 \hline  \hline
1.&\multicolumn{2}{c} {0.16, 0.1597, 0.0104, 0.0104,1.7 \times 10^{-6} }  &365.07,0,0 \\
2.& \multicolumn{2}{c} {0.2056, 0.2054, 0.05, 0.0499, 3.8 \times 10^{-6}} & 365.07,0,0 \\
3.&\multicolumn{2}{c} {0.3021, 0.302, 0.0781, 0.0781, 4.4 \times 10^{-7} }& 365.07,0,0 \\
4.&\multicolumn{2}{c} {0.1806, 0.1805, 0.1254, 0.1254, 7.8\times 10^{-7}} & 365.07,0,0 \\
5.&\multicolumn{2}{c} {0.1946, 0.1945, 0.0533, 0.0532 ,7.2\times 10^{-7} } & 365.07,0,0 \\
6.&\multicolumn{2}{c} {0.2837, 0.2836,
0.0129, 0.0128, 6.4\times 10^{-6}  }  & 365.07,0,0 \\
\hline  \hline
   \end{array}$$
   \caption{\small{Yukawa eigenvalues and mixing angles for $N_g$=3 ($Rank[\mathbf{\Xi}]=5$), $f$=0.9+0.7 i. $\lambda$ =.48+.3i , $\eta$ =.25  ,
    $h$= 1.3, $\gamma$ =.05, $\bar\gamma$=-1.2, $\xi$=3.645+.363i . $M_{\nu^c}$ is independent of $M_H$ value chosen.\label{table:rank5}}}
\vspace{.3cm} \end{table}
 \be
 {\bar\nu}_{3}=-\frac{Y^\nu_{3A}\nu_A}{M_{33}^{\bar{\nu}}} \ee
 In a right handed neutrino diagonal basis, the effective superpotential reads :
\be W_{eff}={\bar\nu}^T_{a}
Y^\nu_{aB}\nu_B+\frac{1}{2}{\bar\nu}^T_{a}M_{aa}{\bar{\nu}}_{a}
-\nu_A^T\biggr(\frac{1}{2}\frac{Y^\nu_{3A} Y^\nu_{3B}
 }{M_{33}^{\bar{\nu}}}\biggr)\nu_B
\ee We can write a dimension five ($\kappa$ ) operator for three
left handed neutrinos: \be \kappa_{AB}=-  (Y^{
\nu})^T_{A3}M_{33}^{-1}(Y^{ \nu})_{3B} \ee The light sterile
neutrino will get Dirac mass only, so the mass matrix is given by
: \be M_{light} =\frac{1}{2}
\begin{pmatrix} \kappa_{11} & \kappa_{12} & \kappa_{13} &
Y^{\nu}_{11}& Y^{\nu}_{21}\cr \kappa_{21} & \kappa_{22} &
\kappa_{23} & Y^{\nu}_{12}& Y^{\nu}_{22} \cr \kappa_{31} &
\kappa_{32} & \kappa_{33} & Y^{\nu}_{13}& Y^{\nu}_{23} \cr
Y^{\nu}_{11}& Y^{\nu}_{12} &Y^{\nu}_{13} & 0& 0\cr Y^{\nu}_{21}&
Y^{\nu}_{22} &Y^{\nu}_{23} & 0& 0 \end{pmatrix} \ee Using the
above solution and random superpotential parameters, we
 have calculated the Yukawa structure and neutrino masses for all
 the $M_H$ values as shown in Table \ref{table:rank5}. Notice that in
 this case neutrino masses are larger comparative to the earlier
 case due to the mixing of Dirac coupling. Superheavy spectrum is
 shown in Tables \ref{tracefullrank5spec} and \ref{htrank5-6dim}, it also exhibits
 pseudo-Goldstones in the $G[1,1,0]$ sector.
\subsubsection{\textbf{B. $Rank[\mathbf{\Xi}]= 4$}}
In this case, one can determine 4 $\tilde\sigma$ variables, out of
a total of six. We have additional conditions for vanishing $5
\times 5 $ minors of $\mathbf{\Xi}$ along
 with $Det[\mathbf{\Xi}]$.
By calculating $\tilde\sigma_{11}$, $\tilde\sigma_{22}$,
$\tilde\sigma_{33}$, $\tilde\sigma_{12}$ in
 terms of ($\tilde\sigma_{13}$, $\tilde\sigma_{23}$), we can construct $\tilde\sigma$ as we have discussed earlier for $Rank[\mathbf{\Xi}]= 5$ :
\be \tilde\sigma= A \tilde\sigma_{13}+ B \tilde\sigma_{23} \ee
where the matrices A and B   are functions of the  $\tilde\chi$
elements. Now we can't factorize $Det[\tilde\sigma]$ separating
$\tilde\chi$ elements and $\tilde\sigma_{13}$, $\tilde\sigma_{23}$
as in the previous case. So none of the $Det[\tilde\sigma]$
factors is common with $Det[\mathbf{\Xi}]$. Even in the special
case: $\tilde\sigma_{13}=\tilde\sigma_{23}$
   $Det[\tilde\sigma]$ factors are different from that of $Det[\mathbf{\Xi}]$. Therefore,
  $Rank[\mathbf{\Xi}]=4$  could be a workable scenario with Type I seesaw neutrino masses
   and without light sterile neutrinos. Vanishing of the common factor of $5 \times 5 $ minors
    results in a complicated system. For convenience, we choose two factors to vanish which results in null
     dimension 5 minors. Thus, we get three conditions, one from $Det[\mathbf{\Xi}]$=0  and following two from $5 \times 5 $
     minors:
\begin{table}
 $$
 \begin{array}{ccccc}
 \hline\hline&&&&\vspace{-.3 cm}\\
   {\rm  {\small{S.No.}}  }&  {\rm M_H}& {\rm  Y_u}&  {\rm Y_d} &{\rm {\small{\{\theta_{13}, \theta_{12}, \theta_{23}\}^Q}} }\\
   &&&&(deg.) \vspace{0.1 cm} \\
 \hline\hline
1.&-4.323+1.47i&.0007,.0021,.0215&.001,.0019,.0219&9.,15.9,15.6 \\
2.&.465+3.382i&.0018,.0148,.0182&.0020,.0197,.0222&11.3,1.5,4.9  \\
3.&.76-2.193i&.0029,.0113,.0137&.0054,.0233,.0385&1.5,6.2,7.4 \\
4.&-0.002+0.968i&.0105,.040,.077&.0035,.0174,.0408&5.2,3.7,2.8\\
5.&-.508-.209i&.0077,.053,.1126&.0019,.0159,.0381&1.1,12.1,1.4\\
6.&-.092-.032i&.0041,.0467,.0558&.0035,.0413,.0522&8.7,5.5,2.6\\
 \hline\hline&&&&\vspace{-.3 cm}\\
   {\rm  {\small{S.No.}}  }&{\rm Y_l}&{\rm Y_{\nu}}&{\rm \{\theta_{13}, \theta_{12}, \theta_{23}\}^L }&\tilde{M}_{\nu^c}\\
   &&&(deg.)& \vspace{0.1 cm} \\
 \hline\hline
1.&.0013,.0041,.0517&.0023,.0064,.0468&3.6,20.3,23.3&5.9,5.3,1.5\\
2.&.0034,.0148,.0205&.0032,.0126,.0162&27.5,14.5,47.0&5.9,5.3,1.5\\
3.&.0053,.0121,.0458&.0033,.0102,.020&13.6,11.1,41.1&5.9,5.3,1.5\\
4.&.0048,.0174,.0473&.0092,.0181,.0915&23.9,14.5,17.7&5.9,5.3,1.5\\
5.&.0042,.0224,.0382&.0061,.0584,.0835&23.7,26.1,49.4&5.9,5.3,1.5\\
6.&.0043,.0497,.0621&.0049,.0355,.0518&14.1,37.6,46.3&5.9,5.3,1.5\\
\hline\hline &&&&\vspace{-.3 cm}\\
{\rm  {\small{S.No.}}}&\multicolumn{2}{c} {{\rm
{m_\nu/10^{-4}(meV)}}}&\multicolumn{2}{c} {{\rm {\Delta
{\rm m_\nu^2/10^{-13}(eV^2)} }}}\vspace{0.1 cm}\\
 \hline \hline1.& \multicolumn{2}{c} {0.039 ,0.187,40.543} &\multicolumn{2}{c} {.0033,164.372}\\
2.& \multicolumn{2}{c} {0.079,0.807 ,4.09 }&\multicolumn{2}{c} {0.0645,1.6074}\\
3.& \multicolumn{2}{c} {0.12 ,0.32 ,6.946} &\multicolumn{2}{c} {0.0088,4.8143}\\
4.&\multicolumn{2}{c} {1.33 ,4.503 ,23.439}&\multicolumn{2}{c} {1.851,52.9128}\\
5.&\multicolumn{2}{c} {0.926 ,17.046,34.448}&\multicolumn{2}{c} {28.9722,89.6078} \\
6.&\multicolumn{2}{c} {1.03,4.96,9.806 }&\multicolumn{2}{c} {2.3544,7.1562}\\
 \hline\hline
\end{array}$$
   \caption{\small{Yukawa eigenvalues and mixing angles for $N_g$=3 ($Rank[\mathbf{\Xi}]=4$), $f$= -0.11  + .02 i . $\tilde{M}_{\nu^c}\equiv
   \lambda  {M}_{\nu^c}/m$. $m/\lambda $ is taken to be $10^{16}$ GeV
  to estimate $\Delta m_\nu^2 $. $\lambda$= 0.48 - .05  i, $\eta$= -.18, $h$= .26, $\gamma$= 0.12, $\bar\gamma$ =
  -1.44, $\xi$= 1.7278  - 0.1734i \label{table:tracefulldata}}}
 \end{table}
\[  \tilde\chi_{13}\tilde\chi_{23}-\tilde\chi_{12} \tilde\chi_{3}=0
\] \be
  \tilde\chi_{12}^2 \tilde\chi_{13} + \tilde\chi_{12} \tilde\chi_{2} \tilde\chi_{23} + \tilde\chi_{13}
  \tilde\chi_{23}^2=0\ee
 We solve  these equations for  $\tilde\chi_2$ and $\tilde\chi_3$
\bea \tilde\chi_1
=\frac{\tilde\chi_{12}\tilde\chi_{13}}{\tilde\chi_{23}} \quad ;
\quad \tilde\chi_2
=\frac{-\tilde\chi_{13}(\tilde\chi_{12}^2+\tilde\chi_{23}^2)}{\tilde\chi_{12}\tilde\chi_{23}}
\quad ; \quad \tilde\chi_3
=\frac{\tilde\chi_{13}\tilde\chi_{23}}{\tilde\chi_{12}}  \eea
Besides Eqns. (\ref{dl7},\ref{dl8},\ref{dl10}) we have seven
equations (4 $\tilde\sigma$ equations and 3 above conditions). For
consistency we fix $\xi$ parameter using one extra condition.
\begin{table}
 $$
 \begin{array}{ccccc}
 \hline \hline&&&&\vspace{-.3 cm}\\
   {\rm  S.No.  }&  {\rm M_H}& {\rm  Y_u}&  {\rm Y_d} &\rm {\{\theta_{13}, \theta_{12}, \theta_{23}\}^Q}\\
   &&&& (deg.)\vspace{0.1 cm}\\
 \hline \hline
1.&-4.323+1.47i&0.001,0.003,0.026&0.001,0.003,0.027&0.002,0.008,0.015\\
2.&.465+3.382i&0.002,0.013,0.015&0.002,0.015,0.017&0.014,0.002,0.003\\
3.&.76-2.193i&0.003,0.01,0.014&0.005,0.019,0.028&0.002,0.008,0.013\\
4.&-0.002+0.968i&0.007,0.033,0.079&0.004,0.017,0.042&0.005,0.004,0.002\\
5.&-.508-.209i&0.006,0.047,0.105&0.002,0.017,0.038&0.001,0.013,0.002\\
6.&-.092-.032i&0.003,0.044,0.054&0.003,0.043,0.053&0.009,0.006,0.004\\
\hline \hline &&&&\vspace{-.3 cm}\\
  {\rm  S.No.  }&{\rm Y_l}&{\rm Y_{\nu}}&{\rm \{\theta_{13}, \theta_{12}, \theta_{23}\}^L}&\tilde{M}_{\nu^c}\\
  &&& (deg.)&\vspace{0.1 cm}\\
  \hline \hline
1.&0.001,0.003,0.027&0.001,0.003,0.026&0.27,9.82,2.13&0.006,0.005,0.001\\
2.&0.002,0.015,0.017&0.002,0.013,0.015&2.41,22.56,37.83&0.006,0.005,0.001\\
3.&0.005,0.019,0.028&0.003,0.01,0.014&2.97,28.08,18.45&0.006,0.005,0.001 \\
4.&0.004,0.017,0.042&0.007,0.033,0.079&5.81,7.17,25.52&0.006,0.005,0.001\\
5.&0.002,0.017,0.038&0.006,0.047,0.105&2.7,6.33,54.28&0.006,0.005,0.001 \\
6.&0.003,0.043,0.053&0.003,0.044,0.054&3.38,7.5,58.18&0.006,0.005,0.001\\
\hline \hline &&&&\vspace{-.3 cm}\\
{\rm  S.No.  }&\multicolumn{2}{c} {\rm m_\nu(meV)}
&\multicolumn{2}{c}{ \rm{\Delta m_\nu^2/10^{-5}
(eV^2)}}\vspace{0.1 cm}\\\hline \hline
1.&\multicolumn{2}{c} {0.0006,0.0047,1.2427}&\multicolumn{2}{c} {2.16\times 10^{-6},0.15}\\
2.&\multicolumn{2}{c} {0.0056,0.0284,0.3286}&\multicolumn{2}{c} {7.74\times 10^{-5},0.011}\\
3.&\multicolumn{2}{c} {0.0096,0.0256,0.3686}&\multicolumn{2}{c} {5.64 \times 10^{-5},0.013}\\
4.&\multicolumn{2}{c} {0.0652,0.8982,3.3757}&\multicolumn{2}{c} {0.080 ,1.06} \\
5.&\multicolumn{2}{c} {0.0701,2.2717,3.2128}&\multicolumn{2}{c} {0.516 ,0.516}\\
6.&\multicolumn{2}{c} {0.025,0.7852,1.4609}&\multicolumn{2}{c}
{0.062,0.152}\\  \hline \hline
 \end{array}$$
   \caption{\small{Yukawa eigenvalues and mixing angles for $N_g$=3 ($Rank[\mathbf{\Xi}]=4$) and the same parameter values as in Table \ref{table:tracefulldata} but
  with much smaller $f$= -0.00011  + .00002 i. $\tilde{M}_{\nu^c}\equiv
   \lambda  {M}_{\nu^c}/m$. $m/\lambda $ is taken to be $10^{16}$ GeV
  to estimate $\Delta m_\nu^2 $.\label{table:tracefulldata-fsmall} }}
 \end{table}

We calculate  Yukawa eigenvalues (Eq. \ref{yukawas3d}) along with
quark and lepton mixing angles for large and small values of $f$
and a random illustrative set of superpotential parameters. The
results are given as Tables \ref{table:tracefulldata} and
\ref{table:tracefulldata-fsmall} and superheavy spectrum in Tables
\ref{spec6dimrank4} and \ref{htrank4-6dim}.
\subsubsection{\textbf{C. Using symmetric irrep of O(3)}}
Another alternative that we have investigated is considering only
the 5-dimensional irrep of O(3). We write the traceless symmetric
$3 \times 3 $ representation as
\[ \hat \phi_{AB}= \phi_{11} \frac{(\lambda_3)_{AB}}{\sqrt{2}} +\phi_{22} \frac{(\lambda_8)_{AB}}
{\sqrt{2}}+ \frac{\phi_{KL}}{\sqrt{2}} \delta^K_{(A} \delta^L_{B)}
\] Here $\lambda_3$ and $\lambda_8$ are the usual  diagonal $3\times 3$ Gell-Mann
matrices. Matrix $\mathbf{\Xi}$ is given by \be
\mathbf{\Xi}=\begin{pmatrix} -\tilde\chi_{22} + 2 \xi &
-\tilde\chi_{11} - \tilde\chi_{22} + \xi & \tilde\chi_{12} &
  0 & -\tilde\chi_{23} \cr -\tilde\chi_{11} - \tilde\chi_{22} + \xi & -\tilde\chi_{11} + 2 \xi & \tilde\chi_{12} & -\tilde\chi_{13} &
  0 \cr \frac{\tilde\chi_{12}}{2} & \frac{\tilde\chi_{12}}{2} & \frac{(\tilde\chi_{11} + \tilde\chi_{22} + 2 \xi)}{2} & \frac{\tilde\chi_{23}}{2} & \frac{\tilde\chi_{13}}{
  2} \cr 0 & -\frac{\tilde\chi_{13}}{2}& \frac{\tilde\chi_{23}}{2} &  \frac{-\tilde\chi_{22} + 2 \xi}{2}& \frac{\tilde\chi_{12}}{2} \cr -\frac{\tilde\chi_{23}}{2} &
   0 & \frac{\tilde\chi_{13}}{2} & \frac{\tilde\chi_{12}}{2} &  \frac{-\tilde\chi_{11} + 2 \xi}{2} \end{pmatrix} \ee
$\hat\Sigma=\{\tilde\sigma_{11},\tilde\sigma_{22},
\tilde\sigma_{12}, \tilde\sigma_{13}, \tilde\sigma_{23}\}$.  In
the present scenario matrix function $\Omega'_3[V]$ has the
following form : \be \Omega'_3[V] =
\begin{pmatrix} \frac{V_{11}+V_{22}}{2} & \frac{V_{11}-V_{22}}{2 \sqrt{3}} & 0 & \frac{V_{13}}{2} & -\frac{V_{23}}{2} \cr
\frac{V_{11}-V_{22}}{2 \sqrt{3}} & \frac{-V_{11}-V_{22}}{2} &
\frac{V_{12}}{\sqrt{3}} & -\frac{V_{13}}{2\sqrt{3}} &
-\frac{V_{23}}{2\sqrt{3}}\cr 0& \frac{V_{12}}{\sqrt{3}} &
\frac{V_{11}+V_{22}}{2} &\frac{V_{23}}{2} & \frac{V_{13}}{2}\cr
\frac{V_{13}}{2}&-\frac{V_{13}}{2\sqrt{3}} &\frac{V_{23}}{2}&
-\frac{V_{22}}{2} &\frac{V_{12}}{2} \cr -\frac{V_{23}}{2} &
-\frac{V_{23}}{2\sqrt{3}}&\frac{V_{13}}{2}&\frac{V_{12}}{2}&-\frac{V_{11}}{2}
\end{pmatrix}
 \ee Higgs mass matrix can be obtained by using $\Omega'_3$ and
rows and columns are labelled by $\{\frac{(\bar H_{11}-\bar H
_{22})}{\sqrt{2}}, \sqrt{\frac{3}{2}}({\bar H}_{11}+{\bar
H}_{22}),\sqrt{2} \bar H_{12},\sqrt{2} \bar H_{13},\sqrt{2} \bar
H_{23}\}$ and $\{\frac{( H_{11}- H _{22})}{\sqrt{2}},
\sqrt{\frac{3}{2}}({ H}_{11}+{ H}_{22}),\sqrt{2}  H_{12},\sqrt{2}
H_{13},\sqrt{2}  H_{23}\}$. Up and down quark Yukawas are given as
: {\footnotesize{\[ Y_u =
\begin{pmatrix} \hat h (\frac{\hat V_1}{\sqrt{2}}+\frac{\hat V_2}{\sqrt{6}} )+ \hat f( \frac{\hat V_6}{\sqrt{2}}+\frac{\hat V_7}{\sqrt{6}}) &
 \hat h \frac{\hat
V_3}{\sqrt{2}} + \hat f \frac{\hat V_8}{\sqrt{2}} &  \hat h
\frac{\hat V_4}{\sqrt{2}} + \hat f \frac{\hat V_9}{\sqrt{2}}   \cr
\hat h \frac{\hat V_3}{\sqrt{2}} + \hat f \frac{\hat
V_8}{\sqrt{2}} & \hat h (-\frac{\hat V_1}{\sqrt{2}}+\frac{\hat
V_2}{\sqrt{6}} )+ \hat f( -\frac{\hat V_6}{\sqrt{2}}+\frac{\hat
V_7}{\sqrt{6}}) &\hat h \frac{\hat V_5}{\sqrt{2}} + \hat f
\frac{\hat V_{10}}{\sqrt{2}}  \cr \hat h \frac{\hat V_4}{\sqrt{2}}
+ \hat f \frac{\hat V_9}{\sqrt{2}}    & \hat h \frac{\hat
V_5}{\sqrt{2}} + \hat f \frac{\hat V_{10}}{\sqrt{2}} & -2 \hat h
\frac{ \hat V_2}{\sqrt{6}} -2 \hat f \frac{ \hat V_7 }{\sqrt{6}}
\end{pmatrix} \] }} {\footnotesize{ \be
Y_d = \begin{pmatrix} \hat h (\frac{\hat W_1}{\sqrt{2}}+\frac{\hat
W_2}{\sqrt{6}} )+ \hat f( \frac{\hat W_{11}}{\sqrt{2}}+\frac{\hat
W_{12}}{\sqrt{6}}) &
 \hat h \frac{\hat
W_3}{\sqrt{2}} + \hat f \frac{\hat W_{13}}{\sqrt{2}} &  \hat h
\frac{\hat W_4}{\sqrt{2}} + \hat f \frac{\hat W_{14}}{\sqrt{2}}
\cr \hat h \frac{\hat W_3}{\sqrt{2}} + \hat f \frac{\hat
W_{13}}{\sqrt{2}} & \hat h (\frac{\hat W_2}{\sqrt{6}}-\frac{\hat
W_1}{\sqrt{2}} )+
 \hat f( \frac{\hat W_{12}}{\sqrt{6}}-\frac{\hat W_{11}}{\sqrt{2}}) &\hat h \frac{\hat
W_5}{\sqrt{2}} + \hat f \frac{\hat W_{15}}{\sqrt{2}}  \cr \hat h
\frac{\hat W_4}{\sqrt{2}} + \hat f \frac{\hat W_{14}}{\sqrt{2}} &
\hat h \frac{\hat W_5}{\sqrt{2}} + \hat f \frac{\hat
W_{15}}{\sqrt{2}} & -2 \hat h \frac{ \hat W_2}{\sqrt{6}} -2 \hat f
\frac{ \hat W_{12} }{\sqrt{6}}\end{pmatrix}
 \label{5dimyuk}\ee}}

\begin{table}
 $$
 \begin{array}{ccccc}
 \hline \hline&&&&\vspace{-.3 cm}\\
   {\rm  S.No.  }&  {\rm M_H}& {\rm  Y_u}&  {\rm Y_d} &\rm {\{\theta_{13}, \theta_{12}, \theta_{23}\}^Q}\\
     &&&& (deg.)\vspace{0.1 cm}\\
 \hline \hline
1.&135.29+11.98  i&.054,.068,.1718& .0001,.0012,.0019&8.23,34.35,32.04\\
2.&25.4+1.72  i& 0.0,.0588,.0938&0.0,.0021,.0034&(.1,1.7,1.2)\times 10^{-6}\\
3.&24.6+1.16  i& 0.0,.0614,.1325&0.0,.0016,.0036&8.12\times 10^{-8},0,0\\
4.&18.41+1.4  i&.001,.0555,.125&.0035,.011,.0264&3.48,6.01,8.67\\
5.&18.32+1.23  i&0.0,.0584,.1322&0.0,.0112,.0255&4.86\times
10^{-9},0,0\\ \hline \hline &&&&\vspace{-.3 cm}\\
  {\rm  S.No.  }&{\rm Y_l}&{\rm Y_{\nu}}&\rm {\{\theta_{13}, \theta_{12}, \theta_{23}\}^L}&\tilde{M}_{\nu^c}\\
    &&& (deg.)&\vspace{0.1 cm}\\
 \hline \hline
1.&.0029,.0066,.013&.174,.2183,.551&5.92,32.15,9.6&22.39,8.89,7.05\\
2.&0.0,.0177,.0283&0.0,.2272,.3629&(.11,8.5)\times 10^{-7},28.7&22.39,8.89,7.05\\
3.&0.0,.0068,.0147& 0.0,.236,.5094&(.03,1.2)\times 10^{-6},19.4&22.39,8.89,7.05\\
4.&.0098,.0136,.038&.0028,.122,.275&14.33,20.53,35.13&22.39,8.89,7.05\\
5.&0.0,.015,.034& 0.0,.1284,.291&5.7\times 10^{-9},0,42.03&22.39,8.89,7.05\\
  \hline \hline &&&&\vspace{-.3 cm}\\
  {\rm  S.No.  } &\multicolumn{2}{c}{\rm {m_\nu(meV)}}&\multicolumn{2}{c} {\Delta m_\nu^2(eV^2)} \vspace{0.1 cm}\\\hline \hline
  1.&\multicolumn{2}{c} {(1.29 ,1.62 ,4.1)\times 10^{-2}} &\multicolumn{2}{c} {(.096,1.42)\times 10^{-9} }\\
  2.&\multicolumn{2}{c} {6.5\times 10^{-11},.012,.028}&\multicolumn{2}{c} {(1.49,6.18)\times 10^{-10}}\\
  3.&\multicolumn{2}{c} {2.0\times 10^{-10},.017,.045}&\multicolumn{2}{c} {(2.75,17.9)\times 10^{-10}}\\
  4.&\multicolumn{2}{c} {1.5\times 10^{-6},.010,.011}&\multicolumn{2}{c} {(1.08,.23)\times 10^{-10}}\\
  5.&\multicolumn{2}{c} {1.1\times 10^{-10},.012,.013}&\multicolumn{2}{c} {(1.33,.29)\times 10^{-10}}\\ \hline \hline
   \end{array}$$
   \caption{\small{Yukawa eigenvalues and mixing angles for
  traceless case
  $N_g$=3 ($Rank[\mathbf{\Xi}]=4$), $f$ = 0.23 +.04 i . $\tilde{M}_{\nu^c}\equiv
   \lambda  {M}_{\nu^c}/m$. $m/\lambda $ is taken to be $10^{16}$ GeV
  to estimate $\Delta m_\nu^2 $. $\lambda$= 0.18 - .03i,
$\eta$ = .034, $\gamma$ = -0.53, $\bar\gamma$ = -2.60, $h$ = .14,
$\xi$= 7.677  + 0.15772i. $M_{\nu^c}$ is independent of $M_H$
value chosen. \label{table:tracelessdata}}} \vspace{.2cm}
\end{table} \vspace{.2cm} We have solved the least degenerate
($Rank[\mathbf{\Xi}]=$ 4) case. This option has relatively fewer
parameters so it is easier to perform numerical searches. We
solved some equations analytically and remaining numerically to
find solution. Yukawa eigenvalues and mixing angles are given in
Table \ref{table:tracelessdata}, superheavy spectrum in Tables
\ref{5dimspec} and \ref{htrank4-5dim}.
\section{Discussion and Outlook}
We proposed \cite{Aulakhkhosa} dynamical generation of flavour
based upon the Susy SO(10) and family gauge group. In literature
O(3) family symmetry with traceful representation is considered
for non-renormalizable and non-GUT Yukawa-on models \cite{koide},
however our model is renormalizable and GUT based. Yukawon fields
break flavour and GUT symmetry spontaneously. Emergence of light
Higgs of the effective MSSM among the large number of YUMGUT MSSM
type Higgs multiplets is ensured by the consistency condition
$Det(\mathcal{H})$=0. SM fermion and neutrino Yukawa couplings are
generated by the VEV of the Yukawon field. SO(10) Yukawa couplings
are just single complex number thus parameter reduction is one of
the main virtue of YUMGUTs. Consistent SSB is achieved with the
introduction of ($O(N_g)$ symmetric- two field $S$, $\phi$) BM
(hidden sector) superpotential. $O(N_g)$ singlet $(S_s)$ breaks
Susy and traceless part $\hat{S}$ is fixed against visible sector
fields contribution to $O(N_g)$ D terms and thus facilitates
YUMGUTs.

We have analyzed the toy model ($N_g$=2) and realistic case
($N_g$=3) without any optimization. As explained earlier the rank
of the coefficient matrix $\mathbf{\Xi}$ of the
$F_{\sigma,\bar{\sigma}}$=0 equation is crucial for determining
the SSB. For $N_g$=2 $Rank[\mathbf{\Xi}]<$ 2 $\Rightarrow$
$Rank[\mathbf{\Xi}]=0$, so to avoid problematic pseudo-Goldstone
non-degenerate $Rank[\mathbf{\Xi}]=2$ is the only viable scenario.
Using random set of superpotential parameters we find \emph{Yukawa
eigenvalues different by a factor of about 10, small quark and
large lepton mixing angle}.

 In the realistic case $(N_g=3)$ we have
considered several possibilties like $Rank[\mathbf{\Xi}]=$ 5,
$Rank[\mathbf{\Xi}]=$ 4. If we consider the reducible
6-dimensional symmetric representation of O(3) with equal
superpotential couplings for traceless and singlet part then
non-degenerate case ($Rank[\mathbf{\Xi}]=5$) give rise to light
sterile neutrino. This motivates reconsideration of the no-go
\cite{blmdm} in the MSGUT using light sterile  neutrinos. Rank
reduction of homogeneous system provides a possible route to find
non-zero eigenvalues of $\sigma(\bar\sigma)$ VEV. We have also
studied the case using traceless -dimensional representation. We
can't use traceless representation for $N_g$=2 because cubic
invariants in the superpotential do not contribute. We have
calculated the complete superheavy spectrum for all the cases
considered (given in Appendix B) to check the existence of
pseudo-Goldstones which may be present when there is Higgs
duplication. Spectra do not contain pseudo-Goldstone except the SM
singlet G[1,1,0] sector which does not affect unification.

In all the cases studied acceptable Yukawa hierarchy and mixing is
achieved but neutrino masses generated are too small. Type I
contribution can be raised by suppressing $f$ which provides
unacceptable Yukawa structure. We have not considered the
contribution of Type II seesaw generated by the VEV of symmetric
multiplet $O^-$. Although we expect it to be small as compared to
Type I as in MSGUTs, but some special points may yield significant
contribution. Addition of $\bf{120}$-plet, which along with
$\bf{10}$-plet is mainly responsible for generating charged
fermion masses, is another way to get neutrino masses in
experimentally measured range. The $Rank[\mathbf{\Xi}]=5$ case
(considering 6 dim symmetric representation)
 phenomenology needs to be investigated because this provides sterile neutrino. With optimization,
one can expect to find the flavour blind parameters of YUMGUT
which can produce actual MSSM Yukawas. To completely demonstrate
this idea we need to  produce realistic SM mass mixing data
respecting NMSGUT fitting features which will require a huge
computational effort.

A number of experimental signals such as light moduli fields and
singlet pseudo-Goldstones \cite{BMsugry} which can also be DM
candidates are associated with our proposal. These fields also
cause cosmological problems. Thus our work has laid the basis for
an extensive program of future studies in unification and
cosmology.
\newpage\section*{Appendix A : YUMGUT VEVs } The values of
the VEVs of the YUMGUT Higgs fields responsible for breaking
$SO(10)\rightarrow \rm{MSSM} $ in units of $m/\lambda\sim 10^{16}$
GeV are :
\subsection*{1. $N_g$=2, $Rank[\mathbf{\Xi}]=$2} \be
 W= 
 \ee
\bea \bar D_X &=& 2(|p_+|^2-|p_-|^2
+3(|a_+|^2-|a_-|^2)+6(|w_+|^2-|w_-|^2) \nonumber\\&&+
{\frac{1}{2}}|\sigma_+|^2-|\sigma_-|^2 +
{\frac{1}{2}}|\bar{\sigma}_+|^2-|\bar{\sigma}_-|^2) =-8.94\eea
\subsection*{2. $N_g$=3}
VEVs are written in prime basis where D-terms point in third
direction.
\subsubsection{A. $Rank[\mathbf{\Xi}]=$5 (tracefull symmetric representation)}
\be \tilde\sigma'=\tilde{\bar\sigma}' = %
\ee \vspace{.3cm}
\be {D'}_{X}^a = 3319.0 \delta^a_3  \ee

\subsubsection{B. $Rank[\mathbf{\Xi}]=$4 (tracefull symmetric representation) }
\vspace{.2cm} \be \tilde\sigma'=\tilde{\bar\sigma}'  =
%
 \ee\vspace{.3cm}
\be {D'}_{X}^a = 47.04 \delta^a_3  \ee
\subsubsection*{C. $Rank[\mathbf{\Xi}]=$4 (traceless symmetric
representation) } \vspace{.3cm} \be
\tilde\sigma'=\tilde{\bar\sigma}' =%
  \ee\vspace{.3cm}
\be {D'}_{X}^a = 275.15 \delta^a_3  \ee \vspace{.4cm}
\section*{Appendix B : Superheavy Spectra}
\vspace{.3cm} We present here superheavy spectra (in units of the
MSGUT scale parameter $m/\lambda$) in the toy model ($N_g=2$)  and
realistic case ($N_g=3$):
\begin{table}\vspace{-.4cm}
 $$
 \begin{array}{cc}
 \hline \hline&\vspace{-.3 cm}\\
 {\rm  Field }&\hspace{10mm} {\rm Masses}\\
 {\rm [SU(3),SU(2),Y]}&\hspace{10mm}
 \vspace{0.1 cm}\\ \hline  \hline
       A[1,1,4]&  {4.093, 3.321, 0.137}    \\
       B[6,2,{5/3}]&   {0.106, 0.099, 0.091}        \\
          C[8,2,1]&{1.727, 1.727, 1.224, 1.224, 0.614, 0.614}\\
          D[3,2,{7/ 3}]&{ 1.919, 1.433, 1.191, 0.810, 0.205, 0.134     }\\
               E[3,2,{1/3}]&{{1.475, 1.043, 0.716, 0.716, 0.677, 0.594, 0.506}} \\
\hline \hline
 \end{array}
 $$
 \label{table a} \caption{\small{Mass spectrum of superheavy fields in units of $m/\lambda \sim 10^{16}$ GeV for $N_g$=2. Only
  the spectra of $h[1,2,\pm 1],$ $t[3,1,\mp2/3]$
   (given in colors S.No. 1 (Blue), S.No. 2 (Green), S.No. 3 (Red)) depend on the value of
      $M_H$ chosen (see Table \ref{table:ng=2} for the values of $M_H$).
   {\label{Ngeq2PGoldies}}}}\end{table}

 \begin{table}
 $$
 \begin{array}{cc}
 \hline \hline&\vspace{-.3 cm}\\
 {\rm  Field }&\hspace{10mm} {\rm Masses}\\
 {\rm [SU(3),SU(2),Y]}&\hspace{10mm}
 \vspace{0.1 cm}\\ \hline \hline
       A[1,1,4]&  {9.93, 8.08, 7.34, 4.06, 2.53, 0.84}    \\
       B[6,2,{5/3}]&   {6.46, 6.38, 5.8, 5.18, 4.83, 2.88}        \\
          C[8,2,1]&{4.62, 4.62, 4.36, 4.36, 4., 4.,2.2, 2.2, 1.88, 1.88, 0.35, 0.35}\\
          D[3,2,{7/ 3}]&{ 7.55, 4.64, 4.25, 3.95, 2.71, 2., 1.92, 1.61, 1.39, 1.1, 1.08, 0.53 }\\
               E[3,2,{1/3}]&{{ 28.48, 28.48, 20.62, 19.48, 18.45,  18.19,14.85, 13.38}} \\
\hline \hline
 \end{array}
 $$
\caption{\small{Mass spectrum of superheavy fields in units of
$m/\lambda \sim 10^{16} $ GeV in
 6-dim symmetric tracefull ($Rank[\mathbf{\Xi}]=5$) scenario with  $N_g$=3.
 Only the spectra of $h[1,2,\pm 1],$ $t[3,1,\mp2/3]$ depend on the value of
      $M_H$ chosen (see Table \ref{htrank5-6dim} for the  spectra for each of the 6 values of $M_H$ ). \label{tracefullrank5spec} }}\end{table}

\begin{table}
 $$
 \begin{array}{ccc}
 \hline\hline&&\vspace{-.3 cm}\\ {\mbox {$M_H$ } }&h[1,2,1] &t[3,1,-2/3]\vspace{0.1 cm}\\
 \hline\hline
2.55 + 0.13 i& 34.81, 32.15, 27.12, 21.08& 39., 36.27, 35.17, 27.86, 27.6, 24.06 \\
& 20.53 ,18.99,12.6, 12.06& 21.67, 18.22, 13.95, 12.81,10.72, 8.94\\
&11.62, 10.96, 10.38, 9.51, 7.29 & 8.28, 7.46, 6.5, 5.07, 4.49, 4.09\\
&6.87, 3.57, 2.7, 2.28, 2.06 & 3.44, 3.08, 2.23, 1.52, 1.36,0.78 \\
&0.74,0.59, 0.32,0.13, 0.05&  0.49, 0.43, 0.26, 0.21,0.11,
0.04\\\hline

1.44 - 0.61 i& 34.77, 32.11, 27.08, 21.06 &38.97, 36.25, 35.14, 27.88, 27.49, 24.01\\
\hline\hline
 \end{array}
 $$
 \caption{\small{Mass spectrum of superheavy fields $h[1,2,\pm 1],$ $t[3,1,\mp2/3]$ which depend on the value of
      $M_H$ chosen in units of $m/\lambda \sim 10^{16} $ GeV in
 6-dim symmetric tracefull ($Rank[\mathbf{\Xi}]= 5$) scenario with  $N_g$=3 for all values of $M_H$.\label{htrank5-6dim}}}\end{table}

\begin{table}
 $$
 \begin{array}{cc}
 \hline \hline&\vspace{-.3 cm}\\
 {\rm  Field }&\hspace{10mm} {\rm Masses}\\
 {\rm [SU(3),SU(2),Y]}&\hspace{10mm}
 \vspace{0.1 cm}\\ \hline \hline
       A[1,1,4]&  {1.531, 1.462, 1.282, 1.265, 0.379, 0.072}    \\
       B[6,2,{5/3}]&   {{2.885, 2.445, 2.409, 1.901, 1.219, 1.001}}        \\
          C[8,2,1]&{{1.195, 1.195, 0.911, 0.911, 0.889, 0.889}}\\
& {0.625, 0.625, 0.604, 0.604, 0.594, 0.594} \\
          D[3,2,{7/ 3}]&{{1.269, 0.896, 0.862, 0.719, 0.667, 0.625} }\\
\hline \hline
 \end{array}
 $$
 \caption{\small{Mass spectrum of superheavy fields in units of $m/\lambda \sim 10^{16} $ GeV
  in 6-dim symmetric tracefull scenario ($Rank[\mathbf{\Xi}]=4$).\label{spec6dimrank4}}}\end{table}

\begin{table}
 $$
 \begin{array}{ccc}
 \hline\hline &&\vspace{-.3 cm}\\{\mbox {$M_H$ } }&h[1,2,1] &t[3,1,-2/3]\vspace{0.1 cm}\\
 \hline\hline
-4.323+1.47i&10.09, 9.86, 8.22, 7.93, 7.31 &9.96, 9.82, 7.84, 7.49, 7.12, 6.57 \\
& 6.46, 5.85, 3.74, 3.49, 3.26&  5.94,3.43, 3.29, 3.05, 2.22, 1.84 \\
& 2.39, 2.04, 1.76, 1.41, 0.89& 1.72, 1.49,1.32, 1.16,1.05, 1.0\\
&   0.75, 0.63, 0.61 ,0.48 &  0.99, 0.79, 0.74, 0.65, 0.63, 0.61\\
&0.33, 0.27, 0.1, 0.03 &0.52, 0.4, 0.32, 0.28, 0.2, 0.12\\\hline

0.465+3.382i&9.71, 9.48, 7.76, 7.42, 6.71 & 9.63, 9.39, 7.35, 7.03, 6.59, 6.28\\
\hline\hline
 \end{array}
 $$
 \caption{\small{Mass spectrum of superheavy fields $h[1,2,\pm 1],$ $t[3,1,\mp2/3]$ which depend on the value of
      $M_H$ chosen in units of $m/\lambda \sim 10^{16} $ GeV in
 6-dimensional symmetric tracefull ($Rank[\mathbf{\Xi}]= 4$) scenario with  $N_g$=3 for all values of $M_H$.
 \label{htrank4-6dim}}}\end{table}

\begin{table}
 $$
 \begin{array}{cc}
 \hline \hline&\vspace{-.3 cm}\\
 {\rm  Field }&\hspace{10mm} {\rm Masses}\\
 {\rm [SU(3),SU(2),Y]}&\hspace{10mm}
 \vspace{0.1 cm}\\ \hline \hline
       A[1,1,4]&  {{0.602, 0.596, 0.515, 0.461, 0.446}}   \\
       B[6,2,{5/3}]&  {1.099, 0.861, 0.826, 0.248, 0.244}  \\
          C[8,2,1]&{{0.563, 0.563, 0.552, 0.552, 0.532}}\\&{{ 0.532, 0.483, 0.483, 0.482, 0.482}}\\
          D[3,2,{7/ 3}]&{{0.584, 0.571, 0.563, 0.561, 0.531}}\\
\hline \hline
 \end{array}
 $$
 \caption{\small{Mass spectrum of superheavy
  fields in units of $m/\lambda \sim 10^{16} $ GeV in 5-dimensional symmetric traceless case ($Rank[\mathbf{\Xi}]=4$).\label{5dimspec}} }\end{table}

\begin{table}
 $$
 \begin{array}{ccc}
 \hline\hline &&\vspace{-.3 cm}\\{\mbox {$M_H$ } }&h[1,2,1] &t[3,1,-2/3]\vspace{0.1 cm}\\
 \hline\hline
135.29+11.98  i&136.697, 136.68, 136.528&136.517, 136.501, 136.392, 135.942 \\
&135.959,135.936,2.939 & 135.923,2.293,2.282, 1.407,0.697 \\
&2.925,1.921,.78, .77,0.655 &0.687, 0.652, 0.609,0.593,0.573, .559 \\
&0.55,0.62,0.557,0.505& 0.527, 0.522, 0.52, 0.483,0.477\\
&0.481 , 0.466, 0.266,0.217 & 0.391, 0.387, 0.169,0.139,
0.072\\\hline

25.4+1.72  i&29.744, 29.664, 28.968&28.908, 28.83, 28.323, 26.094 \\
&26.179,26.063,3.287, 3.252 & 25.996, 2.715,2.688, 1.506, 1.195 \\
&1.821, 1.52,0.992, 0.927 & 0.86, 0.809, 0.679,0.633, 0.599, .572\\
&0.658, 0.605, 0.579, 0.517  &0.556, 0.534, 0.526, 0.398,0.392 \\
&0.284, 0.115, 0.103, 0.008& 0.258, 0.177, 0.161,0.026,
0.007\\\hline

24.6+1.16  i&29.034, 28.952, 28.237&28.176, 28.096, 27.575, 25.282 \\
&25.37,25.25, 3.299, 3.263 & 25.181, 2.732, 2.703, 1.533, 1.202\\
&1.828, 1.55,1.003, 0.937 & 0.868, 0.816, 0.682 ,0.635, 0.599, .572 \\
& 0.659,0.606, 0.585, 0.518& 0.556, 0.534, 0.526, 0.398, 0.392\\
& 0.285,0.104, 0.092, 0.008  & 0.259, 0.17, 0.153,0.031, 0.012
\\\hline

18.41+1.4  i&24.01, 23.909, 23.03&22.965, 22.865, 22.217, 19.321 \\
&19.435,19.281,3.413, 3.353 & 19.19, 2.89, 2.841, 1.831, 1.228 \\
&2.055, 1.669, 1.115, 1.043 & 0.942, 0.885, 0.734, 0.644,0.604, .577\\
&0.697, 0.633, 0.602,0.528 & 0.562, 0.537, 0.528, 0.398 ,0.392\\
& 0.291, 0.076, 0.061, 0.004 & 0.276, 0.109, 0.095,0.069,
0.061\\\hline

18.32+1.23  i&23.926, 23.825, 22.943& 22.878, 22.777, 22.127, 19.218\\
&19.332,19.177,3.416, 3.355 & 19.086,2.893, 2.844, 1.838, 1.228\\
&2.062, 1.669, 1.118, 1.046&  0.943, 0.887, 0.735,0.645, 0.604, .577\\
& 0.699, 0.633, 0.602 , 0.528 & 0.562, 0.537, 0.528, 0.398,0.392\\
&0.291, 0.078, 0.062, 0.004 &0.276, 0.108, 0.095,0.069,
0.061\\\hline\hline

 \end{array}
 $$
 \caption{\small{Mass spectrum of superheavy fields $h[1,2,\pm 1]$, $t[3,1,\mp2/3]$ which depend on the value of
      $M_H$ chosen in units of $m/\lambda \sim 10^{16} $ GeV in
 5-dim symmetric traceless ($Rank[\mathbf{\Xi}]=4$) scenario with  $N_g$=3 for all values of $M_H$.\label{htrank4-5dim}}}\end{table}

\chapter{Summary and Outlook} \label{ch:method}
This thesis is based upon a particular GUT, called New Minimal
Supersymmetric
 SO(10) Grand Unified Theory (NMSGUT) \cite{nmsgut}, which
is capable of producing realistic fits of basic fermion mass
mixing data. Baryon decay is a peculiarity of GUTs and various
extensions of SM predict lepton flavour violation. The main motive
of our study is to check the compatibility of the model (NMSGUT)
with experimental data, particularly constraints from baryon
number and lepton flavour violation, and on the basis of realistic
NMSGUT parameter sets to further refine the NMSGUT predictions and
thus subject it to stringent falsification tests. Apart from this
we aimed to improve the NMSGUT fitting process by inclusion of
loop effects on Susy spectrum, consideration of heavy right handed
neutrino thresholds and RG improvements in the large NMSGUT
FORTRAN code.

This work emphasizes the importance of GUT scale threshold
corrections for baryon number violation rates.
 MSSM is the effective theory of the GUT and its light Higgs is a combination of different  Higgs multiplets
  from all the Higgs irrep of the NMSGUT. So, light Higgs can have wave function
   corrections from all the heavy fields at SO(10) Yukawa vertex. Wavefunction renormalizion constant
    of Higgs line can have very small value $Z_{H, \bar{H}} \approx 10^{-2}$. This lowers the tree level SO(10) Yukawas required to match the GUT derived effective MSSM fermion Yukawas with MSSM data. Since the same Yukawas determine the $d=5$ $\Delta B \neq 0$ operator we get
      suppressed B violation rates. We have shown that instead of being problematic the large number of superheavy
       particles at GUT scale can cure the long standing problem of fast baryon decay
        rates in Susy GUTs. We have extended the already available FORTRAN and Mathematica codes of NMSGUT calculations \cite{nmsgut}. Including
        GUT scale threshold corrections, we have searched for the set of NMSGUT superpotential parameters
         and mSUGRY NUHM soft parameters, respecting RG constraints, which accurately fit the fermion mass mixing data and
          experiment compatible B-decay rates (i.e. $<$ $10^{-34}$ yrs$^{-1}$). These fits have smaller values of
           all the superpotential parameters as compared to the tree level fits. Soft
            parameters prefer the same range as found before and provide mini-split Susy spectrum with heavy third s-generation.

The other pressing issue on which we have focussed is computation
of one loop corrected Susy spectrum. Direct inclusion of one loop
 self energies to the Susy spectrum of NMSGUT solution (fits which produce realistic fermion
  data and acceptable B-decay rates) drives slepton and squark masses to negative values. Heavy CP odd pseudoscalar
   Higgs provides huge corrections. Fresh searches were performed to get positive loop corrected sfermion masses
    by implementing a penalty on the ratio $(\frac{\mu}{M_A}
   )^2$(0.3-2.7). This ratio is crucial for the Higgs sector and Higgsino loop correction
    to the scalars and solutions found have this ratio close to the upper limit applied. We have not yet found light smuon solution (which is very desirable for dark matter phenomenology and to resolve muon g-2 anomaly ) after including loop corrections. We perhaps require either more searches or deeper RG analysis of the soft parameters.  The refined NMSGUT fits have large $A_0$, $M_{H,\bar H}$, $\mu$
and $B$ parameters with heavy third s-generation like the tree
level fits. These distinct predictions of the model will be tested
at LHC  with the discovery of Susy particles.

Branching ratio for different lepton flavour violating ($l_i$
$\rightarrow$ $l_j \gamma$  and $l_i$ $\rightarrow$ 3 $l_j
\gamma$) processes is calculated. These processes do not provide
additional constraints on the soft mass matrices since the
calculated BR is much smaller than the experimental upper bound
because of the large sfermion masses and negative soft Higgs mass
square parameters.
 We have calculated the $\Delta a_{\mu}$ for the loop
corrected fits presented in Chapter 4 and 5. Since smuon is not
light, this contribution is also very small. We have not yet
estimated charged lepton electric dipole moments but can and will
do so in upcoming studies.

Another puzzle for the NMSGUT has been the necessity of using non
universal Higgs doublets mass squared values that  are negative :
which is difficult when the soft terms come from supergravity. We
calculated the NMSGUT RG equations and used them to study
parameter evolution between $M_{Planck}$ and $M_{GUT}$. We found
that the universal positive scalar mass squared parameters
provided by SUGRY easily become negative due to the RG flow thus
removing the problem in principle. Future fits will thus include
this third stage of RG evolution.

Finally, we considered a new scenario based upon the SO(10) and
$O(N_g)$(family) gauge symmetry. In this framework, Higgs irreps
of SO(10) also carry family indices whose VEVs break GUT and
family symmetry thus generating matter Yukawas with enough
structure to account for the observed hierarchy. So the
 number of SO(10) Yukawa couplings reduce dramatically.
 Therefore this scenario is called Yukawon ultra minimal grand
unified theory (YUMGUT). Consistent SSB requires introduction of a
special type (`Bajc-Melfo') of superpotential. Study of toy model
($N_g$=2) and realistic three generation case show that the
realistic MSSM data can be produced dynamically by the VEV of
Yukawon field. Consideration of $\boot$-${\bf{126}}$  VEV
homogeneous
 equations of different rank provides new directions for model phenomenology such as existence of sterile neutrinos. YUMGUT also offers novel dark matter candidates from
a SM singlet sector as well as from the hidden sector fields.

This study shows that NMSGUT is quite compatible with B and L
violation experimental data. All these observations make NMSGUT a
leading candidate for physics beyond SM and a mature theory of
particle physics.

\newpage\thispagestyle{empty}\mbox{}
\newpage\thispagestyle{empty}\mbox{}\newpage

\newpage


\newpage
\addcontentsline{toc}{chapter}{Bibliography}
\begin{thebibliography}{}
\bibitem{Aad:2012tfa}
  G.~Aad {\it et al.}  [ATLAS Collaboration],
  Phys.\ Lett.\ B {\bf 716}, 1 (2012)
  [arXiv:1207.7214 [hep-ex]].


\bibitem{Chatrchyan:2012ufa}
  S.~Chatrchyan {\it et al.}  [CMS Collaboration],
  Phys.\ Lett.\ B {\bf 716}, 30 (2012)
  [arXiv:1207.7235 [hep-ex]].

\bibitem{osi1}
  R.~Davis, Jr., D.~S.~Harmer and K.~C.~Hoffman,
  Phys.\ Rev.\ Lett.\  {\bf 20}, 1205 (1968).


\bibitem{osi2}
  B.~T.~Cleveland, T.~Daily, R.~Davis, Jr., J.~R.~Distel, K.~Lande, C.~K.~Lee, P.~S.~Wildenhain and J.~Ullman,
  Astrophys.\ J.\  {\bf 496}, 505 (1998).





\bibitem{Hirata}
  K.~S.~Hirata {\it et al.}  [KAMIOKANDE-II Collaboration],
  Phys.\ Rev.\ Lett.\  {\bf 63}, 16 (1989);
  K.~S.~Hirata {\it et al.}  [KAMIOKANDE-II Collaboration],
  Phys.\ Rev.\ Lett.\  {\bf 65}, 1297 (1990);
  K.~S.~Hirata {\it et al.}  [KAMIOKANDE-II Collaboration],
  Phys.\ Rev.\ Lett.\  {\bf 65}, 1301 (1990).


\bibitem{Ambrosio}
  M.~Ambrosio {\it et al.}  [MACRO Collaboration],
  Phys.\ Lett.\ B {\bf 434}, 451 (1998)
  [hep-ex/9807005].

\bibitem{SKC}
  S.~Fukuda {\it et al.}  [Super-Kamiokande Collaboration],
  Phys.\ Rev.\ Lett.\  {\bf 86}, 5651 (2001)
  [hep-ex/0103032]; Phys.\ Rev.\ Lett.\  {\bf 86}, 5656 (2001)
  [hep-ex/0103033].



\bibitem{Ahmad}
  Q.~R.~Ahmad {\it et al.}  [SNO Collaboration],
  Phys.\ Rev.\ Lett.\  {\bf 87}, 071301 (2001)
  [nucl-ex/0106015];
  Q.~R.~Ahmad {\it et al.}  [SNO Collaboration],
  Phys.\ Rev.\ Lett.\  {\bf 89}, 011301 (2002)
  [nucl-ex/0204008]; Q.~R.~Ahmad {\it et al.}  [SNO Collaboration],
  Phys.\ Rev.\ Lett.\  {\bf 89}, 011302 (2002)
  [nucl-ex/0204009].







\bibitem{weinbergd5}
  S.~Weinberg,
  Phys.\ Rev.\ Lett.\  {\bf 43}, 1566 (1979).


\bibitem{MohapatraSenjanovicSeesaw} P.~Minkowski,
  Phys.\ Lett.\ B {\bf 67}, 421 (1977).
 M.~Gell-Mann, P.~Ramond and R.~Slansky,
in {\it Supergravity}, eds. P.~van~Niewenhuizen and D.Z.~ Freedman
(North Holland 1979); T.~Yanagida, in Proceedings of {\it Workshop
on Unified Theory and Baryon number in the Universe}, eds.
O.~Sawada and A. Sugamoto (KEK 1979); R.N.~Mohapatra and
G.~Senjanovi{\'c}, Phys. Rev. Lett. {\bf 44}, 912 (1980);
R.~N.~Mohapatra and G.~Senjanovi\'c, Phys. Rev. {\bf D23},165
(1981).



\bibitem{TypeII}
  M.~Magg and C.~Wetterich,
  Phys.\ Lett.\ B {\bf 94}, 61 (1980);
  G.~Lazarides, Q.~Shafi and C.~Wetterich,
  Nucl.\ Phys.\ B {\bf 181}, 287 (1981);
  R.~N.~Mohapatra and G.~Senjanovic,
  Phys.\ Rev.\ D {\bf 23}, 165 (1981).










\bibitem{TypeIII}
  R.~Foot, H.~Lew, X.~G.~He and G.~C.~Joshi,
  Z.\ Phys.\ C {\bf 44}, 441 (1989).




\bibitem{Wess}
  J.~Wess and B.~Zumino,
  Phys.\ Lett.\ B {\bf 49}, 52 (1974).

\bibitem{gaugehierarchy}
E. Witten, Nucl. Phys. B{\bf 188}, 513 (1981); S. Dimopoulos and
H. Georgi, Nucl. Phys. B{\bf 193}, 150 (1981); N. Sakai, Z. Phys.
C11, 153(1981); R. K. Kaul, Phys Lett. B {\bf 109}, 19(1982); R.
K. Kaul and P. Majumdar, Nucl. Phys. B {\bf 199}, 36(1982).


\bibitem{martin}
S.~P.~Martin,``A supersymmetry primer'' [arXiv:hep-ph/9709356].


\bibitem{godbole}Theory and Phenomenology of Sparticles, World Scientific Publishing Co. Pte. Ltd., M. Drees, R. Godbole and P. Roy.

\bibitem{Dimopoulos105}
  S.~Dimopoulos and D.~W.~Sutter,
  Nucl.\ Phys.\ B {\bf 452}, 496 (1995)
  [hep-ph/9504415].

\bibitem{Gabbiani:1996hi}
  F.~Gabbiani, E.~Gabrielli, A.~Masiero and L.~Silvestrini,
  Nucl.\ Phys.\ B {\bf 477}, 321 (1996)
  [hep-ph/9604387].


\bibitem{gravity-mediation1}
  H.~P.~Nilles,
  Phys.\ Rept.\  {\bf 110}, 1 (1984).
\bibitem{gravity-mediation2}
  A.~B.~Lahanas and D.~V.~Nanopoulos,
  Phys.\ Rept.\  {\bf 145}, 1 (1987).

\bibitem{gauge-mediation}
  G.~F.~Giudice and R.~Rattazzi,
  Phys.\ Rept.\  {\bf 322}, 419 (1999)
  [hep-ph/9801271].


\bibitem{anomaly1}
  L.~Randall and R.~Sundrum,
  Nucl.\ Phys.\ B {\bf 557}, 79 (1999)
  [hep-th/9810155].

\bibitem{anomaly2}
  G.~F.~Giudice, M.~A.~Luty, H.~Murayama and R.~Rattazzi,
  JHEP {\bf 9812}, 027 (1998)
  [hep-ph/9810442].


\bibitem{pmssm}
  A.~Djouadi {\it et al.}  [MSSM Working Group Collaboration],
  [hep-ph/9901246].

\bibitem{dimopoulos}
  S.~Dimopoulos, S.~Raby and F.~Wilczek,
  Phys.\ Rev.\ D {\bf 24}, 1681 (1981);
  M.~B.~Einhorn and D.~R.~T.~Jones,
  Nucl.\ Phys.\ B {\bf 196}, 475 (1982);
  W.~J.~Marciano and G.~Senjanovic,
  Phys.\ Rev.\ D {\bf 25}, 3092 (1982).





\bibitem{amaldi}
  U.~Amaldi, W.~de Boer and H.~Furstenau,
  Phys.\ Lett.\ B {\bf 260}, 447 (1991);
  J.~R.~Ellis, S.~Kelley and D.~V.~Nanopoulos,
  Phys.\ Lett.\ B {\bf 260}, 131 (1991);
  P.~Langacker and M.~x.~Luo,
  Phys.\ Rev.\ D {\bf 44}, 817 (1991).


\bibitem{PS}
J.~C.~Pati and A.~Salam, ``Lepton Number as the Fourth Color'',
Phys. Rev. D {\bf{10}}, 275 (1974).


\bibitem{GG}
H.~Georgi and S.~L.~Glashow, ``Unity Of All Elementary Particle
Forces'',
 Phys. Rev. Lett. {\bf{32}}, 438 (1974).


\bibitem{Bajc:2006ia}
  B.~Bajc and G.~Senjanovic,
  JHEP {\bf 0708}, 014 (2007)
  [hep-ph/0612029].

\bibitem{Bajc:2007zf}
  B.~Bajc, M.~Nemevsek and G.~Senjanovic,
  Phys.\ Rev.\ D {\bf 76}, 055011 (2007)
  [hep-ph/0703080].




\bibitem{Dorsner:2005ii}
  I.~Dorsner, P.~Fileviez Perez and R.~Gonzalez Felipe,
  Nucl.\ Phys.\ B {\bf 747}, 312 (2006)
  [hep-ph/0512068].

\bibitem{Dorsner:2006dj}
  I.~Dorsner and P.~Fileviez Perez,
  Phys.\ Lett.\ B {\bf 642}, 248 (2006)
  [hep-ph/0606062].

  \bibitem{FM}
  H.~Georgi,
  AIP Conf.\ Proc.\  {\bf 23}, 575 (1975);
H.~Fritzsch, P.~Minkowski,``Unified Interactions of Leptons and
Hadrons'' , Annals Phys. {\bf93},  193 (1975).
\bibitem{E6}
F.~G\"{u}rsey, P. Ramond, and P. Sikivie, Phys. Lett. B 60, 177
(1976).



\bibitem{wilczekzee}
  F.~Wilczek and A.~Zee,
  Phys.\ Rev.\ D {\bf 25}, 553 (1982).

\bibitem{ag1}
C.~S.~Aulakh and A.~Girdhar, Int. J. Mod. Phys. A {\bf{20}},
865(2005) [arXiv:hep-ph/0204097].



\bibitem{LR1}
  R.~N.~Mohapatra and J.~C.~Pati,
  Phys.\ Rev.\ D {\bf 11}, 2558 (1975).

\bibitem{LR2}
  G.~Senjanovic and R.~N.~Mohapatra,
  Phys.\ Rev.\ D {\bf 12}, 1502 (1975).


\bibitem{rparso10.1}
  C.~S.~Aulakh, K.~Benakli and G.~Senjanovic,
  Phys.\ Rev.\ Lett.\  {\bf 79}, 2188 (1997)
  [hep-ph/9703434].

\bibitem{rparso10.2}
  C.~S.~Aulakh, A.~Melfo and G.~Senjanovic,
  Phys.\ Rev.\ D {\bf 57},  4174 (1998)
  [hep-ph/9707256].



\bibitem{LR3}
  C.~S.~Aulakh, A.~Melfo, A.~Rasin and G.~Senjanovic,
  Phys.\ Rev.\ D {\bf 58}, 115007 (1998)
  [hep-ph/9712551].



\bibitem{tbtyuk}
  B.~Ananthanarayan, G.~Lazarides and Q.~Shafi,
  Phys.\ Rev.\ D {\bf 44}, 1613 (1991);
  S.~Dimopoulos, L.~J.~Hall and S.~Raby,
  Phys.\ Rev.\ D {\bf 45}, 4192 (1992);
  G.~W.~Anderson, S.~Raby, S.~Dimopoulos and L.~J.~Hall,
  Phys.\ Rev.\ D {\bf 47}, 3702 (1993)
  [hep-ph/9209250];
  G.~Anderson, S.~Raby, S.~Dimopoulos, L.~J.~Hall and G.~D.~Starkman,
  Phys.\ Rev.\ D {\bf 49}, 3660 (1994)
  [hep-ph/9308333].





\bibitem{typeIIdom}B.~Bajc, G.~Senjanovic and F.~Vissani,
Phys.\ Rev.\ Lett.\  {\bf 90},  051802 (2003)
[arXiv:hep-ph/0210207];
  B.~Bajc, G.~Senjanovic and F.~Vissani,
  Phys.\ Rev.\ D {\bf 70}, 093002 (2004)
  [hep-ph/0402140].


\bibitem{rparso10}C.~S.~Aulakh, B.~Bajc, A.~Melfo, A.~Rasin and G.~Senjanovic,
 Nucl.\ Phys.\ B {\bf 597}, 89 (2001)
 [arXiv:hep-ph/0004031].



\bibitem{smallrep}
  K.~S.~Babu and S.~M.~Barr,
  Phys.\ Rev.\ D {\bf 51}, 2463 (1995)
  [hep-ph/9409285];
  K.~S.~Babu and R.~N.~Mohapatra,
  Phys.\ Rev.\ Lett.\  {\bf 74}, 2418 (1995)
  [hep-ph/9410326];
  S.~M.~Barr and S.~Raby,
  Phys.\ Rev.\ Lett.\  {\bf 79}, 4748 (1997)
  [hep-ph/9705366];
  Z.~Chacko and R.~N.~Mohapatra,
  Phys.\ Rev.\ D {\bf 59}, 011702 (1999)
  [hep-ph/9808458].


\bibitem{aulmoh}C.~S.~Aulakh and R.~N.~Mohapatra, CCNY-HEP-82-4 April 1982,
  CCNY-HEP-82-4-REV,  Jun 1982 , Phys. Rev. D {\bf 28}, 217 (1983).

\bibitem{ckn}T.~E.~Clark, T.~K.~Kuo and N.~Nakagawa, Phys. lett. B {\bf{115}}, 26 (1982).

\bibitem{abmsv}C.~S.~Aulakh, B.~Bajc, A.~Melfo, G.~Senjanovic and F.~Vissani,
Phys.\ Lett.\ B {\bf 588}, 196 (2004) [arXiv:hep-ph/0306242].


 \bibitem{nathsyed}
P.~Nath and R.~M.~Syed, Nucl. Phys. B {\bf{618}}, 138 (2001).


\bibitem{144-1}
  K.~S.~Babu, I.~Gogoladze, P.~Nath and R.~M.~Syed,
  Phys.\ Rev.\ D {\bf 72}, 095011 (2005)
  [hep-ph/0506312].

\bibitem{144-2}
  K.~S.~Babu, I.~Gogoladze, P.~Nath and R.~M.~Syed,
  Phys.\ Rev.\ D {\bf 74}, 075004 (2006)
  [hep-ph/0607244].


\bibitem{Babumohapatra}
  K.~S.~Babu and R.~N.~Mohapatra,
  Phys.\ Rev.\ Lett.\  {\bf 70}, 2845 (1993)
  [hep-ph/9209215].




\bibitem{allferm}K.~Y.~Oda, E.~Takasugi, M.~Tanaka and
M.~Yoshimura,
  Phys.\ Rev.\ D {\bf 59}, 055001 (1999)
  [arXiv:hep-ph/9808241];
 K.~Matsuda, Y.~Koide and T.~Fukuyama,
  Phys.\ Rev.\ D {\bf 64}, 053015 (2001)
  [arXiv:hep-ph/0010026].
H.~S.~Goh, R.~N.~Mohapatra and S.~P.~Ng,
  Phys.\ Lett.\ B {\bf 570}, 215 (2003)  [arXiv:hep-ph/0303055].
   H.~S.~Goh, R.~N.~Mohapatra and S.~P.~Ng,
  Phys.\ Rev.\ D {\bf 68}, 115008 (2003)
  [arXiv:hep-ph/0308197].
  K.~S.~Babu and C.~Macesanu,
  Phys.\ Rev.\ D {\bf 72}, 115003 (2005)
  [arXiv:hep-ph/0505200].

\bibitem{gmblm}C.~S.~Aulakh, \emph{``From germ to bloom"},
 [arXiv:hep-ph/0506291].



\bibitem{blmdm}C.~S.~Aulakh and S.~K.~Garg,
  Nucl.\ Phys.\ B {\bf 757}, 47 (2006)
 [arXiv:hep-ph/0512224]. 

\bibitem{bert}
S.~Bertolini, M.~Frigerio and M.~Malinsky,
Phys.\ Rev.\ D {\bf 70}, 095002 (2004)
[arXiv:hep-ph/0406117]; 
S.~Bertolini and M.~Malinsky,
  Phys.\ Rev.\ D {\bf 72}, 055021 (2005)
  [hep-ph/0504241].






  \bibitem{nmsgut}C.~S.~Aulakh and S.~K.~Garg,
 [arXiv:hep-ph/0612021v1];
 [arXiv:hep-ph/0612021v2].
 [arXiv:hep-ph/0807.0917v1];
 [arXiv:hep-ph/0807.0917v2]; Nucl. Phys. B {\bf 857}, 101 (2012) [arXiv:hep-ph/0807.0917v3].




\bibitem{moh120}
  B.~Dutta, Y.~Mimura and R.~N.~Mohapatra,
  Phys.\ Rev.\ Lett.\  {\bf 94}, 091804 (2005)
  [arXiv:hep-ph/0412105];
  B.~Dutta, Y.~Mimura and R.~N.~Mohapatra,
  Phys.\ Lett.\ B {\bf 603}, 35 (2004)
  [arXiv:hep-ph/0406262].

\bibitem{Oshimo-120}
  N.~Oshimo,
  Phys.\ Rev.\ D {\bf 66}, 095010 (2002)
  [hep-ph/0206239];
  N.~Oshimo,
  Nucl.\ Phys.\ B {\bf 668}, 258 (2003)
  [hep-ph/0305166].




 \bibitem{lepto1}M.~Fukugita and T.~Yanagida, Phys. Lett. B {\bf 174},
45(1986).


\bibitem{Barbieri:1}
  R.~Barbieri and L.~J.~Hall,
  Phys.\ Lett.\ B {\bf 338}, 212 (1994)
  [hep-ph/9408406].

\bibitem{Barbieri:2}
  R.~Barbieri, L.~J.~Hall and A.~Strumia,
  Nucl.\ Phys.\ B {\bf 445}, 219 (1995)
  [hep-ph/9501334].



\bibitem{hisano}J.~Hisano, T.~Moroi, K.~Tobe and M.~Yamaguchi, Phys. Rev. D {\bf  53}, 2442 (1996)
[arXiv:hep-ph/9510309].


\bibitem{csaigckkRG}C.~S.~Aulakh, I.~Garg and C.~K.~Khosa, \emph{``Two Loop Renormalization Group Evolution Equations of the NMSGUT''}, to appear.



\bibitem{ilathesis} Ila Garg, ``NEW MINIMAL SUPERSYMMETRIC SO(10) GUT PHENOMENOLOGY AND ITS COSMOLOGICAL IMPLICATIONS '', Ph.D. Thesis, Panjab Uni. Chandigarh, 2014.


\bibitem{MEG}
J.~Adam {\it et al.}  [MEG Collaboration],
  Phys.\ Rev.\ Lett.\  {\bf 110}, 201801 (2013)
  [arXiv:1303.0754 [hep-ex]].

\bibitem{BaBar}
  B.~Aubert {\it et al.}  [BaBar Collaboration],
  Phys.\ Rev.\ Lett.\  {\bf 104}, 021802 (2010)
  [arXiv:0908.2381 [hep-ex]].

\bibitem{SINDRUM}
  U.~Bellgardt {\it et al.}  [SINDRUM Collaboration],
  Nucl.\ Phys.\ B {\bf 299}, 1 (1988).


\bibitem{Stockinger}
D.~St\"{o}ckinger, J. Phys. G {\bf 34}, R45 (2007)
[arXiv:hep-ph/0609168].

\bibitem{ag2} C.~S.~Aulakh and A.~Girdhar,
  Nucl.\ Phys.\ B {\bf 711}, 275 (2005) [arXiv:hep-ph/0405074].


\bibitem{aulakhgarg}
 C.~S.~Aulakh and S.~K.~Garg, Private Communication.

\bibitem{bmsv}B.~Bajc, A.~Melfo, G.~Senjanovic and F.~Vissani,
Phys.\ Rev.\ D {\bf 70}, 035007 (2004) [arXiv:hep-ph/0402122].

\bibitem{fuku04}T.~Fukuyama, A.~Ilakovac, T.~Kikuchi, S.~Meljanac and N.~Okada,
 Eur.\ Phys.\ J.\ C {\bf 42}, 191 (2005)
 [arXiv:hep-ph/0401213v1.,v2];
  T.~Fukuyama, A.~Ilakovac, T.~Kikuchi, S.~Meljanac and N.~Okada,
  J.\ Math.\ Phys.\  {\bf 46}  033505 (2005)
  [arXiv:hep-ph/0405300].



\bibitem{hall}
L.~J.~Hall, Nucl. Phys. B {\bf{178}}, 75 (1981).


\bibitem{weinberg-RGT}S.~Weinberg, Phys. Lett. B {\bf 91}, 5 (1980).


\bibitem{csaskga3}
  C.~S.~Aulakh and S.~K.~Garg,
  Mod.\ Phys.\ Lett.\ A {\bf 24}, 1711 (2009)
  [arXiv:0710.4018 [hep-ph]].



\bibitem{dixitsher}V.~V.~Dixit and M.~Sher,
 Phys. Rev. D {\bf{40}}, 3765 (1989).




\bibitem{downsim}
Numerical Recipes in Fortran 90, Second Edition(1996), Cambridge
Univ. Press, W. H. Press, S. A. Teukolsky, W. T. Vetterling and
B.P.Flannery.




\bibitem{antuschspinrath}
 S.~Antusch and M.~Spinrath,
 Phys.\ Rev.\ D {\bf 78}, 075020 (2008)
 [arXiv:0804.0717 [hep-ph]].


\bibitem{Castano}
  D.~J.~Castano, E.~J.~Piard and P.~Ramond,
  Phys.\ Rev.\ D {\bf 49}, 4882 (1994)
  [hep-ph/9308335].


\bibitem{MartinVaughn}
S.~P.~Martin and M.~T.~Vaughn,
  Phys.\ Rev.\  D {\bf 50}, 2282 (1994)
  [arXiv:hep-ph/9311340];
S.~P.~Martin,
  Phys.\ Rev.\  D {\bf 66},  096001 (2002)
  [arXiv:hep-ph/0206136].


\bibitem{porod}   W.~Porod,
  Comput.\ Phys.\ Commun.\  {\bf 153}, 275 (2003)
  [hep-ph/0301101].

\bibitem{piercebagger}
 D.~M.~Pierce, J.~A.~Bagger, K.~T.~Matchev and R.~J.~Zhang,
 Nucl.\ Phys.\ B {\bf 491}, 3 (1997)
 [arXiv:hep-ph/9606211].


\bibitem{nmass}
  E.~K.~Akhmedov, Z.~G.~Berezhiani, G.~Senjanovic and Z.~j.~Tao,
  Phys.\ Rev.\ D {\bf 47}, 3245 (1993)
  [hep-ph/9208230].


\bibitem{gutupend}
  C.~S.~Aulakh,
  [arXiv:1107.2963 [hep-ph]].

\bibitem{minisplit1}
  H.~E.~Haber and Y.~Nir,
  Phys.\ Lett.\ B {\bf 306}, 327 (1993)
  [hep-ph/9302228];
  H.~E.~Haber,
  [hep-ph/9505240];
  A.~Arbey, M.~Battaglia, A.~Djouadi, F.~Mahmoudi and J.~Quevillon,
  Phys.\ Lett.\ B {\bf 708}, 162 (2012)
  [arXiv:1112.3028 [hep-ph]];
  A.~Arbey, M.~Battaglia, A.~Djouadi and F.~Mahmoudi,
  JHEP {\bf 1209}, 107 (2012)
  [arXiv:1207.1348 [hep-ph]].

\bibitem{Bajc:SO(10)fitting}
  B.~Bajc,
  AIP Conf.\ Proc.\  {\bf 805}, 326 (2006)
  [hep-ph/0602166].


\bibitem{core}   C.~S.~Aulakh,
 \emph{``Fermion mass hierarchy in the Nu MSGUT. I: The real core"}
  [arXiv:hep-ph/0602132].

 \bibitem{msgreb}  C.~S.~Aulakh, \emph{``MSGUT Reborn ?"} [arXiv:hep-ph/0607252].


\bibitem{Grimus:1}
  W.~Grimus and H.~Kuhbock,
  Phys.\ Lett.\ B {\bf 643}, 182 (2006)
  [hep-ph/0607197].

\bibitem{Grimus:2}
  W.~Grimus and H.~Kuhbock,
  Eur.\ Phys.\ J.\ C {\bf 51}, 721 (2007)
  [hep-ph/0612132].


\bibitem{minisplit2}
  A.~Arvanitaki, N.~Craig, S.~Dimopoulos and G.~Villadoro,
  JHEP {\bf 1302}, 126 (2013)
  [arXiv:1210.0555 [hep-ph]].





\bibitem {kuslangseg}
  A.~Kusenko, P.~Langacker and G.~Segre,
  Phys.\ Rev.\  D {\bf 54} (1996) 5824
  [arXiv:hep-ph/9602414].

\bibitem{superk1}
  H.~Nishino {\it et al.}  [Super-Kamiokande Collaboration],
  Phys.\ Rev.\ Lett.\  {\bf 102}, 141801 (2009)
  [arXiv:0903.0676 [hep-ex]];
  K.~Kobayashi {\it et al.}  [Super-Kamiokande Collaboration],
  Phys.\ Rev.\ D {\bf 72}, 052007 (2005)
  [hep-ex/0502026].

\bibitem{barondecay1}T.~Goto and T.~Nihei, Phys. Rev. D {\bf 59}, 115009 (1999)
[arXiv:hep-ph/9808255].

\bibitem{barondecay2}V.~Lucas and S.~Raby, Phys. Rev. D {\bf 55}, 6986 (1997)
[arXiv:hep-ph/9610293].

\bibitem{barondecay3}
  P.~Nath and P.~Fileviez Perez,
  Phys.\ Rept.\  {\bf 441}, 191 (2007)
  [hep-ph/0601023].


\bibitem{Sakai}
  N.~Sakai and T.~Yanagida,
  Nucl.\ Phys.\ B {\bf 197}, 533 (1982).


\bibitem{Weinberg-proton}
  S.~Weinberg,
  Phys.\ Rev.\ D {\bf 26}, (1982) 287.

\bibitem{Murayama}
  H.~Murayama and A.~Pierce,
  Phys.\ Rev.\ D {\bf 65}, 055009 (2002)
  [hep-ph/0108104].


\bibitem{Chiraltech}
  M.~Claudson, M.~B.~Wise and L.~J.~Hall,
  Nucl.\ Phys.\ B {\bf 195}, 297 (1982).


\bibitem{wright}B.D.~Wright, [arXiv:hep-ph/9404217] (1994).

\bibitem{nonrenorth}
  J.~Wess and B.~Zumino,
  Phys.\ Lett.\ B {\bf 49}, 52 (1974);
  J.~Iliopoulos and B.~Zumino,
  Nucl.\ Phys.\ B {\bf 76}, 310 (1974);
  S.~Ferrara, J.~Iliopoulos and B.~Zumino,
  Nucl.\ Phys.\ B {\bf 77}, 413 (1974);
  B.~Zumino,
  Nucl.\ Phys.\ B {\bf 89}, 535 (1975);
  S.~Ferrara and O.~Piguet,
  Nucl.\ Phys.\ B {\bf 93}, 261 (1975);
  M.~T.~Grisaru, W.~Siegel and M.~Rocek,
  Nucl.\ Phys.\ B {\bf 159}, 429 (1979).

\bibitem{aulakhgargkhosa}C.~S.~Aulakh, I.~Garg and C.~K.~Khosa,
 Nucl.\ Phys.\ B {\bf 882}, 397 (2014) [arXiv:1311.6100[hep-ph]].

\bibitem{langpol}P.~Langacker and N.~Polonsky,
 Phys.\ Rev.\ D {\bf 47}, 4028 (1993)
 [arXiv:hep-ph/9210235]; P.~Langacker and N.~Polonsky,
 Phys.\ Rev.\ D {\bf 52}, 3081 (1995)
 [arXiv:hep-ph/9503214].

\bibitem{Zwirner}
  J.~R.~Ellis, G.~Ridolfi and F.~Zwirner,
  Phys.\ Lett.\ B {\bf 262}, 477 (1991).


\bibitem{deltar}
  P.~H.~Chankowski, A.~Dabelstein, W.~Hollik, W.~M.~Mosle, S.~Pokorski and J.~Rosiek,
  Nucl.\ Phys.\ B {\bf 417}, 101 (1994).


\bibitem{higgs}
  P.~H.~Chankowski, S.~Pokorski and J.~Rosiek,
  Nucl.\ Phys.\ B {\bf 423}, 437 (1994)
  [hep-ph/9303309].


\bibitem{gluino1}
  S.~P.~Martin and M.~T.~Vaughn,
  Phys.\ Lett.\ B {\bf 318}, 331 (1993)
  [hep-ph/9308222].


\bibitem{gluino2}
  D.~Pierce and A.~Papadopoulos,
  Nucl.\ Phys.\ B {\bf 430}, 278 (1994)
  [hep-ph/9403240].


\bibitem{gluino3}
  D.~Pierce and A.~Papadopoulos,
  Phys.\ Rev.\ D {\bf 50}, 565 (1994)
  [hep-ph/9312248].


\bibitem{DRbar1}
  W.~Siegel,
  Phys.\ Lett.\ B {\bf 84}, 193 (1979);
  D.~M.~Capper, D.~R.~T.~Jones and P.~van Nieuwenhuizen,
  Nucl.\ Phys.\ B {\bf 167}, 479 (1980).


\bibitem{higgstwoloop1}
  G.~Degrassi, P.~Slavich and F.~Zwirner,
  Nucl.\ Phys.\ B {\bf 611}, 403 (2001)
  [hep-ph/0105096].


\bibitem{higgstwoloop2}
  A.~Brignole, G.~Degrassi, P.~Slavich and F.~Zwirner,
  Nucl.\ Phys.\ B {\bf 631}, 195 (2002)
  [hep-ph/0112177].



\bibitem{higgstwoloop}
  A.~Dedes and P.~Slavich,
  Nucl.\ Phys.\ B {\bf 657} 333 (2003)
  [hep-ph/0212132].







\bibitem{2loopHiggs}
  A.~Dedes, G.~Degrassi and P.~Slavich,
  Nucl.\ Phys.\ B {\bf 672}, 144 (2003)
  [hep-ph/0305127].


\bibitem{XTata}
  A.~D.~Box and X.~Tata,
  Phys.\ Rev.\ D {\bf 77}, 055007 (2008)
  [Erratum-ibid.\ D {\bf 82}, 119904 (2010)]
  [arXiv:0712.2858 [hep-ph]];
  Phys.\ Rev.\ D {\bf 79}, 035004 (2009)
  [Erratum-ibid.\ D {\bf 82}, 119905 (2010)]
  [arXiv:0810.5765 [hep-ph]].




\bibitem{suspect}
  A.~Djouadi, J.~L.~Kneur and G.~Moultaka,
  Comput.\ Phys.\ Commun.\  {\bf 176}, 426 (2007)
  [hep-ph/0211331].

\bibitem{suspectweb}http://www.coulomb.univ-montp2.fr/perso/jean-loic.kneur/Suspect/



\bibitem{BNL}
  G.~W.~Bennett {\it et al.}  [Muon G-2 Collaboration],
  Phys.\ Rev.\ D {\bf 73}, 072003 (2006)
  [hep-ex/0602035].

\bibitem{Hisano2}
  J.~Hisano and D.~Nomura,
  Phys.\ Rev.\ D {\bf 59}, 116005 (1999)
  [hep-ph/9810479].

\bibitem{lfv1}
  S.~T.~Petcov, W.~Rodejohann, T.~Shindou and Y.~Takanishi,
  Nucl.\ Phys.\ B {\bf 739}, 208 (2006)
  [hep-ph/0510404].

\bibitem{lfv2}
  D.~F.~Carvalho, J.~R.~Ellis, M.~E.~Gomez and S.~Lola,
  Phys.\ Lett.\ B {\bf 515}, 323 (2001)
  [hep-ph/0103256].

\bibitem{lfv3}
  A.~Kageyama, S.~Kaneko, N.~Shimoyama and M.~Tanimoto,
  Phys.\ Lett.\ B {\bf 527}, 206 (2002)
  [hep-ph/0110283].

\bibitem{lfv4}
  T.~Fukuyama, T.~Kikuchi and N.~Okada,
  Phys.\ Rev.\ D {\bf 68}, 033012 (2003)
  [hep-ph/0304190].


 \bibitem{lfv5}
  A.~Masiero, S.~K.~Vempati and O.~Vives,
  Nucl.\ Phys.\ B {\bf 649}, 189 (2003)
  [hep-ph/0209303];
  A.~Masiero, S.~Profumo, S.~K.~Vempati and C.~E.~Yaguna,
  JHEP {\bf 0403}, 046 (2004)
  [hep-ph/0401138];
  L.~Calibbi, A.~Faccia, A.~Masiero and S.~K.~Vempati,
  Phys.\ Rev.\ D {\bf 74}, 116002 (2006)
  [hep-ph/0605139].


\bibitem{nuflow}
S.~Antusch, M.~Drees, J.~Kersten, M.~Lindner and M.~Ratz,
  Phys.\ Lett.\ B {\bf 519}, 238 (2001)
  [arXiv:hep-ph/0108005]; S.~Antusch, M.~Drees, J.~Kersten, M.~Lindner and M.~Ratz,
  Phys.\ Lett.\ B {\bf 525}, 130 (2002)
  [arXiv:hep-ph/0110366];
  S.~Antusch, J.~Kersten, M.~Lindner, M.~Ratz and M.~A.~Schmidt,
  JHEP {\bf 0503}, 024 (2005)
  [arXiv:hep-ph/0501272].


\bibitem{rhn} A.~Ibarra and C.~Simonetto, JHEP 0804, 102 (2008)
[arXiv:hep-ph/0802.3858].

\bibitem{Babu:1993qv}
  K.~S.~Babu, C.~N.~Leung and J.~T.~Pantaleone,
  Phys.\ Lett.\ B {\bf 319}, 191 (1993)
  [hep-ph/9309223].

\bibitem{PDG}
  J.~Beringer {\it et al.}  [Particle Data Group Collaboration],
  Phys.\ Rev.\ D {\bf 86}, 010001 (2012).


\bibitem{mkst}
M.~Endo, K.~Hamaguchi, S.~Iwamoto, T.~Yoshinaga
[arXiv:hep-ph/1303.4256].

\bibitem{Iwamoto:2013mya}
  S.~Iwamoto,
  [arXiv:1304.5171 [hep-ph]].

\bibitem{tm}
T.~Moroi, Phys. Rev. {\bf D53}, 6565 (1996)
[arXiv:hep-ph/9512396].



\bibitem{detail}
G.~C.~Cho, K.~Hagiwara, Y.~Matsumoto and D.~Nomura, JHEP {\bf
1111}, 068(2011) [arXiv:hep-ph/1104.1769].


\bibitem{Sakharov}
  A.~D.~Sakharov,
  JETP Lett.\  {\bf 5}, 27 (1967)
  [Pisma Zh.\ Eksp.\ Teor.\ Fiz.\  {\bf 5}, 36 (1967)].

\bibitem{Luty:1992un}
  M.~A.~Luty,
  Phys.\ Rev.\ D {\bf 45}, 455 (1992).


\bibitem{Davidson}
  S.~Davidson and A.~Ibarra,
  Phys.\ Lett.\ B {\bf 535}, 25 (2002)
  [hep-ph/0202239].

\bibitem{LFVSO10}
  L.~Calibbi, D.~Chowdhury, A.~Masiero, K.~M.~Patel and S.~K.~Vempati,
  JHEP {\bf 1211}, 040 (2012)
  [arXiv:1207.7227].




\bibitem{Susyno}
  R.~M.~Fonseca,
  Comput.\ Phys.\ Commun.\  {\bf 183}, 2298 (2012)
  [arXiv:1106.5016 [hep-ph]].

\bibitem{spurion} 
  N.~Cabibbo and L.~Maiani,
  ``Weak interactions and the breaking of hadron symmetries'',
  in \emph{Evolution of Particle Physics}, A Volume Dedicated to Eduardo Amaldi on his Sixtieth Birthday,
  pages 50-80, Edited by M.Conversi, Academic Press, New York (1970).
   For a useful recent pedagogical review see R.A. de Pablo,
arXiv:1307.1904v1[hep-ph].


\bibitem{Altarelli:2010gt}
  G.~Altarelli and F.~Feruglio,
  Rev.\ Mod.\ Phys.\  {\bf 82}, 2701 (2010)
  [arXiv:1002.0211 [hep-ph]].

\bibitem{King:2013eh}
  S.~F.~King and C.~Luhn,
  Rept.\ Prog.\ Phys.\  {\bf 76}, 056201 (2013)
  [arXiv:1301.1340 [hep-ph]].

\bibitem{King:2014nza}
  S.~F.~King, A.~Merle, S.~Morisi, Y.~Shimizu and M.~Tanimoto,
  New J.\ Phys.\  {\bf 16}, 045018 (2014)
  [arXiv:1402.4271 [hep-ph]].


\bibitem{koide}
  Y.~Koide,
  Phys.\ Rev.\ D {\bf 78}, 093006 (2008)
  [arXiv:0809.2449 [hep-ph]];  
  Phys.\ Rev.\ D {\bf 79}, 033009 (2009)
  [arXiv:0811.3470 [hep-ph]];
  Phys.\ Lett.\ B {\bf 665}, 227 (2008).

\bibitem{Aulakhkhosa}
  C.~S.~Aulakh and C.~K.~Khosa,
  Phys.\ Rev.\ D {\bf 90}, 045008 (2014)
  [arXiv:1308.5665 [hep-ph]].



\bibitem{BMsugry}
  C.~S.~Aulakh,
  Phys.\ Rev.\ D {\bf 91}, 055012 (2015)
  [arXiv:1402.3979 [hep-ph]].




\bibitem{BM}
  B.~Bajc and A.~Melfo,
  JHEP {\bf 0804}, 062 (2008)
  [arXiv:0801.4349 [hep-ph]].




\bibitem{ovrab}
  B.~A.~Ovrut and S.~Raby,
  Phys.\ Lett.\ B {\bf 125}, 270 (1983).

\bibitem{hilo}
  C.~S.~Aulakh,
   ``Local Supersymmetry And Hi-lo Scale Induction,''
  CCNY-HEP-83/2;  C.S.~Aulakh,
 PhD Thesis, CCNY, 1983, UMI-84-01477.


\bibitem{cremmer}
  E.~Cremmer, S.~Ferrara, L.~Girardello and A.~Van Proeyen,
  Nucl.\ Phys.\ B {\bf 212}, 413 (1983).

  \bibitem{Ohta:1982wn}
  N.~Ohta,
  Prog.\ Theor.\ Phys.\  {\bf 70}, 542 (1983).

\bibitem{nath}
A.~H.~Chamseddine, R.~L.~Arnowitt and P.~Nath, Phys. Rev. Lett.
\textbf{49}, 970 (1982).
  History and exhaustive original  referennces in : R.~Arnowitt, A.~H.~Chamseddine and P.~Nath,
  Int.\ J.\ Mod.\ Phys.\ A {\bf 27}, 1230028 (2012)
  [Erratum-ibid.\ A {\bf 27}, 1292009 (2012)]
  [arXiv:1206.3175 [physics.hist-ph]].

\bibitem{halllykkwein}
  L.~J.~Hall, J.~D.~Lykken and S.~Weinberg,
  Phys.\ Rev.\ D {\bf 27} 2359 (1983).


\bibitem{damalibra}
  R.~Bernabei, P.~Belli, S.~d'Angelo, A.~Di Marco, F.~Montecchia, F.~Cappella, A.~d'Angelo and A.~Incicchitti {\it et al.},
  Int.\ J.\ Mod.\ Phys.\ A {\bf 28}, 1330022 (2013)
  [arXiv:1306.1411 [astro-ph.GA]].

\end{thebibliography}
\end{document}